\documentclass[12pt,a4wide,twoside]{book}
\frontmatter
\usepackage{subfigure}
\usepackage{latexsym}
\usepackage{graphicx}
\usepackage{fancyhdr}
\usepackage{psfig}
\usepackage{epsfig}
\usepackage{graphics}
\usepackage{astrobib,amssymb}
\def\he2{He${\rm II}$ }
\def\h1{H${\rm I}$ }

\def\lesssim{\mathrel{\hbox{\rlap{\hbox{\lower4pt\hbox{$\sim$}}}\hbox{$<$}}}}
\def\gtrsim{\mathrel{\hbox{\rlap{\hbox{\lower4pt\hbox{$\sim$}}}\hbox{$>$}}}}

\def\gtsima{$\; \buildrel > \over \sim \;$}
\def\ltsima{$\; \buildrel < \over \sim \;$}
\def\prosima{$\; \buildrel \propto \over \sim \;$}
\def\gsim{\lower.5ex\hbox{\gtsima}}
\def\lsim{\lower.5ex\hbox{\ltsima}}
\def\simgt{\lower.5ex\hbox{\gtsima}}
\def\simlt{\lower.5ex\hbox{\ltsima}}
\def\simpr{\lower.5ex\hbox{\prosima}}

\def\Msun{{M_\odot}}
\def\Zsun{{Z_\odot}}

\def\Lya{Ly$\alpha$~}

\def\HI{\hbox{H$\scriptstyle\rm I\~$}}
\def\HII{\hbox{H$\scriptstyle\rm II\~$}}

\def\HeIII{\hbox{He$\scriptstyle\rm III\~$}}

\def\HI{\hbox{H~$\scriptstyle\rm I\ $}}
\def\HII{\hbox{H~$\scriptstyle\rm II\ $}}

\def\HeIII{\hbox{He~$\scriptstyle\rm III\ $}}

\newcommand{\be}{\begin{eqnarray}}
\newcommand{\ee}{\end{eqnarray}}
\newcommand{\beq}{\begin{equation}}
\newcommand{\eeq}{\end{equation}}

\newcommand{\clearemptydoublepage}{\newpage{\pagestyle{empty}\cleardoublepage}}

\endquotation

\begin{document}
\thispagestyle{empty}
\begin{center}
\vspace{3.0in}  
\begin{flushleft}
\hspace{1.7in}{\large \textbf{SISSA}}
\end{flushleft}
\vspace{-0.375in}
\begin{flushright}
{\large \textbf{ISAS}} \ \ \ \ \ \ \ \ \ \ \ \ \ \ \ \ \ \ \ \ \ \ \ \ \ \ \ \ \ \ \ \ \ \ \
\end{flushright}
\vspace{-0.7in}
\begin{figure}[!h]     
\begin{center}
\includegraphics[width=0.17 \textwidth]{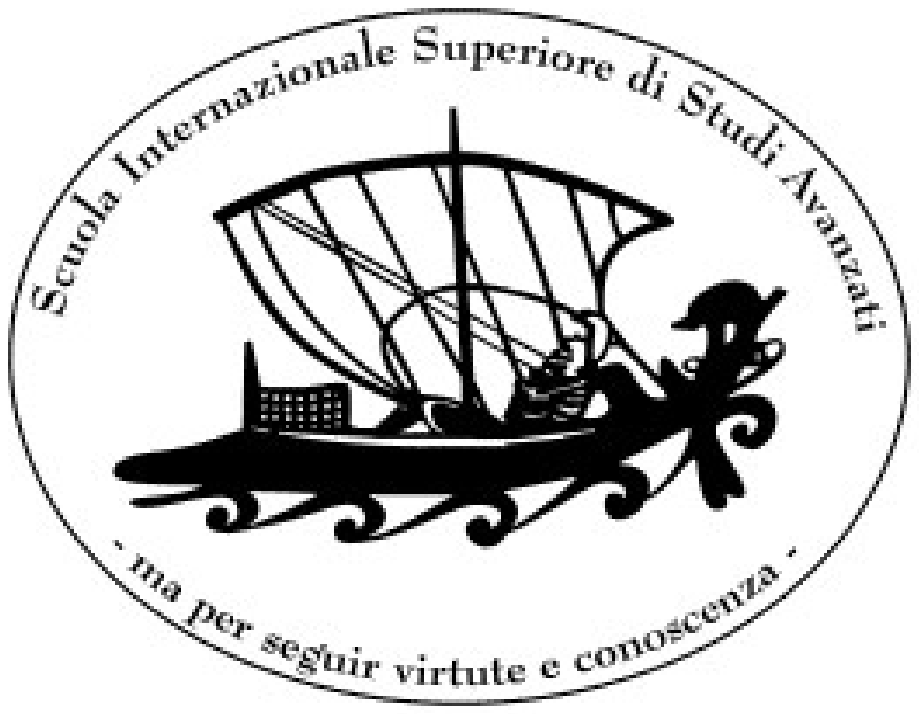}
\end{center}
\end{figure}
\vspace{-0.4in}    
SCUOLA INTERNAZIONALE SUPERIORE DI STUDI AVANZATI \\
INTERNATIONAL SCHOOL FOR ADVANCED STUDIES
\end{center}
\vspace{0.7in}
\begin{center}
{\LARGE \textbf{Cosmic Lighthouses: unveiling the nature of high-redshift galaxies}}
\end{center}
\vspace{0.5in}
\begin{center}
\large{Thesis submitted for the degree of \\ Doctor Philosophi\ae}
\end{center}
\vspace{1.8in}
\begin{flushleft}
\hspace{0.62cm}CANDIDATE \\ 
~~~~~Pratika Dayal
\end{flushleft}
\vspace{-0.785in}
\begin{flushright} 
SUPERVISOR \ \ \ \ \ \ \ \ \ \\ 
Prof.~Andrea~Ferrara\\
\end{flushright}
\vspace{0.5in}
\begin{center} 
October 2010
\end{center}

\newpage
\thispagestyle{empty}

\thispagestyle{empty}

\normalsize

\clearemptydoublepage

\thispagestyle{empty}
\newenvironment{dedication}
{\cleardoublepage \thispagestyle{empty}
\vspace*{\stretch{1}}
\begin{center} \em}
{\end{center}
\vspace*{\stretch{3}} \clearpage}
\begin{dedication}
To all those amber days, in my city of the azure sea and the jade mountains, to all those that passed me by like shining ships in the night, leaving me with thoughts as breathtaking as meteor trails...and most to those that watched, and walked every step with me.
\end{dedication}
\clearemptydoublepage
\pagenumbering{roman}
\tableofcontents
\clearemptydoublepage
\newpage
\newpage
\section{List of acronyms}
Several acronyms will be used in this Thesis. In order to facilitate
the reading we provide their complete list in the following.
 \begin{center}
    \begin{tabular}[h]{cccccc}\hline
      {\bf acronym} & & & & & {\bf extended name} \\
      \hline
      \hline
\\
{CDM} & & & & & Cold Dark Matter \\
\\
{CRASH} & & & & & Cosmological RAdiative Scheme for Hydrodynamics\\
\\
{DM} & & & & & Dark Matter \\
\\
{EW} & & & & & Equivalent Width \\
\\
{ERM} & & & & & Early Reionization Model \\
\\
{GADGET} & & & & & GAlaxies with Dark matter and Gas intEracT \\
\\
{GAMETE} & & & & & GAlaxy MErger Tree \& Evolution \\
\\
{IGM} & & & & &  Intergalactic Medium\\
\\
{IMF} & & & & &  Initial Mass Function\\
\\
{IRA} & & & & &  Instantaneous Recycling Approximation\\
\\
{ISM} & & & & &  Interstellar Medium\\
\\
{LAE} & & & & & Lyman Alpha Emitter \\
\\
{LBG} & & & & & Lyman Break Galaxy \\
\\
{LF} & & & & & Luminosity Function \\
\\
{LRM} & & & & & Late Reionization Model \\
\\
{MDF} & & & & & Metallicity Distribution Function \\
\\
{MW} & & & & & Milky Way \\
\\
 \end{tabular}
  \end{center}

\newpage
\newpage

 \begin{center}
    \begin{tabular}[h]{cccccc}\hline
      {\bf acronym} & & & & & {\bf extended name} \\
      \hline
      \hline
\\
{PopII} & & & & & Population II stars \\
\\
{PopIII} & & & & & Population III (metal-free) stars \\
\\
{SDSS} & & & & & Sloan Digital Sky Survey \\
\\
{SED} & & & & & Spectral Energy Distribution \\
\\
{SFR} & & & & & Star Formation Rate \\
\\
{SNII} & & & & & Type II Supernova \\
\\
{SPH} & & & & & Smoothed Particle Hydrodynamics \\
\\
{QSO} & & & & & Quasi-Stellar Object \\
\\
{UV} & & & & & Ultraviolet \\
\\
{UVB} & & & & & Ultraviolet Background \\
\\
{WMAP} & & & & & Wilkinson Microwave Anisotropy Probe \\
\end{tabular}
  \end{center}

\newpage
\newpage
\section{List of symbols}
Several symbols have been used extensively in this Thesis. In order to facilitate
the reading we provide their complete list in the following.
 \begin{center}
    \begin{tabular}[h]{cccccc}\hline
      {\bf symbol} & & & & & {\bf Definition } \\
      \hline
      \hline
\\
$c$ & & & & & Speed of light  \\
\\
$\chi_{HI}$ & & & & & Fraction of neutral Hydrogen  \\
\\
$EW^{int}$ & & & & & Intrinsic Ly$\alpha$ equivalent width \\
\\
$EW$ & & & & & Observed Ly$\alpha$ equivalent width \\
\\
$f_\alpha$ & & & & & Escape fraction of Ly$\alpha$ ionizing photons from the galaxy \\
\\
$f_{c}$ & & & & & Escape fraction of continuum photons from the galaxy \\
\\
$f_{esc}$ & & & & & Escape fraction of \HI ionizing photons from the galaxy \\
\\
$\Gamma_B$ & & & & & Photoionization rate from UVB \\
\\
$\Gamma_G$ & & & & & Photoionization rate from clustering \\
\\
$\Gamma_L$ & & & & & Photoionization rate from the emitter \\
\\
$h $ & & & & & Planck's constant \\
\\
$k $ & & & & & Boltzmann's constant  \\
\\
$H_0 $ & & & & & Hubble constant (100 h ${\rm km\, s^{-1} \, Mpc^{-1}}$) \\
\\
$\lambda_\alpha$ & & & & & Rest frame Ly$\alpha$ wavelength (1216 \AA)  \\
\\
$L_\alpha^*$ & & & & & Intrinsic stellar powered Ly$\alpha$ luminosity\\
\\
$L_\alpha^g$ & & & & & Intrinsic Ly$\alpha$ luminosity from cooling \HI \\
\\
$L_\alpha^{int}$ & & & & & Total intrinsic Ly$\alpha$ luminosity\\
\\
$L_\alpha^{em}$ & & & & & Total Ly$\alpha$ luminosity emerging from the galaxy\\
\\

    \end{tabular}
  \end{center}

\newpage
\newpage 
\begin{center}
    \begin{tabular}[h]{cccccc}\hline
      {\bf symbol} & & & & & {\bf Definition } \\
      \hline
      \hline \\
$L_\alpha$ & & & & & Total observed Ly$\alpha$ luminosity \\
\\
$L_c^*$ & & & & & Intrinsic stellar powered continuum luminosity\\
\\
$L_c^g$ & & & & & Intrinsic coninuum luminosity from cooling \HI \\
\\
$L_c^{int}$ & & & & & Total intrinsic continuum luminosity\\
\\
$L_c^{em}$ & & & & & Total continuum luminosity emerging from the galaxy\\
\\
$L_c$ & & & & & Total observed continuum luminosity \\
\\
$M_b$ & & & & & Baryonic mass in a galaxy  \\
\\
$M_{dust}$ & & & & & Dust mass in a galaxy  \\
\\
$M_h$ & & & & & Halo mass of a galaxy  \\
\\
$M_*$ & & & & & Stellar mass of a galaxy  \\
\\
$M_g$ & & & & & Gas mass of a galaxy  \\
\\
$\dot M_*$ & & & & & Constant star formation rate of a galaxy  \\
\\
$\nu_\alpha$ & & & & & Rest frame Ly$\alpha$ frequency ($2.5 \times 10^{15}$ Hz)  \\
\\
$Q$ & & & & & \HI ionizing photon production rate  \\
\\
$R_I$ & & & & & Str\"omgren sphere radius \\
\\
$\dot\rho_*$ & & & & & Star formation rate density \\
\\
$\tau_\alpha$ & & & & & Optical depth to Ly$\alpha$ photons  \\
\\
$\tau_c$ & & & & & Optical depth to continuum photons  \\
\\
$T_\alpha$ & & & & & Transmissivity of Ly$\alpha$ photons due to \HI attenuation \\
\\
$t_*$ & & & & & Age of star formation of a galaxy  \\

    \end{tabular}
  \end{center}

\newpage 
\begin{center}
    \begin{tabular}[h]{cccccc}\hline
      {\bf symbol} & & & & & {\bf Definition } \\
      \hline
      \hline \\

\\
$V_I$ & & & & & Str\"omgren region volume \\
\\
$Z_*$ & & & & & Stellar metallicity  \\
\\
$Z_g$ & & & & & Gas metallicity  \\
\\

  \end{tabular}
  \end{center}

\mainmatter
\pagenumbering{arabic}

\chapter{Introduction}\label{ch1_intro}

One of the most compelling questions in Astrophysics today is to understand how the complexity of structures we see around us today arose from the remarkably simple conditions of the early Universe. Our current understanding of the Universe is based on the {\it Hot Big Bang model}, in which, 
the Universe began about 13.7 billion years ago from a singularity of infinite temperature and density. This was immediately followed by a period of accelerated expansion called {\it Inflation}; although inflation lasted for only about $10^{-34}$ seconds, it expanded the Universe by about 60 orders of magnitude in size (see Mazumdar \& Rocher 2010 for a recent review on this topic). At the end of this period, the Universe was highly homogeneous on large scales but locally perturbed as a result of quantistic fluctuations. At this time, the radiation component (consisting of all relativistic particle species, including photons) dominated over the matter component; matter and radiation were thermally coupled through Compton scattering and free-free interactions. Due to the expansion driven by the Big Bang, the Universe was also expanding at the Hubble rate and, cooling as a result of this expansion; the temperature ($T$) of the Universe was decreasing with the redshift $z$, as $T\propto (1+z)$.

About three minutes after the Big Bang, the temperature dropped to $10^9$ K and light nuclei were finally able to synthesize as a result of the strong interactions between neutrons and protons. While Hydrogen nuclei were easily formed, the high temperatures lead to the Deuterium bottleneck; Deuterium nuclei were destroyed immediately after being formed. This process delayed the formation of ${\rm ^4 He}$ nuclei till about 20 minutes after the Big Bang, at which point this nuclei formed in an explosive process, along with trace amounts of ${\rm ^7 Li}$ and ${\rm ^7 Be}$ (Peacock 1993). However, by that time, the Universe was too cool to form heavier nuclei by nuclear fusion. The result of this {\it Big Bang Nucleosynthesis} was the following nuclei: 75\% {\rm H}, 24.8\% ${\rm ^4 He}$, ${\rm ^2 D/H} = 2.75 \times 10^{-5}$ and ${\rm ^3 He/H = 9.28 \times 10^{-6}}$.

The transition from radiation to matter domination occurred at $z \sim 3200$, when the temperature dropped to about $10^4$ K. However, the Universe was still hot enough so that it consisted of an ionized plasma of matter and radiation that were coupled by Compton scattering. Finally, at $z \sim 1100$, the temperature dropped to about $3000$ K, and protons and electrons were finally able to recombine to form hydrogen atoms. At the same time, radiation {\it decoupled} from matter at the so-called last scattering surface, leading to the origination of the Cosmic Microwave Background (CMB) which is one of the pillars of the Big Bang model. This redshifted fossil radiation was first detected by Penzias and Wilson in 1965 and earned them a Nobel prize; it has a blackbody spectrum with a temperature of $2.725 \pm 0.01$ K and temperature anisotropies of the order of $10^{-5}$. This tells us that as the Universe was extremely homogeneous at the last scattering surface; the quantistic density fluctuations that originated in the very early Universe and were expanded to cosmic scales by Inflation had energy density and gravitational potential perturbations of the order of only $10^{-5}$.

After recombination, the Universe was composed of fully neutral hydrogen (\HI) and helium with a residual free electron fraction of about $10^{-4}$. Gravitational instability allowed the density perturbations to grow, leading to the formation of a web-like, filament-dominated structure, called the `cosmic web'. The first stars (and later galaxies) originated predominantly at the intersections of these filaments; the period between $z \sim 1100$ and the formation of the first stars at $z \sim 20-30$ is known as the `Dark Ages' of the Universe due to a lack of any sources of light existing between these times. 

The formation of the first stars and galaxies ended the dark ages and ushered in an era of cosmic enlightenment. The \HI ionizing photons (with energy is larger then $13.6$ eV) from these sources started ionizing the hydrogen in the surrounding fluctuating (lower) density background matter the sources were immersed in, called the `Intergalactic Medium (IGM)'. As these ionized regions started overlapping, they drastically changed both the thermal and the ionization state of the IGM. This evolutionary phase, known as {\it Cosmic reionization}, remains the subject of much debate; while Quasi-stellar object (QSO) data at $z\gsim 6$ seems to indicate that reionization ended (also known as the epoch of reionization, EoR) at $z \approx 6$ (Fan et al. 2006), the data from the Wilkinson Microwave Anisotropy Probe (WMAP) seems to favour a value of $z >10$ (Spergel et al. 2003, Komatsu et al. 2009), as explained further in Sec. \ref{obs_reio_ch1}. Further, the nature and relative contributions of the sources of reionization remain only poorly understood, as discussed further in Sec. \ref{cosmic_reio_ch1}.

We are enormously fortunate to be at a time when the the search for the most distant galaxies, located at the beginning of the cosmic dawn, is now entering its maturity. The last few years have witnessed a tremendous increase in the data available, and the number of candidates at redshifts as high as $z=10$, corresponding to only half a billion years after the Big Bang. This has been made possible by a combination of new technologies and refined selection methods. In the first class of triggers, it is easy to acknowledge the role of the Hubble Space Telescope (HST) and its predecessor, the Hubble Deep Field; follow-up experiments performed with the newly installed Wide Field Camera (WFC3) have allowed to push the exploration to very faint galaxies as remote as $z=10$. The standard selection method applied to these survey data sets is based on the dropout technique introduced by Steidel et al. (1996) and later constantly refined and improved by several authors (e.g. Giavalisco et al. 2004, Bouwens et al. 2007). Though this method has proved to be very solid in identifying high-redshift sources, it has the drawback that the exact source redshift cannot be determined with complete confidence. Another technique that has become increasingly important is narrow-band spectroscopy, aimed at detecting the Ly$\alpha$ line; these searches (Malhotra et al. 2005; Shimasaku et al. 2006; Taniguchi et al. 2005; Kashikawa et al. 2006; Iye et al. 2006; Tilvi et al. 2010; Hibon et al. 2010) have yielded hundreds of $z > 5$ galaxies detected in the Ly$\alpha$, which are therefore called Lyman Alpha Emitters (LAEs, complete details follow in Sec. \ref{laes_ch1}). Finally, another series of experiments involve searching for remote galaxies behind foreground galaxy clusters acting as magnification lenses (Schaerer \& Pell\'o 2005; Richard et al. 2008; Bradley et al. 2008). Although these searches result in deeper magnitudes, their interpretation is hampered by the lens modelling and by the extremely narrow field of views, rendering it difficult to keep cosmic variance under control. As a final remark, we note that the most distant, spectroscopically confirmed, cosmic object is a Gamma Ray Burst (GRB 090423 at $z=8.2$, Salvaterra et al 2009b; Tanvir et al. 2009). Although not a galaxy, the presence of this indicator implies that star formation was already well under way at those early epochs, thus further encouraging deeper galaxy searches.

By virtue of their rapidly growing data sets, LAEs are now one of the most important probes of both reionization and early galaxy evolution, as discussed in more detail in Sec. \ref{imp_laes_ch1}. The aim of this thesis is to present a self-consistent theoretical model that unveils the nature of these high-z galaxies, in addition to explaining the various observed data sets; the hope is that future instruments such as the JWST (James Webb Space TElescope), MUSE (Multi Unit Spectroscopic Explorer) and ALMA (Atacama Large Millimeter Array) will confirm our predictions, thereby strengthening the validity of our work.

In this chapter, we start with an introduction to both the linear and non-linear regimes of structure formation,
in Sec. \ref{struct_form_ch1}. Staring with an introduction to reionization (Sec. \ref{cosmic_reio_ch1}), we discuss the nature of the reionization sources (Sec. \ref{nature_reio_ch1}), which is followed by a brief discussion on the observational constraints on reionization (Sec. \ref{obs_reio_ch1}). We then introduce LAEs in Sec. \ref{laes_ch1} and discuss the types of LAEs, their importance as probes of reionization and early galaxy evolution (Sec. \ref{imp_laes_ch1}), present some of the observed data sets (Sec. \ref{obs_laes}) and discuss some of LAE models exisiting in the literature (Sec. \ref{model_laes_ch1}). We end by summarizing the purpose behind, and the layout of the thesis in Sec. \ref{thesis_purpose}.

We end by observing that data from the CMB, from galaxy clusters and their redshift evolution, and high-z supernovae (SN) have led to an era of precision cosmology and the development of the $\Lambda$CDM cosmological model. In this `standard' cosmlogical model, the Universe is spatially flat and contains cold, weakly interacting, massive dark matter (DM) particles, and a vacuum energy (represented by $\Lambda$) component, in addition to the ordinary baryonic matter. This concordance cosmological model is completely defined once the following parameters are specified: (a) the density of matter, $ \Omega_m = \Omega_{dm} + \Omega_b$, where $\Omega_{dm}$ and $\Omega_b$ represent the density parameters for dark matter and baryons respectively, (b) the density of the vacuum energy, $\Omega_\Lambda$, (c) the Hubble constant, $H_0 = 100 {\rm h\, km \, s^{-1} \, Mpc^{-1}}$, (d) the baryonic density parameter, $\Omega_b$, (e) the root mean square mass fluctuations in a sphere of size $8 {\rm h^{-1}}$ Mpc at $z=0$, $\sigma_8$ and (f) the spectral index of the primordial fluctuations, $n_s$. Their values from the latest observations (see Komatsu et al. 2009) are: $\Omega_m = 0.27\pm0.01$, $\Omega_b = 0.0456\pm 0.0015$, $\Omega_\Lambda= 0.726 \pm 0.015$, $h = 0.705 \pm 0.013$, $\sigma_8=0.812 \pm 0.026$ and $n_s = 0.960 \pm 0.013$. 

\section{Structure formation}
\label{struct_form_ch1}
In this section, we present the commonly adopted theory for structure formation, the {\it gravitational 
instability scenario}, in which primordial density perturbations grow through gravitational Jeans instability to form the complex structures we observe today. For the discussion that follows, we adopt the $\Lambda$CDM cosmological model explained above.

\subsection{Linear regime}
To understand how matter behaves under gravity in an expanding Universe, the linear perturbation theory can be adopted as long as the density fluctuations are small; observations of the CMB ensure that this condition is fully satisfied at $z\approx 1100$. The Universe can be described in terms of a fluid made of collisionless dark matter and collisional baryons, with an average mass density $\overline{\rho}$. Then, the mass density, $\rho({\bf{x}},t)$, at 
any space and time co-ordinates (${\bf x}$ and $t$ respectively) can be written as $\rho({\bf{x}},t)=\overline{\rho}(t)[1+\delta(\mathbf{x},t)]$. During the linear regime ($\delta \ll 1$), the time evolution equation for $\delta$ reads (Peebles 1993):
\be
{\ddot\delta}({\mathbf{x}},t)+2H(t){\dot\delta}({\mathbf{x}},t)=4{\pi}G{\overline{\rho}(t)}{\delta({\mathbf{x}},t)}+\frac{{{c_{s}}^{2}}}{a(t)^{2}}{\nabla^{2}}\delta({\mathbf{x}},t), 
\label{delta_ch1}
\ee
where $c_s$ is the sound speed, $a\equiv (1+z)^{-1}$ is the scale factor describing the expansion of the Universe and $H(t)=H_0[\Omega_m(1+z)^3+\Omega_{\Lambda}]^{1/2}$. The growth of the perturbations due to gravitational collapse (first term on the right hand side) is counteracted by both the cosmological expansion (second term on the left hand side), and by pressure support (second term on the right hand side). While the pressure in the baryonic gas is provided by collisions, it arises from a readjustment of the particle orbits for the collisionless DM. This equation can be used to calculate the evolution of different components in case of a multi-component medium; then, $c_s$ would be the sound speed of the perturbed component providing the pressure support and $\overline \rho$ the density of the gravitationally dominant component, responsible for driving the collapse.

The total density contrast at any spatial location can be described in the Fourier space as a superposition of modes with different wavelength:
\be
\delta({\mathbf{x}},t)=\int\frac{d^3\mathbf{k}}{(2\pi)^3}\delta_{\mathbf{k}}(t)exp(i\mathbf{k}\cdot \mathbf{x}),
\label{fourier_exp_ch1}
\ee
where $\bf k$ is the comoving wave number. Hence, the evolution of any single Fourier component is given by:
\be
{\ddot\delta}_{\mathbf{k}}+2H(t){\dot\delta}_{\mathbf{k}}=\Big (4{\pi}G{\overline{\rho}}-\frac{k^2 c_s^2}{a^2}\Big )\delta_{\mathbf{k}}.
\label{fluct_k_ch1}
\ee
The above equation implicitly defines a critical scale, $\lambda_J$, called the Jeans length (Jeans 1928), at which the competitive pressure and gravitational forces cancel:
\be
\lambda_J=\frac{2\pi a}{k_J}=\Big (\frac{\pi c_s^2}{G{\overline{\rho}}}\Big )^{1/2}.
\label{lambda_j_ch1}
\ee
For $\lambda \gg \lambda_J$, the time-scale needed for the pressure force to act is much larger than the gravitational one and the zero-pressure solution applies. On the other hand, for $\lambda < \lambda_J$ the pressure force is able to counteract gravity and the
density contrast oscillates as a sound wave. Once the Jeans wavelength is fixed, we can also calculate the Jeans mass, $M_J$, which is defined as the mass within a sphere of radius $\lambda_J/2$, such that
\be
M_J={\frac{4\pi}{3}}{\overline{\rho}}\Big ({\frac{\lambda_J}{2}}\Big )^3.
\label{jeans_mass_ch1}
\ee
In a perturbation with mass greater than $M_J$, gravity is much stronger than the pressure force and the structure collapses. This sets a limit on the scales that are able to collapse at each epoch and has a different value for each perturbed component considered, reflecting the differences in their velocity.

Given the initial power spectrum of the perturbations, $P(k)\equiv \langle |\delta^2_{\mathbf k}|\rangle$, the evolution of 
each mode, can be followed using Eq. \ref{fluct_k_ch1}; these can then integrated to recover the global spectrum at any 
time. Inflationary models predict that $P(k) \propto k^n$, with $n=1$, i.e., a scale invariant power spectrum. Although the initial power spectrum is scale invariant, the growth of perturbations result in a modified final value; while on large scales the power spectrum follows a simple linear evolution, its shape on small scales changes due to gravitational growth (and collapse), resulting in a spectrum such that $P(k) \propto k^{n-4}$. Note that in the CDM model of structure formation, most of the power is on small scales, which are therefore the first to become non-linear.

\subsection{Non-linear regime: dark matter halos}
Since cold DM is made of collision-less particles that interact very weakly with the rest of the matter and with the radiation field, density perturbations in this component start growing at early epochs. However, as soon as the density contrast, $\delta \approx 1$, the linear perturbation theory does not apply anymore, and the full non-linear gravitational problem must be considered. Zel'dovich (1970) developed a simple approximation to describe the non-linear stage of gravitational evolution. According to this approach, sheet-like structures called `pancakes' are the first non-linear structures to form from collapse along one principal axis; the observed filaments and knots form from the simultaneous collapse along two and three axes respectively. However, since the probability of collapse along only one axis is the largest, pancakes are the dominant structures formed in such a model. Although the results of such an approximation agree very well with simulations at the beginning of the non-linear collapse, the results are largely inaccurate at the later stages. 

This problem can be overcome to some extent by analytically following the dynamical collapse of a DM halo assuming it to be spherically symmetric and having a constant density. In this case, at some point, this region reaches the maximum radius of expansion, after which it turns around and starts collapsing to a point. However, much before such a situation is reached, the DM 
experience a violent relaxation process, quickly reaching virial equilibrium. A dark matter halo of mass $M$ virializing at redshift $z$, can be fully described in terms of its virial (physical) radius $r_{vir}$, circular 
velocity $v_h=\sqrt{GM/r_{vir}}$, and virial temperature $T_{vir}=\mu m_p v_h^2 (2k_B)^{-1}$, whose expressions are (Barkana \& Loeb 2001):
\be\label{r_vir_ch1}
r_{vir}=0.784\;\Big (\frac{M}{10^8{\rm h^{-1}}M_{\odot}} \Big)^{1/3}\Big [\frac{\Omega_m}{\Omega_m(z)}\frac{\Delta_c}{18\pi^2}\Big ]^{-1/3}\Big ( \frac{1+z}{10}\Big )^{-1} {\rm h^{-1}}\rm{kpc},
\ee
\be\label{v_h_ch1}
v_h=23.4\;\Big (\frac{M}{10^8 {\rm h^{-1}}M_{\odot}} \Big)^{1/3}\Big [\frac{\Omega_m}{\Omega_m(z)}\frac{\Delta_c}{18\pi^2}\Big ]^{1/6}\Big ( \frac{1+z}{10}\Big )^{1/2}\rm{km}\;\rm{s}^{-1},
\ee
\be\label{T_vir_ch1}
T_{vir}=2\times 10^4 \;\Big(\frac{\mu}{0.6}\Big )\Big (\frac{M}{10^8{\rm h^{-1}}M_{\odot}} \Big)^{2/3}\Big [\frac{\Omega_m}{\Omega_m(z)}\frac{\Delta_c}{18\pi^2}\Big ]^{1/3}\Big ( \frac{1+z}{10}\Big )\;\rm{K},
\ee
where $m_p$ is the proton mass, $\mu$ the mean molecular weight, $G$ is the gravitational constant, $k_B$ is the Boltzmann constant and according to the results of Bryan \& Norman (1998),
\be\label{D_c_ch1}
\Delta_c = 18\pi^2+82(\Omega_m(z)-1)-39(\Omega_m(z)-1)^2,
\ee
\be
\Omega_m(z) = \frac{\Omega_m(1+z)^3}{\Omega_m(1+z)^3 + \Omega_\Lambda}.
\ee

Although the spherical collapse approximation is important to derive individual halo properties, it does not yield any information regarding either the inner structure or the abundance of halos. We start by discussing the results available on the inner density profile of DM halos: using numerical simulations of DM halo formation a number of authors (Navarro, Frenk \& White 1996, 1997; Moore et al. 1999; Del Popolo et al. 2000; Ghigna et al. 2000; Jing 2000; Fukushige, Kawai \& Makino 2004; Power et al. 2003) have found the density profile to have a universal shape with an internal cusp, which is independent of the halo mass, the initial density fluctuation spectrum and cosmological parameters; this is explained to be the result of a violent relaxation process which produces an equilibrium independent of the initial conditions. On the other hand, authors including Kravtsov et al. (1998), Avila-Reese et al. (1999), Jing \& Suto (2000), Subramanian, Cen \& Ostriker (2000) and Ricotti (2003) have found the profile to depend upon a number of factors such as the environment the halo is immersed in, the halo mass, the presence/absence of gas etc. The net result is that as of yet, there is no consensus on the existence of a universal shape of the DM halo density profile.

The abundance of halos is much better known; Press \& Schechter (1974) developed an analytic model in which the number density of halos (as a function of the mass) at any $z$ can be calculated from a linear density field by applying a model of spherical collapse to associate peaks in the field with virialized objects in a full non-linear treatment. However, this model can not predict the spatial distribution of the halos. As an alternative approach, Sheth \& Tormen (2002) developed the excursion set formalism that can be used to infer the adundance and distribution of DM halos, in addition to being applicable for both spherical and elliptical collapse, of which, the latter is in better agreement with numerical simulations. Further details of this model are presented later in Sec. \ref{ST_mf}.

\subsection{Non-linear regime: the formation of protogalaxies}
\label{Msf_ch1}
In contrast to DM, as long as the gas is fully ionized, the radiation 
drag on free electrons prevents the formation of gravitationally bound systems. 
It is only after decoupling that perturbations in the baryonic component are 
finally able to grow in the pre-existing dark matter halo potential wells, 
eventually leading to the formation of the first bound objects. The 
virialization process of the gas component is similar to the DM one; 
in this case however, the gas develops shocks during the contraction that 
follows the turnaround, and gets reheated to a temperature at which pressure 
support can prevent further collapse.

The minimum mass of the first bound objects, i.e. the Jeans mass, can be 
derived using Eqns.~\ref{lambda_j_ch1}-\ref{jeans_mass_ch1}, where $c_s$ is the sound 
velocity of the baryonic gas such that $c^2_s=dp/d\rho=RT/\mu$ (here, $R$ is the ideal gas constant). Since the residual 
ionization of the cosmic gas keeps its temperature locked to that of the CMB ($T \propto (1+z)$) 
through different physical processes, down to a redshift, $1+z_t\approx 1000 (\Omega_b h^2)^{2/5}$ (Peebles 1993), this results in a Jeans mass that is time-independent for $z>z_t$. Instead, for $z<z_t$, the gas temperature declines 
adiabatically ($T\propto (1+z)^{2}$), and $M_J$ decreases with decreasing 
redshift such that:
\be\label{M_j}
M_J=3.08 \times 10^3 \left ( \frac{\Omega_m {\rm h^2}}{0.13} \right )^{-1/2} 
\left ( \frac{\Omega_b h^2}{0.022} \right )^{-3/5} \left ( \frac{1+z}
{10}\right )^{3/2} \; {\rm M}_\odot.
\ee
Again, as the determination of the Jeans mass is based on a perturbative 
approach, it can only describe the initial phase of the collapse. Moreover we 
have to stress that $M_J$ only represents a {\it necessary but not sufficient} 
condition for collapse: the gas cooling time, $t_{cool}$, has to be shorter 
than the Hubble time, $t_{H}$, in order to allow the gas to condense. 
Therefore, the efficiency of gas cooling is crucial in determining the minimum 
mass of protogalaxies.

As already pointed out, in standard $\Lambda$CDM models, the first collapsing 
objects are predicted to be the least massive ones, i.e. those with the lowest 
virial temperatures (Eq.~\ref{T_vir_ch1}). For $T_{vir}<10^4$~K and in gas of 
primordial composition, molecular hydrogen, H$_2$, represents the main 
available coolant (see Fig.~12 of Barkana \& Loeb 2001). These first, metal-free stars, are also called Population III (PopIII) stars, to differentiate them from objects that are metal enriched (PopII or PopI) objects where the formation and cooling is governed by metal cooling. PopIII stars are 
predicted to have formed in such H$_2$ cooling halos, usually called 
{\it minihalos}. The cooling ability of these objects essentially depends on 
the abundance of molecular hydrogen: the gas cool by radiative de-excitation 
if the H$_2$ molecule gets rotationally or vibrationally excited through 
a collision with an H atom of another H$_2$ molecule.
Primordial H$_2$ forms with a fractional abundance of $\approx 10^{-7}$ at $z>400$ 
via the H$_2^+$ formation channel. At $z<110$, when the CMB radiation intensity 
becomes weak enough to allow for a significant formation of H$^-$ ions, more 
H$_2$ molecules can be formed:
\be\label{eq:H2}
\nonumber
{\rm H} + {\rm e}^- \to {\rm H}^- + {\rm h\nu}\\
{\rm H}^- + H \to {\rm H}_2 + {\rm e}^-
\nonumber
\ee
Assuming that the H$^-$ channel is the dominant mechanism for H$_2$ formation can lead to 
a typical primordial H$_2$ fraction of $f_{{\rm H}_2}\approx 2\times 10^{-6}$ 
(Anninos \& Norman 1996). However, this is lower than the H$_2$ fraction of $\approx 5\times 10^{-4}$ required to efficiently cool the gas and trigger the SF process  
(Tegmark et~al. 1997). However, during the collapse, the H$_2$ content can be 
significantly enhanced; the fate of a virialized clump conclusively depends on 
its ability to rapidly increase its H$_2$ content during such a collapse phase. 
Tegmark et~al. (1997) investigated the evolution of the H$_2$ abundance for 
different halo masses and initial conditions finding that only the larger 
halos reache the critical molecular hydrogen fraction for the collapse. This 
implies that for each virialization redshift there exists a critical mass, 
$M_{sf}(z)$, such that $M>M_{sf}(z)$ halos will be able to collapse and form 
stars while those with $M<M_{sf}(z)$ will fail. However, the value and 
evolution of $M_{sf}(z)$ is highly debated as it strongly depends on the H$_2$ 
cooling functions used and chemical reactions included (Fuller \& Couchman 
2000). Nevertheless, many of the most complete studies (e.g. Abel, Bryan \& Norman 2000; Machacek, Bryan \& Abel 2001) agree that the absolute minimum mass allowed to collapse is as low as $10^5\Msun$. Note that as soon as more massive halos with $T_{vir}>10^4$~K become non-linear, the gas cooling proceeds unimpeded in these hot objects through atomic line cooling.

 \section{Cosmic reionization}
\label{cosmic_reio_ch1}

As mentioned above, reionization begins when the first sources of \HI ionizing radiation form and start building \HII regions around themselves. We now discuss the basic stages of reionization, these being the {\it pre-overlap}, {\it overlap} and {\it post-overlap} stages, using the terminology introduced by Gnedin (2000). In the initial `pre-overlap' stage, individual sources of ionizing radiation turn on and start ionizing their surroundings. Since the first galaxies form in the most massive halos at high redshifts, they are preferentially located in high density regions; the \HI ionizing photons that escape the galaxy (discussed later) then have to pass this high density region which is characterised by large recombination rates. Once the photons emerge from this region, the ionization fronts propagate more easily into the low density voids. At this stage, the IGM is a two-phase medium, with highly ionized \HII regions separated from neutral \HI regions by ionization fronts. The ionizing intensity too, is very inhomogeneous; even in the ionized regions, its value depends both on the ionizing luminosity of the source as well as the distance from it.

The `overlap' stage begins when neighbouring \HII regions start overlapping. At this stage, each point in the common region ionized by `n' sources is affected by the ionization fronts from all those sources. Therefore, the intensity in the \HII regions rises rapidly, allowing them to expand into the high density gas that had been able to recombine in the presence of a lower ionization intensity. By the end of this stage, most of the IGM regions are able to see many sources, making the ionization intensity both higher and more homogeneous. As more and more galaxies form, this results in a state in which the low-density IGM is ionized and the ionizing radiation reaches everywhere, except for the gas inside self-shielding, high-density clouds. This marks the end of the overlap phase and is referred to as the `moment of reionization'. Although there are many definitions of when reionization completes, we use the most general one in which reionization ends when the volume weighted neutral hydrogen fraction, $\chi_{HI} < 10^{-3}$. 

However, even at the end of the overlap stage, there remains some \HI in high density structures, such as Lyman Limit Systems (LLS) and Damped Lyman Alpha systems (DLAs), which can be seen in absorption at lower $z$. These too, get gradually ionized as galaxies form and the mean ionization intensity increases; the ionization intensity becomes more uniform as the number of sources visible to each point in the IGM increase. This post-overlap phase then continues indefinitely, since collapsed objects retain \HI even at the present time.

Although many different semi-analytic (Shapiro \& Giroux 1987; Fukugita \& Kawasaki 1994; Miralda-Escud\'e \& Rees 1994; Tegmark, Silk \& Blanchard 1994; Haiman \& Loeb 1997; Madau, Haardt \& Rees 1999; Chiu \& Ostriker 2000; Wyithe \& Loeb 2003) and numerical approaches (Gnedin \& Ostriker 1997; Gnedin 2000; Ciardi et al. 2000b; Razoumov et al. 2002; Ciardi, Ferrara \& White 2003; Ricotti \& Ostriker 2004; Sokasian et al. 2003) have been used to model reionization, both the nature of the reionization sources and the reionization history remain unclear and highly debated. 

This is mostly due to uncertainties in modelling several physical issues, which include the properties of the first stars and quasars, the ionizing photon production and radiative transfer, and the IGM clumping, to mention a few. One of the most crucial parameters needed for modelling reionization is the escape fraction of ionizing radiation; whatever the nature of a source, only a fraction, $f_{esc}$, of the ionizing photons emitted escape the production site and reach the IGM. However, the value of $f_{esc}$ is largely unconstrained and ranges between $f_{esc}=0.01-0.73$ ($0.01-0.8$) observationally (theoretically) as is shown from these works: Ricotti \& Shull (2000) found $f_{esc} \geq 0.1$ only for halos of total mass $\leq 10^7 {\rm M_\odot}$ at $z \geq 6$, with the value dropping sharply with increasing halo mass. Using ZEUS-3D simulations, Fujita et al. (2003) found $f_{esc} \leq 0.1$ from disks or dwarf starburst galaxies with total mass $\sim 10^{8-10} {\rm M_\odot}$. While Razumov \& Larsen (2006) concluded that $f_{esc}=0.01-0.1$ in several young galaxies with $M_{vir}= 10^{12-13} {\rm M_\odot}$ at $z \sim 3$, in a recent paper, Razoumov \& Larsen (2010) found $f_{esc} \sim 0.8-0.1$ for galaxies of halo mass $\sim 10^{9-11.5} {\rm M_\odot}$. Using detailed hydrodynamic simulations with radiative transfer, Gnedin et al. (2008) found $f_{esc}=0.01-0.03$ for galaxies with mass $\geq 10^{11} {\rm M_\odot}$ at $z \sim 3-5$. Using high resolution adaptive mesh refinement simulations, Wise \& Cen (2009) found $f_{esc}=0.25-0.8$ for dwarf galaxies between $3 \times 10^{6-9} {\rm M_\odot}$.  Finally, using radiative transfer and SPH simulations, Yajima et al. (2010) found that $f_{esc}$ decreases from $\sim 0.4$ to $\sim 0.07$ as the halo mass increases from $10^{9}$ to $10^{11} {\rm M_\odot}$ at $z \sim 3-6$. The observational results are equally inconclusive: while $z \sim 3$ galaxies have been used to obtain values of $f_{esc} < 0.04$ (Fernandez-Soto et al. 2003) to $f_{esc} \sim 0.1$ (Steidel et al. 2001), observations in the local Universe range between $f_{esc} < 0.01$ (Deharveng et al. 1997) to $f_{esc} \sim 0.1-0.73$ (Catellanos et al. 2002).

To alleviate this problem, Choudhury \& Ferrara (2005, 2006) proposed a self-consistent semi-analytic reionization model that is compatible with a number of different observed data sets. The main features of their model can be summarized along the following points: the model accounts for IGM inhomogeneities by adopting a lognormal distribution, as proposed by Miralda-Escud\'e, Haehnelt \& Rees (2000) and reionization is said to be complete when the low density regions are ionized. The thermal and ionization histories of \HI, \HII and 
\HeIII regions are followed 

\begin{figure}
\center{\includegraphics[angle=270,scale=1.15]{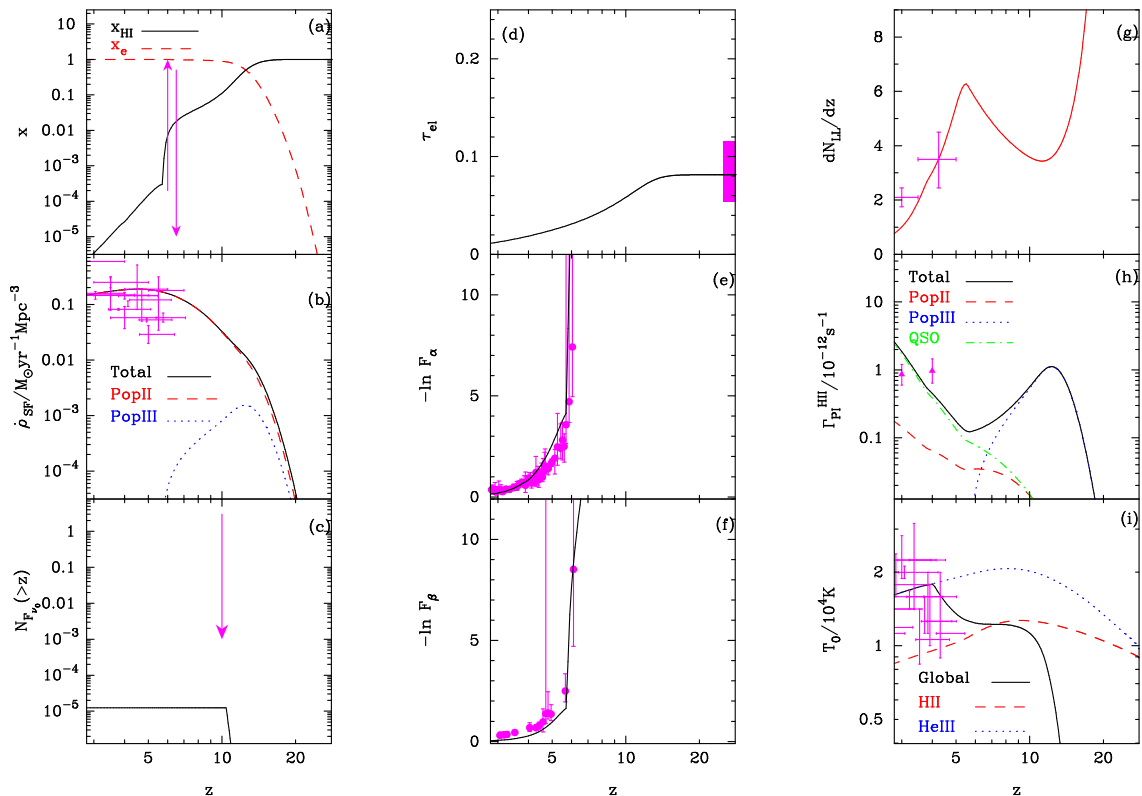}}
\caption{Comparison of the model predictions from Choudhury \& Ferrara (2006) to observations. The panels indicate: (a) The volume averaged \HI fraction, $\chi_{HI}$, with the observational lower limit from QSO absorption lines at $z=6$ and upper limits from Lyman Alpha Emitters at $z=6.5$ (shown with arrows). The ionized fraction, $\chi_e$ is shown with a dashed line. (b) The SFR density for different stellar populations. The points with error bars indicate low-$z$ observations taken from the compilation of Nagamine et al. (2004), (c) the number counts above a certain redshift, with the observational upper limit from the NICMOS Hubble Ultra Deep Field shown by the arrow. The contribution to the source count decreases to zero at low $z$ because of the J-dropout selection criterion. (d) Electron scattering optical depth, with observational constraints from WMAP3, (e) Ly$\alpha$ effective optical depth, with observed points from Songaila (2004), (f) Ly$\beta$ effective optical depth with observed points from Songaila (2004), (g) evolution of the Lyman-Limit systems, with observed points from Storrie-Lombardi et al. (1994), (h) Photoionization rates for hydrogen with estimates from numerical simulations shown by points with error bars (Bolton et al. 2005), and (i) the temperature of the mean density IGM, with observational estimates from Schaye et al. (1999).}  
\label{zch1_cf} 
\end{figure}   
\newpage
simultaneously and self-consistently, treating the IGM as a multi-phase medium. The suppression of star formation in halos due to heating of the IGM (by UV  sources) is taken into account. Three types of sources are considered in order calculate the ionizing flux: Quasi-stellar objects (QSOs), PopII and PopIII stars. The free parameters in their model are the star formation efficiency for PopII, PopIII stars ($f_{II}$, $f_{III}$ respectively) and the associated values of $f_{esc}$ ($f_{esc,II}$, $f_{esc,III}$ respectively).

Using $f_{II} = 0.2, f_{III}=0.07, f_{esc,II}=0.003, f_{esc,III}=0.72$, they were able to reproduce a number of data sets as shown in Fig. \ref{zch1_cf}, including the observed evolution of the cosmic star formation rate (SFR) history, the number counts of high-$z$ sources, the electron scattering optical depth, the Gunn-Peterson (GP) Ly$\alpha$, Ly$\beta$ optical depths, the evolution of the Lyman-Limit systems and the temperature of the IGM.

We now move on to a discussion of the nature of the reionization sources (Sec. \ref{nature_reio_ch1}) and then summarize the main observational constraints available on the EoR in Sec. \ref{obs_reio_ch1}.
 
 \subsection{Nature of the reionization sources}
\label{nature_reio_ch1}
We now briefly discuss our knowledge of the main sources of cosmic reionization, which include the first stars, galaxies, QSOs, GRBs and decaying/annihilating DM. 

\subsubsection {(i) The first stars}

The first stars represent the first sources of light and dust in the Universe, and affect all later generations of stars and galaxies through chemical (enrichment of the IGM with metals), mechanical (energy injection from massive stars into winds) and radiative (ionization/dissociation of atoms/molecules) feedbacks. Hierarchical models of structure formation predict the first collapsed objects to have small masses and a primordial, {\it metal-free} composition; their formation and cooling is governed by molecular hydrogen, as discussed in Sec. \ref{Msf_ch1}.

Although it is quite hard to model the primordial star formation process due to a poor understanding of the fragmentation process, advances have been made along this direction using numerical simulations: by means of a smooth particle hydrodynamics (SPH) simulation, Bromm, Coppi \& Larson (1999) have found the first stars to be massive, with $M \geq 10^2 M_\odot$. Using 3D adaptive mesh refinement simulations, Abel, Bryan \& Norman (2000) have reached the same conclusion, and Nakamura \& Umemura (2001) have found similar masses using a 2D hydrodynamic code for filaments of low initial density. On the other hand, as shown by Uehara \& Inutsuka (2000), and Omukai (2000), small mass clumps of $M \sim 0.1, 0.03 M_\odot$ can be produced if the cloud collapse is driven by HD or atomic cooling respectively. However, it is important to note that PopIII star formation is suppressed as soon as the interstellar medium (ISM) from which they form is enriched by metals produced by previous generations of PopIII stars.

Since the metal-free composition restricts the stellar energy source to be proton-proton burning rather than the CNO cycle, PopIII stars are hotter and have a harder spectra than PopII/PopI stars; this results in strong Ly$\alpha$ (1216\AA) and HeII (1640\AA) lines for these objects. In addition, PopIII stars with $M>300 M_\odot$ produce 10 times more \HI ionizing photons (Bromm, Kudritzki \& Loeb 2001) while for PopIII stars with $M<100 M_\odot$, the ionizing photon rate takes twice as long to decay from its peak value, as compared to ordinary metal-enriched stars. Both these facts might mean that their contribution to reionization is highly important. This is exactly the result found by Choudhury \& Ferrara (2007), who have shown (see Fig. 1 of their paper) that reionization is initially driven by metal-free stars in low mass ($M < 10^8 M_\odot$) halos; the conditions for the formation of these objects are soon erased by the combined action of chemical and radiative feedbacks at $z < 10$.  

However, the initial mass function (IMF), which specifies the distribution of mass in a newly formed stellar population, is largely unknown for PopIII objects due to a poor understanding of their fragmentation processes. This is important for reionization studies since the luminosity mostly comes from stars with $M < 1 M_\odot$ (although metal enrichment and feedback are from stars with $M> 10 M_\odot$). For many years, there have been many speculations in the literature suggesting that the IMF was dominated by massive stars in early times (e.g. Schwarzchild \& Spitzer 1953; Larson 1998). Several indirect observations confirm this view, e.g. Hernandez \& Ferrara (2001) have shown that the observational data on metal poor stars in the Milky Way (MW) suggests that the IMF of the first stars was increasing high-mass weighted towards high-$z$. However, starting with the work of 
Salpeter (1955), it has been well established (e.g. Massey 1998) that the present day IMF of stars (with mass $M>1 M_\odot$) in the solar neighbourhood can be well approximated by a power law, such that such that the number of stars in a mass range $M$ to $M+dM$ is $n_*(M)dM \propto M^{-2.35}$; there is no consensus yet for stars with $M<1 M_\odot$. 

This implies the need for a transition from the top heavy IMF to a normal one. One of the most favoured explanations concerns the metal enrichment of PopIII stars; PopIII star formation is suppressed when the metallicity of the gas they form from exceeds a certain `critical' value, enriched by metals from the first supernovae (SN) (see Schneider et al. 2002 and references therein). The physical interpretation is that the fragmentation properties of the collapsing gas cloud change as the mean metallicity of the gas increases above a critical threshold, $Z_{cr} = 10^{-5 \pm1} Z_\odot$. Although cosmological SPH simulations that include chemodynamics (Tornatore et al. 2007) have shown that PopIII star formation continues down to $z =2.5$, using the same simulation, Salvaterra et al. (2010b) have found that the ratio of PopIII to PopII stars is a decreasing function of $z$, as expected, and never exceeds a value of $10^{-3}$ below $z = 10$ (see also Fig. 1, Tornatore et al. 2007). This is discussed in more detail in Sec. \ref{pop3_ch7}.

\subsubsection{(ii) Galaxies} 
According to the bottom-up, hierarchical model of structure formation, it is now well established that small halos form first, which later merge to form larger systems. However, the physics of the baryons inside galaxies still remains one of the unsolved problems in astrophysics. One of the main problems in calculating the contribution of galaxies to reionization concerns modelling their intrinsic \HI ionizing photon rate (which depends on the SFR) as well as the value of $f_{esc}$ for each galaxy. 

Using suitable parameter choices of the SFR and $f_{esc}$, a number of authors have shown that star forming galaxies are capable of reionizing the Universe by $z \sim 6-15$. This includes work done using semi-analytic models (Fukugita \& Kawasaki 1994; Tegmark, Silk \& Blanchard 1994; Haiman \& Loeb 1997; Chiu \& Ostriker 2000; Ciardi et al. 2000b; Choudhury \& Ferrara 2005, 2006) and simulations (e.g. Gnedin \& Ostriker 1997; Gnedin 2000). In particular, using their model, Choudhury \& Ferrara (2007) find that by $z \sim 6$, 70\% of the \HI ionizing photons come from galaxies with halo mass $M_h < 10^9 M_\odot$.

Although the SFR can be modelled using the Schmidt Law (1959, later modified by Kennicutt 1998), according to which the SFR for a given region varies according to the gas density in that region, the contribution of the galaxies to reionization also depends on the value of $f_{esc}$, which remains largely unknown, as explained before. This is why knowledge of the amount of star formation taking place in high-$z$ galaxies would provide precious constraints on reionization. 

As mentioned above, the last few years have witnessed a tremendous increase in the data available, and the number of candidates at redshifts as high as $z=10$, by virtue of state of the art instruments (HST, the WFC3) and techniques (the narrow band spectroscopy for LAEs and gravitational lensing). In the future, the James Webb Space Telescope (JWST), with its exquisite sensitivity of about $1 {\rm nJy}$ in the $1-10 \mu {\rm m}$ band will be ideally suited to probe optical-UV emission from galaxies with $z >10$. All these data sets together will hopefully shed light on questions such as when and how much galaxies of a certain masses contribute to reionization.
 
 \subsubsection{(iii) Quasars} 
QSOs are powered by accretion of gas onto a black hole (BH); the excess rotation of the gas spiralling in towards the BH yields viscous dissipation of heat that makes the gas glow. The seed BH is generally explained to be the remnant of a massive PopIII explosion.

QSOs are more efficient than stars for reionization because of three main reasons: (a) their emission spectrum is harder than that from stellar sources, (b) the radiative efficiency of accretion onto a BH is about 10 times mlarger than the radiative efficiency of stars, and (c) the value of $f_{esc}$ is larger for QSOs than for stars, for a given density distribution of the host galaxy. Estimates of super massive BH (SMBH) powering high-$z$ QSOs indicates that objects with BH masses $>10^9 M_\odot$ exist even at $z \sim 5-6$ (Dietrich \& Hamann 2004). This means that the history of reionization could have been significantly affected by the existence of massive BH in low mass galaxies in the early Universe.

Although Madau et al. (2004) have shown that BHs produced by PopIII stars can accrete and act as miniquasars (which could produce a significant amount of \HI ionizing photons), it is important to see that the recombination rate, and hence the emissivity required to ionize the IGM increase with $z$ (Miralda-Escud\'e, Haehnelt \& Rees 2000). This, and the relatively short QSO lifetimes could effectively reduce the contribution of QSOs to reionization.

 \subsubsection{(iv) Gamma Ray Bursts} 
GRBs are flashes of gamma rays associated with extremely energetic explosions and are amongst the most luminous electromagnetic events known, detectable out to $z>10$ (Ciardi \& Loeb 2000; Lamb \& Reichart 2000). GRBs are believed to originate in the compact remnants (neutron star or BH) of massive stars, and hence give an opportunity to detect the elusive PopIII stars.

At high-$z$, according to the hierarchical structure formation model, galaxies are smaller and less luminous. GRB afterglows, which already produce a peak flux similar to that from QSOs and galaxies at $z \sim 1-2$, will therefore outshine any other sources of radiation, such as the ubiquitous dwarf galaxies. Indeed, upto 10\% of all GRBs should be at high-$z$, as shown by Natarajan et al. (2005), Daigne, Rossi \& Mochkovitch (2006); Bromm \& Loeb (2006); Salvaterra et al. (2007). Therefore, GRBs will provide superb probes of PopIII SF.
 
The satellite {\it SWIFT} has already detected 5 GRBs at $z \geq 5$ (Gehrels et al. 2004) and a GRB has been detected at $z=6.29$, as reported by Tagliaferri et al. (2005) and Kawai et al. (2006). As a final remark, we note that the most distant, spectroscopically confirmed, cosmic object is a Gamma Ray Burst (GRB090423 at $z=8.2$, Salvaterra et al 2009b; Tanvir et al. 2009). Although not a galaxy, the presence of this indicator implies that star formation was already well under way at those early epochs, thus further encouraging deeper galaxy searches.
 
 \subsubsection{(v) Dark Matter}
As proposed by Hansen \& Haiman (2004), decaying DM particles could represent additional sources of \HI ionizing photons. Mapelli et al. (2006) have built a theoretical model, considering 4 types of DM candidates: Light DM (LDM), gravitinos, neutralinos and sterile neutrinos. They find that all 4 particle types have negligible effect on the CMB. However, while LDM (1-10 Mev) and sterile neutrinos (2-8 Mev) can be early ($z \leq 100$) sources of heating and ionization for the IGM, gravitinos and neutralinos do not have any effect on reionization.
 
 \subsection{Observational constraints on reionization}
\label{obs_reio_ch1}
At the present time, there are both cosmological and astrophysical constraints on reionization; while the former are from the CMB, the latter are inferred using spectral data from QSOs, GRBs and LAEs. All of these are now explained in greater detail.

\subsubsection {(i) CMB constraints on reionization}

As mentioned before, at $z \sim 1100$, electrons and protons combine to form \HI atoms, and photons decouple from matter at this last scattering surface. These photons are observed today as the CMB and carry information about the state of the Universe at the decoupling epoch; small fluctuations in density, velocity and gravitational potential lead to anisotropies in the CMB (Sachs \& Wolfe 1967; Bennett et al. 1996). Reionization alters the anisotropy signal by erasing some of the primary anisotropies and generating polarization signals on large scales, that can be used to probe the EoR. If reionization occurs instantaneously at a redshift $z$, then the total electron scattering optical depth in a $\Lambda$CDM model can be expressed as (see Sec. 7.1.1, Hu 1995)
\be
\tau_{es} = 0.041 \frac{\Omega_b {\rm h}}{\Omega_m} {\sqrt{1-\Omega_m + \Omega_m(1+z)^3} -1 }.
\label{tau_es_ch1}
\ee
From the WMAP first year data (Spergel et al. 2003; Kogut et al. 2003), the electron scattering optical depth was calculated to be $\tau_{es} = 0.017^{+0.08}_{-0.07}$, which for an instantaneous reionization translates into a reionization redshift, $z \sim 15$. With a larger data set, the WMAP 5 year data shows a much higher value of the optical depth, $\tau_{es} = 0.084 \pm 0.016$; this naturally leads to a later reionization at $z \sim 10.9 \pm 1.4$ (Komatsu et al. 2009). A complete discussion on this can be found in Sec. 9.2 Barkana \& Loeb (2001).

We now move onto a discussion of the astrophysical (spectral) constraints on reionization; indeed, the spectral data collected for LAEs and their interpretation form the backbone of the work carried out in this thesis.

\subsubsection {(ii) Spectral constraints on reionization}
\label{spec_cons_reio}

\begin{figure}
\center{\includegraphics[scale=0.6]{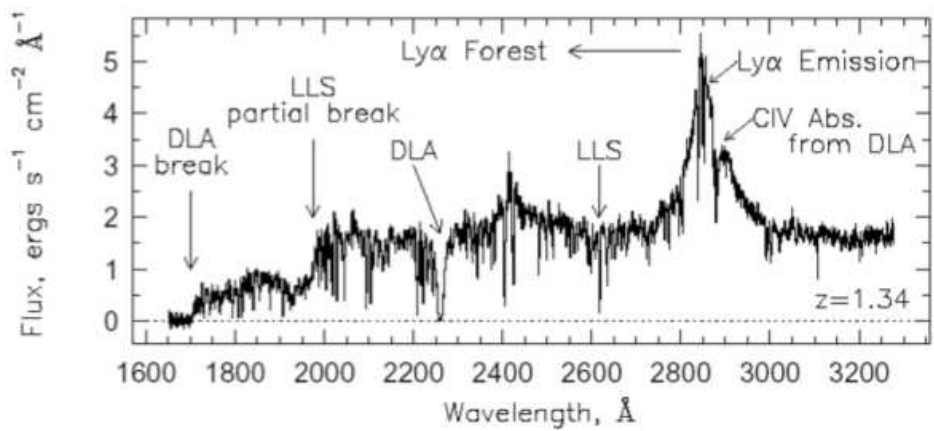}}
\caption{Rest frame spectrum of a QSO at $z=1.34$. The Ly$\alpha$ forest is clearly seen blueward of the Ly$\alpha$ emission line at $1216$ \AA. }  
\label{zch1_eg_spec} 
\end{figure}

The IGM manifests itself in numerous absorption lines along the line of sight (LOS) of observed QSOs, GRBs and galaxies. These absorption lines arise when a LOS intersects a patch of \HI that absorbs the continuum radiation which redshifts into the Ly$\alpha$ (1216 \AA) range. According to the amount of absorption, different kinds of absorbers can be distinguished in the observed spectra. As seen from Fig. \ref{zch1_eg_spec}, the {\it Ly$\alpha$ forest} arises from absorbers that have a column density of \HI such that $N_{HI} \leq 10^{16}\, {\rm cm^{-2}}$; these absorbers lie in shallow DM potential wells, containing gas in various stages of infall and collapse. The {\it Lyman Limit systems} (LLS) have a column density $N_{HI} \geq 10^{17}\, {\rm cm^{-2}}$ and absorption is caused by relatively cool gas associated to star forming galaxies in high density regions. Absorbers that have a column density of $N_{HI} \geq 2 \times 10^{20}\, {\rm cm^{-2}}$ are called {\it Damped Ly$\alpha$ systems} (DLAs) and they are the highest density systems that are believed to be drawn from the bulk of proto-galactic halos which evolve to form present day galaxies such as the MW (Kauffmann 1996).

Most spectral methods of constraining reionization rely on the Ly$\alpha$ line to constrain the amount of \HI in the IGM. This is because specific spectral signatures like the strength of the line and the Gunn-Peterson (1965) absorption bluewards of it, make its detection relatively unambiguous: the Ly$\alpha$ emission line at 1216 \AA \,(which corresponds to the energy of 10.6 eV, emitted when a hydrogen atom makes a transition from the $n=2$ to the $n=1$ ground state) is the strongest spontaneous transition of the hydrogen atom with an Einstein co-efficient of $6.265 \times 10^8 \, {\rm s^{-1}}$; this makes the Ly$\alpha$ line the strongest emission signal. Further, due to a large GP optical depth, in the presence of \HI, the observed spectra shows a sharp cut-off blueward of the Ly$\alpha$ line. The effective GP optical depth to Ly$\alpha$ photons can be expressed as
\be 
\label{tau_tot_ch1}
\tau_{GP}^{eff} = \frac{\pi e^2 f \lambda_\alpha}{m_e c H(z)} n_H \frac{n_{HI}} {n_H}, 
\ee  
where  
\be 
\label{tau_alpha_ch1}
\frac{\pi e^2 f \lambda_\alpha}{m_e c H(z)} n_H = 1.76 \times 10^5 {\rm h^{-1}} \Omega_m^{-1/2} \frac{\Omega_b h^2}{0.022} \bigg(\frac{1+z}{8}\bigg)^{3/2}.  
\ee 
Here $e$ is the electron charge, $f$ is the oscillator strength ($0.4162$), $\lambda_\alpha$ is the wavelength of Ly$\alpha$ in its rest frame (1216 \AA), $m_e$ is the electron mass, $c$ is the speed of light, $H(z)$ is the Hubble parameter at the required redshift, $n_{HI}$ is the global neutral hydrogen density and $n_H$ is the global mean hydrogen density at that redshift. Note that 
\be
\chi_{HI} =\frac{n_{HI}} {n_{H}} , 
\ee 
where $\chi_{HI}$ is the fraction of neutral hydrogen at the redshift under consideration. As is clear from Eq. \ref{tau_alpha_ch1} even a \HI fraction of $10^{-4}$ can lead to a significant attenuation of the Ly$\alpha$ line.

\begin{figure}
\center{\includegraphics[scale=0.65]{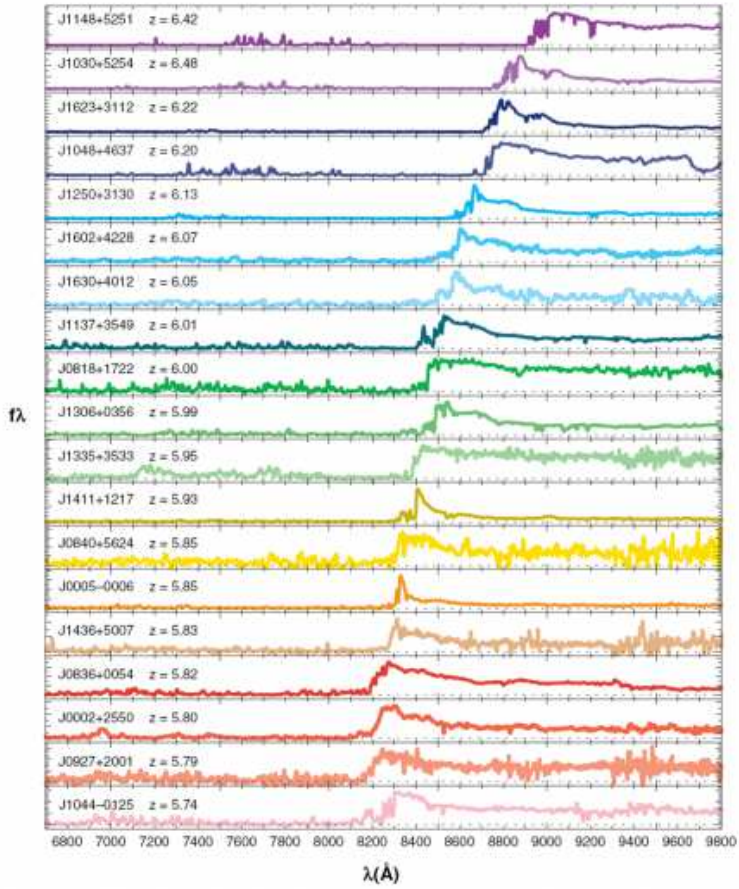}}
\caption{Spectra of 19 SDSS QSOs with $5.74 < z<6.42$ from bottom to top (Fan et al. 2006). }  
\label{zch1_fan_spec} 
\end{figure}

We start by discussing how QSO spectra are used for constraining the ionization state of the IGM. Fan et al. (2006) have obtained spectra for 19 SDSS (Sloan Digital Sky Survey) QSOs between $z \sim 5.7$ to $6.4$, which are shown in Fig. \ref{zch1_fan_spec}. 

\begin{figure}[htb]
\center{\includegraphics[scale=0.5]{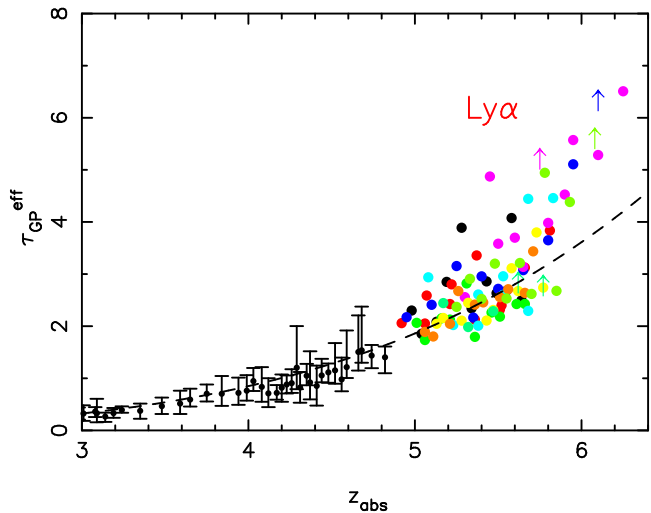}}
\caption{Evolution of the Ly$\alpha$ GP optical depth with redshift. The large circles show the results from Fan et al. (2006) for a sample of 19 SDSS QSOs at $5.74 < z<6.42$. For QSOs showing complete GP troughs, for which no flux is detected, the $2-\sigma$ lower limit on the optical depth is shown with arrows. The small symbols represent low redshift measurements from Songaila (2004).}  
\label{zch1_fan_taualpha} 
\end{figure}

As seen from this figure, as the redshift increases from $z \sim 5.7$ to $6.4$, larger portions of the spectra, blueward of the Ly$\alpha$ line are completely attenuated; the increasing $z$ makes this break shift to longer wavelengths. Fan et al. (2006) have translated this into an optical depth, as shown in Fig. \ref{zch1_fan_taualpha}. 

\vspace{-3.0cm}

This figure clearly shows a sudden rise in the value of $\tau_{GP}^{eff}$, which has been interpreted as an evidence of reionization finishing at $z \sim 6$ (Fan et al. 2006). Indeed, this result is in agreement with the findings of Becker et al. (2001), who detected the first evidence of a complete GP trough in a QSO at $z=6.28$; for this QSO, no flux is detected over 300 \AA\, blueward of the Ly$\alpha$ line.

However, as seen clearly from Eqns. \ref{tau_tot_ch1} and \ref{tau_alpha_ch1}, Ly$\alpha$ photons are highly sensitive to the presence of \HI and a volume averaged value of $\chi_{HI}$ as small as $10^{-4}$ is sufficient to completely depress the Ly$\alpha$ flux. Therefore, it is important to note that the detection of a GP trough only translates into a lower limit of $\chi_{HI}$.  

Using these spectra, many groups (Gallerani et al. 2006; Becker et al. 2007) have tried to check if reionization was over by $z \sim 6$; in particular, Gallerani et al. (2006) find that the QSO data is compatible with a highly ionized IGM at $z\sim 6$. However, it is important to note that reionization is a very patchy process, as a result of which, the Ly$\alpha$ transmission is very different along different LOS; the fact that the IGM seems quite ionized along some LOS, while being neutral along others has been interpreted as signalling the end of reionization (Wyithe \& Loeb 2006).

The distribution of the dark gaps (regions showing no flux) in the QSO spectra have also been used to constrain $\chi_{HI}$; using this information, Fan et al. (2006) have constrained $\chi_{HI} <0.1-0.5$. Using the gap and peak statistics (regions showing no transmission/transmission respectively), Gallerani et al. (2008) have found that the data favours a model where reionization finishes at $z \sim 7$ and robustly constrain $\chi_{HI}<0.36$ at $z=6.3$. Similar conclusions have been reached by Dayal et al. (2008) using LAE data, as will be presented later in Sec. \ref{lf_fit}.

GRB spectra can be used to constrain reionization using the same principles outlined above; indeed, Totani et al. (2006) have used the spectrum of GRB 050904 ($z = 6.3$) to constrain $\chi_{HI}<0.17 \, (0.6)$ at the 68\% (95\%) confidence levels, respectively. We finally summarize the main advantages of using GRB spectra: (a) for standard afterglow lightcurves and spectra, the increase in the luminosity distance is compensated by an increase in the intrinsic luminosity at earlier times (Ciardi \& Loeb 2000), (b) since GRBS originate from stellar sources, their intrinsic luminosity is independent of the DM halo mass, making them outshine the luminosity of their dwarf host galaxy, (c) they have a smooth, power-law afterglow spectra, which makes it easier to extract information about the absorption features, as compared to the spectra of QSOs and galaxies. 

As can be celarly inferred from the above discussion, one of the major challenges of reionization models is to be able to simultaneously account for the considerable, and often apparently conflicting, data accumulated by experiments exploiting QSO absorption line spectra (Fan et al. 2006), GRB spectra (Totani et al. 2006) and the CMB data (Komatsu et al. 2009). Given the many assumptions necessarily made by all reionization models, it has been suggested (Malhotra \& Rhoads 2004, 2005; Santos 2004; Haiman \& Cen 2005; Mesinger, Haiman \& Cen 2004; Dijkstra, Lidz \& Wyithe 2007a; Dijkstra, Wyithe \& Haiman 2007b; Mesinger \& Furlanetto 2007; Dayal et al. 2008; Dayal, Maselli \& Ferrara 2010)  that a class of high redshift galaxies, the LAEs can be suitably used to put additional constraints on the reionization history

\section{Understanding Lyman Alpha Emitters}
\label{laes_ch1}
Lyman Alpha Emitters (LAEs) are objects that have a visible Ly$\alpha$ line, which corresponds to the energy emitted (10.2 eV) when an electron falls from the $n=2$ to the ground state of the hydrogen atom. The first observations of highly redshifted Ly$\alpha$ photons came with the discovery of QSOs: in 1963, Schmidt published a report of the first high-$z$ object (3C 273, $z=0.15$); this was followed soon after (Schmidt, 1965) by a QSO (3C 9) at $z=2.01$, detected by means of its Ly$\alpha$ line. Using the latter QSO (3C 9), which had complete transmission bluewards of the Ly$\alpha$ line, Gunn \& Peterson (1965) became the first researchers to put constraints on the IGM neutral fraction. Anticipating the rise in high-$z$ data, Partridge \& Peebles (1967) made the first predictions for the observability of Ly$\alpha$ emission from young, star forming galaxies. However, two decades would pass before more sources of high-$z$ Ly$\alpha$ emission would be found; after the 1967 Partridge \& Peebles paper, many groups unsuccessfully tried to detect LAEs, which included Partridge (1974), Davis \& Wilkinson (1974), Meier (1976), Hogan \& Rees (1979), Pritchet \& Hartwick (1989, 1990), Rhee, Webb \& Katgert (1989). The first successful detections of LAEs finally took place in the early 1990s, by virtue of data collected by Lowenthal et al. (1991), Wolfe et al. (1992), M$\o$ller \& Warren (1993) and Macchetto et al. (1993). The past two decades have seen a massive rise in the data collected on LAEs, between $z \sim 2.25 - 6.6$, as is discussed later in this section.

There are three main types of astrophysical objects that are visible in the Ly$\alpha$: galaxies, active galactic nuclei (AGN) and Ly$\alpha$ blobs (LABs), each of which is briefly described now. 

(i) {\it Galaxies:} Generally, star formation has been considered to be the main source of Ly$\alpha$ emission in galaxies; star formation produces photons with energy larger than 13.6 eV that ionize the \HI in the ISM. Due to the extremely high density of the ISM, the resulting electrons and protons recombine on short timescales and give rise to a Ly$\alpha$ line. However, as shown by Dijkstra (2009) and Dayal, Ferrara \& Saro (2010), cooling of collisionally excited \HI can also contribute to the Ly$\alpha$ emission of a galaxy. Both these calculations are explained in complete detail in Sec. \ref{intrinsic_lum} and \ref{lum_coolh1_ch4}. 

(ii) {\it AGN:} AGN can also show Ly$\alpha$ emission or have a Ly$\alpha$ halo surrounding them (Schmidt 1965; McCarthy 1993; Villar-Martin et al. 2005). The luminosity of AGNs comes from the accretion of gas onto a BH; the gas spiralling in towards the BH gets heated up due to viscous drag and emits energy, some of which is in the form of \HI ionizing photons. As in galaxies, the recombination of electrons and protons gives rise to Ly$\alpha$ emission (Haiman \& Rees 2001; Weidinger et al. 2005). However, AGN contribute $<1$\% to the data collected for LAEs, as shown by Dawson et al. (2004), Gawiser et al. (2007) and Ouchi et al. (2008). 

(iii) {\it LABs:} Searches for LAES have also revealed a number of extended (tens to $>150$ Kpc) sources emitting copiously in the Ly$\alpha$ with luminosities of the order of $10^{43} - 10^{44}\, {\rm erg \, s^{-1}}$; these are known as LABs. By now, hundreds of LABs have been detected, mostly around $z \sim 3$ (Steidel et al. 2000; Matsuda et al. 2004; Nilsson et al. 2006). Although not the only scenario, the general understanding is that cool IGM gas that flows into DM halos gets heated due to the release of gravitational energy; it then cools by radiating Ly$\alpha$ photons (Haiman et al. 2000; Fardal et al. 2001; Dijkstra et al. 2006).

However, due to the low number densities of both AGN and LABs at high-$z$ ($\geq 4.5$), and the unclear mechanism powering LABs, in this thesis, we concentrate solely on galaxies that are visible by means of their Ly$\alpha$ line; from now on, the term LAEs refers only to such galaxies.

\subsection{The importance of LAEs}
\label{imp_laes_ch1}
Over the past few years, LAEs have rapidly been gaining importance as probes of both reionization and early galaxy evolution. The main reasons for this are now summarized : \\

\begin{figure}
\center{\includegraphics[scale=0.6]{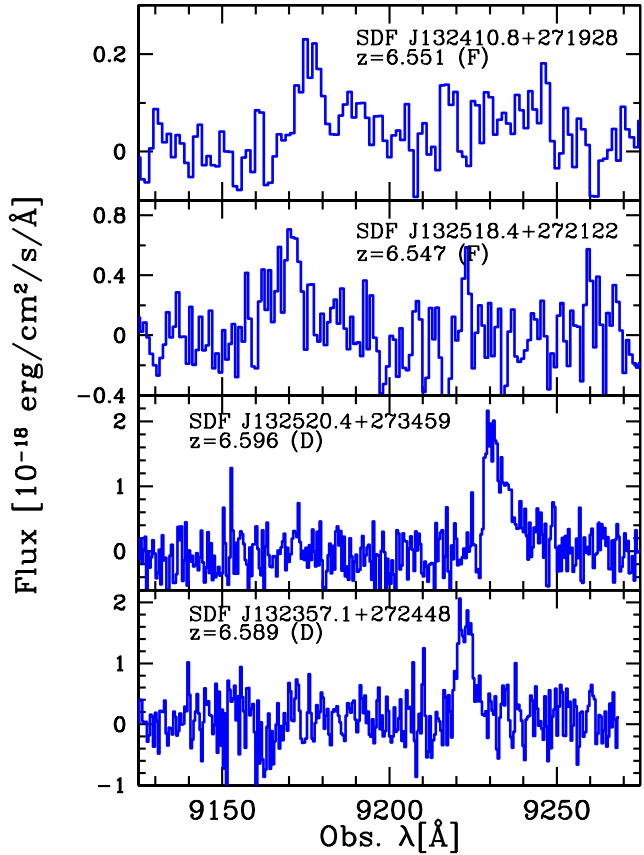}}
\caption{Observed spectra of 4 spectroscopically confirmed LAEs at $z \sim 6.5$ (Kashikawa et al. 2006). }  
\label{zch1_lae_spec_kashi} 
\end{figure}   

(i) The Ly$\alpha$ line shows several specific signatures, including a high strength, a narrow width and the continuum break bluewards of the Ly$\alpha$ line. These can be clearly seen in the spectra of four LAEs at $z \sim 6.5$ shown in Fig. \ref{zch1_lae_spec_kashi}, collected by Kashikawa et al. (2006) using the narrow band technique (explained in Sec. \ref{obs_laes} that follows); even at this high-$z$, the Ly$\alpha$ line signatures mentioned are clearly visible, while most of the continuum signal is hidden under noise. Such clear line signatures make the detection of LAEs unambiguous to a large degree.  \\

(ii) Secondly, since Ly$\alpha$ photons have a large absorption cross-section against \HI, their attenuation can be used to put constraints on the ionization state of the IGM as explained in Sec. \ref{spec_cons_reio}.  \\

(iii) Thirdly, there are hundreds of confirmed LAEs at $z \sim 4.5$ (Dawson et al. 2007), $z \sim 5.7$ (Malhotra et al. 2005; Shimasaku et al. 2006) and $z \sim 6.6$ (Taniguchi et al. 2005; Kashikawa et al. 2006), which exactly probe the redshift range around which reionization is supposed to have ended. Also, since they comprise the largest sample of high-$z$ galaxies known, they are excellent probes of galaxy evolution at these early epochs. \\

\subsection{Observing LAEs}
\label{obs_laes}
There are two main techniques for observing LAEs: these involve the use of weak lensing and narrow band (NB) filters respectively. Weak lensing relies on the presence of a foreground galaxy cluster to magnify the luminosity from a faint high-$z$, background galaxy. Using this technique, Hu et al. (2002) found a LAE at $z=6.56$, with a lensing amplification of 4.5 caused by the cluster Abell 370. Santos et al. (2004) found 11 LAE candidates near 9 clusters, all with magnification factors larger than 10, with $z=2.2-5.6$; of which 3 were confirmed as LAEs at $4.7 <z <5.6$. Stark et al. (2007) also used 9 clusters with magnification factors between 10 and 50 to find 6 LAE candidates between $z=8.7-10.2$; at least 2 of these candidates are confirmed to be at $z>8$.

However, most of the LAEs have been found using the NB technique, which involves searching for Ly$\alpha$ emission using narrow filter bands (i.e. redshift ranges), in windows of low OH sky emission. A number of surveys including the LALA (Large Area Lyman Alpha), MUSYC (MUltiwavelength Survey by Yale-Chile) and SDF (Subaru Deep Field) have successfully used this technique to find high-$z$ LAEs. We now summarize the results of NB searches between $z \sim 3- 6.7$.

\begin{itemize}

\item {$z \sim 3.1$ :} From the MUSYC survey covering an area of about 992 arcmin$^2$, Gawiser et al. (2006) obtained 23 LAE candidates at $z=3.1$, of which 18 were confirmed as LAEs using follow up spectroscopy. Using broad band and narrow band imaging at the VLT (Very Large Telescope), Venemans et al. (2005) obtained 77 LAE candidates in a 49 arcmin$^2$ field, of which 31 were confirmed as LAEs.

\item {$z \sim 4.5$ :} The LALA survey was intended to study large samples of LAEs using imaging from the 4m Mayall telescope on Kitt peak national observatory and spectroscopy from the Keck telescope. Within 7 years of the starting date, by 2007, 97 LAE candidates were found, of which 73 were confirmed as LAEs, using the Deep Imaging Multi-object spectrograph (DEIMOS) on Keck II (Dawson et al. 2007).

\item {$z \sim 5.7$ :} Using the LALA survey, Rhoads et al. (2003) confirmed the presence of 3 LAEs at $z \sim 5.7$. Using the Advanced Camera for Surveys (ACS) for sources in the Hubble Ultra Deep Field (HUDF), Malhotra et al. (2005), confirmed 15 LAEs at this redshift. Using the state of the art imaging instrument, the SuprimeCam, installed on the 8.3m Subaru telescope, and using the spectra from DEIMOS on Keck II, Hu et al. (2004) found 18 confirmed LAEs. Finally, Shimasaku et al. (2006) used the SuprimeCam for imaging an area of 925 arcmin$^2$ in the Subaru Deep Field (SDF) and combined this with spectroscopy from FOCAS (Faint Object Camera and Spectrograph, on Subaru) and DEIMOS on Keck II to obtain 89 LAE candidates; of these 28 were confirmed as LAEs. 

\item {$z \sim 6.6$ :} Taniguchi et al. (2005) detected 58 possible LAEs using Subaru at $z \sim 6.6$ and obtained the spectra for 20 of them using the FOCAS. They found that only 9 of the above objects showed sharp cut-off at the Ly$\alpha$ wavelength, narrow line widths and asymmetric profiles, thus being confirmed as LAEs at $z\sim 6.6$. These included the two LAEs discovered by Kodaira et al. (2005) at $z=6.541$ and $6.578$. Using the same selection criterion and instruments as Taniguchi et al. (2005) and including the LAEs confirmed using the Keck II DEIMOS spectrograph, Kashikawa et al. (2006) added 8 more LAEs at $z \sim6.6$ to this list. Thus, the
Subaru observations have a total of 17 confirmed LAEs at $z\sim 6.6$. 

\item {$z \sim 7.7$ :} Hibon et al. (2010) have photometrically detected 7 possible LAE candidates at $z \sim 7.7$ using the WIRCam (Wide Field near-IR Camera) on the Canadian-French-Hawaian Telescope (CFHT). Tilvi et al. (2010) have detected 4 more candidates using the NEWFIRM (National optical astronomy observatory Extremely Wide Field IR Mosaic) imager; however, none of these candidates has yet been confirmed as a LAE due to lack of spectra.

These observations can then be used to construct the LAE UV and Ly$\alpha$ luminosity functions (LF), the equivalent width (EW) distributions, the line skewness and the correlation functions; using the spectral energy distributions (SEDs), observers have also inferred the ages, stellar masses, SFR and color excess for LAEs, as will be shown in later chapters.

\end{itemize}

\subsection{Modelling LAEs}
\label{model_laes_ch1}
The simple picture of how \HI affects Ly$\alpha$ photons (Eqns. \ref{tau_tot_ch1}, \ref{tau_alpha_ch1}) is complicated by a number of important physical effects. First of all, Ly$\alpha$ photons produced due to stellar processes (as explained above) have to propagate through and escape out of the ISM of the LAE. During their travel they are multiply scattered by \HI atoms (thus being either removed from or added to the LOS) and possibly
absorbed by dust grains (Neufeld 1991; Tasitsiomi 2005; Hansen \& Oh 2006; Finkelstein et al. 2007; Dayal et al. 2008; Dayal et al. 2009; Dayal, Ferrara \& Saro 2010; Dayal, Hirashita \& Ferrara 2010; Dayal, Maselli \& Ferrara 2010). These processes modify both the emerging Ly$\alpha$ luminosity and the shape and equivalent width of the line.  Second, the ionizing radiation from the same stars builds regions of ionized IGM around the emitters, whose size depends on the star formation rate, age, escape of ionizing photons from the galaxy and the stellar IMF (the case of very massive stars has been explored, for example, by Dijkstra \& Wyithe 2007c). As a result, the flux redwards of the Ly$\alpha$ line can escape, attenuated only by the red damping wing of the Gunn-Peterson absorption (Miralda-Escud\'e 1998; Madau \& Rees 2000).  To a first approximation, the spatial scale imposed by the Gunn-Peterson damping wing on the size of the \HII region corresponds to a redshift separation of $\Delta z\approx 0.01$, i.e. about 200 kpc (physical) at $z=10$. The effects of the damping wing fade away if the emitter is powerful enough to create a large enough \HII region and/or if the universe is already reionized when the emitter turns on.  Alternatively, one would observe the damping wing if there were even a small fraction of neutral hydrogen left inside the sphere and/or if a \HI cloud is present along the LOS to the source; all these effects combine to shape the observed LFs

Therefore, understanding and constraining the ionization state of the IGM using LAEs requires: (a) a detailed knowledge of the physical properties of each galaxy, including the SFR, stellar age and stellar metallicity, needed to calculate the intrinsic Ly$\alpha$ and continuum luminosity produced by stellar sources, (b) the intrinsic Ly$\alpha$ and continuum luminosity produced by the cooling of collisionally excited \HI in the ISM, (c) an understanding of the dust formation, enrichment and distribution in each galaxy, necessary to calculate the fractions of escaping Ly$\alpha$ and continuum photons, and (d) a full radiative transfer (RT) calculation to obtain the fraction of Ly$\alpha$ luminosity transmitted through the IGM for each galaxy. All these will be dealt with in complete detail in the following chapters of this thesis.

Since we are interested in using LAEs to probe reionization, we concentrate on the data at $z \sim 5.7, 6.6$ accumulated by Shimasaku et al. (2006) and Kashikawa et al. (2006) respectively. We begin by mentioning that for these dats sets, those galaxies are identified as LAEs that have (a) an observed Ly$\alpha$ luminosity in the currently observable range, $L_\alpha \geq 10^{42.2} \, {\rm erg \, s^{-1}}$ and (b) a value of the observed EW, $EW>20$ \AA. It is interesting to note that while the data accumulated on LAEs shows no evolution in the apparent Ly$\alpha$ LF between $z=3.1$ - $5.7$ (Ouchi et al. 2008), the LF changes appreciably between $z=5.7$ and $6.6$ (Kashikawa et al. 2006) with $L_*$ at $z=6.6$ being about 50\% of the value at $z=5.7$. Surprisingly, however, the UV LF does not show any evolution between these same redshifts.

We now briefly summarize some of the LAE models that have been constructed by different groups to explain this data, and the main results obtained by them.

\subsubsection{(i) Semi-analytic models}
Santos (2004) has presented a LAE model to calculate the Ly$\alpha$ transmission, in the presence of gas infall onto the galaxy. Using this model, he showed that if galactic winds are unimportant, $\chi_{HI} \lsim 0.1$ at $z \sim 6.5$. If, however, galactic winds (inflows / outflows) can not be neglected, he showed that the value of $\chi_{HI}$ can not be constrained using the present LAE data. 

Dijkstra, Lidz \& Wyithe (2007) presented a model that accounts for IGM gas clumping and the fact that high-$z$ galaxies reside in over dense regions, where the velocity fields are different from the Hubble flow. The showed that the observed Ly$\alpha$ line shape depends on a number of factors including the velocity fields, the SFR and the local ultra-violet background (UVB). They further showed that the increase in the UVB at the end of reionization barely affects the Ly$\alpha$ transmission; this implies that the completion of reionization would not lead to a significant change in the observed Ly$\alpha$ luminosity of a galaxy. 

Dijkstra, Wyithe \& Haiman (2007) used a model in which the free parameters were the SFR efficiency of each DM halo and the Ly$\alpha$ transmission; using a search in a parameter space comprised of these two free parameters, they showed that the evolution in the Ly$\alpha$ LF between $z=5.7$ and $6.6$ could be explained solely by an evolution of the underlying mass function. Therefore, according to this model, reionization is over by $z \sim 6.6$.

Kobayashi, Totani \& Nagashima (2007) used a semi-analytic model of galaxy formation, that reproduces a number of properties of low and high-$z$ galaxies, and used the effective escape fraction of Ly$\alpha$ photons (which includes the effects of dust and galactic outflows) as the free parameter to fit the observed LFs. This model requires a change in the ionization state at $z\sim 6$, implying that reionization finishes after $z \sim 6.6$.

Samui et al. (2009) have presented a LAE model, in which, reproducing the data requires galaxies at $z > 5$ to be less dusty, with a more clumpy ISM and complex velocity fields. However, they do not require any change in the ionization state of the IGM, implying that reionization was over before $z \sim 6.6$.

Tilvi et al. (2009) have presented a semi-analytic model of populating DM halos with galaxies such that the SFR of a galaxy is proportional to the cold mass accretion rate. They claim that this model fits the observations without requiring any need of dust or Ly$\alpha$ transmission calculation, implying that the IGM was ionized before $z \sim 6.6$.

\subsubsection{(ii) Models based in numerical simulations}

McQuinn et al. (2007) carried out analytic RT calculations on DM (only) simulations. They found that reionization increases the measured clustering of LAEs and they presented this as a clear signature of the EoR. They too found that the data at $z \sim 6.6$ is easily reproduced by a fully ionized Universe.

Nagamine et al. (2008) used cosmological SPH simulations to obtain the intrinsic properties of each galaxy. Using the escape fraction of Ly$\alpha$ photons or the duty cycle of SF (fraction of LAEs turned on at any time) as two different scenarios, they showed that the duty cycle scenario fits the data better. With this scenario, the data is well fit in the absence of both dust and \HI, i.e., the IGM was highly ionized at $z \sim 6.6$.

Iliev et al. (2008) ran full RT calculations on cosmological DM (only) simulations; they find that assuming a constant mass to light ratio for each galaxy, the data is well reproduced at $z \sim 6$.

Zheng et al. (2010) combined cosmological SPH simulations with a Monte Carlo RT code for LAEs at $z \sim 5.7$.  In their model, the Ly$\alpha$ transmission was the single free parameter that was needed to obtain the observed LF from the intrinsic ones. 

\section{Thesis purpose and plan}
\label{thesis_purpose}
As shown in this chapter (see Sec. \ref{obs_reio_ch1}), one of the major problems for reionization models is to be able to account for the huge amount of data collected using QSOs, GRBS and from WMAP, which are often in conflict. It has therefore become important to use newer and higher-redshift data sets to place constraints on reionization; one of the most important probes in this sense is represented by LAEs. Due to specific spectral signatures, LAEs are relatively unambiguous to identify. Indeed, as mentioned, there are now hundreds of confirmed emitters at $z \sim 4.5 - 6.6$, with tentative detections at $z \sim 7.7$. These high-$z$ redshifts exactly probe the IGM ionization state at the time when reionization is supposed to have ended; in addition, LAEs can also be used to probe galaxy evolution at these early epochs, where little other galaxy data is available. As shown in Sec. \ref{model_laes_ch1}, although a number of models, both semi-analytic and numerical, have been constructed to probe reionization using LAEs, none of them contain all the ingredients necessary for understanding reionization. As mentioned before, these include: (a) a detailed knowledge of the physical properties of each galaxy, needed to calculate the intrinsic Ly$\alpha$ and continuum luminosity produced by stellar sources, (b) the intrinsic Ly$\alpha$ and continuum luminosity produced by the cooling of collisionally excited \HI in the ISM, (c) an understanding of the dust formation, enrichment and distribution in each galaxy, necessary to calculate the fractions of escaping Ly$\alpha$ and continuum photons, and (d) a full radiative RT calculation to obtain the fraction of Ly$\alpha$ luminosity transmitted through the IGM for each galaxy. 

In this thesis, our aim is to construct a physically motivated, self-consistent model for LAEs, containing all these ingredients, so as to be able to understand the importance of the IGM ionization state, dust and peculiar velocities in shaping the LAE Ly$\alpha$ and UV LFs. By doing so, the aim is to gain insight on the nature of LAEs and put constraints on their elusive physical properties. 

We start by building a semi-analytic model for LAEs, in Chapter 2 (Dayal et al. 2008) by populating DM halos with galaxies using assumptions about the ages, SFR and metallicity, and calculating the IGM transmission for each galaxy. This model explains a number of data sets including the Ly$\alpha$ and UV LFs, and the weighted skewness measurements. Although simple, this powerful model favors the early reionization model (ERM, Gallerani et al. 2008), wherein reionization ends at $z \sim 7$. We improve on this model in Chapter 3 (Dayal et al. 2009) by including the effects of clustered sources and coupling our IGM Ly$\alpha$ production/transmission model to state of the art cosmological simulations run using {\tt GADGET-2}. We again find that the ERM gives model results in good agreement with the observed data; however, we find that only a fraction of the Ly$\alpha$ and UV photons must escape out of each galaxy, undamped by dust. The information on the physical properties of each galaxy, obtained from the simulations, allows us to place precious constraints on the nature of LAEs. In addition, it also allows us to build synthetic SEDs to compare to the observed ones; with no free parameters, we find these are in excellent agreement. In Chapter 4 (Dayal, Ferrara \& Saro 2010; Dayal, Hirashita \& Ferrara 2010), we add two important ingredients to the model presented in Chapter 3: we model the amount of dust in each galaxy depending on its intrinsic properties and we include the luminosities produced by cooling of collisionally excited \HI in the ISM. As a consequence of the dust modelling, we are also able to predict the far infrared (FIR) emission expected from LAEs, detectable with future instruments such as ALMA (Atacama Large Millimeter Array). In Chapter 5 (Dayal, Maselli \& Ferrara 2010), a full RT calculation is combined with the models presented in earlier chapters, to build the most complete LAE model, as of date. This model yields a very surprising result: there is a degeneracy between the Ly$\alpha$ photons transmitted through the IGM and those that escape the galaxy undamped by dust. In other words, the IGM ionization state can not be well constrained unless the dusty nature of LAEs is well understood. In Chapter 6 (Salvadori, Dayal \& Ferrara 2010), we couple our LAE model to a semi-analytic code ({\tt GAMETE}), to establish a possible connection between high-$z$ LAEs ($z \sim 5.7$) and the building blocks of the Milky Way (MW). We show that some of the Galactic building blocks could indeed be amongst the LAEs that have been observed at this $z$. In Chapter 7 (Salvaterra, Ferrara \& Dayal 2010), we use a small-size simulation box ($10 \, {\rm h^{-1} \, Mpc}$), obtained by running {\tt GADGET-2}, to explain the observed UV LF at $z \sim 5-10$. Due to the high-resolution, we are able to even resolve dwarf galaxies at these early times; we unveil the nature of these early galaxies and probe their contribution to reionization. We end by summarizing the main results I have obtained during the PhD, in Chapter 8.

\chapter{Modelling high redshift LAEs}\label{ch2_lya_sam}
In this chapter, we introduce a semi-analytic model for LAEs which is used to derive various LAE properties to compare to observations at $z \sim 4.5, 5.7$ and $6.6$. Starting from a simple yet physical model of galaxy formation within dark matter halos coupled with a population synthesis code, we derive the intrinsic luminosities, so as to build the intrinsic Ly$\alpha$ and continuum (Ultra-violet, UV) LFs in Sec. \ref{ST_mf} and \ref{intrinsic_lum}. We then describe the calculation of the attenuation of Ly$\alpha$ photons in the IGM in Sec. \ref{observed_lum}. The principal parameters that affect the Ly$\alpha$ line profile are studied in Sec. \ref{basic_dep}. We then use two different reionization scenarios, so as to obtain the observed Ly$\alpha$ LFs to compare to the observations; comparing model results to the observations requires ISM dust attenuation of both Ly$\alpha$ and continuum photons, as shown in Sec. \ref{lf_fit} and \ref{uv_fit}. Once the model results reproduce the observed data, we compare the SFR density of LAEs to the cosmic value (Sec. \ref{sfr_density}), we see to what extent the predicted EW distributions match the observations in Sec. \ref{ew_ch2} and we compare the Ly$\alpha$ line profile asymmetry values to the observations in Sec. \ref{skewness discussion}. We end the chapter by pointing out the main results and drawbacks of our model in Sec. \ref{conc_ch2}.

Throughout this chapter, we use the best-fit cosmological parameters from the 3-year WMAP data (Spergel et al. 2007), i.e., a flat universe with ($\Omega_b$, $\Omega_m$, $\Omega_{\Lambda}) = (0.041, 0.24, 0.76)$ respectively. Further, the Hubble constant is expressed as $H_0 = 100 \, {\rm h} \, {\rm Km \, s^{-1} \,Mpc^{-1}}$ where ${\rm h} = 0.72$ and $\Omega_b {\rm h}^2 = 0.022$. The parameters defining the linear dark matter power spectrum are $\sigma_8=0.82$, $n_s=0.95$, $dn_s/d\ln k =0$. Here, $n_s$ is the spectral index of the density fluctuations generated by inflation and $dn_s/d\ln k =0$ implies no scale-change of the spectral index. We use a value of $\sigma_8$ much higher that quoted from WMAP3 (0.76) as the combination of WMAP3 and SDSS data give $\sigma_8 \sim 0.78$ (0.86) for low (high) resolution Ly$\alpha$ forest data (Viel et al. 2006). Throughout, Mpc is comoving and pMpc represents values in physical Mpc units.

\section{The semi-analytic model} 
\label{model_ch2} 
We start by building the Sheth-Tormen mass function (Sec. \ref{ST_mf}) to obtain the number density of DM halos at the redshifts of interest. Once the mass function is obtained, a SFR recipe is used to obtain the intrinsic Ly$\alpha$ and continuum luminosity for any halo on the mass function in Sec. \ref{intrinsic_lum}, thereby providing the intrinsic Ly$\alpha$/ continuum LFs. The calculation of the attenuation of the intrinsic Ly$\alpha$ luminosity by \HI in the IGM is explained in Sec. \ref{observed_lum}; ignoring the effects of dust inside the ISM of galaxies on the Ly$\alpha$/ continuum luminosity, this calculation allows the intrinsic LFs to be translated into the observed LFs.

 
\subsection{The Sheth-Tormen mass function} 
\label{ST_mf} 

We start with the well known Sheth-Tormen mass function (Sheth \& Tormen 1999) which is used to calculate the number density of DM halos of mass between $M$ and $M+dM$ at any redshift $z$, represented by $n(M,z) dM$, as 
 
\begin{equation} 
n(M,z) dM = A \bigg(1+\frac{1}{\nu'^{2q}}\bigg) \sqrt\frac{2}{\pi} \frac{\bar\rho}{M} \frac{d\nu'}{dM} e^{-\nu'^2/2} dM, 
\label {stormen} 
\end{equation} 
where $\nu'=\sqrt{a}\nu$ and $\bar\rho$ is the mean cosmic density at the redshift considered. In Eq. \ref{stormen}, $A$, $a$ and $q$ are modifications to the original Press-Schechter mass function (Press \& Schechter 1974) to make it agree better with simulations. Here, $A\approx 0.322$, $q=0.3$ and $a = 0.707$. 
 
As in the Press-Schechter mass function, 
\begin{eqnarray*}  
\nu &=& \frac{\delta_c}{D(z)\sigma(M)}, \\
D(z)&=&g(z)/[g(0)(1+z)], \\ 
g(z) &= &2.5 \Omega_m [\Omega_m^{4/7} - \Omega_\Lambda + (1+\Omega_m/2)(1+\Omega_\Lambda/70)]^{-1}.  
\end{eqnarray*} 
Here, $\nu$ is the number of standard deviations which $\sigma$ represents on a mass scale $M$, $\delta_c$$(=1.69)$ is the critical overdensity for spherical collapse and $D(z)$ is the growth factor for linear fluctuations, Carroll, Press \& Turner (1992). Further, the variance of the mass $M$ contained in a radius $R$ is given by  
\begin{equation} 
\sigma^2(R) = \frac{1}{2\pi^2} \int k^3 P(k) W^2(kR) \frac{dk}{k}.
\label{sig} 
\end{equation} 
In Eq. \ref{sig}, $W(kR)=3(\sin[kR]-kR\cos[kR)]) $ is the window function that represents the Fourier transform of a spherical top hat filter of radius $R$. $P(k)$ is the power spectrum of the density fluctuations, extrapolated to $z=0$ using linear theory, which is expressed as
\begin{equation}
P(k) = A_p k^n T^2(k),
\end{equation}
where $A_p$ is the amplitude of the density fluctuations calculated by normalizing $\sigma(M)$ to $\sigma_8$, and $\sigma_8$ represents the variance of mass in a sphere of size $8{\rm h^{-1}}$ Mpc at $z=0$. The term $T(k)$ is a transfer function which represents differential growth from early times, Bardeen et al. (1986). 
\begin{equation} 
T(k) = \frac{0.43 q^{-1} \ln(1+2.34q)}{[1+3.89q+(16.1q)^2+(5.46q)^3+(6.71q)^4]^{1/4}},  
\end{equation} 
where $q = k(\Omega_m {\rm h^2})^{-1}$. Given that $\delta_c$ decreases with $z$, while for CDM models, $\sigma(M)$ increases with decreasing masses (see Fig. 5, Barkana \& Loeb 2001), the typical mass associated with low-$\sigma$ density fluctuations is higher at lower $z$. In other words, more massive halos become more abundant with decreasing $z$.
 
 
\subsection{The intrinsic and emergent luminosities } 
\label{intrinsic_lum} 

Assuming the ratio of the baryonic and DM halo mass to be the same as the cosmological fraction ($\Omega_b/\Omega_m$) for all halos, the baryonic mass, $M_b$, contained within a DM halo of mass $M_h$ can be expressed as $M_b = (\Omega_b / \Omega_m)M_h$. 

We assume that a fraction $f_*$ of this baryonic matter forms stars over a timescale $t_*=\epsilon_{dc}t_H$,  
where $\epsilon_{dc}$ is the duty cycle and $t_H$ is the Hubble time at $z=0$. Thus, we can write the SFR ($\dot M_*$) as 
\begin{equation} 
\label{sfr2} 
\dot M_* = \frac{f_*}{\epsilon_{dc}} \frac{1}{t_H} \frac {\Omega_b}{\Omega_m} M_h. 
\end{equation} 
 
We then make the following assumptions about the stellar population of the halos on the mass function, so as to obtain their spectra using the population synthesis code {\tt Starburst99} (Leitherer et al. 1999):
\begin{itemize}
\item The stellar metallicity is assumed to be $Z_*=0.05 Z_\odot$. Determining the metallicity of the LAEs proves very challenging, as for most of the cases, only the Ly$\alpha$ line can be detected from these objects. To guess their metallicity, we use the results from studies of LBGs (Lyman Break Galaxies) and DLA (Damped Ly$\alpha$) systems, which indicate values of $0.05-0.10 Z_\odot$, which justifies our assumption, Pettini (2003).

\item The age of the stellar population is taken to be of the order of $t_* =$100 Myr, for continuous star formation. Since the \HI ionizing photon rate, and hence the Ly$\alpha$ luminosity produced, become rapidly independent of age after about 10 Myr of continuous star formation, our results are not highly affected by this assumption, unless LAEs are extremely young objects ($\leq 10$ Myr).  

\item The IMF, which specifies the distribution of mass of a newly formed stellar population, is taken to be a Kroupa (2001) IMF such that the number of stars in a mass range $M$ to $M+dM$ is $n_*(M)dM \propto M^{-\alpha}$ where $\alpha = 1.3, 2.3$ for stellar masses in the range $0.1-0.5 M_\odot$ and $0.5-100 M_\odot$ respectively.
\end{itemize} 

Star formation in LAEs produces photons with energy $> 1$~Ryd. For a given set of values of $Z_*$, $t_*$ and the IMF, the hydrogen ionizing photon rate, $Q$, emitted by a galaxy scales linearly with $\dot M_*$ and can be calculated using the {\tt STARBURST99} spectra as
\begin{equation}
Q = \dot M_* \int_{\nu_L}^\infty \frac{L_\nu} {h \nu} \, d\nu
\label{calq}
\end{equation}
where $L_\nu$ is the specific ionizing luminosity of the emitter [erg s$^{-1}$ Hz$^{-1}$] for a SFR of $1 \, {\rm M_\odot \, yr^{-1}}$, $\nu_L$ is the Lyman limit frequency ($3.28 \times 10^{15}$ Hz) corresponding to the wavelength 912 \AA, $h$ is Planck's constant and $c$ is the speed of light. From our model, for a galaxy with $t_*=100 \, {\rm Myr}$, $Z_* = 0.05 Z_\odot$ and a Kroupa IMF,
\be
Q = 10^{53.41} \frac{\dot M_*}{\rm {M_\odot \, yr^{-1}}}\, [{\rm s^{-1}}]
\label{scale_q}
\ee

These photons ionize the interstellar \HI, leading to the formation of free electrons and protons inside the galaxy. Due to the high density of the ISM, these electrons and protons then recombine on the recombination time scale, giving rise to a Ly$\alpha$ emission line. The \textit{intrinsic} Ly$\alpha$ luminosity, $L_\alpha^{int}$, produced in the galaxy can be expressed as
\begin{equation}
L_\alpha^{int} = \frac{2}{3} \, Q \, h \nu_\alpha \, [{\rm erg \, s^{-1}}],
\label{lya_int2}
\end{equation}
where $\nu_\alpha$ is the frequency of Ly$\alpha$ photons in the rest frame of the galaxy ($2.46 \times 10^{15}$ Hz). The factor two-thirds arises since there is a two-thirds probability of the recombination leading to a Ly$\alpha$ line and a one-third probability of obtaining photons of frequencies different from the Ly$\alpha$ (Osterbrock 1989), for recombinations leading to \HI optically thick to Ly$\alpha$ photons, i.e., a case B recombination. Since $Q$ scales linearly with $\dot M_*$, as shown in Eq. \ref{scale_q}, $L_\alpha^{int}$ can also be expressed as
\begin{equation}
L_\alpha^{int} = 2.8 \times 10^{42} \frac{\dot M_*}{\rm {M_\odot \, yr^{-1}}} \, [{\rm erg\, s^{-1}}].
\label{val_lyaint}
\end{equation}

For each galaxy, the intrinsic continuum luminosity, $L_c^{int}$, is calculated in a band between 1250-1500 \AA\,(centered at 1375 \AA), directly from the spectra as
\begin{equation}
L_c^{int} (1375\,{\rm \AA})=2.14\times10^{40}\frac{\dot M_*}{\rm {M_\odot\,yr^{-1}}}\,[{\rm erg\,s^{-1}\AA^{-1}}]
\label{lc_int2}
\end{equation}

The Ly$\alpha$ line profile that emerges from the galaxy is doppler broadened by the galactic rotation velocity ($v_c$), unlike the continuum.  For quiescent star formation, for realistic halo and disc properties, the galaxy rotation velocity can have values between 1-2 times the halo rotation velocity (Mo, Mao \& White 1998; Cole et al. 2000). We use a value of $1.5$ in this work. We calculate the velocity of the halo, $v_h$, assuming that the collapsed region has an overdensity of roughly 200 times the mean cosmic density contained in a radius $r_{200}$. Then, $v_h$, the halo velocity at $r_{200}$ is expressed as 
\begin{equation} 
v_h^2(z) ={\frac{G M_h}{r_{200}}} = G M_h \left[\frac{100 \Omega_m(z) H(z)^2}{G M_h}\right]^{1/3},  
\label{rot vel2} 
\end{equation} 
where $\Omega_m(z)$, $H(z)$ are the matter density and Hubble parameters, respectively, at the redshift of the emitter and $G$ is the universal gravitational constant.  

Two processes determine the Ly$\alpha$ luminosity which emerges from the galaxy. First, only a fraction of the \HI ionizing photons ionize the ISM, contributing to the Ly$\alpha$ luminosity while the rest ($f_{esc}$) escape the galaxy and ionize the IGM surrounding it. Second, only a fraction, $f_\alpha$, of the Ly$\alpha$ photons produced escape the galaxy, unabsorbed by dust in the ISM. The Doppler broadened Ly$\alpha$ luminosity profile that emerges from the galaxy is expressed as
\begin{equation}
L_\alpha^{em} (\nu) = \frac{2}{3} \, Q \, h \nu_\alpha \, (1-f_{esc}) \, f_\alpha \frac{1}{\sqrt{\pi} \Delta \nu_d} \exp^{-(\nu-\nu_\alpha)^2 / \Delta \nu_d^2} \, [{\rm erg \, s^{-1}}] ,
\label{lya_emm2}
\end{equation}
where $\Delta \nu_d = [v_c/c] \nu_\alpha$. The total emergent Ly$\alpha$ luminosity, $L_\alpha^{em}$, can be determined by summing Eq. \ref{lya_emm2} over the width of the entire dopper broadened Ly$\alpha$ line.

The continuum band is also attenuated by dust in the ISM. The continuum luminosity emerging from the galaxy is expressed as 
\be 
L_c^{em} = f_c \, L_c^{int} \, [{\rm erg \, s^{-1} \AA^{-1}}],
\label{lc_emm2}
\ee
where $f_c$ is the fraction of continuum photons that escape out of the galaxy, undamped by dust in the ISM. It is important to mention that the fraction, $f_c$ ($f_\alpha$) of continuum (Ly$\alpha$) photons which escape the galaxy depends on the total amount of dust in the ISM as well as its distribution (homogeneous or clumped), as will be explained in later chapters.

To summarize, the \textit{intrinsic} Ly$\alpha$ luminosity depends upon: the ionization rate $Q$, which in turn depends on $\dot M_*$ (which is a function of the halo mass), the stellar metallicity $Z_*$, the IMF and the age of the emitter $t_*$, chosen such that the number of ionizing photons emitted per second settles to a constant value. The Ly$\alpha$ luminosity that {\it emerges} from the galaxy further depends on the escape fraction of \HI ionizing photons $f_{esc}$, the escape fraction of Ly$\alpha$ photons ($f_{\alpha}$) and the rotation velocity of the galaxy $v_c$. 

The {\it intrinsic} continuum luminosity depends on $\dot M_*$, $t_*$, $Z$ and the IMF. The continuum luminosity that {\it emerges} from the galaxy only depends on the fraction ($f_c$) of continuum photons that escape the galaxy, undamped by dust. 
\subsection{The observed Ly$\alpha$ and continuum luminosities} 
\label{observed_lum}

The Ly$\alpha$ photons that emerge from the galaxy suffer further attenuation as they travel through the IGM due to their large \HI optical depth; even small amounts of neutral hydrogen in the IGM can attenuate the Ly$\alpha$ luminosity by large amounts as seen in Sec. \ref{spec_cons_reio} (Eqns. \ref{tau_tot_ch1}, \ref{tau_alpha_ch1}). We now describe the calculation of the \HI distribution between the source and the observer, and the attenuation caused by it.

\subsubsection{(i) Global $\chi_{HI}$ calculation}
To calculate the IGM attenuation, we require an estimate of the amount of \HI between the source and the observer. For this, we use the global value of the \HI fraction, $\chi_{HI}$, resulting from the modeling of Gallerani, Choudhury \& Ferrara (2006), further refined in Gallerani et al. (2008), defined as
\begin{equation}
\chi_{HI}=\frac{n_{HI}}{n_H},
\label{define_chi}
\end{equation}
where $n_{HI}, n_H$ are the number densities of \HI and hydrogen respectively, at the redshift of interest. The main features of the model are summarized here. Mildly non-linear density fluctuations giving rise to spectral absorption features in IGM are described by a Log-Normal distribution as proposed by Miralda-Escud\'e, Haehnelt \& Rees (2000). This has been shown to fit the observed probability distribution function of the transmitted flux between redshifts 1.7 and 5.8 by Becker, Rauch \& Sargent (2007). For a given IGM equation of state, this being the temperature-density relation, $\chi_{HI}$ can be computed from photoionization equilibrium as a function of baryonic over-density ($\Delta \equiv \rho/ \bar{\rho}$) and photoionization rate ($\Gamma_B$) due to the ultra-violet background (UVB) radiation field. These quantities must be determined from a combination of theory and observations. Gallerani et al. (2008) included two types of ultraviolet photons: from QSOs and Pop II stars. The free parameters in their model were (i) the SFR efficiency of the PopII stars ($f_{II}$) and (ii) the escape fraction of ionizing photons from the galaxy hosting PopII stars ($f_{esc,II}$). These were calibrated to match the redshift evolution of Lyman-limit systems, Ly$\alpha$ and Ly$\beta$ optical depths, electron scattering optical depth, cosmic SFR history and number density of high redshift sources. From their work, the following reionization scenarios provide a good fit to observational data:
\begin{itemize}
\item {\it Early Reionization Model}: ERM, in which reionization ends at $z=7$, ($f_{II} = 0.1$, $f_{esc,II} = 0.07$), 

\item {\it Late Reionization Model}: LRM, where reionization ends at $z=6$, ($f_{II} = 0.08$, $f_{esc,II} = 0.04$). 
\end{itemize} 

\subsubsection{(ii) Neutral hydrogen profile}

Once that $\chi_{HI}$ is fixed to the the values obtained by Gallerani et al. (2008), we can calculate the photoionization rate ($\Gamma_B$) contributed by the ionizing background light produced by quasars and galaxies. Since the IGM is approximately in local photoionization equilibrium, ionizations are balanced by recombinations, such that 
\begin{equation} 
n_{HI} \Gamma_B = n_e n_p \alpha_B,  
\label{ion_rec}
\end{equation} 
where $n_p$, $n_e$ are the number density of protons and electrons respectively (at the redshift considered) and $\alpha_B = 2.6 \times 10^{-13} \, {\rm cm^3 \, s^{-1}}$ is the hydrogen Case B recombination coefficient. Since $n_p=n_e=(1-\chi_{HI})n_H$, $\Gamma_B$ can be written as: 
\be 
\Gamma_{B} = {(1-\chi_{HI})^2\over \chi_{HI}} n_H \alpha_B. 
\label{gamma_b2}
\ee
  
Further, the radiation from stars inside the galaxy ionizes the region surrounding the emitter, forming the the so-called Str\"omgren sphere. The evolution of the Str\"omgren sphere is given by the following relation, (Shapiro \& Giroux 1987; Madau, Haardt \& Rees 1999)
\begin{equation} 
\label{strom} 
\frac{dV_I}{dt} - 3H(z)V_I = \frac{Qf_{esc}}{\chi_{HI} n_H} - \frac{V_I}{t_{rec}}, 
\label{cal_rs2}
\end{equation} 
where, $V_I$  is the physical volume of the Str\"omgren sphere and $t_{rec}=\left[1.17\alpha_Bn_p\right]^{-1}$ is 
the volume averaged recombination timescale, Madau \& Rees (2000). The physical radius  
$R_I= (3V_I/4\pi)^{1/3}$, identifies a redshift interval $\Delta z$ between the emitter and the edge of the 
Str\"omgren sphere, given by the following: 
\begin{equation} 
\Delta z = 3.33 \times 10^{-4} (\Omega_m {\rm h^2})^{\frac{1}{2}} (1+z)^{\frac{5}{2}} R_I\, [{\rm Mpc}]. 
\end{equation} 
Though this equation is not strictly valid at $z\sim 0$, it is a good approximation at the high redshifts
we are interested in ($z\geq 4.5$). 

If $z_{em}$ is the redshift of the emitter, for redshifts lower than the Str\"omgren sphere redshift ($z_s=z_{em}-\Delta z$), i.e., inside the Str\"omgren sphere, the ionization rate has two contributions : (a) a constant value, $\Gamma_B$, determined by the UVB photoionization rate (which is assumed to be homogeneous and isotropic) and (b) a radius dependent value, $\Gamma_L$, determined by the luminosity that emerges out from the galaxy. The total photoionization rate ($\Gamma_{BL}$) at any distance $r$ from the galaxy can be expressed as
\begin{eqnarray}
\label{gamma_bl}
\Gamma_{BL} (r) & = & \Gamma_B + \Gamma_L (r), \\
\label{gamma_blr2}
\Gamma_{BL} (r)& =  &\Gamma_B + \int _ {\nu_L} ^\infty \frac{L_\nu^{em}}{4 \pi r^2} \frac{\sigma_L}{h \nu} \bigg(\frac{\nu_L}{\nu}\bigg)^3 d\nu,
\end{eqnarray}
where $L_\nu^{em}= L_\nu^{int} f_{esc} \,[\rm erg \, s^{-1} \, Hz^{-1}]$, is the specific ionizing luminosity emerging from the emitter and $\sigma_L$ [$6.3\times 10^{-18}\, {\rm cm^2}$] is the hydrogen photoionization cross-section. 

Thus, inside the ionized region, $\chi_{HI}$ is computed as following: 
\begin{equation} 
\chi_{HI}(r) = \frac{2 n_H \alpha_B+\Gamma_{BL}(r)\pm \sqrt { \Gamma_{BL}^2(r) + 4 n_H \alpha_B \Gamma_{BL}(r) }}{2 n_H \alpha_B }.
\label{chi_r}
\end{equation} 
The solution must be chosen such that $\chi_{HI}<1$, which only happens for a negative sign before the square root. At the edge of the Str\"omgren sphere, we force $\chi_{HI}(r)$ to attain the global value in the IGM.  
 
\subsubsection{(iii) Ly$\alpha$ optical depth and transmitted flux}

Then, if $z_{em}$ and $z_{obs}$ are the redshifts of the emitter and observer respectively, we calculate the total optical depth ($\tau_\alpha$) to the Ly$\alpha$ photons along the LOS as 
\begin{eqnarray}
\label{tau_alpha2}
\tau_\alpha(\nu) &=& \int_{z_{em}}^{z_{obs}} \sigma\, n_{HI}(z)\, \frac{dl}{dz} dz ,\\
& = & \int_{z_{em}}^{z_{obs}} \sigma_0 \phi(\nu) \chi_{HI}(z) n_H(z) \frac{dl}{dz} dz ,
\end{eqnarray}
where $dl/dz = c/[H(z) (1+z)]$. Here, $\sigma$ is the total absorption cross-section and $\phi$ is the Voigt profile. We use $\sigma_0 = \pi e^2 f (m_e c)^{-1}$, where $e$, $m_e$ are the electron charge and mass respectively and $f$ is the oscillator strength (0.4162).

For regions of low \HI density, the natural line broadening is not very important and the Voigt profile can be approximated by the Gaussian core: 
\begin{equation} 
\phi(\nu_i) \equiv \phi_{gauss} = \frac{1}{\sqrt{\pi} (b/\lambda_\alpha)} e^{-({\frac{\nu_i-\nu_\alpha}{\nu_\alpha} \frac{c}{b}})^2},
\label{gauss2} 
\end{equation} 
In Eq. \ref{gauss2}, $\nu_i$ is used since a photon of initial frequency $\nu$ has a frequency $\nu_i = \nu [(1+z_i)/(1+z_{em})]$ at a redshift $z_i$ along the LOS. The Doppler width is expressed as $b/\lambda_\alpha$, where $b=\sqrt {2kT/m_H}$ is the Doppler width parameter, $m_H$ is the hydrogen mass, $k$ is the Boltzmann constant and $T=10^4 K$ is the IGM temperature (Santos 2004; Schaye et al. 2000; Bolton \& Haehnelt 2007b). 

In more dense regions the Lorentzian damping wing of the Voigt profile becomes important.
According to Peebles (1993), this can be approximated as  
\begin{equation} 
\phi_{lorentz}(\nu_i) = \frac{\Lambda (\nu_i /\nu_\alpha)^4}{4\pi^2(\nu_i-\nu_\alpha)^2 + \frac{\Lambda^2}{4} (\nu_i/\nu_\alpha)^6} 
\label{lorentz2}
\end{equation} 
where $\Lambda= 6.25\times 10^8$~s$^{-1}$ is the decay constant for the Ly$\alpha$ resonance. 

Eqns. \ref{tau_alpha2} - \ref{lorentz2} can also be written in terms of photon velocities, such that
\begin{equation}
\label{tau_vel2}
\tau_\alpha({\rm v}) = \int_{z_{em}}^{z_{obs}} \sigma_0 \phi({\rm v}) n_{HI}(z) \frac{dl}{dz} dz ,
\end{equation}
where ${\rm v} = (\lambda - \lambda_\alpha)[\lambda_\alpha c]^{-1}$ is the rest-frame velocity of a photon with wavelength $\lambda$, relative to the line centre (rest-frame wavelength $\lambda_\alpha = 1216$ \AA, velocity ${\rm v}_\alpha =0$) and the other symbols retainn the meanings explained above. Then, the Gaussian core (Eq. \ref{gauss2}) can be written as
\begin{equation} 
\phi({\rm v}_i) \equiv \phi_{gauss} = \frac{\lambda_\alpha}{\sqrt{\pi} b} e^{-(\frac{{\rm v}_i + {\rm v}_p-{\rm v}_\alpha}{b})^2},
\label{gauss_vel2} 
\end{equation} 
where ${\rm v}_i$ is the velocity of a photon of initial velocity ${\rm v}$ at a redshift $z_i$ along the LOS, ${\rm v}_p$ is the peculiar velocity at $z_i$ and  ${\rm v}_\alpha$ is the velocity of the Ly$\alpha$ photons at $z_i$, which is 0 in this expression. The Lorentzian damping wings (Eq. \ref{lorentz2}) can be written as
\begin{equation} 
\phi_{lorentz}({\rm v}_i) = \frac{R_\alpha \lambda_\alpha}{\pi[({\rm v}_i + {\rm v}_p-{\rm v}_\alpha)^2 + R_\alpha^2]} 
\label{lorentz_vel2}
\end{equation} 
where $R_\alpha = \Lambda \lambda_\alpha [4 \pi]^{-1}$. Although computationally more expensive than the above approximations, using the Voigt profile to compute the absorption cross-section gives precise results, and therefore we have implemented it in our code to obtain all the results presented below.  

The Ly$\alpha$ luminosity that is observed at a frequency $\nu_{obs} = \nu [1+z_{em}]^{-1}$ can be expressed as 
\be
L_\alpha(\nu_{obs}) = L_\alpha^{em}(\nu) T_\alpha(\nu) \, [{\rm erg \, s^{-1}}],
\label{la_obs2}
\ee
where a fraction $T_\alpha(\nu) = e^{-\tau_\alpha(\nu)}$ of the Ly$\alpha$ luminosity emerging from the galaxy is transmitted through the IGM. The value of the total observed Ly$\alpha$ luminosity, $L_\alpha$, is then obtained by summing $L_\alpha(\nu)$ over the width of the entire doppler broadened Ly$\alpha$ line. The total value of the IGM transmission, $T_\alpha$ can then be determined as $T_\alpha =  L_\alpha / L_\alpha^{em}$.

Since the continuum is unaffected by transmission through the IGM, the observed continuum luminosity, $L_c$, is calculated to be 
\be
L_c = L_c^{em} \, [{\rm erg \, s^{-1} \, \AA^{-1}}].
\label{lc_obs2}
\ee

We now briefly discuss the effects of the peculiar velocities, ${\rm v}_p$ (Eqns. \ref{gauss_vel2}, \ref{lorentz_vel2}) on $T_\alpha({\rm v})$. If ${\rm v}_p$ is negative (positive) with respect to the velocity, ${\rm v}$, of an emitted Ly$\alpha$ photon, i.e., there is an inflow towards (outflow from) the galaxy, ${\rm v}_i + {\rm v}_p \propto (\lambda_i - \lambda_\alpha)$ decreases (increases). Therefore, $\lambda_i$ is smaller (larger) than it would be in the case of ${\rm v_p}=0$, i.e., the wavelength has been blueshifted (redshifted). Consequently, the values of the $e^{-{\rm v_i+v_p}}$ and $1/[{\rm v_i+v_p}]$ terms (Eqns. \ref{gauss_vel2} and \ref{lorentz_vel2}) increase (decrease) which leads to a higher (lower) value of $\tau_\alpha({\rm v})$, i.e. a lower (higher) value of $T_\alpha({\rm v})$. In brief, using the above Voigt profile, peculiar velocities caused by galactic scale inflows (outflows) of gas blueshift (redshift) the Ly$\alpha$ photons, leading to a lower (higher) value of $T_\alpha$.

 
\section{Basic dependencies of the observed Ly$\alpha$ luminosity} 
\label{basic_dep} 
 
The Ly$\alpha$ optical depth explained above in Sec. \ref{observed_lum} depends on three quantities: the SFR (which fixes the value of $Q$), the ionized region radius, and the global neutral fraction, i.e., 
\be
\tau_\alpha = \tau_\alpha(\dot M_\star, R_I, \chi_{HI}).
\label{tau_fn}
\ee
Once these three parameters are given, the transmissivity of the Ly$\alpha$ line, $T_\alpha$, is uniquely determined. Notice that 
\be
R_I=R_I(f_{esc},t_\star ,\dot M_\star,\chi_{HI}).
\ee
If instead we are interested in the observed Ly$\alpha$ luminosity, a fourth parameter needs to be specified, the ``effective" \Lya photon escape fraction 
\begin{equation} 
f_{esc,\alpha} = (1-f_{esc}) f_\alpha, 
\end{equation} 
which expresses the physical fact that the condition to observed  Ly$\alpha$ photons is that some ionizing photons
are absorbed within the galaxy and only a fraction $f_\alpha$ of produced Ly$\alpha$ photons can escape to infinity. Note that $f_{esc,\alpha}$ does not affect the transmissivity as both the emerging and the observed luminosity depend on it and therefore it factors out. A full exploration of the physical effects of the parameters on the observed luminosity, $L_\alpha$, can be performed by varying only the parameters $\dot M_\star, R_I, \chi_{HI}$ and $f_{esc,\alpha}$. The effects of other parameters (as, for example, metallicity, $Z_*$) can be estimated by simple scaling of the results below.  
 
\begin{figure}[htb] 
  \center{\includegraphics[scale=0.7]{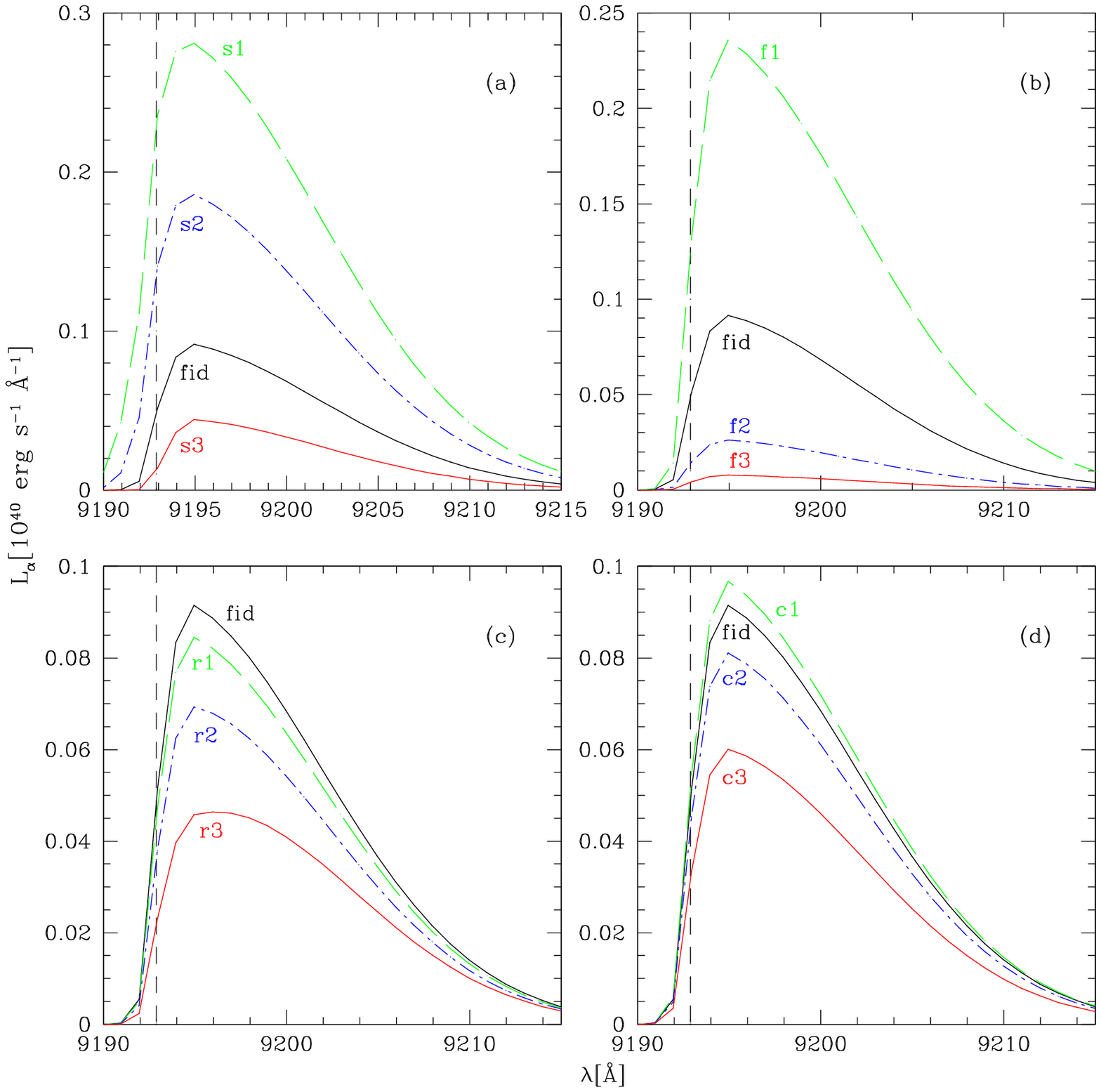}} 
  \caption{Effect of varying (a) $\dot M_*$ (b) $f_{esc,\alpha}$ (c) $R_I$ and (d) $\chi_{HI}$ on $L_\alpha$. Refer to Tab. \ref{table1_ch2} for the parameters used for each of the lines in this plot. The dashed vertical line shows the wavelength of the redshifted (emission redshift $z=6.56$) Ly$\alpha$ line.} 
\label{zch2_line_prof} 
\end{figure}

To understand the impact of each of the three relevant quantities on $L_\alpha$ we have selected a {\it fiducial}
case with parameters broadly similar to those we inferred under realistic (i.e. observationally derived) conditions for LAEs and allow them to vary in isolation taking three different values. We therefore considered $1\times {\rm fiducial} + 4 \times 3 =13$ different cases shown in Fig. \ref{zch2_line_prof} and summarized in detail in Tab. \ref{table1_ch2}.   
 
\begin{table*} 
\begin{center} 
\begin{tabular}{|c|c|c|c|c|c|} \hline 
Model&$\dot M_*$&$f_{esc,\alpha}$&$\chi_{HI}$&$R_I$& $T_\alpha$ \\  
     &$[M_\odot {\rm yr}^{-1}]$&   &        &[pMpc] & \\  
\hline 
Fiducial&$27$&$0.35$&$0.01$&$5.95$& $0.47$\\  
s1&$81$&$-$&$-$&$-$&$0.52$ \\  
s2&$54$&$-$&$-$&$-$&$0.50$  \\  
s3&$13.5$&$-$&$-$&$-$&$0.44$  \\ \hline 
 
f1&$-$&$0.9$&$-$&$-$&$0.47$ \\ 
f2&$-$&$0.1$&$-$&$-$&$0.47$ \\ 
f3&$-$&$0.03$&$-$&$-$&$0.47$ \\ \hline

r1&$-$&$-$&$-$&$2.97$&$0.44$ \\ 
r2&$-$&$-$&$-$&$1.48$&$0.37$ \\ 
r3&$-$&$-$&$-$&$0.74$&$0.27$ \\ \hline 
 
c1&$-$&$-$&$3\times 10^{-4}$&$-$&$0.49$ \\  
c2&$-$&$-$&$0.05$&$-$&$0.42$ \\  
c3&$-$&$-$&$0.15$&$-$&$0.32$  \\ \hline
\end{tabular} 
\end{center} 
\caption {Parameters of the fiducial model as well as for the different cases plotted in Fig. \ref{zch2_line_prof}. For all cases, the halo mass is $10^{11.8} M_\odot$. Dashes indicate that fiducial model values have been used. Note that $R_I$ is expressed in physical Mpc.} 
\label{table1_ch2}
\end{table*} 
 
 
\subsection{Star formation rate} 
\label{sfr}
 
The ionizing photon rate, $Q$, of the emitter is directly proportional to $\dot M_*$. As a result, a larger value of $\dot M_*$ results in (a) an increase of $L_\alpha^{int}$, (b) a larger ionized radius, $R_I$ around the galaxy, (c) a lower value of $\chi_{HI}$ at each point within the Str\"omgren sphere (see Eqns. \ref{gamma_bl}-\ref{chi_r}). The net effect is that as $\dot M_*$ increases, the IGM transmission, $T_\alpha$, of a stronger Ly$\alpha$ line increases due to decreased damping by both the Gaussian core and the red damping wing. This is shown in panel (a) of Fig. \ref{zch2_line_prof}. For the fiducial case we find that 47\% of the intrinsic \Lya luminosity is transmitted; this value increases with $\dot M_*$, reaching 52\% when $\dot M_\star= 81 M_\odot {\rm yr}^{-1}$, as seen from Tab. \ref{table1_ch2}.

\subsection{Effective \Lya photon escape fraction }  
\label{fa}
 
The effective \Lya photon escape fraction  $f_{esc,\alpha}$ scales both $L_\alpha^{em}$ and $L_\alpha$ equally,
without changing either the size of the Str\"omgren sphere or the \HI profile within it. The fraction of Ly$\alpha$ luminosity transmitted is hence, the same in all the cases. The variation of $L_\alpha$ with $f_{esc,\alpha}$ is shown in panel (b) of Fig. \ref{zch2_line_prof}.

\subsection{Ionized region radius} 
\label{ri} 
\begin{figure} 
\center{\includegraphics[scale=0.45]{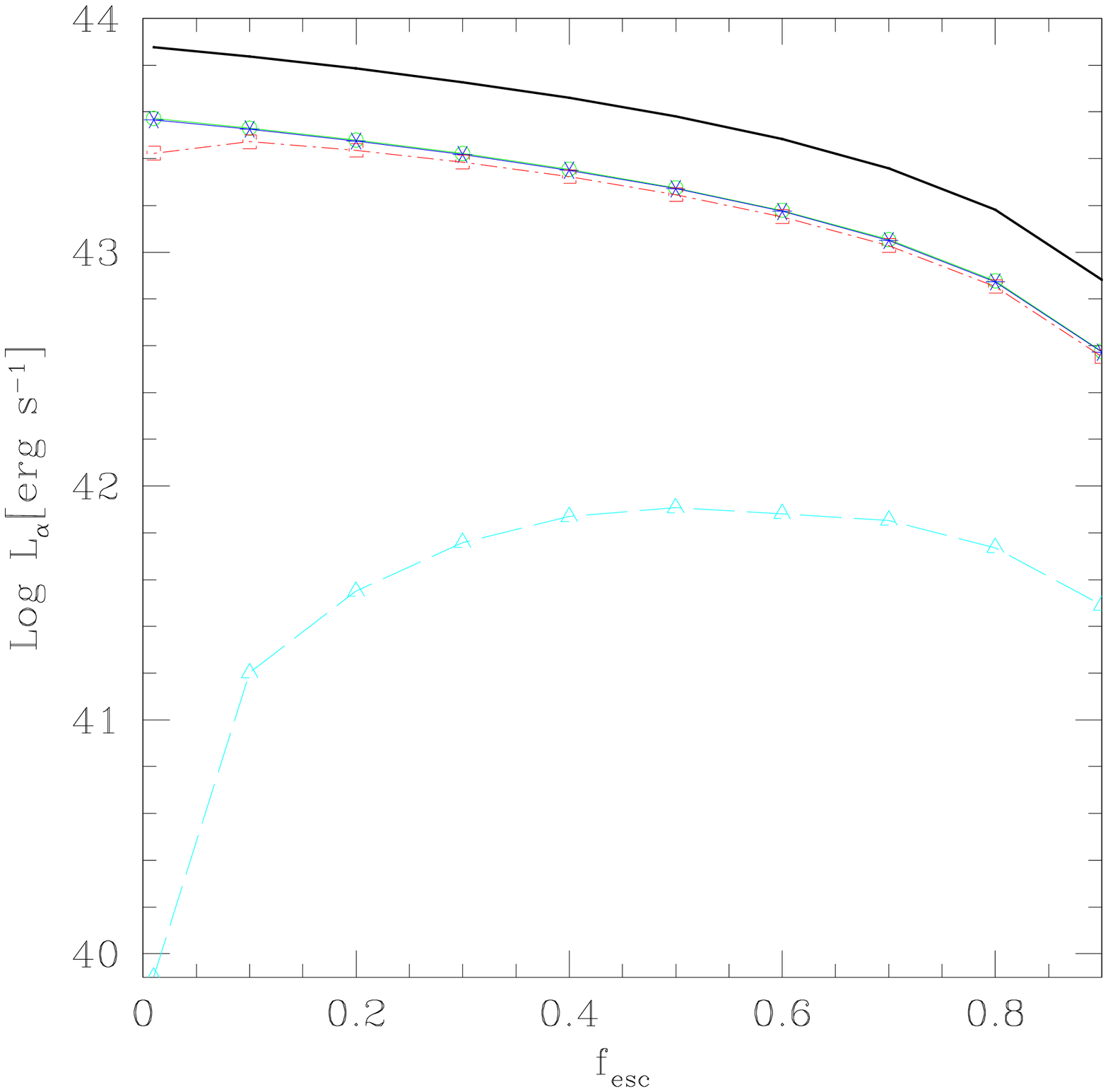}}
\caption{Dependence of $L_{\alpha}$ on $f_{esc}$ for different values of $\chi_{HI}$. Adopted parameters are 
$\dot M_\star=27 M_\odot {\rm yr}^{-1}$, $t_*=10^8$~yr, $f_\alpha=1$. The solid line shows $L_\alpha^{em}$. Curves with symbols refer to different values of $\chi_{HI}=0.15, 0.01, 10^{-3}, 3\times 10^{-4}$, from bottom to top, respectively.}  
\label{zch2_fesc_fractx_lya} 
\end{figure}   

As the ionized region radius, $R_I$, becomes larger, due to a more robust input on ionizing photons from the source, the Ly$\alpha$ photons reach the edge of the sphere more redshifted. Hence, the \HI outside the ionized bubble is less effective in attenuating the flux. The size of the ionized region radius is therefore very important for galaxies in regions of high $\HI$ density  and loses importance as the $\HI$ density decreases. We show the variation of $L_\alpha$ with $R_I $ (in pMpc) in panel (c) of Fig. \ref{zch2_line_prof}. from which we can readily appreciate that as $R_I$ increases (at a fixed $\chi_{HI}$ and $\dot M_*$), $T_\alpha$ increases due to the aforementioned effect. As, to a good approximation,  
 \begin{equation} 
R_I \propto \left(\frac{Q f_{esc} t_*}{\chi_{HI} n_H}\right)^{1/3}, 
\label{ri} 
\end{equation} 
for a fixed value of $Q$ (i.e. $\dot M_*$) and $\chi_{HI}$, $R_I$ can vary either due to $t_*$ or $f_{esc}$. These two parameters play a qualitatively different role. While the age variation can be embedded in a variation of $R_I$ only, changing the value of the escape fraction also affects $L_\alpha^{em}$ (see Eq. \ref{lya_emm2}) giving rise to a physically interesting effect. In Fig. \ref{zch2_fesc_fractx_lya}, for illustration purposes, we fix $\dot M_\star=27 M_\odot {\rm yr}^{-1}$, $t_*=10^8$~yr, $f_\alpha=1$ and study the effect of $f_{esc}$ on $L_{\alpha}$ for different values of $\chi_{HI}$.  
 
The observed Ly$\alpha$ luminosity, $L_\alpha$, decreases monotonically with $f_{esc}$ for low values of $\chi_{HI}$ ($<0.01$), just mirroring the decreasing value of the Ly$\alpha$ line emerging from the galaxy. Here, the fact that the size of the Str\"omgren sphere built increases with increasing $f_{esc}$ has no effect on $L_\alpha$ simply because the $\HI$ density is too low to cause (red) damping wing absorption, irrespective of the size of the ionized region. For $\chi_{HI} \geq 0.01$, the $L_{\alpha}$ trend with $f_{esc}$ in not monotonic anymore (see also Santos, 2004). For example, for $\chi_{HI}=0.15$, $L_\alpha$ reaches a maximum at $f_{esc}\approx 0.5$. This can be explained by the following: for low ($<0.5$) $f_{esc}$ values, as $f_{esc}$ increases, the ionized volume increases, thus leading to larger transmission. When $L_\alpha$ reaches its maximum (for $f_{esc}\approx 0.5$, in our example), a further $f_{esc}$ increase reduces $L_\alpha$, as a consequence of the decreasing value of $L_\alpha^{em}$. This highlights the fact that while for low values of $\chi_{HI}$, $f_{esc}$ affects $L_\alpha$ only through the emerging Ly$\alpha$ line, for high values of $\chi_{HI}$, the effect of $f_{esc}$ on the Str\"omgren sphere size becomes considerably important. 

 
\subsection{Neutral hydrogen fraction} 
\label{vary chi} 
 
In panel (d) of Fig. \ref{zch2_line_prof}, we study the effect of different $\chi_{HI}$ values on the Ly$\alpha$ line. It can be seen from Tab. \ref{table1_ch2} that the Ly$\alpha$ line is quite damped ($T_\alpha \sim 0.32$) for high values of $\chi_{HI}$ ($=0.15$). As the value of $\chi_{HI}$ decreases, the effect of both the Gaussian core and the red damping wing start reducing, allowing more of the line to be transmitted. For $\chi_{HI}=3 \times 10^{-4}$, most of the line redwards of the \Lya wavelength escapes without being damped. This occurs because the emitter is able to (a) strongly ionize the \HI within the Str\"omgren sphere (already ionized to a large extent even outside it) even further, and (b) build a large Str\"omgren sphere such that the Ly$\alpha$ line is not affected by the damping wing of the \HI outside. 
 
We remind the reader that $L_\alpha = L_\alpha (\dot M_\star, f_{esc,\alpha}, R_I, \chi_{HI})$. For a continuous 
star formation mode, the luminosity of the source becomes rapidly independent of age (typically after $\sim 10$ Myr); if, in addition, we adopt the values of $\chi_{HI}$ obtained from Gallerani et al. (2008) by matching the experimental data, we are left with two free parameters, $\dot M_\star$ and $f_{esc,\alpha}$. Recalling that $\dot M_\star \propto f_*/\epsilon_{dc}$, the free parameters in our model reduce to (a) $f_*/\epsilon_{dc}$ and (b) $f_{esc,\alpha}$. 

 
\section{Comparison with observations} 
\label{comp} 
Once that the semi-analytic model, described in Sec. \ref{model_ch2}, is applied to all the galaxies on the mass function, galaxies with $L_\alpha \geq 10^{42.1} (10^{42.2})\, {\rm erg\, s^{-1}}$, which is the current observational detection limit at $z \sim 4.5$ (5.7,6.6), are identified as LAEs comprising the Ly$\alpha$ LFs. Once the LAEs are identified at each redshift of interest, we are in a position to compare the results obtained from our model to observations of the LAE LF, the UV LF, the line profile asymmetries, the EW distribution and the cosmic SFR density. In particular, our aim is to assess to what extent the study of these quantities for LAEs can be used to discriminate between the early (ERM) and late (LRM) reionization scenarios, as deduced from the study of Gallerani et al. (2008), summarized in Sec. \ref{observed_lum}. 
 
First, we briefly summarize the data sets we use to compare our models to, with complete details having been given in Chapter 1 (see Sec. \ref{obs_laes}).
\begin{itemize}
\item {\it $z \sim 4.5$}: Dawson et al. (2007) conducted the Large Area Ly$\alpha$ (LALA) survey to look for LAEs at $z\sim4.5$ and found 97 candidates; 73 of which were confirmed using DEIMOS on KECK II and the Low Resolution Imaging Spectrograph (LRIS). 

\item {\it $z \sim 5.7$}: Shimasaku et al. (2006) identified 89 LAE candidates in the Subaru Deep Field (SDF) at $z\sim 5.7$ by using the 8.2m Subaru Telescope and the following selection criteria: (a) $i'-{\rm NB816} \geq 1$; (b) ${\rm NB816} \leq 26$. By using the Faint Object Camera and Spectrograph (FOCAS) on Subaru and DEIMOS, 28 candidates were confirmed as LAEs. 

\item {\it $z \sim 6.5$}: Taniguchi et al. (2005) detected 58 possible LAEs using Subaru at $z \sim 6.5$ and obtained the spectra for 20 of them using the FOCAS; 9 of these objects were confirmed as LAEs at $z\sim 6.5$.  Kashikawa et al. (2006) added 8 more LAEs, confirmed using the Keck II DEIMOS spectrograph, to this list. Thus, the Subaru observations have a total of 17 confirmed LAEs at $z\sim 6.5$. 
\end{itemize}

\subsection{The Ly$\alpha$ luminosity function}  
\label{lf_fit} 
\begin{figure*}[htb]
\vspace{-0.0cm}
\center{\includegraphics[scale=0.64]{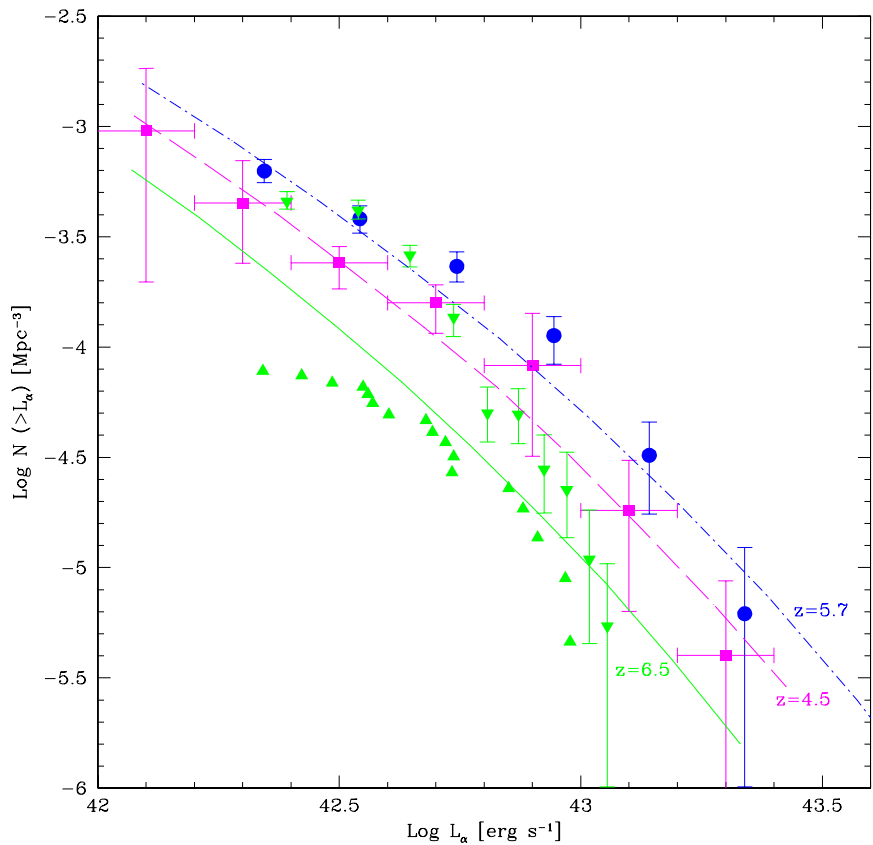}} 
\caption{Cumulative LAE Luminosity Function for the early reionization model (ERM). Points represent the data at 
three different redshifts:  $z=4.5$ Dawson et al. (2007) (squares), $z=5.7$ Shimasaku et al. (2006) (circles), 
$z=6.56$ Kashikawa et al. (2006) with downward (upward) triangles showing the upper (lower) limits. Lines refer to
model predictions at the same redshifts: $z=4.5$ (dashed), $z=5.7$ (dot-dashed), $z=6.56$ (solid).} 
\label{zch2_lyalf} 
\end{figure*}

As a first remark, it is useful to point out that if the Ly$\alpha$ LF evolution were to result purely from the evolution of the dark matter halos predicted by hierarchical structure formation, one would expect the comoving number density of luminous objects to increase with decreasing redshift. Although data errors are still large, it must be noted that instead there is an indication that there is no evolution of the Ly$\alpha$ LF between $z \sim 3-6$ (Dawson et al. 2007; Ouchi et al. 2008). Obviously, a number of different effects could produce this non-monotonic trend, a few examples being, SFR evolution, redshift dependent escape fractions and dust extinction, as we discuss in the following. In Fig. \ref{zch2_lyalf}, we plot the cumulative LFs at $z=4.5, 5.7$ and $6.56$ together with our best fit results. We now discuss the predictions of ERM and LRM separately. 
 
The ERM predicts an evolution of the hydrogen neutral fraction such that $\chi_{HI} = 1.3\times10^{-5}, \, 
8.6\times10^{-5}, \, 3\times 10^{-4}$ for $z=4.5, \, 5.7$ and $6.56$ respectively. Interestingly, a very good fit to the data can be obtained for the two highest redshifts with a single value of the star formation efficiency parameter $f_*/ \epsilon_{dc}=3.5$, thus implying that $\dot M_*$ for any given halo mass is not very much dependent on redshift. While a reasonable fit to the data at $z=5.7$ and $z=6.56$ is obtained for a single value of $f_{esc,\alpha}\approx 0.3$; a better fit is obtained by allowing for a 40\% increase of $f_{esc,\alpha}$ towards larger masses. The typical LAE dark matter halo masses corresponding to the observed luminosities are in the range $M_h=10^{10.7-12.0} M_\odot$ at $z=6.56$; at the same redshift, $\dot M_*$ ranges from $2$ to $43$ $M_\odot{\rm yr}^{-1}$.   

The data at $z=4.5$ instead pose a challenge to the model because, assuming non-evolving values of  $f_*/
\epsilon_{dc}=3.5$ and $f_{esc,\alpha}$, the observed number density of luminous objects is lower than that
predicted by the evolution of the theoretical LF. Given the relative constancy of the star formation efficiency and of the effective \Lya photons escape fraction noted for the two highest redshifts considered, the most natural
explanation is in terms of increasing dust extinction. To reconcile the prediction with the data at $z=4.5$ we then require that the \Lya line suffers an additional damping due to the presence of dust; which we find to be equal to $1/4.0=0.25$, i.e $f_\alpha$ (and hence $f_{esc,\alpha}$) decreases by a factor of 4. A strong increase of the dust content inside galaxies is expected on cosmic time scales larger than 1 Gyr (corresponding to $z\simlt 5$) when evolved stars rather than core-collapse supernovae become the primary dust factories. Such a hypothesis needs to be checked carefully, as the dust would not only affect the \Lya line but also the continuum emission, finally affecting the equivalent width of the line. We will discuss these effects of dust in Secs. \ref{uv_fit} and \ref{ew_ch2}. Hence, it seems that overall, a model in which reionization was completed relatively early ($z=7$) matches the data quite well.  
 
The LRM has a much slower reionization history, as is clear from the values of $\chi_{HI} = 1.4\times10^{-5}, \,
1.3\times10^{-4}, \, 0.15$ for $z=4.5, \,5.7$ and $6.56$ respectively. At the lowest redshifts ($z=4.5$ and $5.7$) this model requires exactly the same value $f_*/ \epsilon_{dc}=3.5$ as the ERM. This does not come as a surprise of course, as $\chi_{HI}$ is so small at these epochs in both the ERM and the LRM that $L_\alpha$ is unaffected. However, as $\chi_{HI}$ is much larger at $z=6.56$ in the LRM as compared to the ERM, a higher star formation efficiency, $f_*/ \epsilon_{dc}=16$ is required to fit the data at $z=6.56$ for the LRM. As a result $\dot M_*$ of LAEs in the LRM are increased by the same amount, ranging from $11$ to $197 \, {\rm M_\odot}$~yr$^{-1}$. As in the ERM, we use the same value of $f_{esc,\alpha}\approx 0.3$ (increasing by 40\% for larger halo masses) for $z=5.7$ and $6.56$, but the data at $z=4.5$ again require $f_\alpha$ to decrease by a factor of 4.

\begin{figure}  
\center{\includegraphics[scale=0.45]{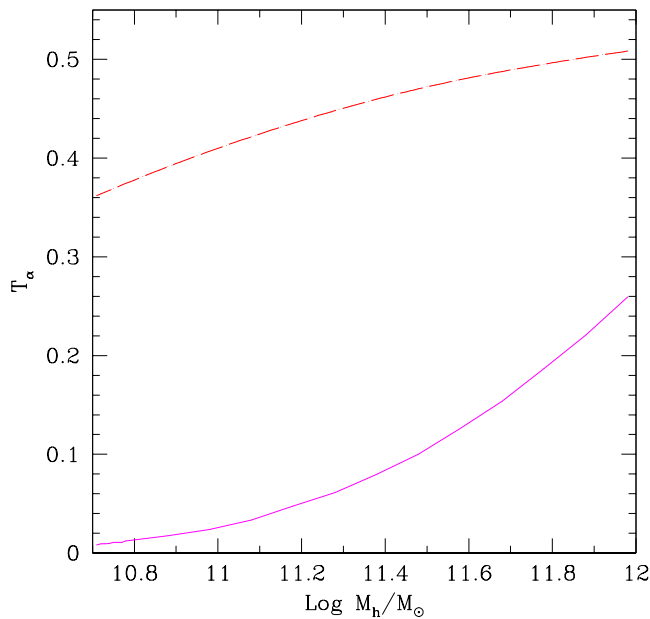}} 
\caption{\Lya transmissivity as a function of the LAE DM halo mass at $z=6.56$ for the LRM (solid line) and ERM (dashed).} 
\label{zch2_mh_tx}   
\end{figure}  

A comparison of the IGM transmissivity, $T_\alpha$, for the two reionization models considered is shown in Fig. \ref{zch2_mh_tx} for $z=6.56$. In both cases, $T_\alpha$ increases towards more massive halos because of their generally larger values of $\dot M_*$; also, at a given halo mass, $T_\alpha$ varies from $0.36$ to $0.51$ for the ERM, while it varies from $0.01$ to $0.26$ for the LRM i.e. it is considerably smaller for the LRM. In the LRM, small LAEs are characterized by a lower $T_\alpha$ with respect to larger ones relative to ERM. This is because even though the values of $\dot M_*$ are higher than in the ERM, the smaller LAEs are not able to build large enough HII regions; as a result, their \Lya line is much more damped as compared to that for the larger LAEs. 

In conclusion, the LF data seem to require a strong increase of the SFR from $z=5.7$ to $6.56$ in the LRM to fit the observed LFs while a SFR that smoothly decreases with increasing redshift fits the observations for the ERM. Looking at the general trend, one finds that SFR densities decrease with increasing redshift. Hence, we find that the LF data favors the reionization scenario described by the ERM, i.e. a highly ionized ($\approx 3\times10^{-4}$) Universe at $z=6.56$. The Best fit parameter values for the ERM are shown in Tab. \ref{table2_ch2}.

A caveat is that this analysis has been done for isolated emitters. As shown by McQuinn et al. 2007, clustering significantly increases the amount of Ly$\alpha$ luminosity that can be transmitted by an emitter by
adding a boost term to the background ionization rate. We find that such a boost factor of $\sim$ 100 boosts the luminosity transmitted by the LAEs at $z=6.56$ with $\chi_{HI}=0.15$ significantly and in that case, the LRM can be fit by the same parameters ($f_{esc,\alpha}$, $f_*/\epsilon_{dc}$) as the ERM. However, an estimate of the boost in the background requires an accurate understanding of the radial dependence of the clustering and the contribution of each emitter to the boost. Hence, we can not rule out the LRM completely till clustering is included and better measurements of SFR densities at $z\geq 6.56$ are obtained. 


\subsection{The UV luminosity function}
\label{uv_fit}

Once that the Ly$\alpha$ LFs are fixes, we can also try to match our model UV LFs to the observed ones. We start by briefly describing how the UV LFs are built up observationally.

\begin{table*} 
\begin{center} 
\begin{tabular}{|c|c|c|c|c|c|} 
\hline 
$z$&$M_h [M_\odot]$&$f_*/\epsilon_{dc}$&$\dot M_*[M_\odot {\rm yr}^{-1}]$&$f_{esc,\alpha}$&$f_c$\\  
\hline 
$4.5$& $10^{11.1-12.5}$&$3.5$&$6-160$&$\sim 0.075$&$\sim0.045$\\
$5.7$& $10^{10.8-12.3}$&$3.5$&$3-103$&$\sim 0.3$&$\sim 0.25$\\
$6.56$& $10^{10.7-12.0}$&$3.5$&$2-43$&$\sim 0.3$&$\sim 0.5$\\ \hline
\end{tabular} 
\end{center}  
\caption {Best fit parameter values for the ERM to fit both the Ly$\alpha$ LF and UV LF. For each redshift (col 1), we mention the DM halo mass range required (col 2), the SFR efficiency (col 3), the associated SFR (col 4), the effective escape fraction of Ly$\alpha$ photons (col 5) and the escape fraction of continuum photons (col 6).}
\label{table2_ch2} 
\end{table*} 

\begin{figure}[htb]
\center{\includegraphics[scale=0.45]{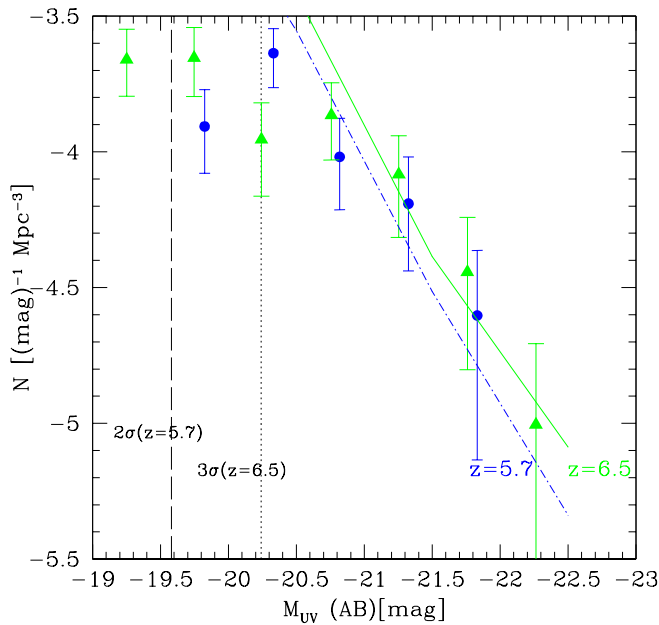}} 
\caption{UV LAE Luminosity Function for the early reionization model (ERM). Points represent the data at 
two different redshifts: $z=5.7$ Shimasaku et al. (2006) (circles), $z=6.56$ Kashikawa et al. (2006) 
(triangles). Lines refer to model predictions at the same redshifts: $z=5.7$ (dot-dashed), $z=6.56$ (solid). The vertical dashed (dotted) lines represent the 2$\sigma$ (3$\sigma$) limiting magnitude for $z=5.7$ ($z=6.56$). } 
\label{zch2_uvlf} 
\end{figure}

\begin{itemize}

\item {$z \sim 5.7$}: Shimasaku et al. (2006) transformed the z' band magnitude from the photometric sample of 89 LAE candidates into the far UV continuum at the rest frame. The UV LF was calculated by dividing the number of LAEs in each 0.5 magnitude bin by the effective volume corresponding to the FWHM of the bandpass filter used (NB816). Objects fainter than the 2$\sigma$ limiting magnitude (27.04 mag) in the z' band were not included in calculating the UV LF and this corresponds to the vertical line at $M_{UV}=-19.58$ in Fig. \ref{zch2_uvlf}. The authors mention that the apparent flattening at $M_{UV}>20.5$ might be due to the incompleteness in the measurement of the far UV LF.

\item {$z \sim 6.5$}: Kashikawa et al. (2006) used the same methodology mentioned above to derive the rest UV continuum from their photometric sample of 58 LAEs. Their LF measurements at magnitudes fainter than $M_{UV}=-20.24$ (3$\sigma$) are uncertain due to the z' band magnitudes no longer being reliable beyond this value.

\end{itemize}

Both the above calculations have accounted for the detection completeness of the narrow band filters. They also find that cosmic variance is not severe for the UV LF. An important point to note is that the UV LFs at $z=5.7$ and $z=6.5$ are in very good agreement and show no evolution between these redshifts, which is in clear contrast to the Ly$\alpha$ LF which shows a deficit of high luminosity LAEs at $z=6.5$ as mentioned before in Sec. \ref{lf_fit}.

We build the UV LF by translating the intrinsic UV LF into the observed one, as explained in Sec. \ref{intrinsic_lum}, \ref{observed_lum} (see Eqns. \ref{lc_int2}, \ref{lc_emm2}, \ref{lc_obs2}), for all galaxies identified as LAEs. However, using these conversions and the best fit parameter values of $f_*/\epsilon_{dc}$ for the ERM as mentioned in Sec. \ref{lf_fit}, we find that the UV LFs for both $z \sim 5.7$ and $6.6$ lie above the observed ones. Hence, additional dust damping of the UV LF is required to match with the observations. 

We quantify this additional damping by including $f_c$, the fraction of continuum photons that escape the galaxy unabsorbed by dust, into our calculations, as explained in Sec. \ref{intrinsic_lum}. Using a single value of $f_c$ for a specific redshift (see Tab. \ref{table2_ch2}), across the entire mass range considered, we find a reasonably good agreement with the observed UV LF for the bright LAEs. However, the model fails to reproduce the bending of the UV LF observed for the low luminosity emitters. This could either be due to detection incompleteness in the observations or due to the lack of a physical effect such as a halo mass dependent $\dot M_*$ or $f_c$. A simple prescription for the former would be a value of $\dot M_*$ that decreases with decreasing halo masses, as will be shown in later chapters. 

It is interesting to note that for this model, while at the highest redshift, continuum photons are less absorbed by dust as compared to the Ly$\alpha$ photons, the trend reverses at lower redshifts. This could hint at dust whose inhomogeneity/ clumpiness evolves with redshift. However, robust estimates of the ages, metallicites, IMF and detailed studies of dust distribution and its evolution inside LAEs are needed before such a strong claim can be made.

\subsection{Cosmic star formation rate density}
\label{sfr_density}

\begin{figure} 
\center{\includegraphics[scale=0.45]{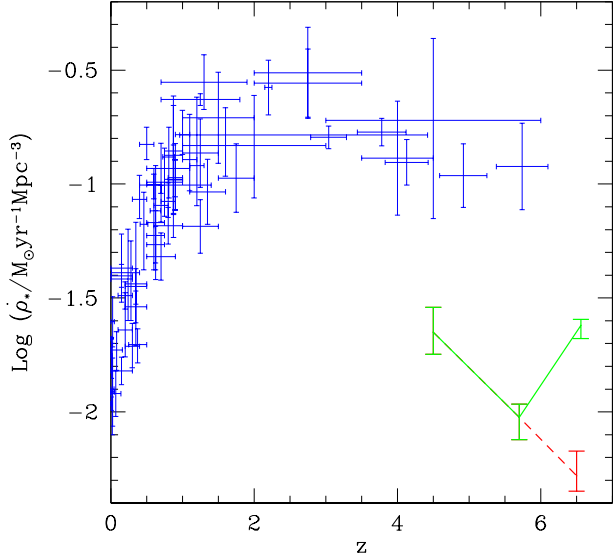}}
\caption{Comparison of the LAE SFR density, $\dot\rho_*$, to the cosmic SFR density evolution from our best fit models. Points show the measurements by Hopkins (2004); the dashed (solid) line is the prediction from ERM (LRM) for all LAEs on the Ly$\alpha$ LFs, shown in Fig. \ref{zch2_lyalf}.}  
\label{zch2_sfr_den} 
\end{figure}  

As a sanity check, using the parameters that best fit the data as discussed in Sec. \ref{lf_fit}, we calculate the contribution of LAEs on the Ly$\alpha$ LFs shown in Fig. \ref{zch2_lyalf}, to the SFR density, $\dot\rho_*$, at $z=4.5,5.7$ and $6.56$. We compare these with the values of the cosmic SFR density observed by Hopkins (2004) (see Tab. \ref{table2_ch2} of their paper) for the common dust-correction case, the results for which are plotted in Fig. \ref{zch2_sfr_den}. We find that for the best-fit parameters, the contribution of LAEs to the cosmic SFR density is redshift-dependent, with $\dot\rho_* \sim 8$\% at $z=5.7$ (SFRs are in the range $3 < \dot M_\star/M_\odot {\rm yr}^{-1} < 103$), and even higher at $z=4.5$, although the data present a large scatter at the latter epoch.

Further, two points are worth noticing about the predicted values of $\dot\rho_*$. First, the value of $\dot\rho_*$ must increase strongly from $z=5.7$ to $6.56$ in the LRM case. Although not impossible, such behavior is certainly puzzling and not easy to interpret. As the dust formation timescale is about 10 Myr, if the latter is copiously produced in supernova ejecta, as pointed out by several authors (Kozasa, Hasegawa \& Nomoto 1991, Todini \& Ferrara 2001, Schneider, Ferrara \& Salvaterra 2004, Bianchi \& Schneider 2007) and recently confirmed by the extinction curves of high redshift quasars (Maiolino et al. 2004), supernova-produced dust would rapidly increase the opacity to both continuum and \Lya photons, thus causing a rapid fading of the emitter. 

Second, the contribution of LAEs is about 8\% of the cosmic star formation rate density at $z=5.7$. Thus, either
the duty cycle of the actively star forming phase in these objects is of the same order, or one has to admit that only a very small fraction ($\sim 1/12$) of high redshift galaxies experience this evolutionary phase. In the first case, the star formation duration would last about 8\% of the Hubble time at $z=5.7$, i.e. 72 Myr.

\subsection{The \Lya equivalent width} 
\label{ew_ch2} 
 
From our model it is easy to derive the intrinsic rest-frame \Lya line equivalent width, $EW^{int}$, which is defined as
\be
EW^{int} = \frac{L^{int}_{\alpha} \,[{\rm erg \, s^{-1}}]}{L_c^{int} \,[{\rm erg \, s^{-1} \AA^{-1}}]}.
\ee 
Since both the $L_\alpha^{int}$ and $L_c^{int}$ scale linearly with $\dot M_*$ (see Eqns. \ref{val_lyaint}, \ref{lc_int2}), the intrinsic EW distribution is a $\delta$-function at $EW^{int}\sim 131$\AA.  
 
From our model, the observed EW in the rest frame of the emitter is calculated as 
\begin{equation} 
EW = EW^{int} (1-f_{esc}) T_\alpha \left(\frac{f_\alpha}{f_c}\right). 
\label{ewint_ch2}
\end{equation} 
Here, the ratio $f_{\alpha}/f_c$ expresses the differential extinction of the \Lya line with respect to continuum radiation due to ISM dust grains. 

At $z=4.5$, we have seen that we require a factor $\approx$ 4 suppression of the \Lya line luminosity by dust, i.e. $f_{esc,\alpha} \approx 0.075$. As dust affects also the continuum, and hence the EW, we need to estimate the value of $f_c$ (calculated at $\lambda=1375$~\AA). We find that for $f_c \approx 0.045$, the mean value of $EW$ from our model (155 \AA) is the same as the observed EW (155 \AA). We then use the following relations to obtain the color extinction: 
\begin{equation} 
A_\lambda(1375{\rm \AA}) = -2.5 \log f_c,  
\end{equation} 
\begin{equation} 
E(B-V) = \frac{A_V}{R_V} \approx \frac{1}{4}\frac {A_\lambda(1375{\rm \AA})}{R_V},  
\end{equation} 
where $R_V\approx 3$ and we have assumed a Galactic extinction curve. From these expressions we obtain $E(B-V) 
= 0.28$. 

\begin{figure} 
\center{\includegraphics[scale=0.45]{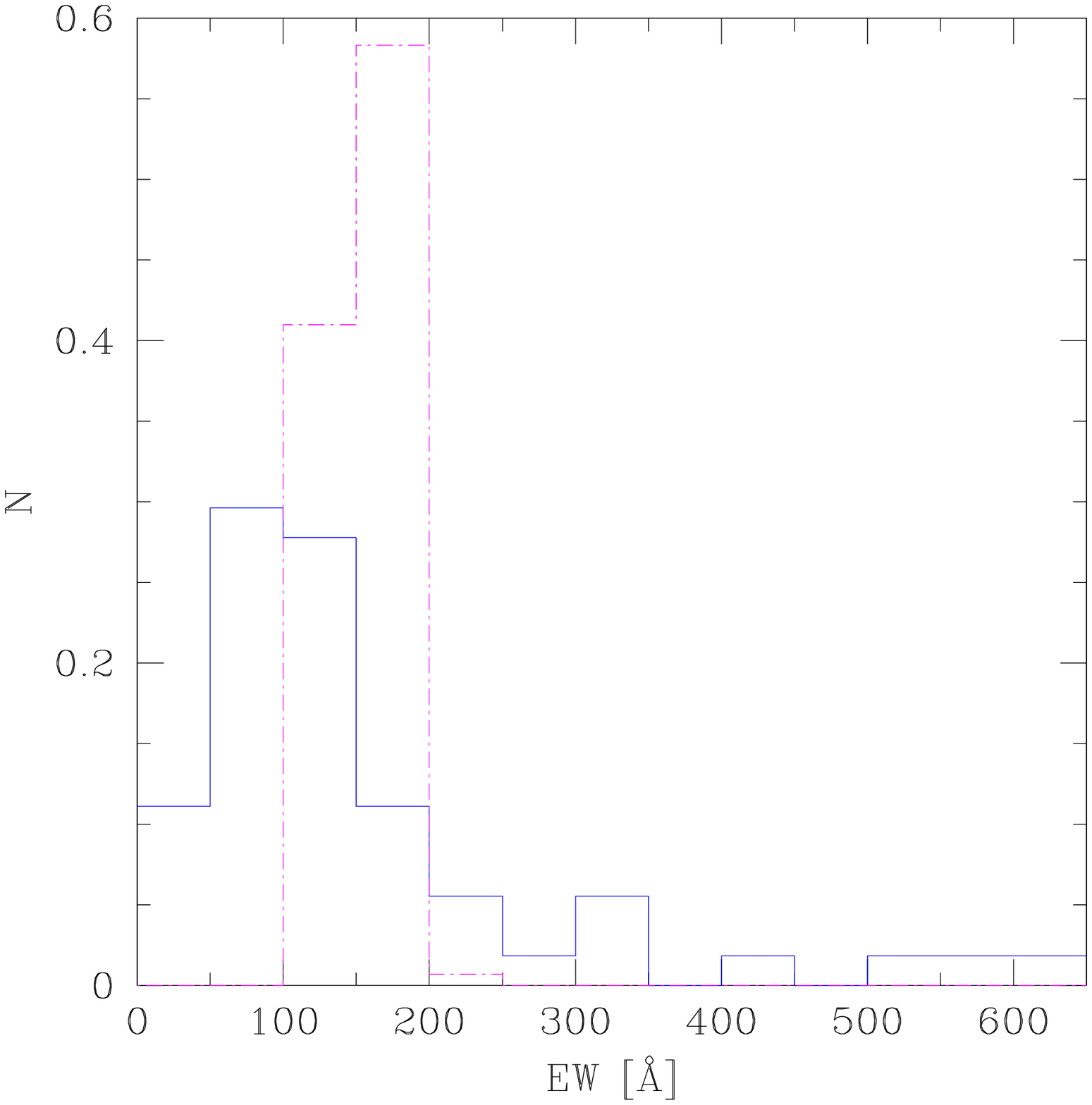}} 
\caption{Normalized distribution of the rest frame EW for LAEs at $z=4.5$. Observed values from Dawson et al. (2007) (model results) are shown by solid (dot-dashed) lines.} 
\label{zch2_ew_4.5} 
\end{figure}  

The value of $f_c$ implies that the continuum is extincted about 1.6 times more heavily than the \Lya line (assuming $f_{esc} \sim 0$ so that $f_\alpha = 0.075$). This is not inconceivable if LAE interstellar dust is inhomogeneously distributed and/or clumped, as showed by Neufeld (1991). With these two values we then derive the predicted EW distribution and compare it with the Dawson et al. (2007) data in Fig. \ref{zch2_ew_4.5}. As mentioned before, for the best fit parameters to the LF at $z=4.5$, $f_{esc,\alpha} \approx 0.075$ and $T_\alpha \approx 0.50$. Note that $f_{esc,\alpha}$ and particularly $T_\alpha$ depend on the LAE luminosity/mass and increase by about 45\% and 20\% respectively towards higher masses. 
 
The predicted EWs are concentrated in a range, $114~{\rm \AA} < EW < 201~{\rm \AA}$ (mean =$155~{\rm \AA}$),
whereas the observed distribution is considerably wider, spanning the range $6-650 \, {\rm\AA}$ with a mean of 
155\, \AA. As explained above, the spread of the predicted EW distribution arises only from the corresponding spread of $\dot M_*$ (6-160 $M_\odot {\rm yr}^{-1}$)  required in order to match the LF at $z=4.5$, via the dependence of $T_\alpha$ on $\dot M_*$. 

Calculating the rest frame EWs is easier at $z=5.7$ since we have an estimate of $f_c$ from the UV luminosity function as mentioned in Tab. \ref{table2_ch2}. We calculate the EWs using $f_{esc,\alpha} \approx 0.3$, $f_c \approx 0.25$ and $T_\alpha \approx 0.37$. As for $z=4.5$, $f_{esc,\alpha}$ and $T_\alpha$ depend on the halo mass and increase by 40\% and 45\% respectively towards higher masses. The calculations then yield EWs that range between 56-127 \AA. The mean from our model ($\sim 92.3$ \AA) is much less than the mean value of 120 \AA, observed by Shimasaku et al. (2006). 

\begin{figure} 
\center{\includegraphics[scale=0.45]{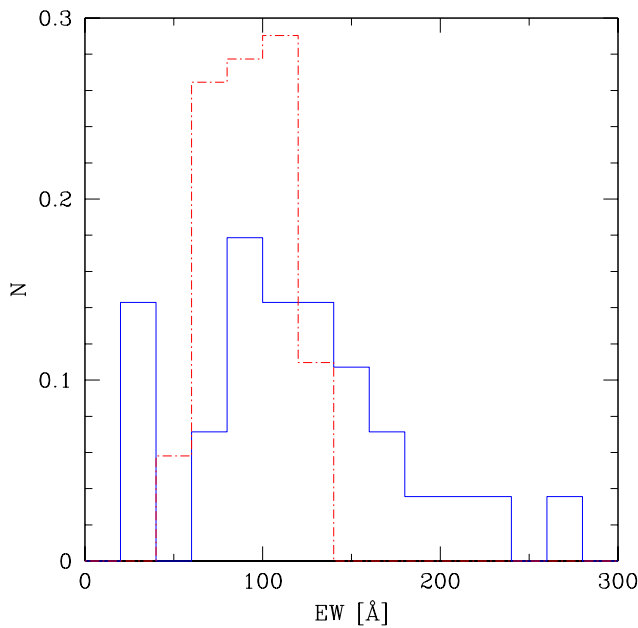}} 
\caption{Normalized distribution of the rest frame EW for LAEs at $z=5.7$. Observed values from Shimasaku et al. (2006) (model results) are shown by solid (dot-dashed) lines.} 
\label{zch2_ew_5.7} 
\end{figure}  

The narrow range (z=4.5) and lower mean (at z=5.7) of EWs calculated from our model can easily be explained by the fact that our model does not include inflows/outflows, assumes an age of about 100 Myr for all the emitters and a metallicity which is 1/20 of the solar value. In reality, a larger spread would be expected from the addition of physical effects lacking in this model, such as (i) gas kinematics (inflow/outflow); (ii) variations of the IMF, metallicity, and stellar populations including Population III (PopIII or metal-free stars), and (iii) young stellar ages. 

While inflows erase the red part of the Ly$\alpha$ line, thereby reducing the EW, outflows shift the line centre redwards, helping more of it to escape. Outflows can also add a bump to the red part of the line due to backscattering of Ly$\alpha$ photons, as shown by Verhamme, Schaerer \& Maselli (2006). A top heavy IMF produces more \HI ionizing photons, as does decreasing the metallicity. Hence, both these effects increase the EW. Further, for very young emitters ($\sim 10$ Myr), the continuum is much less than the continuum at 100 Myr and so, the EW would be much larger for younger emitters. 

\subsection{Line profile asymmetries} 
\label{skewness discussion}

Additional constraints on the model can come from the information embedded in the observed line profiles, as for
example the line profile asymmetry. This can be suitably quantified by the {\it weighted skewness} parameter, $S_W$, introduced by Kashikawa et al. (2006), which we calculate for the best fit parameter values for the ERM mentioned above. We adopt the following definition for such a quantity: 
\begin{equation} 
S_W = S \Delta \lambda = S (\lambda_{10,r}-\lambda_{10,b}), 
\label{weighted_skew} 
\end{equation} 
where $\lambda_{10,r}$ ($\lambda_{10,b}$) is the wavelength redward (blueward) of the Ly$\alpha$ line where the flux  
falls to 10\% of the peak value. In addition, we have that 
\begin{eqnarray} 
I &=& \sum_{i=1}^n f_i, \\ 
\bar x & = & \frac{\sum_{i=1}^n x_if_i}{I}, \\ 
\sigma^2 & = & \frac{\sum_{i=1}^n(x_i-\bar x)^2f_i}{I}, \\ 
S & = & \frac{\sum_{i=1}^n(x_i-\bar x)^3f_i}{I\sigma^3}, \label{skew} 
\end{eqnarray} 
where $f_i$ is the line flux in the wavelength pixel $i$ whose coordinate is $x_i$, and the summations are performed over the pixels covered by the \Lya line. On general grounds one would expect that the observed \Lya line shape would be more symmetric (i.e. low $S_W$) in reionization models characterized by a lower value of $\chi_{HI}$. However, given the above definition, just the opposite is true. In fact, for any reasonable value of the relevant parameters (see Fig. \ref{zch2_line_prof}) the blue part of the line is heavily absorbed, thus yielding a high value of $S_W$; as $\chi_{HI}$ is increased, also the long-wavelength part of the line is affected by the red damping wing, making the line more symmetric around the peak.   

\begin{figure} 
  \center{\includegraphics[scale=0.45]{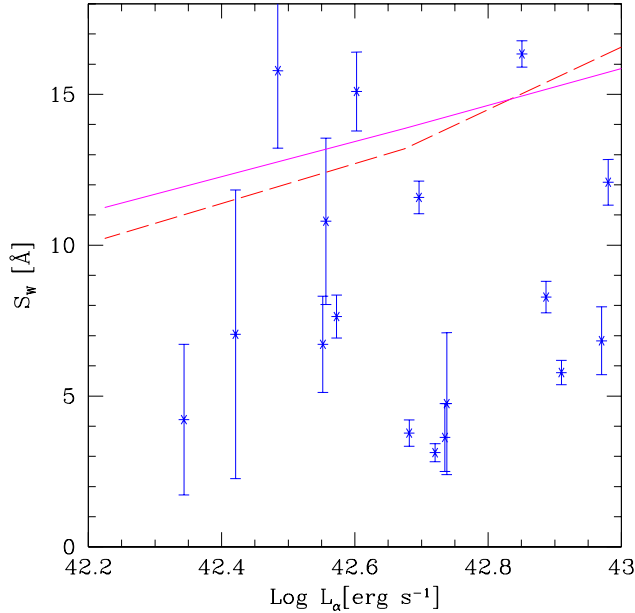}} 
  \caption{Weighted skewness of the observed Ly$\alpha$ line for different models. The asterisks are the data from Kashikawa et al. (2006). The dashed (solid) line correspond to the best fit ERM (LRM) at $z=6.56$.} 
\label{zch2_weighted_skew_2} 
\end{figure}  
 
The predicted trend of $S_W$ with the observed \Lya luminosity at $z=6.56$ is shown in Fig. \ref{zch2_weighted_skew_2}, for the parameters of the ERM and LRM that best fit the LF data (discussed in Secs. \ref{lf_fit}, \ref{uv_fit}). For both models, the weighted skewness of the line increases for more luminous objects; however, such dependence is steeper for the ERM than for the LRM. In general, though, the two reionization scenarios predict $S_W$ values in the range 10-17. The data from Kashikawa et al. (2006) spans the somewhat larger range 3-17, with many of the data points lying around $S_W=5$. Given the paucity of the observed points and the large errors associated to them, it is probably premature to draw any strong claim from these results. However, given the constant increase in the amount and quality of LAE data, it is quite possible that the line skewness could represent a very interesting tool to constrain reionization models in the near future. It has to be noted that the data show a large scatter of $S_W$ at a given value of $L_\alpha$, perhaps indicating that local conditions, including gas infall/outflow, density inhomogeneities and interaction of the Ly$\alpha$ line with the interstellar medium of the galaxy, might play a dominant role. 

From the theoretical point of view it is instructive to summarize the response of the skewness to different physical conditions. As we have seen from Fig. \ref{zch2_weighted_skew_2}, $S_W$ increases with $L_\alpha$ (or, equivalently with $\dot M_*$); this is true for any fixed value of $\chi_{HI}$.  This is because as $\dot M_*$ increases, more of the Ly$\alpha$ line escapes forcing $S_W$ to increase as a result of the larger value of $\Delta \lambda$. Further, the long -wavelength part of the observed Ly$\alpha$ line begins to flatten with increasing $\chi_{HI}$ due to attenuation by the red damping wing. Hence, $\Delta \lambda$ varies slower with $L_\alpha$ ($\dot M_*$) for high $\chi_{HI}$ (LRM) as compared to lower values (ERM); this makes the slope of $S_W$ steeper for the ERM. 
 
\begin{figure*}[htb] 
\center{\includegraphics[scale=0.85]{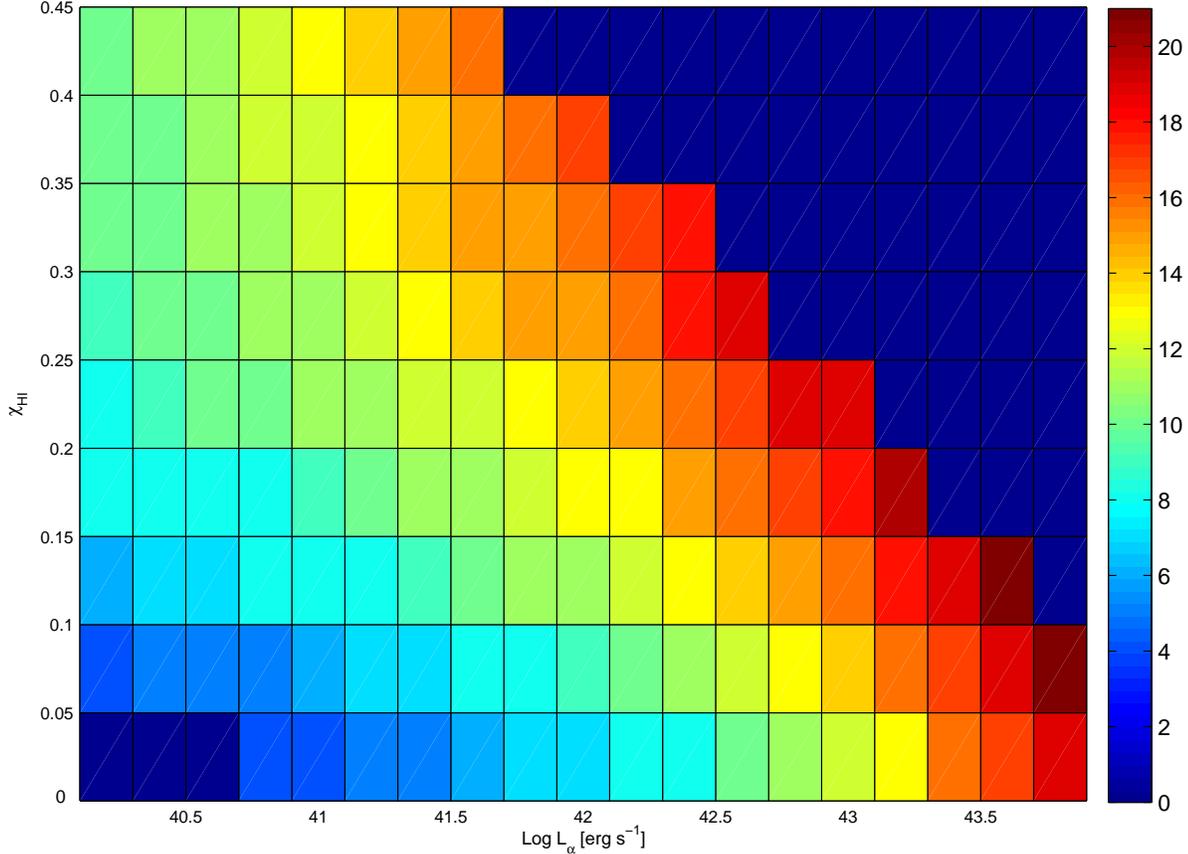}} 
\caption{Dependence of $S_W$ (values are color-coded by the bar on the right) on $\chi_{HI}$ and $L_\alpha$ at 
$z=6.56$ for a set of LAEs with SFR in the range predicted by the two reionization models, i.e. $\dot M_\star=2.7-197 
M_\odot$yr$^{-1}$.} 
\label{zch2_contour_all} 
\end{figure*}

A more general view of the dependence of $S_W$ on $L_\alpha$ (hence on $\dot M_*$) and $\chi_{HI}$ is shown in Fig. \ref{zch2_contour_all}. The plot has been obtained by dividing $L_\alpha$ into bins and averaging the weighted skewness over the number of LAEs in each bin at a given value of $\chi_{HI}$. The regions with weighted skewness values equal to zero represent a lack of LAEs in that bin. 
 
The most intriguing feature of Fig. \ref{zch2_contour_all} is a clear anti-correlation between $L_\alpha$ and $\chi_{HI}$. Given the range of SFR considered ($\dot M_\star=2.7-197 M_\odot$yr$^{-1}$), LAEs populate progressively fainter \Lya luminosity bins as the IGM becomes more neutral. Notice that relatively luminous objects ($L_\alpha \approx 10^{42.5}$~erg~s$^{-1}$) would not be detected if $\chi_{HI}\simgt 0.25$. Within the range in which these objects are visible, the most luminous objects always show the largest $S_W$ at fixed $\chi_{HI}$; however, such maximum value is also seen to increase with decreasing $\chi_{HI}$. 

However, it must be noted that the model does not include important effects such as inflows/outflows and interaction of the Ly$\alpha$ photons with the ISM, which will definitely leave an imprint on the $S_W$ and hence, weaken the $L_\alpha$-$\chi_{HI}$ 
anti-correlation.

\section{Conclusions} 
\label{conc_ch2} 
 
We briefly summarize the main results obtained and the caveats involved in the work discussed in this chapter.

\begin{itemize}
 
\item The LFs observed by Dawson et al. (2007), Shimasaku et al. (2006) and Kashikawa et al. (2006) at $z=4.5$, $5.7$ and $6.56$ respectively, can be reproduced by both the ERM and the LRM. However, we favour the ERM since it requires no redshift evolution or mass dependence of the star formation efficiency, while the fitting the data using the LRM requires a puzzling upturn of the star formation efficiency from $z \sim 5.7$ to $6.6$, which is at odds with the observed cosmic star formation rate density. The ERM being the right reionization scenario implies that the LF evolution can be explained solely by an evolution of the underlying dark matter halo mass function; the ionization state of the IGM does not change between $z \sim 5.7$ to $6.6$. However, more observations of the SFR density and information regarding the boost added to the ionizing background due to clustering at $z \sim 6.5$ are required to completely rule out the LRM. 
 
\item Reproducing the Ly$\alpha$ LFs also requires that only a fraction of the Ly$\alpha$ photons escape the galaxy environment, undamped by dust. A reasonable fit to the data at $z=5.7$ and $z=6.56$ is obtained for a single value of $f_{esc,\alpha}\approx 0.3$ (although a good fit is obtained by allowing for a (40\%) increase of $f_{esc,\alpha}$ towards larger masses). The data at $z=4.5$ require $f_{esc,\alpha}\sim 0.075$, signalling an increase in the dust content in low redshift galaxies. Matching the model UV LFs to the observations also requires that only a fraction of the UV photons escape the galaxy environment, undamped by dust.

\item We calculate the expected EWs at $z=5.7$ and find that the mean ($\sim$ 92 \AA) is much less than the observed value of 120 \AA. At $z=4.5$, a dust extinction of $E(B-V)\approx 0.28$ brings the predicted mean \Lya EWs ($\approx 155$~\AA) in very good agreement to the observed mean ($\approx 155$~\AA). However, additional effects which vary on a galaxy to galaxy basis, such as outflows/inflows or peculiar stellar populations are required to account for the spread of EW seen in the data.

\item For the given IMF and metallicity, while at higher redshifts ($z=6.5$), the continuum photons are less absorbed by dust as compared to Ly$\alpha$ photons, this trend reverses at lower redshifts($z= 4.5$, $5.7$). This could be explained by imhomogeneously distributed/clumped dust. However, the IMF, ages and metallicities of the emitters must be fixed robustly using simulations and infall must be included in the model before such a strong claim can be made.

\item The contribution of LAEs to the cosmic SFR density is small, amounting to roughly 8\% at $z=5.7$. Thus either the duty cycle of the actively star forming phase in these objects is of the same order, or one has to admit that only about one-twelfth of high redshift galaxies experience this evolutionary phase.  
 
\item Although additional useful information can be extracted from the line profile by using indicators like the line weighted skewness and equivalent width, the results presented here (Secs. \ref{ew_ch2}, \ref{skewness discussion}) must be considered as very preliminary due to both the scarce and relatively poor statistical quality data, as well as the simplifications of the model which make the comparison only meaningful at a basic level. Nevertheless, it is encouraging that the model results are broadly in agreement with the data, at least for what concerns mean values.   
 
\end{itemize}

We now discuss some of the main ingredients missing in our model:

\begin{itemize}

\item We use a single stellar population ($t_*=100$ Myr, $Z_* = 0.02 Z_\odot$) for all galaxies and use a value of $\dot M_*$ that scales linearly with $M_h$, i.e, we do not account for SFR dependent metal enrichment, or outflows leading to a decrease in the SFR for low mass galaxies.

\item Since we do not model the dust enrichment of galaxies, we use a single value of both $f_\alpha$ and $f_c$ for all galaxies at any given redshift.

\item We use a homogeneous and isotropic IGM density field which could affect our results substantially;
although the largest galaxies are able to ionize the largest regions around themselves, they also reside in regions of the largest over-density.

\item Due to a lack of the positional information of galaxies and feedback modeelling, we are unable to consider peculiar IGM velocities in our calculations of $T_\alpha$. However, as pointed out by Santos (2004) and Iliev et al. (2008), peculiar gas motions might affect the line profile considerably: while inflows of gas erase the Ly$\alpha$ line, galaxy scale outflows produced by supernova (or AGN) feedback enable more of the Ly$\alpha$ to escape. 

\item Due to a lack of positional information regarding the galaxies, we are unable to account for the boost in the photoionization rate due to clustered sources.

\end{itemize}

\chapter{LAE evolution in the reionization epoch}\label{ch3_lya_sim}
In Chapter \ref{ch2_lya_sam} we introduced a self consistent semi-analytic LAE model that reproduces the observed Ly$\alpha$ and UV LFs, and provides color excess and line skewness values in broad agreement with the observed data. However, there are a number of missing ingredients in the model, the most important of these being: (a) the calculation of the intrinsic Ly$\alpha$ luminosity, UV luminosity and SED (spectral energy distribution) as a function of the SFR, age, metallicity and IMF of the galaxy under consideration and, (b) the boost in the ionization rate imparted by galaxy clustering and its effects on the visibility of galaxies of different masses.

In this chapter, we use state-of-the-art cosmological SPH (smoothed particle hydrodynamic) simulations, presented in Sec. \ref{sim_ch3} to fix the SFR, age, metallicity for each galaxy to obtain the intrinsic Ly$\alpha$ luminosity, UV luminosity and the SED (Sec. \ref{intrinsic_lum_ch3}). The simulations are then coupled to a Ly$\alpha$ transmission model to obtain the observed Ly$\alpha$ and UV LFs (Sec. \ref{lyalf_ch3}, \ref{uvlf_ch3}). Using this model, we are able to quantify the importance/effect of clustering on Ly$\alpha$ luminosity transmission and its contribution to shaping the Ly$\alpha$ LF (Sec. \ref{lyalf_ch3}). By doing so, we gain insight on the nature of LAEs and put constraints on their elusive physical properties in Sec. \ref{nature_ch3}. We discuss the contribution of LAEs to reionization in Sec. \ref{lae_reio_ch3} and end with a summary of the main results and the caveats of the model, in Sec. \ref{Conc_ch3}.

The cosmological model adopted in the simulations used corresponds to the $\Lambda$CDM Universe with $\Omega_m=0.26,\ \Omega_{\Lambda}=0.74,\ \Omega_b=0.0413$, $n_s=0.95$, $H_0 (= 100 \,{\rm h}) = 73$
km s$^{-1}$ Mpc$^{-1}$ and $\sigma_8=0.8$, thus consistent with the 5-year analysis of the WMAP data (Komatsu et al. 2009).

\section{The theoretical model}
\label{theoretical_model_ch3}

We run {\small GADGET-2}, a cosmological SPH code (details follow in Sec. \ref{sim_ch3}), to obtain simulation snapshots at $z \sim 5.7, 6.6, 7.6$. We obtain the intrinsic properties ($t_*, \dot M_*, Z_*$) for each galaxy identified in the snapshots, as explained in Sec. \ref{sim_ch3}; the values of $t_*, \dot M_*$ and $Z_*$ are then input into {\tt STARBURST99} to obtain the intrinsic spectrum for each galaxy, which is then used to calculate the intrinsic Ly$\alpha$/ continuum luminosities and the Ly$\alpha$ transmission, as shown in Sec. \ref{intrinsic_lum_ch3}. The effects of clustered sources and their inclusion into calculating the IGM transmission are discussed in Sec. \ref{clus_ch3}. 

\subsection{The simulations: GADGET2 with chemodynamics}
\label{sim_ch3}

{\tt GADGET} is a freely available code for cosmological N-body/SPH simulations; the name is an acronym of `GAlaxies with Dark matter and Gas intEracT'. It computes gravitational forces with a hierarchical tree algorithm (optionally in combination with a particle-mesh scheme for long-range gravitational forces) and represents fluids by means of smoothed particle hydrodynamics (SPH). The code can be used for studies of isolated systems, or for simulations that include the cosmological expansion of space, both with or without periodic boundary conditions. In all these types of simulations, {\tt GADGET} follows the evolution of a self-gravitating collisionless N-body system, and allows gas dynamics to be optionally included. Both the force computation and the time stepping of GADGET are fully adaptive, with a dynamic range which is, in principle, unlimited. GADGET can therefore be used to address a wide array of astrophysically interesting problems, ranging from colliding and merging galaxies, to the formation of large-scale structure in the universe. With the inclusion of additional physical processes such as radiative cooling and heating, GADGET can also be used to study the dynamics of the gaseous intergalactic medium, or to address star formation and its regulation by feedback processes.

The simulation analyzed in this paper has been carried out using the TreePM-SPH code {\small {GADGET-2}} (Springel 2005), with the implementation of chemodynamics as described by Tornatore et al. (2007). It is part of a larger set of cosmological runs, which are presented and discussed in detail elsewhere (Tornatore et al. 2010, in preparation). The
periodic simulation box has a comoving size of $75 {\rm h^{-1}} {\rm Mpc}$ and contains $512^3$ DM particles and initially, an equal number of gas particles.  As such, the masses of the DM and gas particles are $m_{\rm DM}\simeq 1.7\times 10^8\,{\rm h^{-1}}\, {\rm M}_\odot$ and $m_{\rm gas}\simeq 4.1\times 10^7\,{\rm h^{-1}}\, {\rm M}_\odot$, respectively. The
Plummer--equivalent softening length for the gravitational force is set to $\epsilon_{\rm Pl}=2.5\, {\rm h^{-1}}$ kpc, kept fixed in physical units from $z=2$ to $z=0$, while being $\epsilon_{\rm Pl}=7.5\, {\rm h^{-1}}$ kpc in comoving units at higher redshift. The value of the softening parameter of the SPH kernel for the computation of hydrodynamic forces is allowed to drop at most to half of the the gravitational softening.

The run assumes a metallicity-dependent radiative cooling (Sutherland \& Dopita 1993) and a uniform redshift-dependent UVB produced by quasars and galaxies as given by Haardt \& Madau (1996). The code also includes an effective model to describe star formation from a multi-phase ISM and a prescription for galactic winds triggered by SN explosions (see Springel \& Hernquist 2003 for a detailed description). In their model, star formation occurs due to collapse of condensed clouds embedded in an ambient hot gas. Stars with mass larger then 8${\rm M_\odot}$ explode as supernovae and inject energy back into the ISM. These interlinked processes of star formation, cloud evaporation due to supernovae, and cloud growth caused by cooling, lead to self-regulated star formation. Galactic winds are assumed to have a fixed velocity of 500 km s$^{-1}$, with a mass upload rate equal to twice the local star formation rate. The code includes the description of chemical enrichment given in Tornatore et al. (2007). Metals are produced by Type II SN (SNII), Type I SN (SNIa) and intermediate and low-mass stars in the asymptotic giant branch (AGB).  We assume SNII arise from stars having mass above $8\,{\rm M_\odot}$. As for SNIa, we assume their progenitors to be binary systems, whose total mass lies in the range (3--16)$\,{\rm M_\odot}$. The relative number of stars of different mass is computed for this simulation by assuming the Salpeter (1955) IMF between 1 and 100 ${\rm M_\odot}$ such that the number of stars with masses between $M$ and $M +dM$ is given by $n_*(M) dM \propto M^{-2.35}$. Metals and energy are released by stars of different masses by properly accounting for mass--dependent lifetimes. In this work we assume the lifetime function proposed by Padovani \& Matteucci (1993).  We adopt the metallicity--dependent stellar yields from Woosley \& Weaver (1995) and the yields for AGB and SNIa from van den Hoek \& Groenewegen (1997).

As for the identification of galaxies, they are recognized as gravitationally bound groups of star particles.  For each analyzed snapshot we first run a standard friends-of-friends (FOF) algorithm with a linking length of 0.2 in units of the mean particle separation. Each FOF group is then decomposed into a set of disjoint substructures, which are identified by the SUBFIND algorithm (Springel et al. 2001) as locally overdense regions in the density field of the
background main halo. After performing a gravitational unbinding procedure, only sub-halos with at least 20 bound particles are considered to be genuine structures (see Saro et al. 2006, for further details). For each ``bona-fide'' galaxy, we compute the mass-weighted age\footnote{This method tends to slightly bias the age towards larger values. On the other hand numerical resolution limits the ability to resolve the smallest halos harboring the oldest stars.}, 
the total halo/stellar/gas mass, the SFR, the mass weighted gas/stellar metallicity, the mass-weighted gas temperature
and the half mass radius of the dark matter halo.

We compute the Ly$\alpha$ emission and spectral properties for all the structures identified as galaxies in the simulation boxes at the redshifts of interest ($z\sim 5.7, 6.6, 7.6$). Obviously, not all these galaxies will be necessarily classified as LAEs.

\subsection{The intrinsic and observed luminosities }
\label{intrinsic_lum_ch3}
We start by summarizing the main features of the model used to obtain the intrinsic Ly$\alpha$ and continuum luminosities and their transformation to the observed luminosity values. Complete details of these calculations can be found in Secs. \ref{intrinsic_lum} and \ref{observed_lum}.

Star formation in galaxies gives rise to continuum and \HI ionizing photons, of which the the latter ionize the ISM. Due to the high density of the ISM, recombinations take place on a short timescale and this gives rise to a Ly$\alpha$ emission line. The intrinsic UV continuum and Ly$\alpha$ luminosity depend on the galaxy properties including the IMF, SFR ($\dot M_*$), stellar metallicity ($Z_*$) and age ($t_*$); all these quantities are taken from the simulation outputs, as discussed above. We then use {\rm STARBURST99}, (Leitherer et al. 1999), to obtain the intrinsic emission spectrum for each of the galaxies in the simulation boxes.  This spectrum is used to calculate the values of $L_\alpha^{int}$ and $L_c^{int}$ (in a band between 1250-1500 \AA, centered at 1375 \AA), as explained in Sec. \ref{intrinsic_lum}.

From our model, for a galaxy with $t_*=200 \, {\rm Myr}$, $Z_* = 0.2 Z_\odot$, $ \dot M_* = 1 {\rm M_\odot} \, {\rm yr^{-1}}$, $Q = 10^{53.47} {\rm s^{-1}}$ and the corresponding value of $L_\alpha^{int}=3.25 \times 10^{42} \,{\rm erg \, s^{-1}}$. This is very consistent with the value of $3.3 \times 10^{42} \,{\rm erg \, s^{-1}}$ shown in Tab. 4  of Schaerer (2003) for similar values of the age, metallicity and IMF. For the same galaxy, $L_c^{int} = 3.5 \times 10^{40} \,{\rm erg \, s^{-1}\AA^{-1}}$, which yields a value of $EW^{int} \sim 93$ \AA.

The doppler-broadened Ly$\alpha$ line, $L_\alpha^{em}$, that emerges out of the galaxy is calculated using Eqns. \ref{rot vel2}, \ref{lya_emm2}, while the continuum luminosity that emerges out of the galaxy, $L_c^{em}$, is calculated using Eq. \ref{lc_emm2}.

We use the value of the neutral hydrogen fraction, $\chi_{HI}$, (see Eq. \ref{define_chi}) obtained by Gallerani et al. (2008) from the ERM, to calculate the UVB photoionization rate, $\Gamma_B$ (see Eqns. \ref{ion_rec}, \ref{gamma_b2}) and the Str\"omgren region volume, $V_I$, built by each galaxy around itself (Eq. \ref{cal_rs2}). We then use Eq. \ref{gamma_clus} that accounts for clustered sources (explained in Sec. \ref{clus_ch3} that follows), to obtain the total photoionization rate $\Gamma_{T}$ seen by any galaxy, thus obtaining $\chi_{HI}$ (Eq. \ref{chi_r}) as a function of distance from the galaxy within the  Str\"omgren sphere, assuming a homogeneous and isotropic IGM density field. These equations are then used to calculate the IGM Ly$\alpha$ transmission, $T_\alpha$, as explained in Eqns. \ref{tau_alpha2} to \ref{lorentz2}. The observed values of Ly$\alpha$ and $L_c$ are then obtained as shown in Eqns. \ref{la_obs2} and \ref{lc_obs2}, respectively.

We digress here to stress the uncertainties in our knowledge of $f_{esc}$; it is clearly seen that the escape fraction of \HI ionizing photons, $f_{esc}$ affect $L_\alpha^{em}$ (Eq. \ref{lya_emm2}), $V_I$ (Eq. \ref{cal_rs2}) and $\Gamma_{T}$ (Eq. \ref{gamma_clus}). However, estimating $f_{esc}$ proves to be somewhat of a challenge; though an enormous amount of work has been put into its calculation by various theoretical and observational groups, its value remains largely debated between $f_{esc}=0.01-0.8$ ($0.01-0.73$) theoretically (numerically), as discussed in Sec. \ref{cosmic_reio_ch1}. Since modelling $f_{esc}$ is much beyond the scope of this thesis, we adopt a fiducial value of $f_{esc}=0.02$, based on the results of Gnedin et al. (2008), as a reasonable best estimate of $f_{esc}$ for all the galaxies in the simulation boxes at $z\sim5.7,6.6,7.6$.

\subsection{The effects of clustering}
\label{clus_ch3}
One of the most important drawbacks of modelling LAEs semi-analytically is that only an average value of the UVB photoionization rate, $\Gamma_B$, is usually assumed, making any inference on the additional contribution, $\Gamma_G$, due to galaxy clustering not possible. As McQuinn et al. (2007), have shown, the observed clustering of LAEs is increased by transmission through a patchy IGM as expected during reionization. In addition to being an important tool to probe reionization, clustering will also leave imprints on the Ly$\alpha$ LF.

Thanks to our cosmological SPH simulations, we are able to include the effect of clustering on the Ly$\alpha$ LF and transmission. Clustering of galaxies implies that the mean separation could become smaller than the typical size of their ionized regions. Therefore, more than one galaxy affects the size of the \HII region and the neutral hydrogen profile within it, thereby increasing the visibility of individual galaxies.

\begin{figure*}[htb]
  \center{\includegraphics[scale=0.5]{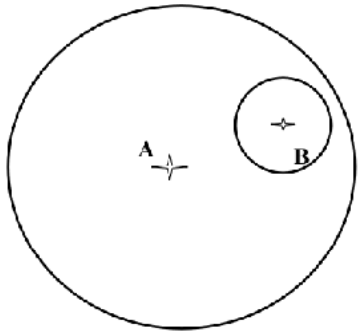}} 
  \caption{A schematic diagram to elucidate the effects of clustering on galaxy visibility (see text). The circles indicate the boundaries of individual \HII regions.}
\label{zch3_sch} 
\end{figure*}

Consider for example the case of a luminous galaxy, A, and a fainter one, B (see Fig. \ref{zch3_sch}). Each one of them will carve an \HII region in the IGM, whose size depends on their luminosity, as shown in the Figure. Depending on the IGM hydrogen neutral fraction, none, one or both galaxies can be seen individually. This is because, to a first approximation, the spatial scale imposed by the Gunn-Peterson damping wing on the size of the \HII region corresponds to a redshift separation of $\Delta z \approx 0.01$, i.e. about 200 kpc (physical) at $z=10$ (Miralda-Escud\'e 1998). However, when the mean separation between galaxies is smaller than the size of the smallest \HII region (i.e. galaxies are considerably clustered) then it can occur that both galaxies can be observed as LAEs due to the increased \HII region size and decreased \HI fraction within it boosting the transmissivity. The importance of such clustering effect on LAE visibility is stronger in the initial reionization phases, when the gas is almost neutral, implying that only clustered galaxies would be visible under those circumstances.

To quantify this effect from simulations, we use a post-processing technique in which we start by calculating the size of the ionized region around each galaxy as explained in Eq. \ref{cal_rs2}, Sec. \ref{observed_lum}. Any two galaxies that are separated by a distance smaller than either of their Str\"omgren radii are then treated as a local enhancement of the photoionization rate for the other, i.e., as shown in Fig. \ref{zch3_sch}, for galaxy A, we add an extra contribution, $\Gamma_G$, evaluated as the local value of the photoionization rate of galaxy B at the position of galaxy A. The same procedure is followed when considering the transmissivity of galaxy B. More generally, the total ionization rate, $\Gamma_T$, seen at the position of galaxy $j$, whose separation from $N$ other galaxies is smaller than either of the Str\"omgren radii can be expressed as
\begin{equation}
\Gamma_{T,j} (r) =  \Gamma_{BL}(r) + \sum_{i=1, i \neq j} ^ N \int_{\nu_L}^\infty \frac{L_{\nu,i}^{em}}{4 \pi r_{ij}^2} \frac{\sigma_L}{h \nu} \bigg(\frac{\nu_L}{\nu}\bigg)^3  d\nu, 
\label{gamma_clus}
\end{equation}
where the third term on the right hand side represents $\Gamma_G$ and $\Gamma_{BL}$ is calculated in the same way as explained in Sec. \ref{observed_lum} (see Eqns. \ref{gamma_bl}, \ref{gamma_blr2}). Further, $L_{\nu,i}^{em}= L_{\nu,i}^{int} \,f_{esc}$ is the ionizing luminosity emerging from the $i^{th}$ galaxy and $r_{ij}$ is the radial distance between galaxies $i$ and $j$.

As a caveat, we point out that we are calculating clustering effects assuming that the sizes of the \HII regions correspond to sources embedded in the IGM at the mean
ionization fraction given by the ERM. However, prior to complete overlap, the ionization field is very patchy and galaxies are more likely to be immersed in either an almost neutral or a highly ionized region. Because the average $\chi_{HI}=0.16$, it means that $1-\chi_{HI}=84$\% of the volume is substantially ionized at $z=7.6$. Hence, in general, our method provides a lower limit to the number of detectable LAEs. Several authors (Zahn et al. 2007; Mesinger \& Furlanetto 2007; Geil \& Wyithe 2008) have presented schemes that avoid detailed radiative transfer (RT) calculations and still provide ionization schemes in good agreement with simulations. A precise calculation of clustering effects can also be done by properly following RT in detail, an approach which we use later in Chapter \ref{ch5_lya_rt}.

To summarize, the boost in the ionization background imparted by clustering is important for all galaxies when the IGM is close to neutral. The importance of this effect decreases as reionization proceeds, however, it decreases faster for more luminous galaxies.

\section{Comparison with observations}
\label{results_ch3}

Once the model presented above has been implemented, we are in a position to compare the model results with observations. In particular, we compare the calculated UV and Ly$\alpha$ LFs to the data obtained by Shimasaku et al. (2006) ($z\sim5.7$) and Kashikawa et al. (2006) ($z\sim6.6$), and make predictions for the LFs at $z \sim 7.6$. In addition, we also explore the effects of clustering on Ly$\alpha$ transmission and present synthetic SEDs to match with Lai et al. (2007) observations at $z\sim5.7$. 

\subsection{The Ly$\alpha$ LF and the effects of clustering}
\label{lyalf_ch3}

We use the procedure explained in Secs. \ref{intrinsic_lum_ch3} and \ref{clus_ch3} to calculate the observed Ly$\alpha$ and continuum luminosity for each of the galaxies in the simulation boxes. Galaxies with (a) an observed Ly$\alpha$ luminosity in the currently observable range, $L_\alpha \geq 10^{42.2} \, {\rm erg \, s^{-1}}$ and (b) a value of the observed EW, $EW>20$ \AA\, are then identified as LAEs, which are used to build the cumulative Ly$\alpha$ LF to compare to observations.

\begin{figure}[htb]
  \center{\includegraphics[scale=0.5]{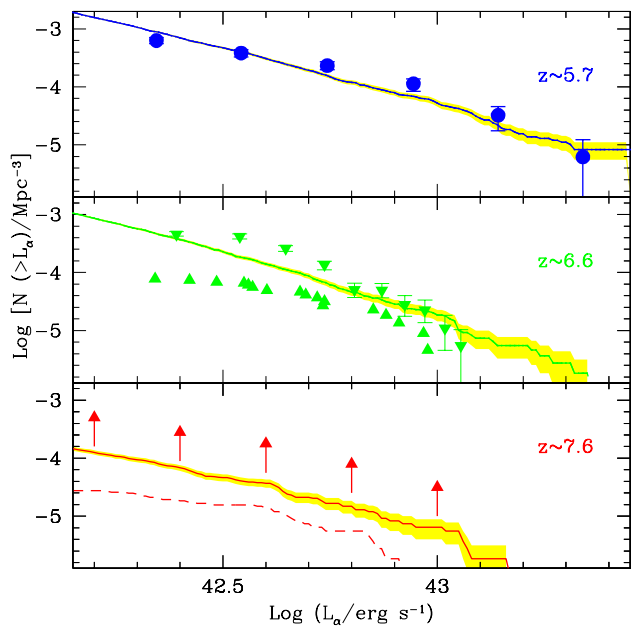}} 
\caption{Cumulative Ly$\alpha$ LF for the ERM. The panels are for $z\sim 5.7$, $6.6$ and $7.6$ from top to bottom. Points represent the data at two different redshifts:  $z\sim 5.7$ (Shimasaku et al. 2006) (circles) and $z \sim 6.6$ (Kashikawa et al. 2006) with downward (upward) triangles showing the upper (lower) limits. Solid (dashed) lines in each panel refer to model predictions at $z\sim 5.7,6.6,7.6$ for the parameter values in Tab. \ref{table1_ch3} including (excluding) clustering effects. Shaded regions in all panels show poissonian errors. Though curves with/without clustering effects are indistinguishable at $z=5.6, 6.6$, the difference is appreciable at $z=7.6$. Because radiative transfer is not accounted for in the simulations, at $z=7.6$ our results represent lower limits on the number density of observed objects.}
\label{zch3_lya3} 
\end{figure}   

The number of objects identified as LAEs from our simulations are (1696, 929, 136) at $z\sim (5.7, 6.6, 7.6)$; since the definition of LAE is an operational one based on the observed Ly$\alpha$ luminosity, the number depends on the adopted values of $f_\alpha$ and $T_\alpha$. The values given above are for the best fit parameters shown in Tab. \ref{table1_ch3} and Fig. \ref{zch3_lya3}. 

\begin{table*} 
\begin{center} 
\begin{tabular}{|c|c|c|c|c|c|c|c|}
\hline 
$z$& $\Gamma_B$ & $\chi_{HI}$ &  $\langle B \rangle$ & $\langle T_\alpha \rangle$ &$f_\alpha$&$f_c$&$E(B-V)$\\  
$$& $(10^{-12} {\rm s^{-1}})$ & $$ &  $$ & $$ &$$ &$$ &$$\\  
\hline 
$5.7$& $0.47$ &$6 \times 10^{-5}$ & $1.04$ & $0.49$& $0.3$ & $0.22$ & $0.15$\\
$6.6$& $0.19$ & $2.3 \times 10^{-4}$ & $1.05$ & $0.49$ &$0.3$ & $0.37$ & $0.1$\\
$7.6$& $0.28\times 10^{-3}$ &$0.16$ & $58.5$ & $0.42$ & $0.3$ & $0.37$ & $0.1$\\
  \hline
\end{tabular} 
\end{center}
\caption {Best fit parameter values to match the Ly$\alpha$ and UV LFs in the ERM, including clustering effects. For each redshift (col. 1), we report the UVB photoionization rate (col. 2), the fraction of neutral hydrogen corresponding to the UVB photoionization rate (col. 3), the average value of the boost parameter (col. 4), the average transmission of the Ly$\alpha$ luminosity (col. 5), the value of the the escape fraction of Ly$\alpha$ photons (col. 6), the escape fraction of UV continuum photons (col. 7) and the color excess calculated using the supernova dust extinction curve (col. 8).}
\label{table1_ch3} 
\end{table*}

We find that, independent of clustering, to match the data at both $z\sim5.7$ and $6.6$, only a certain fraction ($f_\alpha = 0.3$) of the Ly$\alpha$ luminosity {\it must emerge out of each galaxy, unabsorbed by dust within the ISM}. As shown in the uppermost panel of Fig. \ref{zch3_lya3}, within error bars, the theoretical LF nicely matches the data at $z\sim 5.7$. 

With the same value of $f_\alpha=0.3$ at $z \sim 6.6$, the theoretical LF lies close to the upper bound of the data as shown in the central panel. Since the dust attenuation of Ly$\alpha$ photons does not evolve between $z \sim 6.6$ and $5.7$, it is reasonable to assume that the same value also holds at $z \sim 7.6$ (the cosmic time between $z \sim 7.6$ and $6.6$ is only about $0.1$ Gyr, which might be too small for any significant dust evolution) and then use it to predict the LF (bottom panel of Fig. \ref{zch3_lya3}). 

Fig. \ref{zch3_lya3} provides interesting information on the effects of clustering as well: from the overlapping of the LFs including/excluding clustering at $z \sim 5.7$ and $6.6$, we conclude that these are negligible at these epochs. Clustering of sources, however, plays a key role at $z \sim 7.6$. If clustering effects are neglected, very few objects (about 30) would be luminous enough to be detectable in current Ly$\alpha$ surveys; instead, the luminosity boost due to clustered sources leads to about 136 objects  to become visible in our simulation volume. We reiterate that, because we do not follow the RT of ionizing radiation through the IGM, the results presented in Fig. \ref{zch3_lya3} (and in Fig. \ref{zch3_uvlf3}) at $z=7.6$ must be seen as lower limits to the actual number density of LAEs.    

We pause to discuss an issue concerning the LFs at $z\sim 6.6$ and 7.6. We start by comparing the model predictions and the observations at $z \sim 6.6$ where the predictions for the faint end of the LF lie between the upper and lower limits set by the observational data. Although this could be regarded as a success of the model, it is safe to discuss if physics not included in our model or biases in the data might spoil this agreement. While the lower bound of the data is made up of spectroscopically confirmed LAEs, the upper bound is a photometric sample composed of all galaxies identified as LAE candidates at this redshift. Spectroscopic analysis of LAE candidates on the upper bound of the LF could then possibly rule out a number of them as being low-$z$ interlopers and contaminants, thus bringing the upper limit in agreement with our curve. This would require that up to 60\% of the candidates might not be confirmed. Another possibility is that most of the faint objects are not individual galaxies but unresolved groups, a point made by Mori \& Umemura (2006), who have shown that a number of small galaxies undergoing mergers can be identified as a single LAE at high redshifts. At $z \sim 6.6$, the resolution of the data is about $1''$, which corresponds to a physical separation of $5.4$ kpc. Although in our simulations, we do not find any objects that are separated by such small distances, this might also be due to insufficient resolution on scales of about 5 kpc. In this work, we discard this possibility, since, although a small number of such pairs might be found with higher resolution, they would probably not be enough to boost up the low luminosity end of the LF substantially. 

We now briefly discuss the predictions from our model to recently obtained data at $z \sim 7.7$. Hibon et al. (2010) have photometrically detected 7 possible LAE candidates at $z \sim 7.7$ using the WIRCam (Wide Field near-IR Camera) on the CFHT. Tilvi et al. (2010) have detected 4 more candidates using the NEWFIRM (National optical astronomy observatory Extremely Wide Field IR Mosaic) imager and obtained the Ly$\alpha$ LF, which they claim to be the upper limit at $z \sim 7.7$. It is very encouraging that the slope of the curve predicted by our model (including clustering) at $z \sim 7.7$ matches very well to the slope observed by Hibon et al. (2010), although the amplitude of the observed LF lies above the predicted one by a factor of about 4. 
Some of the most plausible explanation for this discrepancy are: (a) spectroscopic confirmation would rule out a number of the candidates as LAEs, thereby shifting the observed LF down towards the predicted one, (b) as mentioned before, our results represent lower limits to the observed objects; including a full RT calculation could boost up the LF to bring it into accordance with the observations, (c) the dust content of galaxies might be much lower at $z \sim 7.7$ as compared to that at $z \sim 6.6$; the resulting increase in $f_\alpha$ could boost up the predicted LF. However, as yet unavailable, $z \sim 7.7$ Ly$\alpha$ and UV LFs comprising spectroscopically confirmed LAEs are required, in order for theoretical models to distinguish between these different explanations.

\begin{figure}[htb]
  \center{\includegraphics[scale=0.5]{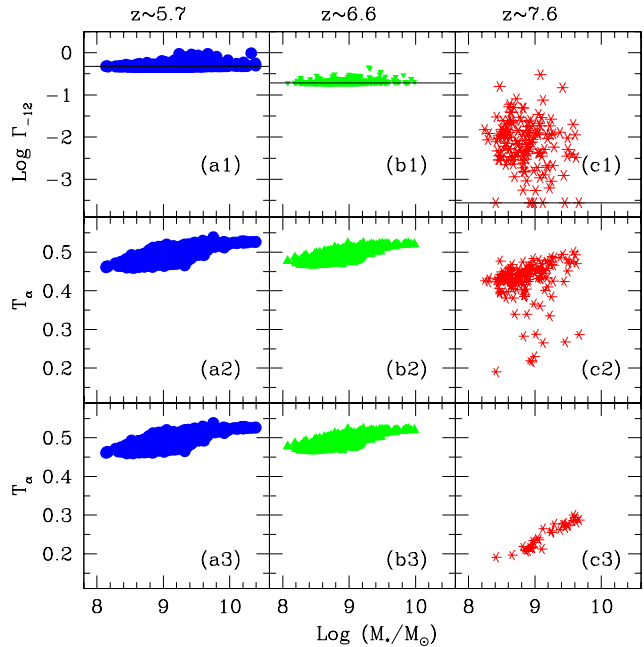}} 
  \caption{{\it Upper row}: Photoionization rates as a function of LAE stellar mass, $M_*$. The horizontal line corresponds to the contribution of the UVB, $\Gamma_B$; symbols in each panel denote the value of $\Gamma_B+\Gamma_G$. {\it Middle}: Ly$\alpha$ transmissivity, $T_\alpha$ including clustering effects. {\it Lower}: Ly$\alpha$ transmissivity without clustering effects. Columns refer to $z\sim 5.7,6.6,7.6$ as indicated.}
\label{zch3_clus} 
\end{figure}

We now come back to clustering. We start by defining the boost parameter, $\langle B \rangle$, such that $\langle B \rangle =  1 + \langle \Gamma_G \rangle / \Gamma_B$, where 
\be
\langle \Gamma_G \rangle = (1/N) \sum_{i=1}^N \Gamma_G(i),
\ee
and $\Gamma_G(i)$ is the photoionization boost due to clustered LAEs seen by the $i^{th}$ LAE of the total $N$ LAEs at the redshift considered. The reason why the effects of clustering become important at $z\sim 7.6$ is easily explained using Fig. \ref{zch3_clus}. First, from the uppermost three panels (a1, b1, c1), it is seen that the value of $\Gamma_B$ is very low at $z \sim 7.6$ ($2.8\times 10^{-16}$ s$^{-1}$). Between $z \sim 7.6$ and $6.6$, it increases rapidly by about 3 orders of magnitude such that at $z \sim 6.6$, $\Gamma_B = 1.9 \times 10^{-13}$ s$^{-1}$. Afterwards, the ionization rate increases only by a factor of about $2.5$ between $z \sim 6.6$ and $5.7$. Second, at $z \sim 5.7, 6.6$ the value of $\langle B \rangle$ is less than a factor of 1.5, while at $z \sim 7.6$, $\langle B \rangle \approx 58$, i.e. the photoionization rate is strongly dominated by the local emission from the clustered LAEs. Since the IGM is already highly ionized at $z < 7$ in the ERM, the extra local contribution from clustered LAEs does not affect the Ly$\alpha$ transmission sensibly, as seen from the comparison of panels (a2,a3) and (b2,b3) of Fig. \ref{zch3_clus}. However, at $z \sim 7.6$, the effects of clustering on the transmissivity, $T_\alpha$, are dramatic. The photoionization rate boost due to clustered LAEs makes the IGM transparent enough that about 136 in our simulation volume become visible as compared to 30 that would be detected as LAEs in the absence of clustering effects. This is clear from the comparison of panels (c2) and (c3), from which we conclude that the transmissivity is increased up to values of $T_\alpha=0.2-0.5$ when clustering is included in the computation as compared to $T_\alpha=0.2-0.3$ when it is not.

A few additional points are worth mentioning here: (a) on a galaxy to galaxy basis, it is not necessarily the environments of the most massive galaxies that experience the largest enhancement of the photoionization rate due to clustering as seen from the upper three panels of Fig. \ref{zch3_clus}. This shows that the boost in photoionization rate seen by a small galaxy within the \HII region of a more massive one is greater than or comparable to the boost seen by a massive galaxy due to the contribution from a large number of small galaxies embedded in its \HII region. (b) If $\chi_{HI}\ll 1$, both including/excluding the contribution of $\Gamma_G$, $T_\alpha$ increases with $M_*$ and hence with star formation rate (panels a2, a3, b2, b3). This leads us to conclude that for a highly ionized IGM, the contribution of $\Gamma_L (\propto \dot M_*)$ always dominates over that of $\Gamma_G$. (c) If $\chi_{HI} \approx 1$, $T_\alpha$ still increases with the star formation rate both including/excluding $\Gamma_G$, although the scatter is much larger in the case including $\Gamma_G$. This is because excluding $\Gamma_G$, very few galaxies are luminous enough to transmit enough of the Ly$\alpha$ luminosity to be visible as LAEs (panel c3). However, including $\Gamma_G$ dramatically increases the $T_\alpha$ such that about 4 times as many galaxies become visible (panel c2). Hence, we conclude that in this case, $\Gamma_G$ contributes significantly in making the environment around LAEs more transparent to Ly$\alpha$ photons.


\subsection{The UV LF}
\label{uvlf_ch3}

\begin{figure}[htb]
  \center{\includegraphics[scale=0.5]{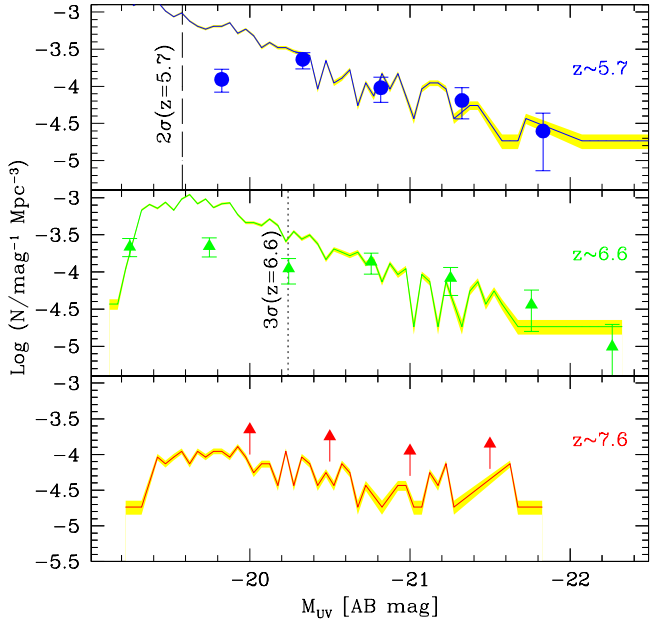}} 
  \caption{UV LAE LF for the ERM. Points represent the data at two different redshifts: $z \sim 5.7$ (Shimasaku et al. 2006) (circles), $z \sim 6.6$ (Kashikawa et al. 2006) (triangles). Lines refer to model predictions including clustering at the redshifts (from top to bottom): $z \sim 5.7,6.6,7.6$, for the parameter values in Tab. \ref{table1_ch3}. The vertical dashed (dotted) lines represent the observational 2$\sigma$ (3$\sigma$) limiting magnitudes for $z = 5.7$ ($z =6.6$). The shaded region in all panels shows the poissonian errors. Because radiative transfer is not accounted for in the simulations, at $z=7.6$ our
results represent lower limits to the observed objects.} 
\label{zch3_uvlf3} 
\end{figure}

For each galaxy identified as a LAE included in the Ly$\alpha$ LF, we calculate its observed continuum luminosity, $L_c$, as explained in Sec. \ref{observed_lum}. We bin the number of galaxies on the basis of the continuum luminosity magnitude and divide by the volume of the simulation box ($75^3 {\rm h^{-3}}$ comoving ${\rm Mpc^3}$) to obtain the UV LF for the LAEs identified in Sec. \ref{lyalf_ch3}. This is shown in Fig. \ref{zch3_uvlf3}. 

We find that to match to the observations, the value of $f_c$ for each LAE must decrease with decreasing redshift, going from $0.37$ at $z \sim 6.6$ to $0.22$ at $z \sim 5.7$, as shown in Tab. \ref{table1_ch3}. This means that while 37\% of all continuum photons escape any LAE at $z \sim 6.6$, only $22$\% escape at $z \sim 5.7$. We interpret this decreasing $f_c$ to be the result of an increase in the dust content of the galaxies. To make predictions at $z \sim 7.6$, we again use the same value of $f_c$ as at $z \sim 6.6$, making the assumption that the dust content of LAEs does not evolve between these two redshifts. The values of $f_c$ for all the three redshifts are shown in Tab. \ref{table1_ch3}. 

An interesting point here is that at $z \sim (5.7, 6.6)$, $f_\alpha/f_c \sim (1.4, 0.8)$, which are in reasonable agreement to the values of $f_\alpha/f_c \sim (1.2, 0.6)$ (see Tab. \ref{table2_ch2}) obtained using the semi-analytic model presented in Chapter \ref{ch2_lya_sam}. Using the Supernova extinction curve, $f_\alpha/f_c \sim 0.8$ for a homogeneous distribution of dust, Bianchi \& Schneider (2007). However, no single extinction curve (Galactic, Small Magellanic Cloud or Supernova) can give a value of $f_\alpha/f_c >1$. The relative damping at $z \sim 5.7$ can only be explained by an inhomogeneous two-phase ISM model in which clumped dust is embedded in a hot ISM, as proposed by Neufeld (1991). The data thus, seem to hint at the fact that the dust distribution in the ISM of LAEs becomes progressively inhomogeneous/clumped with decreasing redshift.

We also calculate the color excess $E(B-V)$ ($ = A_v/R_v$) for LAEs at each of the redshifts as a sanity check. Using the supernova dust extinction curve (Bianchi \& Schneider 2007), we calculate $A_v = A_\lambda (1375\, {\rm \AA}) /  5.38 $ and $R_v=2.06$. The color excess is then calculated as
\begin{equation}
E(B-V)=\frac{A_\lambda (1375\,{\rm \AA})}{5.38 R_v} = - \frac{2.5 \log_{10} f_c}{5.38 R_v}.  \\
\label{colorexs_sn_ch3}
\end{equation}
With the above formulation, we find that $E(B-V) \sim 0.15$ at $z \sim 5.7$, $E(B-V) \sim 0.1$ at $z \sim 6.6$. 

When expressed in terms of the widely used Calzetti extinction law (Calzetti et al. 2000), we again find $E(B-V)=0.15, 0.1$ at $z \sim 5.7,6.6$. The inferred color excess value at $z \sim 5.7$ is in very good agreement with recent experimental determinations: by fitting the SEDs of 3 LAEs at $z = 5.7$, Lai et al. (2007) have inferred $E(B-V) < 0.225-0.425$; in a sample of 12 LAEs at $z = 4.5$, Finkelstein et al. (2009a) have found $E(B-V) = 0.04-0.37$; using 3 galaxies at $z = 4-5.76$, Pirzkal et al. (2007) have found $E(B-V)=0.012-0.15$. Further, using SPH simulations similar to those used here, Nagamine et al. (2008) have inferred a value of $E(B-V)\sim 0.15$.

\subsection{The observed SEDs}

\begin{figure*}[htb] 
  \center{\includegraphics[scale=0.93]{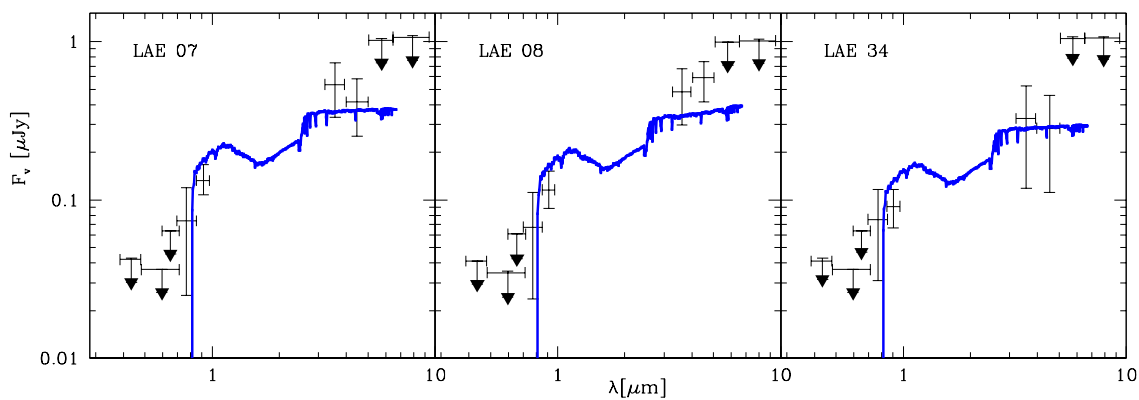}} 
  \caption{Comparison of theoretical SEDs (lines) with observations (points) for LAE \#07, \#08 and \#34,
(from left to right) from Lai et al. (2007). Points with downward pointing arrows represent the $3 \sigma$ upper limits of the data.}
\label{zch3_spec} 
\end{figure*}

Lai et al. (2007) have observed the spectra for three LAEs (\#07, \#08 and \#34) at $z \sim 5.7$ with observed Ly$\alpha$ luminosities, $L_\alpha = (4.9,  4.3, 3.6) \times 10^{42}\, {\rm erg\, s^{-1}}$ (K. Lai, private communication). From amongst the galaxies we identify as LAEs at $z\sim 5.7$, we select those three galaxies whose observed Ly$\alpha$ luminosities match most closely with the three values observed. The spectra of each of the selected galaxies (obtained using {\rm STARBURST99}, as explained in Sec. \ref{intrinsic_lum_ch3}) is then attenuated using the SN dust extinction curve and  $E(B-V)=0.15$, to produce synthetic spectra which are shown in Fig. \ref{zch3_spec}. 

\label{sed_ch3}

As shown by Lai et al. (2007), the spectra can be well fit by different kinds of stellar populations, varying the age, metallicity and color excess. However, we have no free parameters since the values of $t_*, \dot M_*$ and $Z_*$ for each of the galaxies is obtained from the simulation outputs and the color excess value is obtained by comparing the theoretical UV LF to the observed one at $z \sim 5.7$ (Shimasaku et al. 2006) as mentioned above in Sec. \ref{uvlf_ch3}. As seen, a remarkable agreement is found between the synthetic and observed SEDs. Even though we are using a small and biased data set of LAEs selected based on the IRAC $3.6 \mu m$ and $4.5 \mu m$ detections, the agreement between the synthetic and observed SEDs provides a strong consistency test of our model.

We briefly mention the physical properties of the LAEs whose synthetic spectra are shown in Fig. \ref{zch3_spec}. The stellar ages  are 182--220 Myr for the three LAEs (see also Tab. \ref{table2_ch3}). Hence, these LAEs are intermediate age objects, rather than being very old ($t_* \sim 700$ Myr) or very young ($t_* \sim 5$ Myr). Their stellar metallicities are about $0.2-0.3\, Z_\odot$, and the SFR are between 7-10 ${\rm M_\odot} {\rm yr^{-1}}$. Further, the SEDs for all the three LAEs are well reproduced by a single value of the color excess which shows that all these objects possibly contain similar amounts of dust in the ISM.

\begin{table} 
\begin{center} 
\begin{tabular}{|c|c|c|c|c|} 
\hline 
$\# {\rm LAE} $ & $t_* $ & $Z_* $ & $\dot M_* $ & ${\rm E(B-V)}$  \\  
$ $ & ${\rm (Myr)}$ & $ (Z_\odot)$ & $ ({\rm M_\odot} {\rm yr^{-1}})$ & $$  \\  
\hline
$07$ & $191$ & $0.23$ & $9.7$ & $0.15$ \\
$08$ & $182$ & $0.32$ & $9.6$ & $0.15$ \\ 
$34$ & $220$ & $0.23$ & $7.3$ & $0.15$ \\  
 \hline
\end{tabular} 
\end{center} 
\caption {Physical properties of the LAEs from the simulation which match the observed Ly$\alpha$ luminosities most closely. For each LAE observed by Lai et al. (2007) (col. 1), we show the age (col. 2), stellar metallicity (col. 3), SFR (col. 4) and color excess (col. 5) for the corresponding LAE from our simulation.}
\label{table2_ch3}
\end{table}

\section{The nature of LAEs}
\label{nature_ch3}

Having selected the galaxies in the simulated volume that would be experimentally defined as LAEs, and having shown that their LFs and SED are consistent with the observed data, we can take a further step and quantify the physical properties of these objects. This includes the relation between stellar and halo mass, age, metallicity, star formation rate and evolution. We will present these results up to $z \sim 7.6$; however it must be kept in mind that at that redshift, only 136 galaxies are bright enough to be observed using standard narrow-band techniques which makes the scatter much larger as compared to that at lower redshifts.   

\begin{figure}[htb]
\center{\includegraphics[scale=0.5]{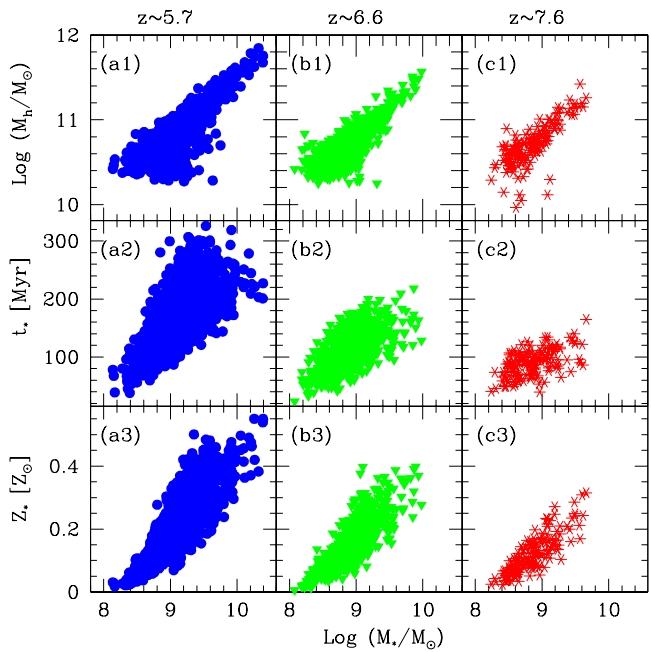}} 
\caption{Physical properties of LAEs at $z\sim 5.7$, $z \sim 6.6$ and $z \sim 7.6$ as referred to by columns. As a function of stellar mass, $M_*$, we show  (a) the halo mass, $M_h$, (a1-c1), (b)  mass-weighted stellar ages, $t_*$, (a2-c2), and (c) mass-weighted stellar 
metallicity, $Z_*$, (a3-c3).}
\label{zch3_phy1} 
\end{figure}  

\begin{table*} 
\begin{center} 
\begin{tabular}{|c|c|c|c|c|c|c|c|} 
\hline 
$z$ & $t_* $ & $\langle Z_* \rangle$ & $\langle \dot M_* \rangle $ & $\langle EW^{int}\rangle$ & $\langle EW \rangle $ & $\langle \cal I \rangle$ & $\dot\rho_* $  \\ 
$$ & $({\rm Myr})$ & $(Z_\odot)$ & $({\rm M_\odot} {\rm yr^{-1}})$ & $({\rm\AA})$ & $({\rm\AA})$ & $$ & $({\rm M_\odot} {\rm yr^{-1} Mpc^{-3}})$  \\ 
\hline
$5.7$ & $38-326$ & $0.22$ & $6.9$ & $94.3$ & $63.5$ & $0.72$ & $1.1 \times 10^{-2}$ \\
$6.6$ & $23-218$ & $0.15$ & $5.7$ & $104.1$ & $41.7$ & $0.72$ & $4.9 \times 10^{-3}$ \\ 
$7.6$ & $39-165$ & $0.12$ & $7.7$ & $108.9$ & $37.8$ & $0.78$ & $9.6 \times 10^{-4}$\\  
 \hline
\end{tabular} 
\end{center}
\caption {For all the LAEs comprising the Ly$\alpha$ LF including clustering effects, at the redshifts shown (col. 1), we show the range of mass weighted ages (col. 2), the average stellar metallicity (col. 3), the average SFR (col. 4), the average intrinsic and observed EWs (col. 5, 6), the average value of the star formation indicator (col. 7) and the SFR density (col. 8) obtained from our simulation.}
\label{table3_ch3} 
\end{table*} 

\subsection{The DM halo mass}
\label{mh_ch3}
LAEs are characterized by dark matter halo masses in the range $10^{10.2-11.8} M_\odot$ at $=5.7$, corresponding to $> 2\sigma$ fluctuations at all redshifts. As seen from Fig. \ref{zch3_phy1} (panels a1, b1, c1), this range becomes progressively narrower at earlier epochs because of two occurrences:
(i) the mass function of halos shifts to lower masses in hierarchical structure formation models; and (ii) smaller halos progressively become invisible at higher redshifts as their stellar mass and hence luminosity is too low to be detected (we recall that all the objects analyzed are part of the LFs shown in Figs. \ref{zch3_lya3} and \ref{zch3_uvlf3}).

\subsection{The stellar mass}
\label{ms_ch3}

In spite of the narrowing of the LAE DM mass range with increasing redshift, there is almost no evidence of an evolution of the stellar to halo relation in the redshift range under examination, as seen from Fig. \ref{zch3_phy1} (panels a1, b1, c1), and we find $M_* \propto M_h^{1.56}$, with the best fit given by the following expression: 
\begin{equation}
\log_{10} (M_h) = (0.64-0.06\Delta z) \log_{10} (M_*)  + (5.0+0.5\Delta z), 
\end{equation}
where $\Delta z = (z-5.7)$. The above relation implies that the stellar mass per unit halo mass is increasing towards larger systems, which is just a restatement of the well-known fact that the star formation is less efficient in small galaxies due to the inhibiting effects of mechanical feedback. The typical stellar masses of LAEs are $< 10^{10.5}$ at all redshifts, i.e. they are smaller than the Milky Way.

\subsection{The stellar ages}
\label{ages_ch3}

Although a considerable spread is present, on average larger systems tend to be older than small ones as shown in Fig. \ref{zch3_phy1} (panels a2, b2, c2). LAEs result from the merging of several sub-halo systems in which star formation was ignited long before. In Tab. \ref{table3_ch3}, we show that the range of ages reduces towards high redshifts from $38-326$ Myr at $z \sim 5.7$ to $39-165$ Myr at $z \sim 7.6$. These ages imply that the oldest stars in these systems formed already at $z=8.5$, thus during the reionization epoch. We reiterate that, because of finite numerical resolution, stars in halos below our resolution limit might have formed even before that epoch, so the previous value must be seen as a lower limit to the onset of star formation activity. From the previous discussion it is clear that LAEs are neither pristine nor very young objects forming their stars for the first time. 

\begin{figure}[htb]
  \center{\includegraphics[scale=0.5]{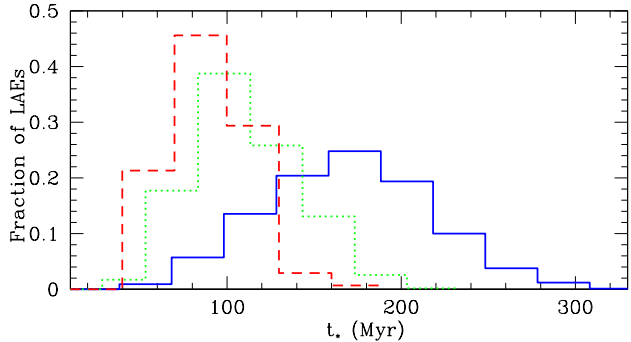}} 
  \caption{Normalized distribution of number of LAEs as a function of the mass weighted stellar age ($t_*$). The lines are for z: 5.7 (solid), 6.6 (dotted) and 7.6 (dashed).  }
\label{zch3_dist_age} 
\end{figure} 

Since the ages of LAE stellar populations are currently hotly debated, we digress briefly to discuss
our results in the framework of the various arguments given in the literature. For example,
Finkelstein et al. (2009a) find a bimodality in their sample of 14 LAEs at $z \sim 4.5$; their 
objects are either very young ($<15$ Myr) or very old ($>400$ Myr). Their results could be 
explained by invoking two different star formation modes in LAEs: a recent strong burst 
or a continuous SF in which the bulk of the population is dominated by old objects.
To better evaluate this possibility, we plot the normalized distribution 
of the mass-weighted stellar ages for all the LAEs identified in the simulation volume at 
$z\sim 5.7,6.6$ and $7.6$ in Fig. \ref{zch3_dist_age}. We do not find any such bimodality from our simulation. LAEs are distributed in age between 39-165 Myr at $z\sim 7.6$ and this range increases with 
decreasing redshift, as already mentioned above.  A possible explanation of the experimental
result could be the presence of dust. LAEs would be visible in the Ly$\alpha$ when 
the age $<15$ Myr so that not enough dust would have formed and at later times $> 400$ Myr 
when the galaxy would have destroyed/ removed most of its dust content. In the intermediate 
periods, the Ly$\alpha$ line would be attenuated below observable limits. 

\subsection{The stellar metallicities}
\label{mets_ch3}

We now return to our discussion of the physical properties of LAEs. The result that LAEs are neither pristine nor very young objects forming their stars for the first time, is confirmed by the range of metallicities found in these systems and shown in panels (a3, b3, c3) of Fig. \ref{zch3_phy1} (see also Tab. \ref{table3_ch3}). For LAEs, $Z_* \sim 0.02-0.55$ $Z_\odot$, with the mean over the sample decreasing with redshift; also, larger objects have higher metallicities as expected from their larger stellar masses. These results are broadly consistent with an enrichment predominantly caused by SNII, and therefore tightly following
the SFR of the galaxy; the increasing scatter seen at the lowest redshift hints at a larger contribution by SNIa. The mass-metallicity relation for LAEs is also an interesting outcome of our study. This can be conveniently expressed as 
\begin{equation}
Z / Z_\odot = (0.25 - 0.05\Delta z)\log_{10} (M_*)  - (2.0-0.3\Delta z),
\end{equation}
where $\Delta z = (z-5.7)$. Differently to the analogous relation observed at lower redshifts (Tremonti et al. 2004; Panter et al. 2008; Maiolino et al. 2008), we do not see the sign of a flattening of metallicity towards larger masses. As the flattening is usually interpreted as a result of a transition from a galactic wind-regulated metal
budget to a close-box evolution in which all metals are retained, we conclude that in our LAEs, winds play an important role because of their relatively low masses, a point already noticed above. 

\subsection{The star formation rates}
\label{sfr_ch3}

The SFR for LAEs are in the range from 2.5-120 ${\rm M_\odot}\,{\rm yr^{-1}}$, i.e. a sustained but not exceptionally large star formation activity, with large objects at later times being the most prominent star factories.

\begin{figure}[htb] 
\center{\includegraphics[scale=0.5]{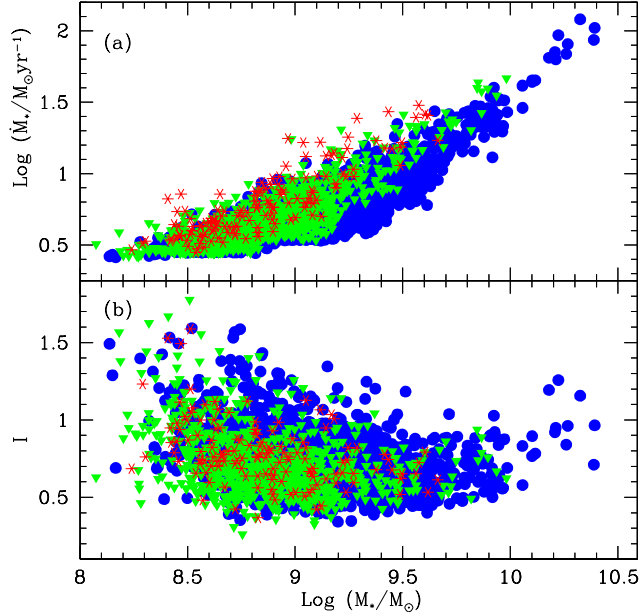}} 
\caption{(a) SFR, $\dot M_*$, and (b) SFR indicator, $\cal I$ = $\dot M_* t_*/M_*$, as a function of stellar mass ($M_*$) for LAEs at three redshifts: $z\sim 5.7$ (circles), $z\sim 6.6$ (triangles) and $z \sim 7.6$ (asterisks).}
\label{zch3_sfr} 
\end{figure}  

 The relation appears to flatten below $M_* \leq 10^{9.3} {\rm M_\odot}$, as is clearly seen. Stated differently, low mass LAEs have star formation rates confined in the narrow range 2.5-10 ${\rm M_\odot}\,{\rm yr^{-1}}$, whereas only (relatively few) larger objects undergo intense star formation events, $\dot M_* > 50 \,{\rm M_\odot} \,{\rm yr^{-1}}$. This might be the result of feedback regulation, which prevents small objects from burning most of their gas fuel at high rates. To investigate this aspect more, we have studied the behavior of a star formation indicator, ${\cal I}=\dot M_* t_*/M_*$. Physically, this is the ratio between the stellar mass produced if a LAE had always formed stars at the rate deduced at the given redshift for a time equal to the mean age of its stars, and the actual total stellar mass of the system. Hence if ${\cal I} > 1$ (${\cal I} < 1$), the star formation rate was lower (higher) in the past. The large majority of LAEs show values of ${\cal I} < 1$, indicating that the star formation rate averaged over the entire history must be larger than the final value. This could be due to huge bursts of star formation, catalyzed by mergers. Some systems (at the low and high ends of the stellar mass distribution at $z \sim 5.7$) have ${\cal I} > 1$, i.e,  they have quietly built up their stellar population at an increasing rate as they grow in mass by subsequent mergings.  
We provide a handy fit for the SFR in terms of the stellar mass:
\begin{eqnarray*}
\log_{10} (\dot M_*) & = & (0.4-0.04\Delta z) [\log_{10} (M_*)]^2 -  \\
& & (6.9-0.8\Delta z ) \log_{10} (M_*) + (30-5\Delta z),
\end{eqnarray*}
where $\Delta z = (z-5.7)$. 

\begin{figure*}[htb] 
  \center{\includegraphics[scale=0.75]{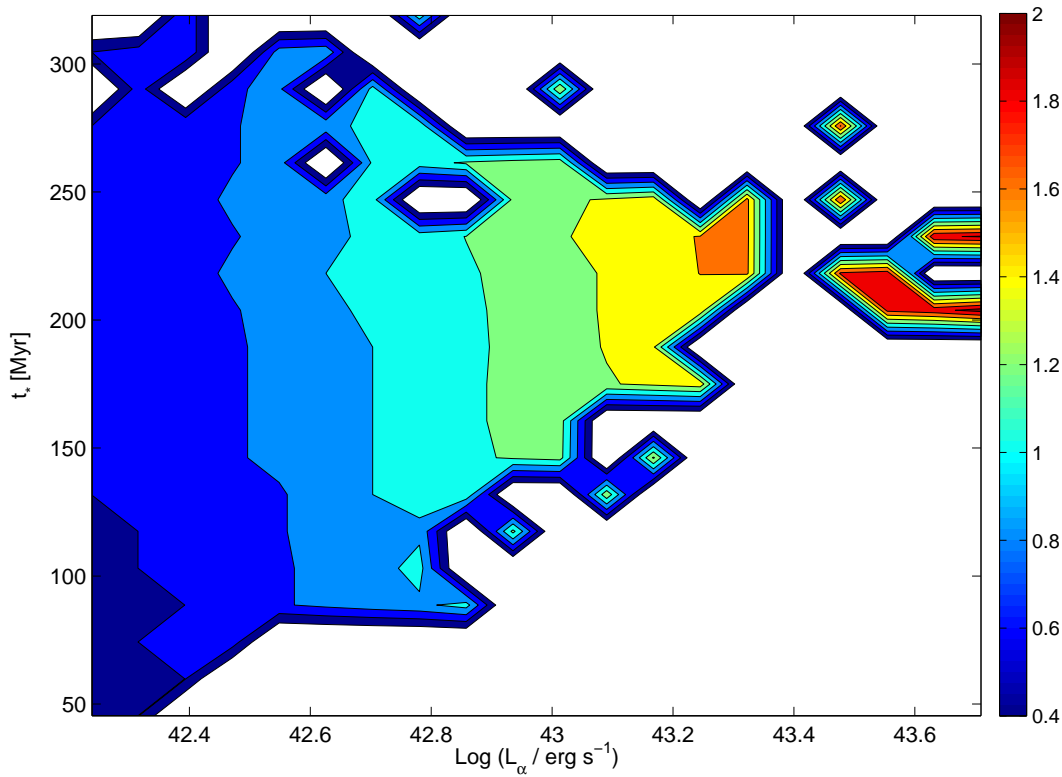}} 
  \caption{Contour plot showing relation between observed Ly$\alpha$ luminosity, age and SFR at $z\sim 5.7$. The SFR values, ${\rm Log} (\dot M_*/{\rm M_\odot} \, {\rm yr^{-1}})$, are color-coded by the bar on the right. }
\label{zch3_contour} 
\end{figure*}  

Finally, we discuss the dependence of the observed Ly$\alpha$ luminosity on age and SFR (Fig. \ref{zch3_contour}). 
The SFR for all galaxies with a given $L_\alpha$ are very similar, even though the ages 
vary between $40-340$ Myr.  As explained before, this is because according to the ERM, at $z \sim 5.7$, 
the Universe is already so ionized ($\chi_{HI} \sim 10^{-5}$) that even the smallest emitters 
are able to build large enough Str\"omgren spheres on short timescales. Further, the 
number of galaxies with small SFR ($\dot M_* \leq 20 \,{\rm M_\odot} {\rm yr^{-1}}$) is quite high, 
after which this number decreases rapidly (see also Fig. \ref{zch3_sfr}), with very few galaxies in the 
highest luminosity bins. Thus, the faint end of the LF samples a mixture 
of young and old objects, while LAEs in the bright end are predominantly massive, 
intermediate (200-250 Myr) age systems.  

\section{LAE contribution to reionization}
\label{lae_reio_ch3}

At the present time, LAEs are among the most distant galaxies known. As the quest for reionization sources 
is struggling to identify the most important populations for this process, it is
worthwhile to assess to what extent LAEs might represent such a long-searched-for
population. 
First we note that (Tab. \ref{table3_ch3}) at $z \sim 5.7$, the SFR density provided by LAEs
is $\dot\rho_* = 1.1 \times 10^{-2}\, {\rm M_\odot\, yr^{-1}\, Mpc^{-3}}$, and this decreases with increasing redshift. 
Comparing this to the cosmic SFR density of $0.12\, {\rm M_\odot\, yr^{-1}\, Mpc^{-3}}$ at the same redshift measured by Hopkins (2004), we find that LAEs contribute only about 9.2\% to the cosmic SFR value, confirming our previous 
results found using a semi-analytic LAE model, as shown in Sec. \ref{sfr_density}. Corresponding to this SFR evolution and $f_{esc}=0.02$, it is easy to derive that the  
\HI ionizing photon rate density ($q_{LAE}$) contributed by LAEs is $q_{LAE}= 3.1\times 10^{49} \, {\rm s^{-1}\, 
Mpc^{-3}}$ at $z=6.6$. We compare this photon rate density to the one necessary to balance recombinations 
given by Madau, Haardt and Rees (1999):

\begin{equation}
q_{rec} = 10^{51.57} \bigg(\frac{C}{30}\bigg) \bigg(\frac{1+z}{7.6}\bigg)^3 
\bigg(\frac{\Omega_b {\rm h^2}}{0.022}\bigg)^2 {{\rm s^{-1}\, Mpc^{-3}}},
\end{equation}
where $C$ is the IGM clumping factor.  For $C=1$, corresponding to a homogeneous IGM, we get 
the minimum \HI ionizing photon rate density necessary to balance recombinations,  i.e. 
$q_{rec} = 1.24 \times 10^{50} \, {\rm s^{-1}\, Mpc^{-3}}$. This means that the LAEs on 
the Ly$\alpha$ LF at $z\sim 6.6$ can contribute at most 25\% of the \HI ionizing photons 
needed to balance recombinations at this redshift. This value decreases to 0.8\% as $C$ increases 
to 30. 

The total \HI ionizing photon rate density contributed by all astrophysical sources ($q_{all}$) 
at $z \sim 6.6$ can be calculated as (Bolton \& Haehnelt, 2007b)
\begin{equation}
q_{all} = 10^{51.13} \Gamma_{-12} \bigg(\frac{1+z}{7.6}\bigg)^{-2} 
{{\rm s^{-1}\, Mpc^{-3}}};
\end{equation}
using $\Gamma_{-12} = 0.19$ (Tab. \ref{table1_ch3}), gives $q_{all}= 2.5 \times 10^{50}\, {\rm s^{-1}\, Mpc^{-3}}$ 
at $z \sim 6.6$.  This implies LAEs contribute about 12.5\% of the total \HI ionizing photon rate, which is consistent with the SFR density contribution of LAEs to the global value. Such a low contribution from LAE, given their halo and stellar masses does not come as a surprise. Indeed, this result is consistent with the previous estimates by Choudhury \& Ferrara (2007) who showed that only a fraction $\simlt 1$\% of the photons required to ionize the IGM come from objects in the LAE range, the bulk being provided at high redshifts by faint (or even Ultra Faint, see Salvadori \& Ferrara 2009) dwarf galaxies.    

\section{Conclusions}
\label{Conc_ch3}
We briefly summarize the main results obtained in this chapter, along with a discussion about the caveats involved.

\begin{itemize}

\item We use a large scale hydrodynamical simulations to derive the physical properties 
(halo, stellar, gas mass, star formation rate, stellar age and metallicity) of 
high redshift galaxies. Using a reionization history consistent with an early 
reionization epoch (ERM), and only {\it two} free parameters ($f_\alpha$ and $f_c$) we obtain the observed Ly$\alpha$ and UV LFs. We find that Ly$\alpha$ photons at both $z \sim 5.7$ and $6.6$ must be attenuated by dust within the galaxy to reproduce the observed LAE luminosity function; $f_\alpha=0.3$ and $f_c=(0.22,0.37)$ at $z =(5.7,6.6)$ for the model results to match the observations. The higher attenuation of the continuum photons relative to the Ly$\alpha$ (Neufeld 1991) at $z \sim 5.7$ hints at an inhomogeneous two-phase ISM with dust clumps embedded in a warm intercloud gas. 

\item Clustering of sources boosts the average value of the total photoionization rate (and 
consequently the ionized hydrogen fraction) in the surroundings of galaxies by $<1.5$ 
times at $z \sim 5.7,6.6$; this value increases by more than a factor of 50 at 
$z \sim 7.6$. 

\item For an almost neutral IGM ($z \sim 7.6, \chi_{HI}=0.16$), if clustering effects are ignored, only those few galaxies that are able to carve out a large enough \HII region are visible. However, about 4 times more galaxies become visible when clustering effects are included. In a highly ionized IGM ($z\sim 6.6, 5.7$, $\chi_{HI} \sim 10^{-4}, 10^{-5}$), however, the effect of clustering is indiscernible on the Ly$\alpha$ LF. WE also find that the clustering boost is not necessarily the highest for the most massive galaxies. 

\item We have discussed in detail the physical properties (and we provide handy fitting 
functions to several relations among them) of galaxies which we identify 
as LAEs at $z=5.6, 6.6$ and $7.6$. The ages of the LAEs 
range between $39-165$ Myr at $z \sim 7.6$ but this range increases to $38-326$ Myr at 
$z\sim 5.7$. Further, the average metallicity of LAEs is $Z_*=0.12 Z_\odot$ at $z\sim 7.6$, 
and increases to $Z_*=0.22 Z_\odot$ at $z\sim 5.7$. Hence, LAEs are more metal-enriched than 
what usually assumed. Star formation is relatively suppressed by feedback in low mass halos 
($M_h \leq 10^{11} M_\odot$) and it rises steeply for the larger halo masses. 

\end{itemize}

We now discuss some of the shortcomings still present in our model. 

\begin{itemize}
\item Since we do not model dust enrichment, we use the same values of $f_\alpha$ and $f_c$ for all galaxies at a given redshift. In reality, however, these values would change on a galaxy to galaxy basis, depending on the amount, topology and distribution of dust inside the galaxy.

\item We do not account for IGM peculiar velocities, which have a large impact on the observed Ly$\alpha$ luminosity value; outflows enable more of the Ly$\alpha$ luminosity to escape, while inflows erase the Ly$\alpha$ line. Further, outflows with a sufficiently high value of the \HI column 
density ($\sim 10^{18-21} {\rm cm^{-2}} $) add a bump to the red part of the Ly$\alpha$ line, 
thereby increasing the observed luminosity (Verhamme et al. 2006).

\item We have carried out an analytic calculation to obtain $T_\alpha$, without accounting for a full RT calculation.
 
\item We have not included Ly$\alpha$ or continuum luminosities from the cooling of collisionally excited \HI in the ISM (Dijkstra 2009). Adding this contribution to the Ly$\alpha$ luminosity from stars could increase the intrinsic Ly$\alpha$ luminosity from the galaxy by large amounts while leaving the intrinsic continuum luminosity value unchanged.

\end{itemize}

\chapter{The cool component of LAEs}\label{ch4_lya_cool}
In this chapter, we build on the model described in Chapter \ref{ch3_lya_sim} by including two physical processes important for LAEs: (a) we calculate the Ly$\alpha$ and continuum luminosities produced by cooling of collisionally excited \HI in the ISM of each galaxy, and (b) we use a detailed model to calculate the dust mass evolution and the corresponding optical depth to continuum photons, enabling us to link $f_c$ to the physical properties of each galaxy, such as its SFR, IMF, age and gas mass. These are described in Sec. \ref{lum_coolh1_ch4} and \ref{dust_model_ch4} respectively. We also introduce a model to calculate the dust-processed FIR emission in Sec. \ref{fir_emm_ch4}. We then discuss the dust enrichment of LAEs in Sec. \ref{dusty_nature_ch4} and compare the dust to gas ratios to observations of local dwarfs. We match the model results to LAE observations, including the Ly$\alpha$ and UV LFs, EW distributions for LAEs at $z \sim5.7,6.6$, in Sec. \ref{match_lf_ch4}. We assess the detectability of the LAE dust-reprocessed sub-millimeter (submm) radiation using telescopes such as ALMA, in Sec. \ref{pred_submm_ch4}. We end by discussing the main results and caveats in our model, in Sec. \ref{conc_ch4}.

As in Chapter \ref{ch3_lya_sim}, the cosmological model adopted in the simulations used corresponds to the $\Lambda$CDM Universe with $\Omega_m=0.26,\ \Omega_{\Lambda}=0.74,\ \Omega_b=0.0413$, $n_s=0.95$, $H_0 (= 100 \,{\rm h}) = 73$ km s$^{-1}$ Mpc$^{-1}$ and $\sigma_8=0.8$, thus consistent with the 5-year analysis of the WMAP data (Komatsu et al. 2009).

\section{The theoretical model}
\label{theoretical_model_ch4}

We start by using the simulation described in Sec. \ref{sim_ch3} to obtain the the mass-weighted age,
the total halo/stellar/gas mass, the SFR, the mass weighted gas/stellar metallicity, the mass-weighted gas temperature and the half mass radius of the DM halo for each bonafide galaxy identified in the simulation snapshots at $z \sim 5.7,6.6$. We then calculate the Ly$\alpha$ and continuum luminosity contributed by cooling of collisionally excited \HI in the ISM (Sec. \ref{lum_coolh1_ch4}) and add these to the stellar-powered Ly$\alpha$ and continuum luminosity calculated in Sec. \ref{intrinsic_lum_ch3}, to obtain the {\it total intrinsic} Ly$\alpha$ and continuum luminosity produced by each galaxy. We calculate the dust enrichment of each galaxy based on its intrinsic properties (SFR, age, gas mass) to obtain the fraction of continuum and Ly$\alpha$ photons that escape the galaxy, undamped by dust in Sec. \ref{dust_model_ch4}; the observed continuum luminosity values are obtained at this point. We calculate the Ly$\alpha$ IGM transmission, as explained in Sec. \ref{intrinsic_lum_ch3}, including the clustering calculation shown in Sec. \ref{clus_ch3}, to obtain the observed Ly$\alpha$ luminosity value for each galaxy. Once these calculations are done, galaxies with an observed Ly$\alpha$ luminosity, $L_\alpha \geq 10^{42.2}\, {\rm erg\, s^{-1}}$ and observed equivalent width, $EW \geq 20$ \AA\, are identified as LAEs, following the current observational criterion for LAE identification. We also calculate the FIR dust emission expected from LAEs in Sec. \ref{fir_emm_ch4}.

\subsection{Intrinsic luminosities from cooling \HI}
\label{lum_coolh1_ch4}

The huge amount of literature on LAEs has generally been focused on stars as the main source of both Ly$\alpha$ and continuum luminosity. However, by virtue of using our SPH simulation, we can consistently calculate the separate stellar and cooling \HI gas contributions to the total Ly$\alpha$ and continuum luminosities. 

We start by mentioning the values of the star-powered intrinsic Ly$\alpha$ ($L_\alpha^*$) and continuum luminosity ($L_c^*$): for a galaxy with $t_*=200 \, {\rm Myr}$, $Z = 0.2\, {\rm Z_\odot}$, $\dot M_* = 1\, {\rm M_\odot} \, {\rm yr^{-1}}$, we find $Q = 10^{53.47} {\rm s^{-1}}$ and the corresponding stellar Ly$\alpha$ luminosity for $f_{esc}=0.02$ is $L_\alpha^*=3.18 \times 10^{42} {\rm erg \, s^{-1}}$. For the same galaxy, the stellar continuum luminosity is $L_c^* = 3.5 \times 10^{40} {\rm erg \, s^{-1}\AA^{-1}}$, which yields a stellar EW of about $91$ \AA. For complete details of this calculation, see Sec. \ref{intrinsic_lum_ch3}.

Our numerical simulations also allow us to precisely track the evolution of the amount of gas present in each galaxy of the studied cosmic volume along with its temperature distribution, so that we are able to calculate the Ly$\alpha$ and UV continuum luminosities from the cooling of collisionally excited \HI in the ISM of LAEs. The Ly$\alpha$ luminosity, $L_\alpha^g$, produced by the cooling of neutral hydrogen gas of total mass $M_{HI}$,  depends both on the number density of electrons and the number density of \HI. We assume the hydrogen gas in each galaxy to be concentrated within a radius $r_g$, such that 
\be
r_g = 4.5 \lambda r_{200},
\ee
where the spin parameter, $\lambda=0.04$ (Ferrara, Pettini \& Shchekinov 2000) and $r_{200}$ is calculated assuming the collapsed region has an overdensity of 200 times the critical density at the redshift considered. Then, the density of electrons can be expressed as $(1-\chi_{HI}) M_{HI}[\mu m_H 4\pi r_g^3]^{-1}$ and the number density of \HI can be expressed as  $\chi_{HI} M_{HI}[\mu m_H 4 \pi r_g^3]^{-1}$, where $\mu$ is the mean molecular weight and $m_H$ is the proton mass. Then using the coefficients of $7.3\times 10^{-19} e^{-118400/T} \chi_{HI}\, [{\rm erg\, cm^3 \, s^{-1}}]$ as given by Dijkstra (2009), the total Ly$\alpha$ luminosity from cooling of collisionally excited \HI from the entire volume in which gas is distributed can be expressed as
\begin{equation}
L_\alpha^g = 7.3\times 10^{-19} \frac{ M_{HI}^2}{4\pi r_g^3 (\mu m_H)^2} G_\alpha(T,\chi_{HI}) \, \, {\rm [erg  \, s^{-1}]},
\label{la_gas_ch4}   
\end{equation}
where 
\begin{equation}
G_\alpha(T,\chi_{HI})=\sum_{i=1}^N\chi_{HI, i}^2 (1-\chi_{HI, i})e^{-118400/T_i} f_{I,i}^2
\end{equation}
is a nondimensional function depending on the \HI temperature, $T$, and the fraction of neutral hydrogen, $\chi_{HI}$. Further, $f_I$ is the \HI mass fraction mass found in the $i$-th bin of the simulated temperature distribution, which is discretized into $N=100$ bins equally spaced in logarithm between $10^{3.4}$~K and $10^{6.8}$~K, to provide a smooth representation of $G_\alpha$. 

\begin{table} 
\begin{center} 
\begin{tabular}{|c|c|c|c|} 
\hline 
$M_h$ & $G_\alpha(T,\chi_{HI})$ & $G_c(T)$ & $EW_i^g$ \\
$[M_\odot]$ & $$ & $$ & $[{\rm \AA}]$  \\ 
\hline
$ <10^{10}$ & $1.31 \times 10^{-7}$ & $0.50$ & $887.1$ \\
$10^{10-10.4}$ & $1.14 \times 10^{-7}$  & $0.45$ & $985.6$\\ 
$10^{10.4-10.8}$ & $7.32 \times 10^{-8}$ & $0.40$ & $1108.8$\\ 
$10^{10.8-11.2}$ & $9.18 \times 10^{-8}$ & $0.37$ & $1198.8$\\ 
$>10^{11.2}$ & $1.18 \times 10^{-7}$ & $0.34$ & $1304.5$\\
\hline
\end{tabular} 
\end{center}
\caption {Values of the functions $G_\alpha(T,\chi_{HI})$, $G_c(T)$ and the corresponding intrinsic EW from cooling of collisionally excited \HI gas in the ISM for different halo mass ranges.}
\label{table1_ch4} 
\end{table} 

\begin{figure}[htb] 
\center{\includegraphics[scale=0.5]{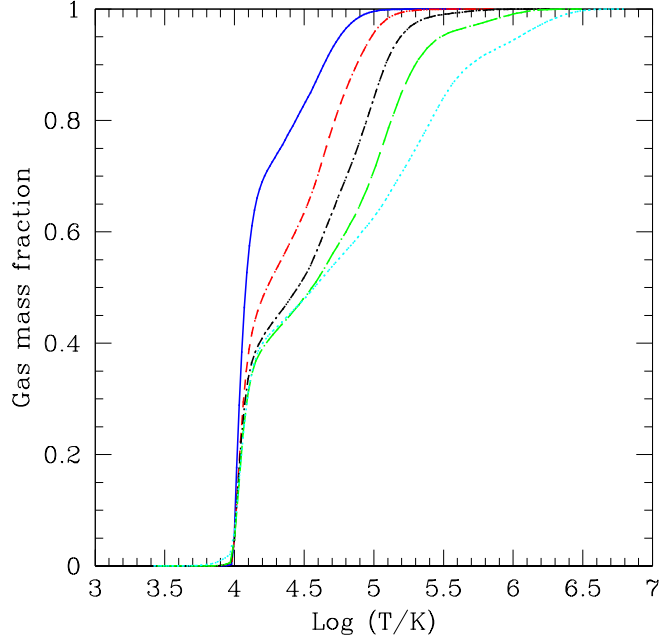}} 
\caption{Cumulative temperature distribution for interstellar gas averaged over all galaxies having dark matter halo masses in the range (in solar mass): $<10^{10}$ (solid line), $10^{10-10.4}$ (short-dashed), $10^{10.4-10.8}$ (dash-dotted), $10^{10.8-11.2}$ (long-dashed), $>10^{11.2}$ (dotted).}
\label{zch4_temp_dist} 
\end{figure} 

The value of the \HI fraction ($\chi_{HI,i}$) in each bin is calculated by inserting the temperature ($T_i$) of that bin into the collisional ionization-recombination rate equation for \HI (Cen 1992). Each galaxy in the simulation box is first assigned to one of the five halo mass ranges shown in Tab. \ref{table1_ch4}. Then we add its gas mass to the corresponding temperature bin and weigh over the total mass in that mass range. The values of $G_\alpha(T,\chi_{HI})$ and galaxy-averaged temperature distributions are shown in Tab. \ref{table1_ch4} and Fig. \ref{zch4_temp_dist}, respectively. From the Figure, it appears that massive galaxies tend to have a larger fraction of their interstellar gas in a hot component. In Eq. \ref{la_gas_ch4}, $M_{HI}$ is the total \HI mass of the galaxy assuming a primordial gas composition (76\% H, 24\% He).

Having determined the value of $L_\alpha^g$, the continuum luminosity, $L_c^g$, produced by cooling of collisionally excited \HI is calculated as 
\begin{equation}
L_c^g  = L_\alpha^{g}\, \phi(\nu_c/\nu_\alpha) \bigg(\frac{\nu_c}{c }\bigg)  G_c(T) \, \,{\rm [erg \, s^{-1} \, \AA^{-1}]},
\end{equation}
where we consider the $2\gamma$ ($2s-1s$) transition to be the only source of the continuum and
\begin{equation}
G_c(T)= \sum_{i=1}^N \frac{\Omega(1s,2s)}{\Omega(1s,2p)} \bigg \vert_i f_{I,i}.
\end{equation}
Here $\nu_c$ is the frequencies corresponding to the continuum (1375 \AA) wavelength and $\phi(\nu/\nu_\alpha) d\nu/\nu_\alpha$ is the probability that a photon is emitted in the frequency range $\nu \pm d\nu/2$, Spitzer \& Greenstein (1951). Again, $f_I$ is the \HI mass fraction mass found in the $i$-th bin of the simulated temperature distribution, which is discretized into $N=100$ bins equally spaced in logarithm between $10^{3.4}$~K and $10^{6.8}$~K. We use $\phi(\nu_c/\nu_\alpha)=3.1$, for the continuum frequency corresponding to 1375 \AA. The collisional excitation rates, $\Omega(1s, 2s)$ and $\Omega(1s, 2p)$ are calculated for each temperature bin by using the results of Anderson et al. (2000). Again using the temperature distribution for all galaxies as shown in Fig. \ref{zch4_temp_dist}, the mass-weighted value of this ratio varies with the halo mass range as shown in Tab. \ref{table1_ch4}. The intrinsic EW from cooling of collisionally excited gas alone decreases from 1449 to 681 \AA, as the continuum wavelength increases from 1250 to 1500 \AA. However, since the stellar continuum always dominates that from the gas, as will be shown later, we show the results at a wavelength corresponding to 1375 \AA, for convenience. Using these conversion factors, the calculated value of intrinsic EW of the Ly$\alpha$ line from gas alone has a value between 887-1304 \AA, depending on the halo mass, as shown in Tab. \ref{table1_ch4}. 

From now on, the Ly$\alpha$/continuum luminosity produced by a galaxy are the sum of the contribution from both the stellar sources and the cooling of \HI in the ISM, such that
\be
L_\alpha^{int}= L_\alpha^* + L_\alpha^g
\label{int_la_ch4}
\ee
and
\be
L_c^{int}= L_c^* + L_c^g
\label{int_lc_ch4}
\ee
Once that the values of $L_\alpha^{int}$ and $L_c^{int}$ have been determined, the total intrinsic equivalent width, $EW^{int}$, can be calculated as
\begin{equation}
EW^{int} = \frac{L_\alpha^{int}}{L_c^{int}} = \frac{L_\alpha^* + L_\alpha^g}{L_c^* + L_c^g}.
\end{equation}
The {\it observed} equivalent width, $EW$, is affected by dust in the ISM (modifying both the continuum and Ly$\alpha$ luminosity) and transmission through the IGM, which only damps the Ly$\alpha$ luminosity. The observed EW in the rest frame of the galaxy is again calculated using Eq. \ref{ewint_ch2}. While we use the Ly$\alpha$ IGM transmission model that includes the effects of clustered sources (see Secs. \ref{intrinsic_lum_ch3}, \ref{clus_ch3}), the determination of $f_\alpha$ and $f_c$ follows in Sec. \ref{dust_model_ch4}.

\begin{figure*}[htb]
\center{\includegraphics[scale=0.9]{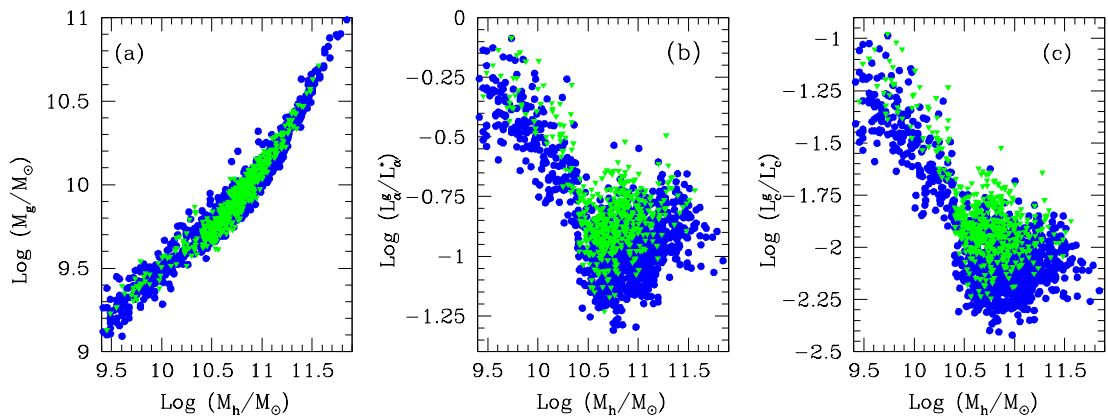}} 
\caption{(a) The total gas mass $M_{g}$; (b) ratio of cooling Ly$\alpha$ luminosity from collisionally excited \HI gas, $L_\alpha^g$, to that from stars, $L_\alpha^*$; (c) same ratio for the continuum luminosity at 1375 \AA, for z=5.7 (circles) and z=6.6 (triangles) shown as a function of the halo mass, $M_h$, for the galaxies identified as LAEs at these redshifts.}
\label{zch4_gal_prop} 
\end{figure*} 

We now discuss the relative importance of the luminosity produced by cooling \HI to that from stellar sources. We find that the contribution of $L_\alpha^g$ to the total intrinsic Ly$\alpha$ luminosity is not negligible; the average value of $L_\alpha^g / L_\alpha^* =(0.16,0.18)$ at $z=(5.7,6.6)$ respectively, as seen from Panel (b), Fig. \ref{zch4_gal_prop}. The ratio $L_c^g/L_c^*$, on the other hand is about a factor ten smaller, i.e. (0.01,0.02) at $z=(5.7,6.6)$, respectively (Panel (c) of same Figure). Adding $L_\alpha^g$ and $L_c^g$, therefore, has the effect of boosting up the intrinsic EW. As shown in Sec. \ref{intrinsic_lum_ch3}, the average intrinsic EW produced by stars only at $z\sim 5.7 = 95$ \AA. However, as shown in Tab. \ref{table2_ch4}, we find that adding $L_\alpha^g$ boosts the average intrinsic EW to $106$ \AA, i.e., by a factor of about 1.12 as expected.

It is interesting to see that even though the gas mass (Panel (a), Fig. \ref{zch4_gal_prop}) is almost similar for a given halo mass at $z \sim 5.7$ and $6.6$, both $L_\alpha^g$ and $L_c^g$ are slightly larger at $z \sim 6.6$. This can be explained noting that the virial radius decreases with increasing redshift, which leads to a corresponding increase in the \HI density. The higher density leads to an increase in the collisional excitation rate and hence, the cooling rate, producing more $L_\alpha^g$ and $L_c^g$.

Further, as has been shown before shown in Sec. \ref{sfr_ch3}, the SFR rises steeply with increasing halo masses, which leads to $L_\alpha^*$ and $L_g^*$ becoming increasingly dominant as compared to  $L_\alpha^g, L_c^g$. Hence, the contribution of ISM gas towards both the Ly$\alpha$ and continuum luminosity decreases with increasing halo mass. The dip in $L_\alpha^g$ (and the corresponding dip in $L_c^g$) for halo masses between $10^{10.4}$ and $10^{10.8}\, M_\odot$ is caused by the lower value of $G_\alpha(T,\chi_{HI})$ for these masses, as seen from Tab. \ref{table1_ch4}.

\subsection{The dust enrichment model}
\label{dust_model_ch4}

We now describe the dust enrichment model used to calculate $f_c$ and $f_\alpha$. Although dust is produced both by supernovae and evolved stars in a galaxy, several authors (Todini \& Ferrara 2001; Dwek et al. 2007) have shown that the contribution of AGB stars becomes progressively less important and at some point negligible towards higher redshifts ($z \gsim 5.7$). This is because the typical evolutionary time-scale of these stars ($\geq 1$ Gyr) becomes longer than the age of the Universe above that redshift. However, under certain conditions thought to hold in quasars, in which extremely massive starbursts occur, the contribution of AGB can become important somewhat earlier (see Valiante et al. 2009). Therefore for the ages and SFR of the galaxies in our simulation boxes (see also Secs. \ref{ages_ch3}, \ref{sfr_ch3}), we make the hypothesis that the dust seen in LAEs at $z \ge 5.7$ is produced solely by type II supernovae. We compute the amount of dust in each LAE by applying the evolutionary model described below by post-processing the simulation outputs.

As mentioned before, for each simulated galaxy, the values of $\dot M_*, M_*$ and $t_*$ are available. We then use the average SFR over the SF history of the galaxy as $\langle \dot M_* \rangle = M_* / t_*$ to calculate the dust mass. This is because the SNII rate, and hence the dust content of the galaxy depend on the entire SFR history and not just the final values obtained from the simulation. The SNII rate is estimated to be $\gamma \langle \dot M_* \rangle$, where $\gamma \sim (54 \,{\rm M_\odot})^{-1}$ for a Salpeter IMF between the lower and upper mass  limits of $1$ and ${100\, \rm M_\odot}$ respectively (Ferrara, Pettini and Shchekinov 2000). We have assumed that the progenitors of SNII have a mass larger than ${8\, \rm M_\odot}$ and in order to simplify the calculation, we have adopted the instantaneous recycling approximation (IRA), i.e., the lifetime of stars with mass larger than $8\, {\rm M_\odot}$ is zero.

Using this SNII rate, we can calculate the evolution of the total dust mass, $M_{dust}(t)$, in the galaxy as
\begin{equation}
\label{md_ch4}
\frac{d M_{dust}(t)}{dt} = y_d \gamma \langle \dot M_* \rangle - \frac{M_{dust}(t)}{\tau_{dest}(t)} - \frac{M_{dust}(t)}{M_{g}(t)} \langle \dot M_* \rangle,
\end{equation}
where  the first, second and third terms on the RHS represent the rates of dust production, destruction and astration (assimilation of a homogeneous mixture of gas and dust into stars) respectively; the initial dust mass is assumed to be zero. Further, $y_d$ is the dust yield per SNII, $\tau_{dest}$ is the timescale of dust destruction due to SNII blast waves and $M_{g}(t)$ is the gas mass in the galaxy at time $t$. For $y_d$, the average dust mass produced per SNII, we adopt a value of ${0.5\, \rm M_\odot}$ (Todini \& Ferrara 2001; Nozawa et al. 2003, 2007; Bianchi \& Schneider 2007). The amount of gas left in the galaxy, $M_g(t)$, at any time $t$ is calculated as ${d M_g(t)}/{dt} = -\langle \dot M_* \rangle$ and each galaxy is assumed to have an initial gas mass given by the cosmological baryon to dark matter ratio $\Omega_b/\Omega_m$. The final gas mass so obtained for each galaxy is very consistent with the value obtained from the simulation snapshot. This is primarily because LAEs are much larger than dwarfs and hence, are not expected to have very strong outflows. This also justifies neglecting the term accounting for the dust lost in outflows, due to galactic winds in Eq. (\ref {md_ch4}).

The destruction timescale, $\tau_{dest}(t)$ is estimated following the results of McKee (1989), Lisenfeld \& Ferrara (1998) as:
\begin{equation}
\tau_{dest} (t) = \frac{M_g(t)}{\gamma \langle \dot M_* \rangle \epsilon M_s {\rm (100\, km \, s^{-1})} } .
\label{tau_dust_ch4}
\end{equation}
Here, $\epsilon$ is the efficiency of dust destruction in a SN-shocked ISM, for which we adopt value $\sim 0.4$; this is a reasonable estimate between the values of 0.1 and 0.5 found by McKee (1989) and Seab \& Shull (1983) for varying densities of the ISM and magnetic field strengths. $M_s {\rm(100\,  km\, s^{-1}) }$ is the mass accelerated to 100 km ${\rm s^{-1}}$ by the SN blast wave and has a value of $6.8 \times 10^3 \, {\rm M_\odot}$ (Lisenfeld \& Ferrara 1998).

Once the final dust mass, $M_{dust}(t_*)$, is calculated for each galaxy in the simulation, we can translate this into an optical depth, $\tau_c$, for continuum photons as
\begin{equation}
\tau_c = \frac{3\Sigma_{d}}{4 a s},
\label{tauc_ch4}
\end{equation}
where $\Sigma_d$ is the dust surface mass density, $a$ and $s$ are the radius and material density of graphite/carbonaceous grains, respectively ($a=0.05 \mu m$, $s = 2.25\, {\rm g\, cm^{-3}}$; Todini \& Ferrara 2001; Nozawa 2003). The dust surface mass density is calculated assuming that grains are concentrated in a radius proportional to the stellar distribution, $r_e$, following the results of Bolton et al. (2008), who have derived fitting formulae to their observations of massive, early type galaxies between $z=0.06-0.36$. 

We digress briefly to explain the relation used to obtain the stellar distribution radius, $r_e$, from the observed V band luminosity, $L_V$ (Bolton et al. 2008),
\begin{eqnarray}
\label{reff_bolton}
\log r_e &=& 1.28 \log \sigma_{e2} -0.77 \log I_e -0.09, \\
(M_{dim} / 10^{11} M_\odot) & = & 0.691 (L_V / 10^{11} L_\odot)^{1.29}, \\
\sigma_{e2}& = &(2 G M_{dim} / r_e)^{0.5} \,[100 \, {\rm km/s}], \\
I_e &= & L_V/ (2\pi r_e^2) \,[10^9 L_V/ (L_\odot {\rm kpc}^2)].
\label{reff_bolton_end}
\end{eqnarray}
Here, $\sigma_{e2}$ is the projected velocity dispersion in one-half effective radius, $I_e$ is the surface brightness and $L_V$ is the luminosity between 4644 and 6296 \AA\, in the observer's frame, which is obtained using the {\rm STARBURST99} (Leitherer et al. 1999) spectrum, for the SFR, age, metallicity and IMF of the galaxy under consideration. For illustration, for a galaxy at $z \sim 5.7$ which has $t_*=200$ Myr, $Z_*=0.2\, {\rm Z_\odot}$, $\dot M_*=1\, {\rm M_\odot yr^{-1}}$, the V band luminosity is 5.05$\times \, 10^{42} \, {\rm erg \, s^{-1}}$ and the corresponding value of $r_e = 0.35$ kpc. Although we have used the V band luminosity in the observer's frame in this work, using the rest frame V band luminosity does not affect our results in any way, using the same values of the free parameters. Though it is not an entirely robust estimate, we extend this result to galaxies at $z\sim 5.7,6.6$, due to the lack of observations about the stellar distribution in high-redshift galaxies.

We now return to the calculation of the continuum optical depth; we calculate $\Sigma_d = M_{dust}(t_*)/(\pi r_d^2)$, where $r_d$ is the effective radius of dust. The best fit to the observed UV LFs requires the dust distribution scale to be quite similar to that of the stellar distribution, such that $r_d =(0.6,1.0) \times r_e$ at $z \sim (5.7,6.6)$. 

The optical depth to continuum photons (Eq. \ref{tauc_ch4}) can be easily converted into a value for the escape fraction of continuum photons assuming a slab-like distribution of dust, such that
\begin{equation}
f_c = \frac{1-e^{-\tau_c}}{\tau_c}.
\end{equation}
We are hence able to determine $f_c$ for each individual galaxy, depending on its intrinsic properties, i.e. the SFR, metallicity, age, gas mass and IMF. 

Using this model, we find that $\langle f_c \rangle \sim (0.23,0.38)$ at $z \sim 5.7, 6.6$ respectively. This is consistent with the values of $(0.22,0.37)$ obtained in Sec.  \ref{results_ch3} (see Tab. \ref{table1_ch3}). The Ly$\alpha$ contribution from the gas allows smaller galaxies to become visible as LAEs: 
the minimum halo mass of LAEs shifts from  $10^{10.2} \, {\rm M_\odot}$ (see Sec. \ref{mh_ch3}) to $(10^{9.0},10^{9.4})$ at $z \sim(5.7,6.6)$. Since smaller galaxies ($M_h \leq 10^{10.2} {\rm M_\odot}$) have lower SFR and dust content, they tend to have a higher $f_c$ value. However, only about 10\% of the galaxies we identify as LAEs in the simulation have $M_h < 10^{10.2} M_\odot$ and hence, they are too few to vary the mean $f_c$.

\begin{figure}[htb] 
  \center{\includegraphics[scale=0.5]{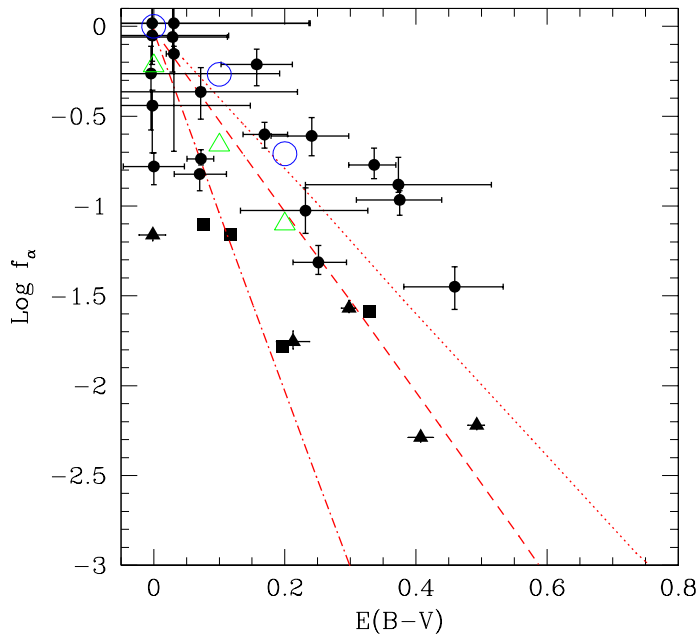}} 
  \caption{The escape fraction of Ly$\alpha$ photons, $f_\alpha$ as a function of the dust color excess, $E(B-V)$. Filled points (circles, triangles, squares) show the observed values for Ly$\alpha$ galaxies from the GALEX LAE sample (Deharveng et al. 2008), IUE local starburst sample (McQuade et al. 1995; Storchi-Bergmann et al. 1995) and Atek et al. (2008) sample respectively. The dashed line is the best fit to all the observations (see Atek et al. 2009), the dot dashed line shows the best fit from Verhamme et al. (2008) obtained from the spectral fitting of $z \sim 3$ LBGs and the dotted line is obtained following the Seaton law (1979). The relations obtained in this chapter using $p=(1.5,0.6)$ for LAEs are shown with empty circles and triangles for $z \sim (5.7,6.6)$, respectively. Details in Sec. \ref{dust_model_ch4}.}
\label{zch4_fesc_clrexs} 
\end{figure} 

As $f_c$ is fully determined from the above procedure, we are left with a single free model parameter, $f_\alpha$, to match the Ly$\alpha$ LF obtained from our calculations to the data. 
The relation between $f_\alpha$ and $f_c$ can be written as
\begin{equation}
f_\alpha  = p (A_\lambda, C) f_c,
\end{equation}
where $p$ depends not only on the adopted extinction curve, $A_\lambda$, but also on the differential radiative transfer effects acting on both Ly$\alpha$ and UV continuum photons in an inhomogeneous medium described by a clumping factor $C=\langle n^2 \rangle/\langle n \rangle^2$. While for the former dependence, some pieces of evidence exist that the SN dust we assume here can successfully be used to interpret the observed properties of the most distant quasars (Maiolino et al. 2006) and gamma-ray bursts (Stratta et al. 2007), the way in which inhomogeneities differentially enhance Ly$\alpha$ with respect to UV continuum luminosities through the so-called "Neufeld effect" (Neufeld 1991; Hansen \& Oh 2006; Verhamme et al. 2008; Finkelstein et al. 2009a; Kobayashi et al. 2009) is still highly debated. A report of a decreasing trend of $f_\alpha$ with $E(B-V)$ has been recently published (Atek et al. 2009);  our results nicely match that empirical relation, as shown in Fig. \ref{zch4_fesc_clrexs}. We therefore tentatively assume $C=1$ and compute $p$ simply from our SN extinction curve, which corresponds to a value of $R_V=2.06$, obtaining $p (A_\lambda, C=1)=0.8$, independently of the LAE mass or luminosity. Although the best fit to the observed Ly$\alpha$ LF favors values that vary around 0.8 ($p=1.5,0.6$ at $z=5.7,6.6$)\footnote{ We set $p=1$ when $f_\alpha \ge 1$.}, it is unclear if these deviations can be interpreted as a genuine imprint of an evolving clumpy ISM.  

\subsection{Calculating the FIR luminosity}
\label{fir_emm_ch4}

To calculate the total FIR luminosity for all the galaxies identified as LAEs at the redshifts of interest, i.e. $z \sim (5.7,6.6)$, we assume that the dust is predominantly heated by UV radiation from stars with wavelengths 912--4000 \AA\,, Buat \& Xu (1996); we use the notation $UV$ for non-ionizing continuum in this wavelength range. 

Assuming that the escape fraction of $UV$ photons is the same as that for the continuum photons (1375 \AA), we calculate the $UV$ luminosity, $L_{UV}$, escaping from the galaxy as 
\be
L_{UV} = L_{UV}^{0}f_c,
\ee
where $L_{UV}^0$ is the intrinsic $UV$ luminosity, calculated from the {\tt STARBURST99} spectra, for each LAE.

Because of the large cross section of dust against $UV$ light and the intense $UV$ field in a star forming galaxy, we can assume that the emitted FIR luminosity, $L_{FIR}$, is equal to the $UV$ luminosity which is absorbed and heats up the dust, such that 
\begin{equation}
L_{FIR} = L_{{UV}}^{0} - L_{{UV}} = (1-f_c)L_{{UV}}^{0}.
\end{equation}

In order to examine the detectability of dust emission, the observed flux is predicted to be
\begin{eqnarray}
f_\nu =\frac{(1+z)L_{\nu (1+z)}}{4\pi d_{L}^2},
\end{eqnarray}
where $d_{L}$ is the luminosity distance (Carroll, Press \& Turner 1992), and $L_\nu$ is the monochromatic luminosity at the chosen frequency of observation.

If dust grains emit thermally with a single dust temperature, $T_{dust}$, $L_\nu$ can be written as
\begin{equation}
L_\nu  =  4\pi M_{dust}\kappa_\nu B_\nu (T_{dust}),
\label{lnu}
\end{equation}
where $\kappa_\nu$ is the mass absorption coefficient, $B_\nu (T_{dust})$ is the Planck function with frequency $\nu$ and temperature $T_{dust}$.

If we assume a power law form for $\kappa_\nu$ such that $\kappa_\nu = \kappa_{\nu_0} (\nu /\nu_0)^\beta$, the total FIR luminosity can be expressed as 
\begin{eqnarray}
L_{FIR} & \hspace{-2mm}= & \hspace{-2mm}\int_0^\infty L_\nu\,d\nu\nonumber\\
& & \hspace{-13mm}=4\pi M_{dust}\kappa_{\nu_0}
\nu_0^{-\beta}\left(\frac{kT_{dust}}{h}\right)^{4+\beta} \left(\frac{2h}{c^2}\right)
\int_0^\infty\frac{x^{3+\beta}}{e^x-1}dx,
\label{Lnu_integ}
\end{eqnarray}
where $x = h\nu(k T_{dust})^{-1}$. We use $\nu_0 = 3.00\times 10^{12}$ Hz (corresponding to a wavelength of 100 $\mu$m), which is an arbitrary frequency in FIR used for normalization.

Since we assume all the dust grains to be spherical with a single size $a = 0.05 \mu$m (appropriate for a SN-produced dust, see Todini \& Ferrara 2001; Nozawa 2003, 2007) and material density $s$, the mass absorption coefficient $\kappa_{\nu_0}$ can be written as $\kappa_{\nu_0} =3Q_\nu[4as]^{-1},$ where $Q_\nu$ is the optical absorption cross section normalized to the geometrical cross section ($\pi a^2$). For graphite/carbonaceous grains, $Q_\nu\,  a^{-1}=1.57\times 10^2$ cm$^{-1}$ at a wavelength of 100 $\mu$m, $s=2.25$ g cm$^{-3}$ and $\beta =2$ (Draine \& Lee 1984). This results in $\kappa_{\nu}=52.2\, (\nu /\nu_0)^\beta$ cm$^2$ g$^{-1}$.
\footnote{The corresponding values for silicates are $Q_\nu \, a^{-1}=1.45\times 10^2$ cm$^{-1}$, $s=3.3$ g cm$^{-3}$ and $\beta =2$, which results in $\kappa_{\nu}=32.9\, (\nu /\nu_0)^\beta$ cm$^2$ g$^{-1}$.} Even though in this work, we assume the dust grains to be graphites, the flux is insensitive to the specific grain material for a given value of $L_{FIR}$ since a small $\kappa_\nu$ is compensated by a high $T_{dust}$ and vice versa. Then, Eq. \ref{Lnu_integ} can be simplified to the following expression
\begin{eqnarray}
T_\mathrm{dust}= 6.73 \left(
\frac{L_\mathrm{FIR}/L_\odot}{M_\mathrm{dust}/M_\odot}
\right)^{1/6}~\mathrm{K}.
\end{eqnarray} 
This temperature is put in Eq. (\ref{lnu}) to obtain $L_\nu$ and hence the dust emission flux detectable in the submm bands in the observer's frame. 


\section{The dusty nature of LAEs}
\label{dusty_nature_ch4}

Once that the model is in place, we can discuss the physical properties and dust enrichment of LAEs at $z \sim 5.7, 6.6$. We start by observing that when the luminosity from cooling \HI and a dust enrichment model that depend on the physical properties of the galaxy are considered, the DM halo mass of LAEs ranges between $10^{9.0-11.8}\, M_\odot$ at $z \sim 5.7$ (see Tab. \ref{table2_ch4}) as compared to the values of $10^{10.2-11.8}\, M_\odot$ obtained at the same redshift, when only stellar-powered luminosity and a single $f_c, f_\alpha$ value was used for all galaxies at a given redshift (see Sec. \ref{mh_ch3}). The decrease in the halo mass also leads to a decrease in the SFR of LAEs; the lower limit of $\dot M_*$ decreases to $0.8-120 \, {\rm M_\odot \, yr^{-1}}$ (Tab. \ref{table2_ch4}) from $2.5-120 \, {\rm M_\odot \, yr^{-1}}$. However, due to the merger-history induced stellar age dispersion, the stellar ages do not show any appreciable difference for the two different scenarios; for both cases they range between $40-326$ Myr (See Tab. \ref{table2_ch4} and Sec. \ref{ages_ch3}). 

\begin{figure}[htb]
  \center{\includegraphics[scale=0.5]{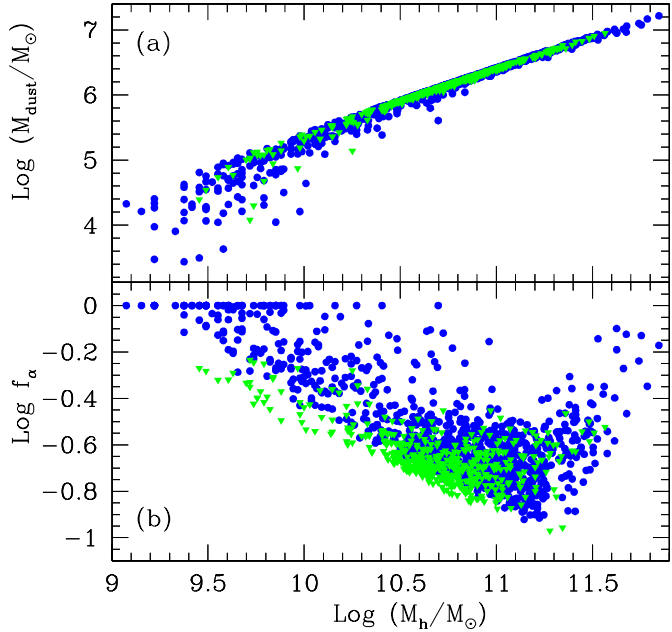}} 
  \caption{(a) Dust mass and (b) $f_\alpha$ as a function of halo mass, $M_h$, for z=5.7 (circles) and z=6.6 (triangles) for the galaxies identified as LAEs at these redshifts using $p=(1.5,0.6)$ for all galaxies at $z\sim(5.7,6.6)$. For details, refer to text in Sec. \ref{dust_model_ch4}.}
\label{zch4_dust_prop} 
\end{figure}

Since we consider a scenario in which the dust amount is regulated only by the SFR, which scales with the halo mass, the dust amounts too scale in the same way, as seen from Panel (a) of Fig. \ref{zch4_dust_prop}. However, while the dust mass increases linearly with the halo mass for most of the objects, the scatter at the low mass end arises from feedback regulation of the star formation in these objects. 

\begin{table*} 
\begin{center} 
\begin{tabular}{|c|c|c|c|c|c|c|c|c|} 
\hline 
$z$ & $M_h$ & $\dot M_*$ & $M_g$  & $E(B-V)$ & $T_\alpha$ & $EW^{int}$ & $EW^{emer}$ & $EW $ \\
$$ & $[M_\odot]$ & $[M_\odot yr^{-1}]$ & $[M_\odot]$ & $$ & $$ & $[{\rm\AA}]$ & $[{\rm\AA}]$ & $[{\rm\AA}]$ \\ 
\hline
$5.7$ & $10^{9.0-11.8}$  & $0.8-120$ & $10^{9.0-11.0}$ & $0.14$ & $0.48$ & $106.0$ & $157.2$ & $75.1$  \\
$6.6$ & $10^{9.4-11.6}$ &  $1.6-46.4$ & $10^{9.1-10.7}$ & $0.09$ & $0.48$ & $117.2$ & $70.3$ & $34.6$ \\ 
\hline

\end{tabular} 
\end{center}
\caption {For all the LAEs comprising the Ly$\alpha$ LF at the redshifts shown (col. 1), we show the range of halo mass (col. 2), the range of mass weighted ages (col. 3), the range of SFR (col. 4), the range of gas mass (col. 5), the average color excess (col. 6), the average transmission of Ly$\alpha$ photons through the IGM (col. 7), the average intrinsic EW (col. 8), the average EW of the line emerging from the galaxy (col. 9) and the average value of the observed EW (col. 10).}
\label{table2_ch4}
\end{table*} 

From Panel (b) of the same figure, we see that the escape fraction of Ly$\alpha$ photons, $f_\alpha$, decreases by a factor of 10, from 1 to about 0.1 as the halo mass increases from $10^{9.0}$ to $10^{11.2}\, {\rm M_\odot}$ at $z \sim 5.7$, consistent with the results obtained by Laursen et al. (2009). However, for the few galaxies between $10^{11.2}$ and $10^{11.8}\, {\rm M_\odot}$, $f_\alpha$ increases again. This is because the optical depth to continuum photons, $\tau_c \propto M_{dust}/r_e^2$, where $r_e$ scales with SFR (see Eq. \ref{reff_bolton} - \ref{reff_bolton_end}). Since the SFR rises steeply with halo mass for $M_h \gsim 10^{11.2}\, {\rm M_\odot}$, the optical depth decreases, even though the total dust mass increases. The decreased optical depth then leads to the larger escape fraction of continuum and hence, Ly$\alpha$ photons seen. When $f_c$ is translated into a color excess using the SN extinction curve (see Eq. \ref{colorexs_sn_ch3}), we find $E(B-V) \sim 0.14$, averaged over all LAEs at $z \sim 5.7$. This is in very good agreement to the values inferred by Lai et al. (2007), Pirzkal et al. (2007), Nagamine et al. (2008), Dayal et al. (2009) and Finkelstein et al. (2009a) (See Sec. \ref{uvlf_ch3} for more details).

\begin{figure}[htb] 
  \center{\includegraphics[scale=0.5]{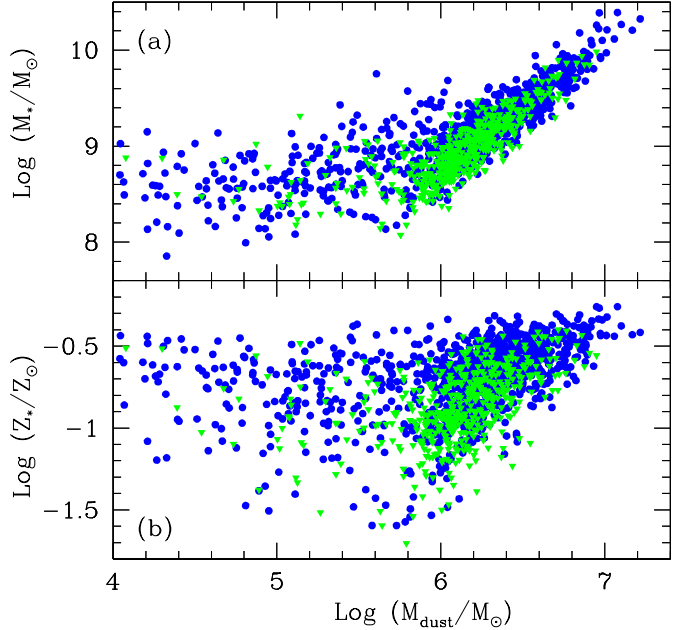}} 
  \caption{Relation between (a) stellar mass, $M_*$, (b) stellar metallicity, $Z_*$ and dust mass, $M_{dust}$, for z=5.7 (circles) and z=6.6 (triangles) for the galaxies identified as LAEs at these redshifts.}
\label{zch4_md_met} 
\end{figure} 

Since more massive galaxies generally preserve more of their gas content, we find that the galaxies with the largest stellar mass are more dust rich, as shown in Panel (a) of Fig. \ref{zch4_md_met}; for the bulk of the galaxies, the dust mass, $M_{dust}$ is related to the stellar mass, $M_*$, as $M_{dust} \propto M_*^{0.7} $. However, the low stellar mass end shows a large scatter in the dust content, likely arising from  feedback regulation leading to a scatter in the star formation rates for smaller halos/galaxies. As expected, galaxies with the highest stellar metallicities have the largest dust masses; the dust mass is related to the stellar metallicity, $Z_*$, as $M_{dust} \propto Z_*^{1.7}$ as shown in Panel (b) of Fig. \ref{zch4_md_met}. Again, the scatter for the low metallicity galaxies is explained by the corresponding scatter in the star formation rates at this end.

Since the simulation also provides us with the mass-weighted gas metallcity of each galaxy, we can also study the relation between the dust-to-gas ratio ($D=M_{dust}/M_g$) and the gas metallicity, $Z_g$, to discuss the dust enrichment scenario and compare it to results obtained by other authors, for the local Universe. We normalize the dust-to-gas ratio to the one measured in the  solar neighborhood, $D_\odot= 1/150 = 6\times 10^{-3}$ (Hirashita \& Ferrara 2002).  

\begin{figure}[htb] 
  \center{\includegraphics[scale=0.5]{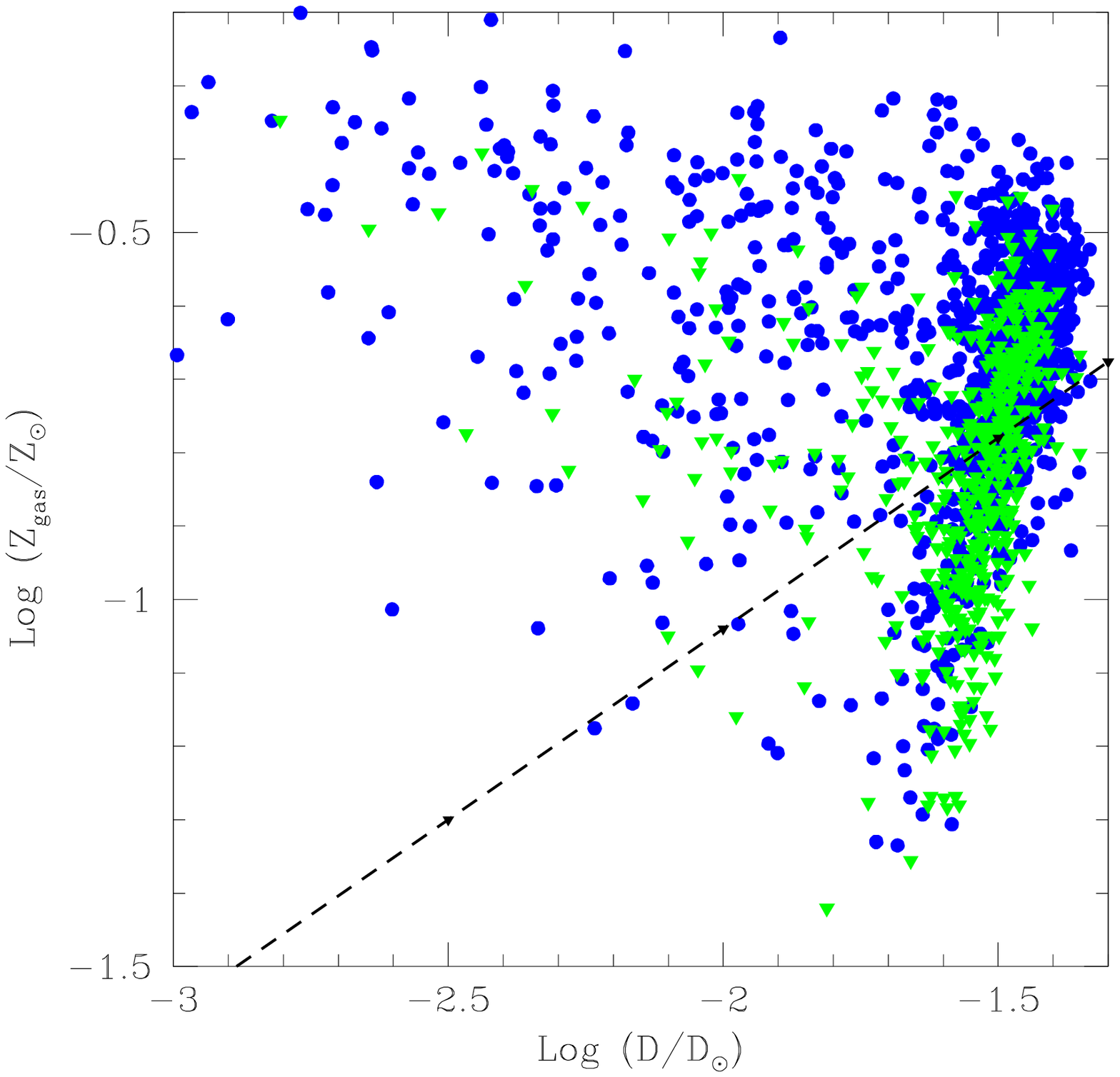}} 
  \caption{Relation between dust-to-gas ratio and gas metallicity for LAEs at $z=5.7$ (circles), 6.6 (triangles). The dashed line corresponds to $Z_g \propto {D}^{0.52}$ (Lisenfeld \& Ferrara 1998). }
\label{zch4_dust_gas_ratio} 
\end{figure}
 
The relation between $D$ and $Z_g$ has been investigated by several authors. In particular, Lisenfeld \& Ferrara (1998) have shown that for nearby dwarf galaxies the following relation holds: $Z_g \propto {D}^q$, with $q=0.52\pm 0.25$. In principle, there is no reason to expect that such low-redshift determination applies equally well to LAEs, due to the very different physical conditions and shorter evolutionary timescales allowed by the Hubble time at $z\approx 6$. Using the above relation for a gas metallicity value equal to $0.2 Z_\odot$, appropriate for our calculation of LAEs, yields $D = 0.003-0.12 D_\odot$; from our result we get a mean value in reasonable agreement with this expectation, $D = 0.001-0.045 D_\odot$ for $Z_g = 0.2 Z_\odot$. 

However, although we confirm a correlation between dust abundance and metallicity, the $D-Z_g$ relation we find, shown in Fig. \ref{zch4_dust_gas_ratio}, shows considerable deviations from the above power-law. First, the slope of the relation is considerably steeper (i.e. $q >1$); second, in spite of the relatively high metallicity of LAES (about $0.1-0.5 Z_\odot$), the dust abundance is relatively depressed with essentially all objects having $D < 0.05 D_\odot$. Taken together, these two points imply that LAEs are relatively poorly efficient dust producers with respect to local galaxies. The physical explanation of this fact is straightforward. Because of their young ages ($< 500$~Myr), LAEs can only have their dust produced by SNII rather than by evolved stars, which dominate the dust production in local galaxies. However, since dust is both produced and destroyed by SN in LAEs, $D$ saturates at a relatively low value; in local galaxies, instead, SN shocks cannot counteract the rate of dust production by evolved stars and $D$ grows to 
larger values.   

Finally, Fig. \ref{zch4_dust_gas_ratio} shows the presence of a considerable number of outliers to the left of the power-law relation, i.e. the low $D$-high  $Z_g$ region, at both redshifts. Since both the dust and gas mass scale with the halo mass (See Panels (a) of Fig. \ref{zch4_gal_prop}, \ref{zch4_dust_prop}) this physically means that even some galaxies with a very small halo mass retain a large metal fraction in the ISM gas. This can possibly be explained by the fact that though galaxies with a small halo mass have very feeble star formation rates and hence produce less amount of metals as compared to larger galaxies, the mechanical feedback that would eject their ISM metal content is comparatively more depressed with respect to larger galaxies.

\section{Observational implications}
\label{match_obs_ch4}

We now compare the model results to the observed Ly$\alpha$ and UV LFs at $z \sim 5.7, 6.6$ and to the observed EW distribution at $z \sim 5.7$. We also make predictions for the sub-millimeter flux that could be detected by upcoming missions such as ALMA, for these high-z galaxies.

\subsection{The LFs and EWs}
\label{match_lf_ch4}

\begin{figure}[htb]
  \center{\includegraphics[scale=0.5]{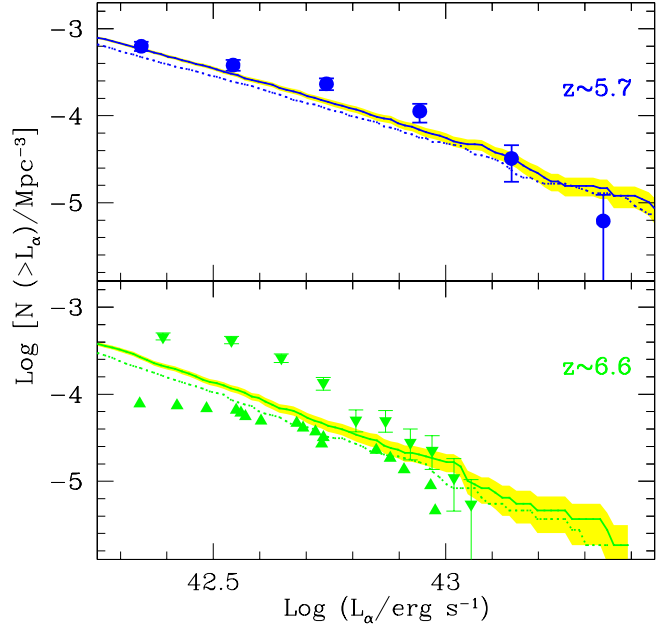}} 
  \caption{Cumulative Ly$\alpha$ LF for the ERM. The panels are for $z\sim 5.7$ and $6.6$. Points represent the data at two different redshifts:  $z\sim 5.7$ (Shimasaku et al. 2006) (circles), $z \sim 6.6$ (Kashikawa et al. 2006) with downward (upward) triangles showing the upper (lower) limits. Solid (dashed) Lines refer to model predictions at $z\sim 5.7,6.6$ for the parameter values in text including (excluding) the contribution from cooling of collisionally excited \HI in the ISM. Shaded regions in both panels show poissonian errors. }
\label{zch4_lya} 
\end{figure}

As shown in Sec. \ref{lum_coolh1_ch4}, cooling of collisionally excited \HI can produce a significant amount of Ly$\alpha$ luminosity in the ISM. However, since galaxies become more compact and denser towards high redshift, this leads to a corresponding increase in $L_\alpha^g$. This can be clearly seen from the cumulative Ly$\alpha$ LF (Fig. \ref{zch4_lya}) where ignoring the gas contribution leads to a slightly larger under estimation of the LF at $z \sim 6.6$ compared to $z\sim 5.7$. 

\begin{figure}[htb] 
  \center{\includegraphics[scale=0.5]{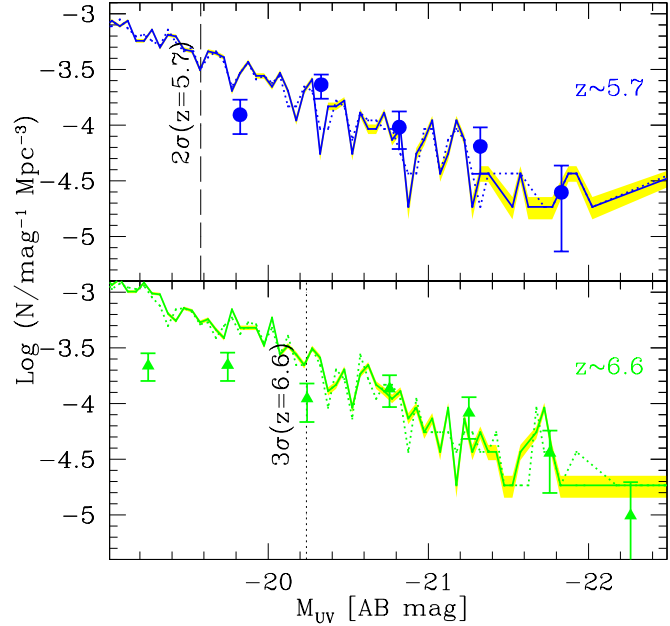}} 
  \caption{UV LAE LF for the ERM. Points represent the data at two different redshifts: $z \sim 5.7$ (Shimasaku et al. 2006) (circles), $z \sim 6.6$ (Kashikawa et al. 2006) (triangles). Solid (dashed) lines refer to model predictions including (excluding) the contribution from cooling \HI in the ISM at the redshifts (from top to bottom): $z \sim 5.7,6.6$, for the parameter values in text. The vertical dashed (dotted) lines represent the observational 2$\sigma$ (3$\sigma$) limiting magnitude for $z = 5.7$ ($z =6.6$). The shaded regions in both panels show the poissonian errors.} 
\label{zch4_uv} 
\end{figure}

As mentioned in Sec. \ref{lum_coolh1_ch4}, cooling of collisionally excited \HI does not contribute much to the continuum luminosity, with the average contribution to the total continuum luminosity being $\leq 10$ \% at either of the redshifts considered. Adding $L_c^g$, therefore,  produces a negligible effect on the UV LF, as seen clearly from Fig. \ref{zch4_uv}, where the UV LFs, both including and excluding the gas contribution overlap. 

The free parameters in this work are the the radius of the dust distribution, $r_d$ relative to the stellar distribution radius, $r_e$ and the escape fraction of Ly$\alpha$ photons relative to continuum photons. Once that these two parameters are obtained by calibrating our results to the UV and Ly$\alpha$ LFs, we have no more free parameters left. Using the values of $f_\alpha/f_c$, $T_\alpha$ and the intrinsic EW for each galaxy we identify as a LAE, we obtain the emergent and observed EWs at $z \sim 5.7$. Due to the cooling of collisionally excited \HI, the total Ly$\alpha$ luminosity increases by a factor of about 1.12 while the continuum almost remains unchanged. This leads to intrinsic EWs in the range 88.8-188.8 \AA \, for the galaxies we identify as LAEs; the value of the intrinsic EW at $z \sim 5.7$, averaged over all LAEs increases to 106.0 \AA\, as shown in Tab. \ref{table2_ch4}, compared to 95 \AA\, found in Dayal et al. (2009) (See Tab. \ref{table3_ch3}), not including gas emission.

The EWs emerging from the galaxy at $z \sim 5.7$ are further enhanced and have a mean value of 157.2 \AA\, (Tab. \ref{table2_ch4}) because of the larger escape fraction of Ly$\alpha$ photons with respect to continuum photons ($f_\alpha/f_c=1.5$). However, since the Ly$\alpha$ luminosity is attenuated by transmission through the IGM, with $\langle T_\alpha \rangle \sim 0.48$, the observed EWs are finally reduced to the range 44.4-106.9 \AA; with an average value of 75.1 \AA as shown in the same table.

\begin{figure}[htb] 
\center{\includegraphics[scale=0.45]{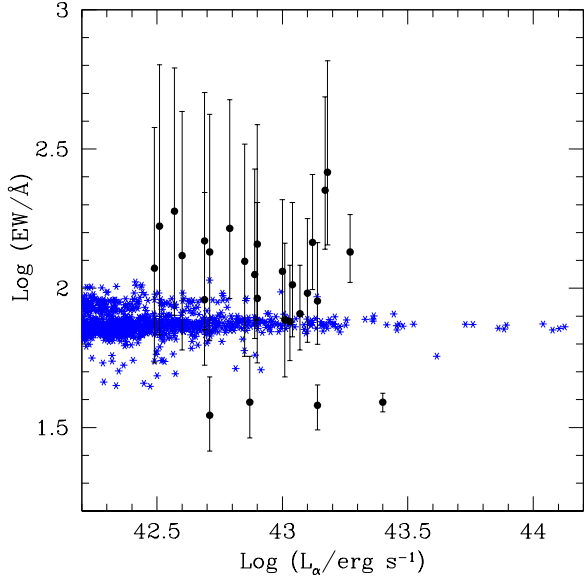}} 
\caption{Observed EWs from Shimasaku et al. (2006) (circles) and model values of the observed EWs (astrexes) as a function of the observed Ly$\alpha$ luminosity.}
\label{zch4_ew_5pt7} 
\end{figure}  

We show the EWs as a function of the observed Ly$\alpha$ luminosity in Fig. \ref{zch4_ew_5pt7}. Though the numerical data is concentrated at ${\rm Log}(EW/{\rm \AA}) \sim 1.9$, it shows some scatter, mostly induced by the different physical properties of galaxies as age and metallicity. However, we are unable to reproduce some of the largest and smallest EW values. This could be due to some of the assumptions made in the modelling of $L_\alpha^g$ including:  (a) we have calculated $L_\alpha^g$ for a homogeneous ISM. This assumption seems to represent a fairly good description of the low-$z$ data discussed in Fig. \ref{zch4_fesc_clrexs}. However, we cannot exclude the possible presence of gas inhomogeneities which would increase the collisional excitation rates and hence increase $L_\alpha^g$. Since we are investigating a large cosmological volume, resolving the small scale structure of the ISM is beyond the possibilities of the present work, and our results in this sense should be considered as a lower limit to the true emissivity of the gas. (b) $L_\alpha^g$ depends on the radius of gas distribution and the simple relation we have assumed might be different for these high redshift galaxies and (c) we have assumed a very simple scaling between $f_\alpha$ and $f_c$. In reality, this should depend on the geometry and distribution of dust within the ISM and this would induce scatter in the EW distribution.

\subsection{Prediction for sub-millimeter fluxes}
\label{pred_submm_ch4}

\begin{figure*}[htb] 
  \center{\includegraphics[scale=1.0]{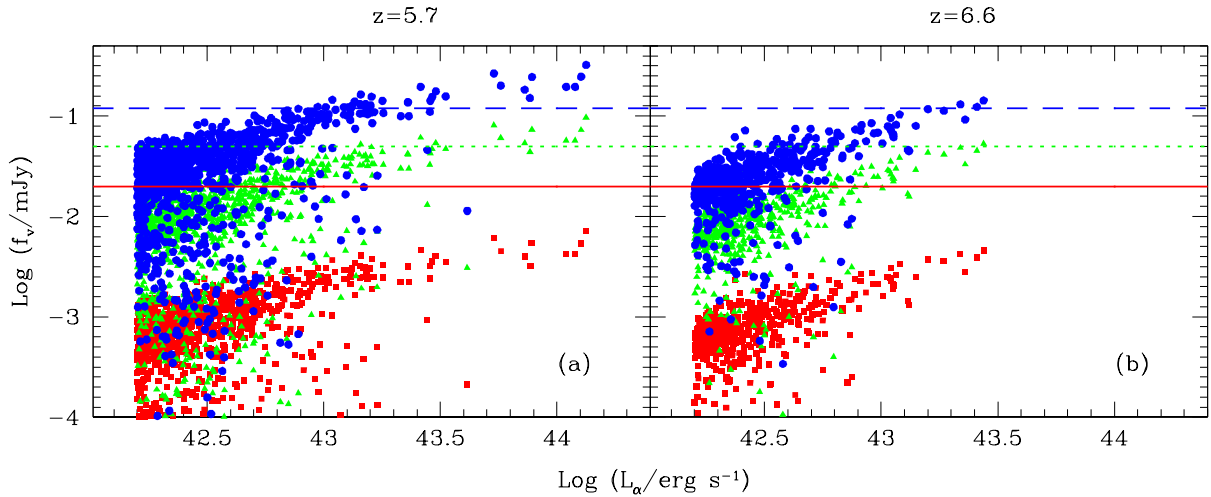}} 
 \caption{Correlations between Ly$\alpha$ luminosity and submm flux for $z=5.7$ (left) and $6.6$ (right). The horizontal solid, dotted and dashed lines show the $5\sigma$ detection limit of ALMA with 1 hour integration for the bands corresponding to $100$ Ghz (3mm), $220$ Ghz (1.4 mm) and $353$ GHz (850$\mu$m) respectively. Symbols (squares, triangles, circles) in both panels show the results obtained from this work for the same bands (100, 220, 353 GHz).}
\label{zch4_lae_visibility} 
\end{figure*} 

ALMA could be a strong tool to directly detect the dust emission from a large number of LAEs. Then, the question of how to identify the sources of submm emission arises. We find that the observed Ly$\alpha$ luminosity at both $z \sim (5.7,6.6)$ (as calculated in Secs. \ref{lum_coolh1_ch4}, \ref{dust_model_ch4}) and the fluxes at $3$mm, $1.4$ mm and $850 \mu$m of LAEs are correlated, as shown in Fig. \ref{zch4_lae_visibility}; in general, the galaxies with the largest observed Ly$\alpha$ luminosities also show the largest value of the FIR fluxes. The reason for this can be explained as follows: the largest galaxies have the largest star formation rates and hence the largest values of $L_\alpha^{int}$ since this scales with the star formation rate. As shown in Sec. \ref{lyalf_ch3}, their $T_\alpha$ is also the largest. Further, the largest galaxies are also the most dust-enriched by SNII since the SNII rate scales with the star formation rate. However, due to large radii of dust distribution (which scales with the SFR), the largest galaxies do not have the smallest $f_\alpha$ as expected. This implies that the largest galaxies show both the largest $L_\alpha$ and $f_\nu$, hence the trend seen. Fig. \ref{zch4_lae_visibility} shows that  the $850 \mu$m band is the optimum one to look for submm emission; (32,\,5) LAEs at $z=(5.7,\,6.6)$ in our simulation volume have an 850$\mu$m flux larger than the $5\sigma$ 1 hour integration limit of ALMA which is $0.12$ mJy (Morita \& Holdaway 2005). This corresponds to about $(3\%,1\%)$  of 
the LAEs in our simulated samples at these redshifts.

\begin{figure}[htb] 
  \center{\includegraphics[scale=0.5]{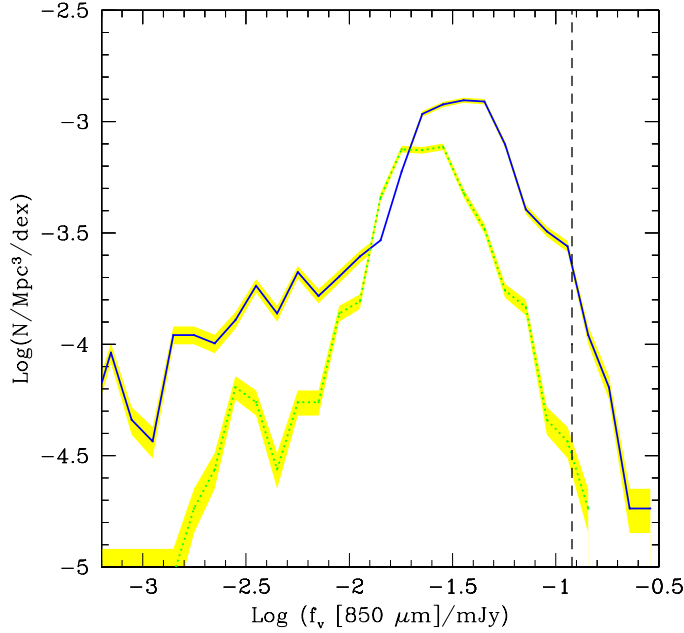}} 
  \caption{The distribution function of the observed $850\mu$m flux of LAEs at $z=5.7$ (solid line) and $6.6$ (dotted line). The vertical dashed line shows the $5\sigma$ 1 hour integration detection limit of ALMA ($=0.12$ mJy) and the shaded regions show the poissonian errors.}
\label{zch4_lf_flux} 
\end{figure} 

In Fig. \ref{zch4_lf_flux}, we show the FIR luminosity and submm flux luminosity functions for our LAE sample at $850 \mu$m since this seems to be the most optimum band to look for FIR emission, as mentioned above. We see that selecting LAEs with $L_\alpha \geq 10^{42.2} {\rm erg\, s^{-1}}$, can efficiently detect galaxies whose submm fluxes are bright enough to be detected by ALMA with an integration time of only 1 hour. A longer integration time drastically increases the number of LAEs detected by ALMA because of the steep rise of distribution function toward faint submm fluxes as shown in the same figure.
 
Thus, we show that we can simply point ALMA towards bright LAEs that have already been identified at high-z to detect LAEs in the sub-mm band at $5\sigma$ with an integration time of only 1 hour.

\section{Conclusions}
\label{conc_ch4}

We now summarize the main results presented in this chapter.
\begin{itemize}

\item We find that the Ly$\alpha$ luminosity from cooling ISM gas ($L_\alpha^g$) constitutes a non-negligible addition to the value from stars ($L_\alpha^*$); the average value of $L_\alpha^g / L_\alpha^* \sim (0.16,0.18)$ at $z \sim (5.7,6.6)$. However, the continuum from stellar luminosity ($L_c^*$) always dominates over that from cooling gas ($L_c^g$); the average value of $L_c^g / L_c^* \sim (0.01,0.02)$ at $z \sim (5.7,6.6)$. 

\item Galaxies with the largest stellar mass, $M_*$ have also the largest dust content, with a scaling relation given by $M_{dust} \propto M_*^{0.7}$. The same trend of increasing dust content is also observed in terms of the stellar metallicity, $M_{dust} \propto Z_*^{1.7}$.

\item The increase in the dust mass with increasing halo mass leads to a decrease in the escape fraction of Ly$\alpha$ photons ($f_\alpha$) from the galaxy, with the value dropping from $\sim 1$ to $0.1$ as the halo mass increases from $10^{9.0}$ to $10^{11.2}\, {\rm M_\odot}$ at $z \sim 5.7$. 

\item Adding Ly$\alpha$ from the cooling \HI, the average intrinsic EW at $z\sim 5.7$ increases by a factor of about 1.12 and has a mean vale of $106.0$ \AA. The emergent EW is further enhanced to an average value of about 157.2 \AA\, at $z \sim 5.7$ since $f_\alpha/f_c=1.5$ at this redshift. However, attenuation of the Ly$\alpha$ luminosity \HI in the IGM reduces the mean observed EW to 75.1 \AA.

\item We find that the observed Ly$\alpha$ luminosity and the FIR fluxes are correlated for LAEs in all the three ALMA bands (3000, 1363 and 850 $\mu$m); the LAEs with the largest observed Ly$\alpha$ luminosity also show the largest fluxes. This means that pointing surveys of the brightest LAEs already identified would yield large fluxes in ALMA. We find that the $850 \mu$m band seems to be the optimum one for detecting the FIR emission from high-z LAEs. Although a few tens of objects would be detectable with a 1 hour integration, the number density of objects detectable would rise steeply with the integration time.

\end{itemize}

In the same spirit, we now summarize some of the main ingredients still missing in this model.

\begin{itemize}

\item The IGM transmission calculation does not include either RT, peculiar velocity fields or density inhomogeneities, all of which could have a large effect on the amount of Ly$\alpha$ transmitted through the IGM. 

\item Due to the large simulation box size, we are unable to resolve individual galaxies; we are therefore forced to assume a homogeneous \HI distribution in the ISM. Considering a clumpy ISM could increase $L_\alpha^g$ as compared to the value we get assuming a homogeneous distribution by many factors.

\item we assume SNII to be the primary sources of dust production/destruction at the redshifts considered. While this is justified for ages $\leq 1$ Gyr, we can not completely rule out the contribution from evolved stars as shown by Valiante et al. (2009). Further, the dust yield per SNII and the dust destruction efficiency due to supernova shocks are uncertain, and only known well enough to within a factor of a few.

\end{itemize}

\chapter{Radiative transfer effects on LAE visibility}\label{ch5_lya_rt}
In this chapter, we build on the model introduced in Chapter \ref{ch4_lya_cool} by using a complete RT calculation; we combine runs of SPH and RT simulations, carried out using ${\tt GADGET-2}$ and {\tt CRASH} respectively, to the previously developed LAE model (see Chapters \ref{ch2_lya_sam}, \ref{ch4_lya_cool}) as explained in Sec. \ref{model_ch5} which follows. We then explore the effects of peculiar velocities and dust on the LAE visibility in sec. \ref{visibility_ch5}; we start by ignoring both these effects in Sec. \ref{nodust_novel_ch5}, we then include the effects of dust in Sec. \ref{incdust_novel_ch5} and finally we consider both these effects in Sec. \ref{incdust_incvel_ch5}. All these calculations reveal a degeneracy between the effects of  dust and the IGM ionization state, as shown in Sec. \ref{degen_ch5}. We discuss the physical properties of LAEs in Sec. \ref{phy_prop_ch5} and compare the EWs obtained from the model to the observations in Sec. \ref{ew_ch5}, ending with a brief summary of the results and the caveats in the model, in Sec. \ref{conc_ch5}.

\section{The model}
\label{model_ch5}
We post-process a {\tt GADGET-2} snapshot at $z \sim 6.1$ with a RT code called {\tt CRASH}, to obtain snapshots of the ionization state of the IGM at different time intervals, as described in Sec. \ref{rt_ch5}. Each of these {\tt CRASH} snapshots is then post-processed separately with the Ly$\alpha$, dust and transmission models which are explained in Sec. \ref{post_proc_ch5} that follows. 

\subsection{ Hydro and radiative transfer}
\label{rt_ch5}
The Ly$\alpha$ and UV LFs presented in this work are based on the results from combined runs of SPH and RT simulations, carried out using {\tt GADGET-2} (Springel 2005)\footnote{http://www.mpa-garching.mpg.de/gadget/} and {\tt CRASH} (Maselli, Ferrara \& Ciardi 2003; Maselli, Ciardi \& Kanekar 2009) respectively, which are coupled to the LAE model explained in Chapters \ref{ch2_lya_sam} and \ref{ch4_lya_cool}. {\tt GADGET-2} generates the redshift evolution of the density field, the baryonic density distribution and the velocity fields in a 100${\rm h^{-1}}$ Mpc comoving volume; we obtain a snapshot of the simulation at $z \sim 6.1$. The specific simulation used in this work is the G5 run described in Springel and Hernquist (2003), which is part of an accurate study focused on modelling the star formation history of the universe. We have chosen the G5 run since it contains several physical ingredients necessary for our investigation. Firstly, the volume is large enough so that cosmic variance is minimized in the determination of the LFs. Secondly, the value of $\dot M_*$ is self-consistently inferred from physically motivated prescriptions that convert gas particles into stellar particles and that properly take account of the mechanical/chemical feedback associated to supernovae and galactic
winds. This point is particularly relevant since the intrinsic luminosity of the galaxies, both in the Ly$\alpha$ and UV, depends sensitively on $\dot M_*$. Also, since the physics governing star formation and galaxy evolution is modelled accurately, the simulation gives us a reliable representation of the galaxy population. As expected, the large simulation volume naturally leads to a relatively coarse mass resolution with the resolution mass being $2.12 \times 10^9$ $M_\odot$ ($3.26\times 10^8$ $M_\odot$) for DM (gas) particles. Running a FOF algorithm on the SPH particle distribution, we identify galaxies and obtain their intrinsic properties, including $\dot M_*$, $Z_*$, $M_g$ and $M_*$. These are then used to calculate the intrinsic Ly$\alpha$/continuum luminosity, the dust enrichment and the escape fraction of continuum/Ly$\alpha$ photons from the galaxy. All these are presented and discussed in greater detail in Sec.~\ref{post_proc_ch5}  and \ref{visibility_ch5}.

The RT calculations have been carried out by post-processing the {\tt GADGET-2} SPH snapshot at $z \sim 6.1$ using the {\tt CRASH} (Cosmological RAdiative Scheme for Hydrodynamics) code. {\tt CRASH} is a 3D, grid based algorithm which combines ray tracing and Monte Carlo sampling techniques to describe the propagation of ionizing radiation through an arbitrary distribution of hydrogen/helium gas and to follow the evolution of the gas properties affected by the matter-radiation interactions, i.e. ionization fractions and temperature. The algorithm implementation is fully described in Maselli, Ferrara \& Ciardi (2003) and Maselli, Ciardi \& Kanekar (2009). In order to use the SPH {\tt GADGET-2} outputs as inputs for the gas distribution in space, we need to map the SPH particles onto a grid; for this work we have chosen a $128^3$ grid, resulting in a spatial resolution of 0.78${\rm h^{-1}}$ comoving Mpc. Although this degrades the original SPH spatial resolution, it allows feasible computational times for the RT simulations. The main assumptions made for running {\tt CRASH} are the following: (a) Using RT codes coupled with the Q5 (10 ${\rm h^{-1}} \, {\rm Mpc}$) run (see Springel and Hernquist 2003), Sokasian et al. (2003) studied the effects of small sources ($10^9 M_\odot$) on Reionization. They found that including these sources, they required $f_{esc}=0.2$ for
Reionization to end at $z\sim 7-8$. They also showed that if these small
sources are neglected, the number of HI ionizing photons required scales
up by about 60\%. However, precisely
because we do not resolve galaxies below $10^9 M_\odot$, we initially assume, at $z\sim 6.1$,  the gas to be in photoionization equilibrium with a uniform value of the UVB produced by sources below the simulation resolution scale, corresponding to an average neutral hydrogen fraction, $\langle \chi_{\rm HI} \rangle= 0.3$, (b) we assume that galaxies have formed stars in the past at an average rate $\dot M_*$, equal to the one derived at $z\sim 6.1$ and (c) since the RT simulation is carried out on a SPH snapshot which contains about 2500 galaxies, to reduce the computational running time, we run the RT calculation in the monochromatic mode, with $h\nu=13.665$ eV. This is a reasonable assumption since the Ly$\alpha$ LFs depend mostly on the average ionization fraction and on the 3D topology of the fully ionized
regions; we are therefore not interested in the detailed profiles of the ionization front or a highly accurate estimate of the IGM temperature, to probe which the ionizing radiation of each galaxy would have to be computed from its spectrum. 

The specific photoionization rate is calculated using the second term on the right hand side of Eq. \ref{gamma_blr2} (Sec. \ref{observed_lum}) and then averaging this value over $Z_*$ and $t_*$ of the galaxies in the the SPH simulation.  The
contribution from all the galaxies determined under the above-mentioned assumptions is then super-imposed on the average UVB value. Finally, we use $f_{esc}=0.02$ (Gnedin et al. 2008), as a characteristic value
of  the escape fraction of \HI ionizing photons from each galaxy. The total ionizing radiation field, which is the sum of that produced by all galaxies identified in the simulation and the UVB, is then evolved with {\tt CRASH} up to the Hubble time corresponding to $z \sim 6.1$. However, as we are interested in assessing the visibility of LAE as a function of the mean ionization fraction which could correspond to different epochs along any assigned reionization history, no particular meaning should be attached to the redshift parameter. 

As mentioned in Sec. \ref{cosmic_reio_ch1}, there are huge discrepancies in the value of $f_{esc}$ obtained by different groups: the value ranges between $f_{esc}=0.01-0.73$ ($0.01-0.8$) observationally (theoretically). For this work, we adopt a fiducial value of $f_{esc}=0.02$, based on the results of Gnedin et al. (2008), since a complete modelling of $f_{esc}$ is much beyond of the scope of this thesis. Although it is clear that varying the value of $f_{esc}$ would indeed affect the progression of Reionization within the box, thereby affecting the visibility of LAEs, as will be shown in Sec. \ref{visibility_ch5}, this problem is alleviated by a degeneracy between the fraction of Ly$\alpha$ photons that escape the galaxy (undamped by dust) and the fraction transmitted through the IGM, as shown in Sec. \ref{degen_ch5}.

\begin{figure*}[htb] 
 \center{\includegraphics[scale=0.65]{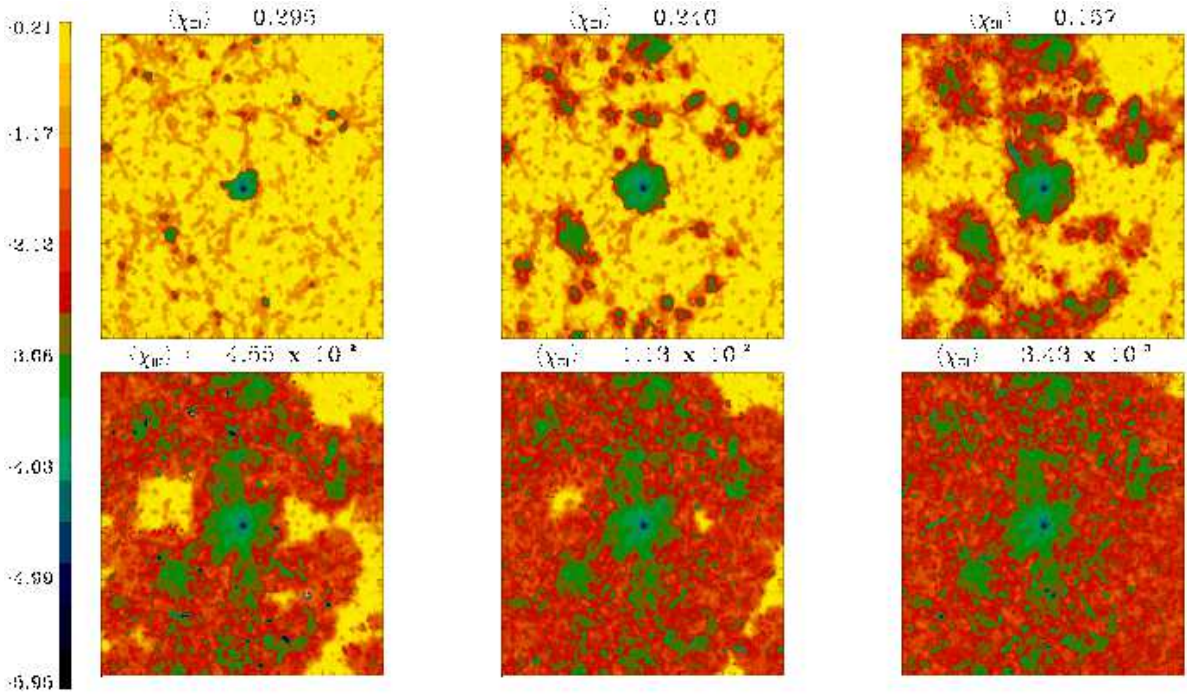}} 
  \caption{Maps showing the time evolution of the spatial distribution of \HI in a 2D plane cut through a 3D box, obtained by running {\tt CRASH} on a {\tt GADGET-2} snapshot (100${\rm h^{-1}}$ Mpc on a side) at $z \sim 6.1$. The colorbar shows the values (in log scale) of the fraction of neutral hydrogen, $\chi_{HI}$. The average decreasing neutral hydrogen fractions marked above each panel, $\langle \chi_{HI} \rangle = 0.295, 0.24, 0.157,4.45\times 10^{-2},1.13 \times 10^{-2} ,3.43 \times 10^{-3}$, correspond to increasing {\tt CRASH} runtimes of $10, 50, 100, 200, 300, 500$ Myr, respectively.  Details in text of Sec.~\ref{rt_ch5}.}
\label{ch5_mapsEvol6}
\end{figure*} 

We now return to the results obtained by the {\tt CRASH} runs. In Fig.~\ref{ch5_mapsEvol6} we show maps of the \HI fraction across a 2D cut through the RT simulation box, at different simulation times. This time sequence then represents a time-line of the reionization process, due to the galaxy population present within the box. Though an approximation, this description catches many of the important features of the reionization process, including the complex topology produced
by the inhomogeneities in the density field, by the galaxy properties and by their spatial distribution. The main aim of this work is to study how the visibility of LAEs ($z \sim 6.1$ here) changes as the ionization field they are embedded in evolves from being highly neutral ($\langle \chi_{HI}\rangle =0.3$) to completely ionized ($\langle \chi_{HI}\rangle \sim 3 \times 10^{-3}$). This requires using a single galaxy population, as obtained from the {\tt GADGET-2} snapshot and only evolving the ionization field; if the galaxy population itself were to change, it would not be possible to disentangle the effects of reionization from the effects of galaxy evolution.

Fig.~\ref{ch5_mapsEvol6} clearly shows the reionization process in its main three phases. First, isolated \HII regions start growing around point sources which are preferentially located along the over-dense filaments of matter. In this stage the size of the isolated \HII regions grows differentially according to the production rate of \HI ionizing
photons from the specific source; the few galaxies with large $\dot M_*$ (e.g. the
largest galaxy located at the center having $\dot M_* \sim 726 \, {\rm M_\odot yr^{-1}}$)
can build an ionized region of about 20 Mpc, in a time as short as 10 Myr (see upper-left panel), which 
grows at a very fast rate due to the vigorous \HI ionizing photon output. If observed at the same evolutionary timescale, due to the smaller \HI ionizing photon budget, the \HII regions of smaller galaxies are confined to sizes of the order of few Mpc ($<4 \, {\rm Mpc}$), which grow slowly. After 10 Myr of continuous star formation activity, the average neutral hydrogen
fraction decreases only to $\langle \chi_{HI} \rangle \sim 0.295$. At this early stage, galaxies with $\dot M_* \leq 25 \, {\rm M_\odot \, yr^{-1}}$, transmit only about 20\% of their Ly$\alpha$ luminosity through the IGM. Consequently the low luminosity end of the LF is depressed with respect to the intrinsic emissivity of these sources. On the other hand, galaxies with larger $\dot M_*$ are able to transmit a larger amount ($\sim 30$\%) of their luminosity through the IGM.

As the \HI ionizing photon production from galaxies continues, the sizes of the \HII regions increase with time. At about $100$ Myr (upper right panel), the isolated \HII regions start overlapping, resulting in an enhancement of the local photo-ionization
rate close to the smaller sources. This allows the ionized regions associated to the latter to grow faster, and results in the
build-up of large \HII regions also around smaller galaxies. This results in an enhancement in the transmission of the Ly$\alpha$ luminosity for these sources, such that this value increases to $\sim 40$\%, consequently leading to a boost of the low-luminosity end of the Ly$\alpha$ LF. As expected, the overlap phase speeds-up the reionization process; for a star formation time $50,100,200, 300$ Myr, $\langle \chi_{HI} \rangle $ decreases to $\sim 0.24,0.16,4.5\times 10^{-2}, 1.1\times 10^{-2}$, respectively. 

Finally at about 300 Myr (lower center panel), the simulation volume is almost completely ionized. This configuration corresponds to the post-overlap phase which leads to a further acceleration of the reionization process. For about 400 Myr of star formation activity, $\langle \chi_{HI} \rangle$ drops to a value of about $4.3 \times 10^{-3}$. This is a consequence of the fact that, as the IGM becomes more ionized, even the ionizing radiation from the numerous smaller galaxies contributes to the overall radiation field. It is very important to point out that, even in this final stage, the ionization field is highly inhomogeneous, with regions of higher ionization fraction corresponding to the environment of point ionizing sources (green regions in the maps). This topology is of fundamental importance in the estimation of the Ly$\alpha$ LFs, as the Ly$\alpha$ transmissivity associated to each source depends on the extent of the highly ionized regions as well as on the residual neutral hydrogen within them; a very tiny fraction of residual neutral hydrogen is sufficient to suppress the intrinsic Ly$\alpha$ luminosity significantly. Even for an ionization fraction, $\langle \chi_{HI} \rangle \sim 3.4 \times10^{-3}$, which results after 500 Myr of star formation, the transmission value ranges between $42-48$\% for galaxies with increasing $\dot M_*$; the blue part of the line can be significantly attenuated even by the tiny residual fraction of about $10^{-6}$, close to the source. As a final note, we again stress that varying $f_{esc}$ would strongly affect both the sizes of the ionized regions and the \HI fraction within them, thereby affecting the progression of ionization within the chosen box.

We then postprocess each of the {\tt CRASH} snapshots shown in Fig. \ref{ch5_mapsEvol6} and explained above, separately with the Ly$\alpha$, dust and transmission models which are explained in Sec. \ref{post_proc_ch5} that follows.

It is important to mention that a single {\tt GADGET-2} snapshot suffices for our work because: (a) the Ly$\alpha$ line is a single, narrow line, which redshifts out of resonance with \HI extremely quickly; to a first approximation, the spatial scale imposed by the Gunn-Peterson damping wing on the size of the \HII region corresponds to a redshift separation of $\Delta z \approx 4.4 \times 10^{-3}$, i.e. about 280 kpc (physical) at $z=6$ (Miralda-Escud\'e 1998), and (b) the decreasing values of both the density of hydrogen and $\chi_{HI}$ with decreasing redshift render the Ly$\alpha$ line transparent to \HI attenuation.

\subsection{Post-processing with the LAE model}
\label{post_proc_ch5}
                                                                                                             
As mentioned in Sec.~\ref{rt_ch5}, for each galaxy in the SPH simulation snapshot, we obtain the values of $\dot M_*$, $Z_*$, $M_g$, and $M_*$; the stellar age is then calculated as $t_* = M_*/\dot M_*$, i.e. assuming  a constant SFR. This assumption has been made for consistency with the RT calculation (see Sec.~\ref{rt_ch5}). These properties are then used to calculate the values of $L_\alpha^{int}$ and $L_c^{int}$ for each galaxy, as explained in Sec. \ref{lum_coolh1_ch4}. These intrinsic values of the Ly$\alpha$ and continuum luminosities so obtained can then be used to calculate the observed luminosity values. 

While it is comparatively easy to calculate the observed continuum luminosity, $L_c$, which is only affected by dust in the ISM (see Eq. \ref{lc_obs2}, Sec. \ref{observed_lum}), the observed Ly$\alpha$ luminosity depends both on the transmission through the ISM dust and the IGM, as explained in Sec. \ref{observed_lum}.

To calculate $f_c$, we again use the SNII dust enrichment model explained in Sec. \ref{dust_model_ch4}, taking into account the processes of SNII dust production, dust destruction and astration. We assume that the average dust mass produced per SNII, $y_d={0.54\, \rm M_\odot}$ (Todini \& Ferrara 2001; Nozawa et al. 2003, 2007; Bianchi \& Schneider 2007). Further, for the efficiency of dust destruction in a SN-shocked ISM, we adopt value $\sim 0.12$; this is a reasonable estimate compared to the value of 0.1 found by McKee (1989). Once the final dust mass, $M_{dust}(t_*)$, is calculated for each galaxy in the simulation, we translate this into an optical depth, $\tau_c$, for continuum photons using Eq. \ref{tauc_ch4}, where we again assume the dust grains to be graphite/carbonaceous ($a=0.05 \mu m$, $s = 2.25\, {\rm g\, cm^{-3}}$; Todini \& Ferrara 2001; Nozawa et al. 2003). We assume the dust distribution scale, $r_d$, to be proportional to the gas distribution scale, $r_g$. As in Sec. \ref{lum_coolh1_ch4}, we calculate $r_g = 4.5 \lambda r_{200}$, where the spin parameter, $\lambda=0.04$ (Ferrara, Pettini \& Shchekinov 2000) and $r_{200}$ is calculated assuming the collapsed region has an overdensity of 200 times the critical density at the redshift considered. The ratio of $r_d$ to $r_g$, is adjusted so as to reproduce the UV LF data, as shown further in Sec.~\ref{visibility_ch5}, and complete details of the calculations mentioned here can be found in Sec. \ref{dust_model_ch4}.

Different extinction curves, including that for the Milky Way, Small Magellanic Cloud and for supernovae give different relations between the extinction of the Ly$\alpha$ and continuum photons, for homogeneously distributed dust. In this spirit, and to get a hint of the dust inhomogeneity, we combine $f_\alpha$ and $f_c$, and use the escape fraction of Ly$\alpha$ photons relative to the continuum ones, $f_\alpha/f_c$, to be the free parameter to calculate $L_\alpha$ in our model. A more detailed explanation of the determination of $f_c$ and $f_\alpha$, required to reproduce the observations, follows later in Sec.~\ref{visibility_ch5}.

The total Ly$\alpha$ optical depth is calculated for each galaxy identified in the {\tt CRASH} outputs corresponding to $\langle \chi_{HI} \rangle \sim 0.29, 0.24, 0.16, 4.5\times 10^{-2}, 1.1 \times 10^{-2}, 4.3\times 10^{-3}, 3.4 \times 10^{-3}$ as mentioned in Sec.~\ref{rt_ch5}, using Eqs. \ref{tau_vel2}-\ref{lorentz_vel2}. The transmission along any LOS is calculated as $T_\alpha{\rm (v)} =e^{-\tau_\alpha ({\rm v})}$, integrating from the position of each galaxy to the edge of the box. To get a statistical estimate of the transmission, we construct 6 LOS starting from the position of each galaxy, one along each of the box axes, for each of the mentioned ionization states. The average value of the transmission for each galaxy is then obtained by averaging the transmission over the 6 LOS. This is done for all the ionization state configurations resulting from the RT calculation.

Once that these calculations are made, all galaxies with $L_\alpha \geq 10^{42.2} {\rm erg \, s^{-1}}$ and an observed equivalent width, $EW = (L_\alpha / L_c) \geq 20$ \AA \,are identified as LAEs and used to construct the cumulative Ly$\alpha$ and UV LFs which can then be compared to the observations. In principle, our results can be compared to both the observed LFs at $z \sim 5.7$ and $6.6$. However, there exists a huge uncertainty between the complete photometric and confirmed spectroscopic sample at $z \sim 6.6$; we limit the comparison between the theoretical model and observations to the data accumulated at $z \sim 5.7$ by Shimasaku et al. (2006), in this work.

It is important to see that once that the value of $f_{esc}$ is fixed (to 0.02 in this calculation), the only {\it two free parameters} of our model to match the theoretical Ly$\alpha$ and UV LFs to the observed ones are: the dust distribution radius, $r_d$, relative to the gas radius, $r_g$ and the relative escape fraction of Ly$\alpha$ photons as compared to the continuum photons, $f_\alpha/f_c$. Both these parameters remain quite poorly understood due to a lack of observational data about the dust distribution/topology in high-redshift galaxies; they must therefore be inferred by comparing the theoretical LFs to the observed ones. Details on how these parameters are determined follow in Sec.~\ref{visibility_ch5}.

\section{LAE visibility during reionization}
\label{visibility_ch5}

Once the combined SPH+RT calculations are carried out and the LAE
model implemented, we are in a position to quantify the importance of
reionization, peculiar velocities and the dust enrichment on the
observed LFs. Physically, the reionization process leads to a decrease in $\langle \chi_{HI} \rangle$, thereby increasing $T_\alpha$; on the
other hand, peculiar velocities caused by galactic scale outflows (inflows), redshift (blueshift) the Ly$\alpha$ photons, thereby leading to a higher (lower) value of $T_\alpha$ as explained in Sec. \ref{observed_lum}. A handle on the dust enrichment is necessary since dust grains absorb both Ly$\alpha$ and continuum photons, thereby affecting their escape fractions from the galaxy.   

\subsection {The LFs for dust-free galaxies and a stationary IGM}
\label{nodust_novel_ch5}

\begin{figure}[htb]
  \center{\includegraphics[scale=0.5] {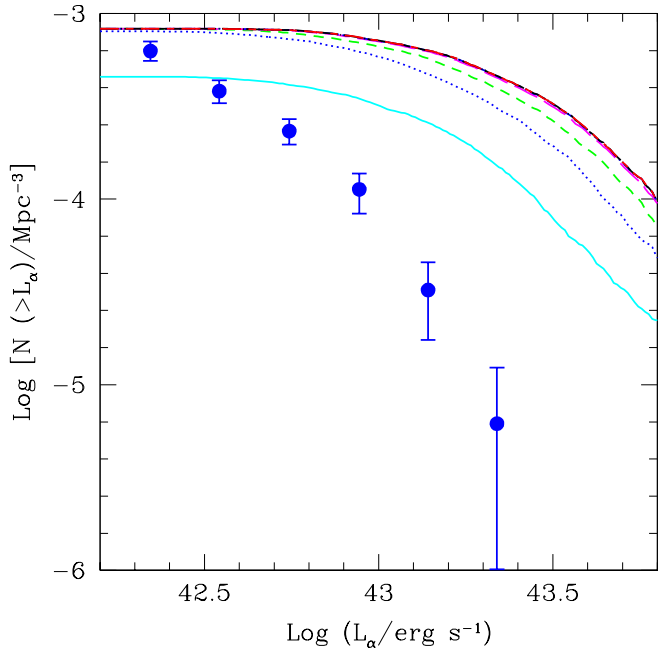}} 
 \caption{Cumulative Ly$\alpha$ LF for galaxies identified as LAEs from the SPH simulation snapshot including the full RT calculation and density fields but ignoring velocity fields (${\rm v}_p=0$) and dust ($f_\alpha=f_c = 1$). The lines from bottom to top correspond to {\tt CRASH} outputs with decreasing values of $\langle \chi_{HI} \rangle$ such that: $\langle \chi_{HI} \rangle \sim$ = 0.29 (solid), 0.24 (dotted), 0.16 (short dashed), $4.5 \times 10^{-2}$ (long dashed), $1.1 \times 10^{-2}$ (dot-short dashed), $4.3 \times 10^{-3}$ (dot-long dashed) and $3.4 \times 10^{-3}$ (short-long dashed). Points show the observed LF at $z\sim 5.7$ (Shimasaku et al. 2006). }
\label{zch5_lya_novel_nodust}
\end{figure}

We begin our study by ignoring the effects of the peculiar velocities and assuming all the galaxies to be dust free, to  quantify how each of these two parameters shapes the observed Ly$\alpha$ and UV LFs. The former assumption means that ${\rm v}_p = 0$ in Eqns. \ref{gauss_vel2}, \ref{lorentz_vel2}; the latter implies that all photons produced inside the galaxy escape into the IGM, i.e. $f_\alpha = f_c=1$ in Eqns. \ref{lya_emm2}, \ref{lc_emm2}. We then use the prescriptions detailed in Sec.~\ref{post_proc_ch5} to identify the galaxies that would be visible as LAEs in each {\tt CRASH} output and build their Ly$\alpha$ LF, as shown in Fig.~\ref{zch5_lya_novel_nodust}. 

Since both peculiar velocities and dust are neglected, in this case, while the UV LF is simply the intrinsic LF, the Ly$\alpha$ LF is shaped solely by the transmission through the IGM. As mentioned in Sec.~\ref{rt_ch5}, after a star formation time scale as short as 10 Myr ($\langle \chi_{HI} \rangle \sim 0.29$), the galaxies in the simulation snapshot are able to form \HII regions, with the region size and the ionization fraction inside it increasing with $\dot M_*$; this results in a $T_\alpha$ which increases with $\dot M_*$. As the galaxies continue to form stars, the sizes of the \HII regions built by each source increase with time, leading $\langle \chi_{HI} \rangle$ to decrease to $0.24,0.16$ at  $t=50, 100$ Myr respectively. This leads to an increase in the transmission of the red part of the Ly$\alpha$ line for all the sources, in turn yielding a corresponding increase of the Ly$\alpha$ luminosity of each source, as depicted in Fig.~\ref{zch5_lya_novel_nodust}. However, after about 200 Myr, $\langle \chi_{HI} \rangle$ reduces to $\sim 0.04$; the \HII regions built by each source are large enough so that almost all of the red part of the Ly$\alpha$ line is transmitted and the value of $T_\alpha$ saturates for all LAEs. This results in very similar LFs for $\langle \chi_{HI} \rangle \leq 0.04 $, as seen from the same figure. 

The key point here is that, if peculiar velocities and dust effects are neglected, at no stage of the reionization history, either the slope or the amplitude of the observed Ly$\alpha$ LF can be reproduced; analogous problems arise also when the UV LFs are considered. 

\subsection {The LFs for dusty galaxies and a stationary IGM}
\label{incdust_novel_ch5}

The above discussion implies the need of one or more physical effects attenuating the Ly$\alpha$ and continuum photons, the most obvious of which is the presence of dust in the ISM of these galaxies, which would absorb both Ly$\alpha$ and continuum photons, simultaneously reducing $f_\alpha$ and $f_c$ (models $S1-S7$, Tab. \ref{table1_ch5}). We then include the dust model described in Sec.~\ref{post_proc_ch5} into our calculations. 

\begin{figure*}[htb]
  \center{\includegraphics[scale=0.95]{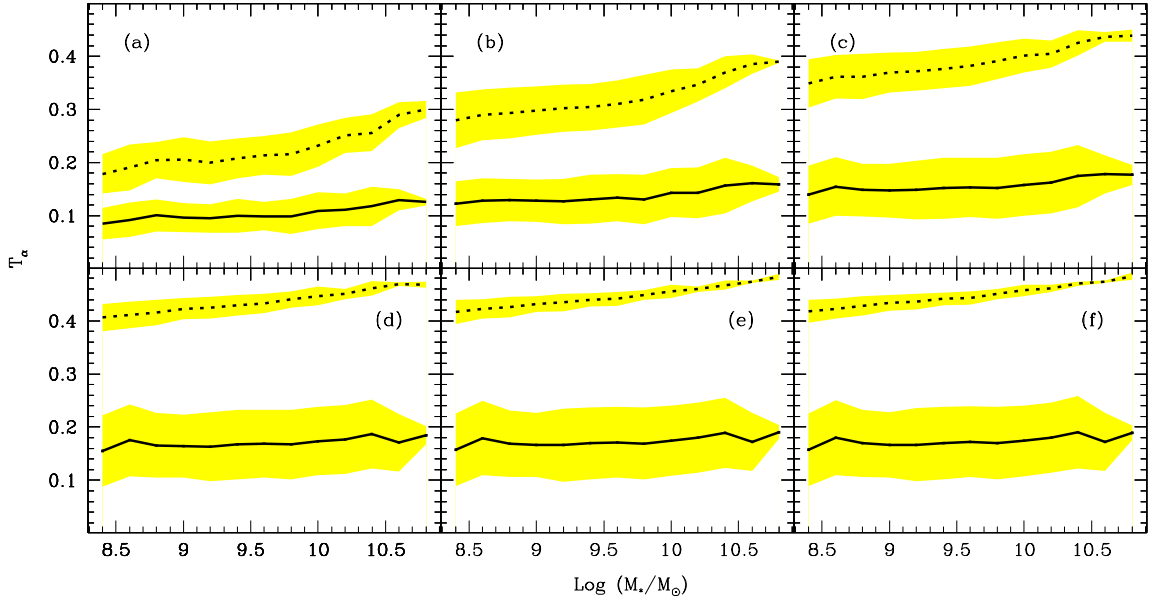}} 
  \caption{$T_\alpha$ as a function of $M_*$ for galaxies identified as LAEs from the simulation snapshot including the full RT calculation, density fields, dust and ignoring (including) velocity fields shown by dotted (solid) lines in each panel. The solid/dotted lines in each panel correspond to the following models given in Tab. \ref{table1_ch5}: (a) ${\rm M1}/{\rm S1}$ ($\langle \chi_{HI} \rangle  \sim 0.29$), (b) ${\rm M2}/{\rm S2}$ (0.24), (c) ${\rm M3}/{\rm S3}$ (0.16), (d) ${\rm M4}/{\rm S4}$ ($4.5 \times 10^{-2}$), (e) ${\rm M5}/{\rm S5}$ ($1.1 \times 10^{-2}$) and (f) ${\rm M7}/{\rm S7}$ ($3.4 \times 10^{-3}$). In each panel, the stellar mass bins span 0.2 dex and the shaded regions represent the $1\sigma$ error bars in each mass bin. }
\label{zch5_fractx_ms} 
\end{figure*}

Since the UV is unaffected by the \HI in the IGM as mentioned in Sec.~\ref{post_proc_ch5} (see also Sec. \ref{observed_lum}), the same value of $r_d = 0.48 r_g$ reproduces the UV LF for all of the ionization states of the IGM, ranging from $\langle \chi_{HI} \rangle \sim 0.29$ to $3.4 \times 10^{-3}$. This assumption of the dust distribution scale leads to an average escape fraction of continuum photons, $f_c \sim 0.12$ for the galaxies we identify as LAEs.

\begin{figure}[htb]
  \center{\includegraphics[scale=0.5]{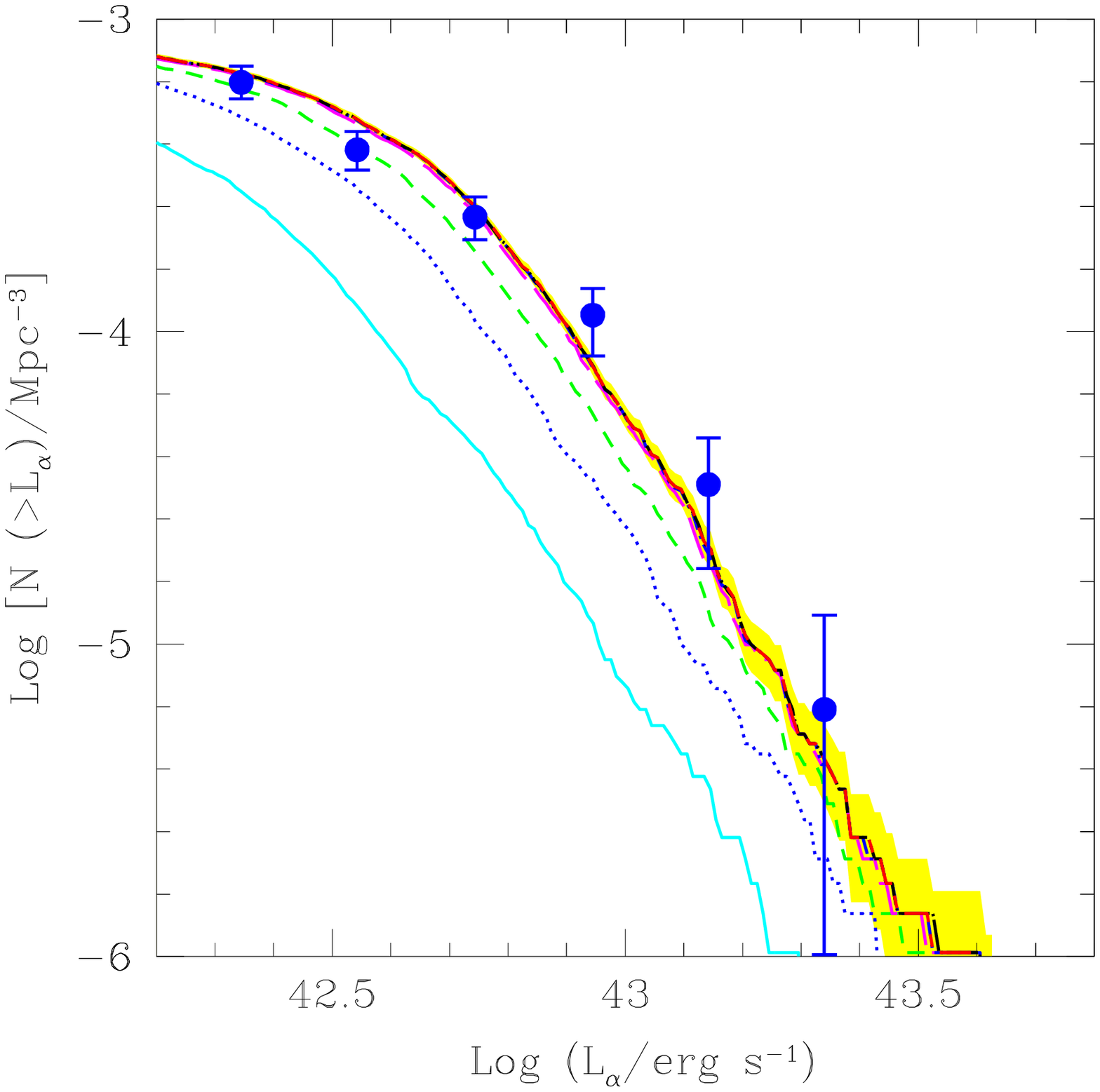}} 
  \caption{Cumulative Ly$\alpha$ LF for galaxies identified as LAEs from the SPH simulation snapshot including the full RT calculation, density fields and dust but ignoring velocity fields (${\rm v}_p=0$), corresponding to models ${\rm S1-S7}$ in Tab. \ref{table1_ch5}. The lines from bottom to top correspond to {\tt CRASH} outputs with decreasing values of $\langle \chi_{HI} \rangle$ such that: $\langle \chi_{HI} \rangle \sim$  0.29 (solid), 0.24 (dotted), 0.16 (short dashed), $4.5 \times 10^{-2}$ (long dashed), $1.1 \times 10^{-2}$ (dot-short dashed), $4.3 \times 10^{-3}$ (dot-long dashed) and $3.4 \times 10^{-3}$ (short-long dashed). The shaded region shows the poissonian error corresponding to $\langle \chi_{HI} \rangle \sim 3.4 \times 10^{-3}$, i.e., model ${\rm S7}$ (Tab. \ref{table1_ch5}) and points show the observed LF at $z\sim 5.7$ (Shimasaku et al. 2006).}
\label{zch5_lya_novel_incdust} 
\end{figure} 

\begin{table*}[htb] 
\begin{center} 
\begin{tabular}{|c|c|c|c|c|c|c|c|} 
\hline 
${\rm Model}$ & $T$ & ${\rm Features}$ & $\langle \chi_{HI} \rangle$ & $\Gamma_{tot}$ & $\langle T_\alpha \rangle$ & $\langle f_c \rangle $ & $f_\alpha/f_c$ \\
$$ & $[{\rm Myr}]$ & $$ & $$ & $[10^{-14}\, {\rm s^{-1}}]$ & $$ & $$ & $$ \\
\hline
${\rm S1}$ & $10$ & ${\rm RT + dust}$ & $0.295$ & $2.8 \times 10^{-3}$  & $0.21$ & $0.12$ & $1.3$\\
${\rm S2}$ & $50$ & ${\rm RT + dust}$ & $0.240$ & $4.0 \times 10^{-3}$  &  $0.30$ & $0.12$ & $1.3$\\
${\rm S3}$ & $100$ & ${\rm RT + dust}$ & $0.157$ & $7.6 \times 10^{-3}$   & $0.37$ & $0.12$ & $1.3$\\
${\rm S4}$ & $200$ & ${\rm RT + dust}$ & $4.55\times 10^{-2}$ & $3.3 \times 10^{-2}$  & $0.42$ &  $0.12$ &$1.3$\\
${\rm S5}$ & $300$ & ${\rm RT + dust}$ & $1.13 \times 10^{-2}$ & $0.14$  & $0.43$ & $0.12$ &$1.3$\\
${\rm S6}$ & $400$ & ${\rm RT + dust}$ & $4.33 \times 10^{-3}$ & $0.33$ & $0.44$ & $0.12$ & $1.3$\\
${\rm S7}$ & $500$ & ${\rm RT + dust}$ & $3.43 \times 10^{-3}$ & $0.48$   & $0.44$ & $0.12$ & $1.3$\\

$$ & $$ & $$ & $$ & $$  & $$ & $$ & $$ \\

${\rm M1}$ & $10$ & ${\rm RT + VF + dust}$ & $0.295$ & $2.8 \times 10^{-3}$   & $0.10$ & $0.12$ & $3.7$\\
${\rm M2}$ & $50$ & ${\rm RT + VF +dust}$ & $0.240$ & $4.0 \times 10^{-3}$   & $0.13$ & $0.12$ & $3.7$\\
${\rm M3}$ & $100$ & ${\rm RT + VF +dust}$ & $0.157$ & $7.6 \times 10^{-3}$  & $0.15$ & $0.12$ &$3.7$\\
${\rm M4}$ & $200$ & ${\rm RT + VF +dust}$ & $4.55 \times 10^{-2}$ & $3.3 \times 10^{-2}$  & $0.16$ & $0.12$ &$3.7$\\
${\rm M5}$ & $300$ & ${\rm RT + VF +dust}$ & $1.13 \times 10^{-2}$ & $0.14$  & $0.17$ & $0.12$ & $3.7$\\
${\rm M6}$ & $400$ & ${\rm RT + VF +dust}$ & $4.33 \times 10^{-3}$ & $0.33$ & $0.17$ & $0.12$ & $3.7$\\
${\rm M7}$ & $500$ & ${\rm RT + VF +dust}$ & $3.43 \times 10^{-3}$ & $0.48$ & $0.17$ & $0.12$ & $3.7$\\
\hline
\end{tabular} 
\end{center}
\caption {The model designation (col 1), the star formation timescale to ionize the IGM (col 2), the features included in the model (VF stands for peculiar velocity fields) (col 3), the average neutral hydrogen fraction (col 4), the value of the total photoionization rate (sum of the contribution from the UVB and from all the galaxies in the snapshot) corresponding to this neutral hydrogen fraction (col 5), the average transmission for all the galaxies identified as LAEs (col 6), the average escape fraction of continuum photons for LAEs for the model presented in Sec.~\ref{post_proc_ch5} (col 7) and the relative escape fraction of Ly$\alpha$ and continuum photons to best fit the observations for $\langle \chi_{HI} \rangle \leq 0.04$ (col 8).}
\label{table1_ch5} 
\end{table*}

We now momentarily digress to discuss the effect of reionization on $T_\alpha$ for different stellar mass ranges. However, since $\dot M_* $ scales with $M_*$ (details in Sec. \ref{phy_prop_ch5}), we discuss $T_\alpha$ in terms of $\dot M_*$. As mentioned in Sec.~\ref{rt_ch5}, we initialize the {\tt CRASH} runs with $\langle \chi_{HI} \rangle \sim 0.3$. In a star formation timescale of 10 Myr (model $S1$, Tab. \ref{table1_ch5}), by virtue of the \HII regions that already start growing, the total photoionization rate (sum of contributions from the UVB and all the galaxies in the simulation) has a value of $\Gamma_{tot}$\footnote{ $\Gamma_{tot}$, the total photoionization rate, is calculated assuming ionization-recombination equilibrium over average values of density and ionization fraction in the simulation volume, so that $$\Gamma_{tot} = \frac{(1-\langle \chi_{HI} \rangle)^2 \langle n_H \rangle \alpha_B}{\langle \chi_{HI} \rangle}.$$ Here, $\langle n_H \rangle$ is the average hydrogen number density in the simulation volume and $\alpha_B$ is the case B recombination co-efficient. } $= 2.8 \times 10^{-17} {\rm s^{-1}}$ and $\langle \chi_{HI} \rangle$  decreases slightly to 0.295. At this point, due to their smaller \HI ionizing photon output, galaxies with $\dot M_* \leq 25 \, {\rm M_\odot \, yr^{-1}}$,  have $T_\alpha \sim 0.2$; galaxies with larger $\dot M_*$, have $T_\alpha \sim 0.3$, as shown in Panel (a) of Fig.~\ref{zch5_fractx_ms}. As the star formation continues, for 50 (100) Myr of star formation (model $S2,S3$ respectively, Tab. \ref{table1_ch5}), the \HII region sizes increase and $\Gamma_{tot}$ increases slightly to $\sim 4.0 \times 10^{-17}$ $(7.7 \times 10^{-17}) \, {\rm s^{-1}}$; $T_\alpha$ increases and ranges between $0.28-0.4$ (0.34-0.44) for $\dot M_* \sim 8-200 \, {\rm M_\odot \, yr^{-1}}$, Panel b (c), Fig.~\ref{zch5_fractx_ms}. Finally for $\Gamma_{tot} \sim 3.3 \times 10^{-16} \, {\rm s^{-1}}$, corresponding to $\langle \chi_{HI} \rangle \sim 0.04$ (model $S4$, Tab. \ref{table1_ch5}), the transmission settles to $T_\alpha \sim 0.4-0.48$ for $\dot M_* \sim 8-200 \, {\rm M_\odot \, yr^{-1}}$ (Fig. \ref{zch5_fractx_ms} d-f, models $S5-S7$, Tab. \ref{table1_ch5}); in about 200 Myr from the ignition of star formation, the Str\"omgren spheres built by these LAEs are large enough so that the redshifted Ly$\alpha$ photons are no longer affected by the \HI outside this region. However, the residual \HI inside this ionized region leads to an absorption of the photons blueward of the Ly$\alpha$ line, and hence, about half of the line is transmitted. The values of $\Gamma_{tot}$, $\langle \chi_{HI} \rangle$ and the $T_\alpha$ values averaged for all LAEs in each {\tt CRASH} output are shown in Tab. \ref{table1_ch5}.

Using the above $T_\alpha$ values and dust model fixed by the UV LF, the only free parameter we are left with, to match the theoretical and observed Ly$\alpha$ LFs is $f_\alpha/f_c$; this only scales the Ly$\alpha$ LF without affecting its shape. As mentioned above, the $T_\alpha$ value for galaxies identified as LAEs settles for $\langle \chi_{HI} \rangle \leq 0.04$, which leads to a corresponding saturation in the Ly$\alpha$ LF. We find that $f_\alpha/f_c \sim 1.3$ reproduces the slope and the magnitude of the observed Ly$\alpha$ LF quite well for all the {\tt CRASH} outputs where $\langle \chi_{HI} \rangle \leq 0.04$ (models $S4-S7$, Tab \ref{table1_ch5}), as shown in Fig.~\ref{zch5_lya_novel_incdust}. However, due to decreasing values of $T_\alpha$, the Ly$\alpha$ LF is progressively under-estimated for increasing values of $\langle \chi_{HI} \rangle$, as seen from the same figure.

\subsection {The LFs including dust and IGM peculiar velocities}
\label{incdust_incvel_ch5}

\begin{figure}[htb] 
  \center{\includegraphics[scale=0.5]{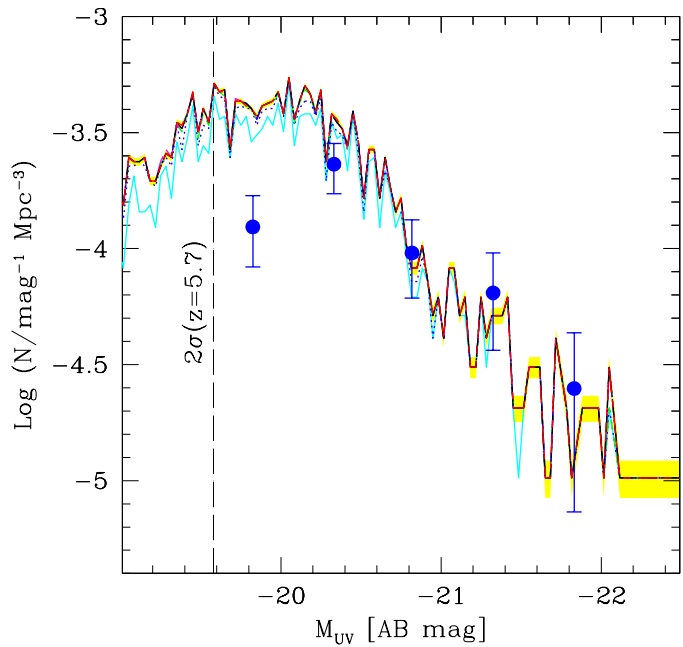}} 
  \caption{UV LF for galaxies identified as LAEs from the simulation snapshot including the full RT calculation, density fields, velocity fields and dust, corresponding to models ${\rm M1}$ to ${\rm M7}$ in Tab. \ref{table1_ch5}. The lines from bottom to top correspond to decreasing values of $\langle \chi_{HI} \rangle$ such that: $\langle \chi_{HI} \rangle \sim$ 0.29 (solid), 0.24 (dotted), 0.16 (short dashed), $4.5 \times 10^{-2}$ (long dashed), $1.1 \times 10^{-2}$ (dot-short dashed), $4.3 \times 10^{-3}$ (dot-long dashed) and $3.4 \times 10^{-3}$ (short-long dashed). The shaded region shows the poissonian error corresponding to $\langle \chi_{HI} \rangle \sim 3.4 \times 10^{-3}$, i.e., model ${\rm M7}$ (Tab .\ref{table1_ch5}) and points show the observed UV LF at $z\sim 5.7$ (Shimasaku et al. 2006).}
\label{zch5_uv_incvel_incdust} 
\end{figure}

\begin{figure}[htb] 
  \center{\includegraphics[scale=0.5]{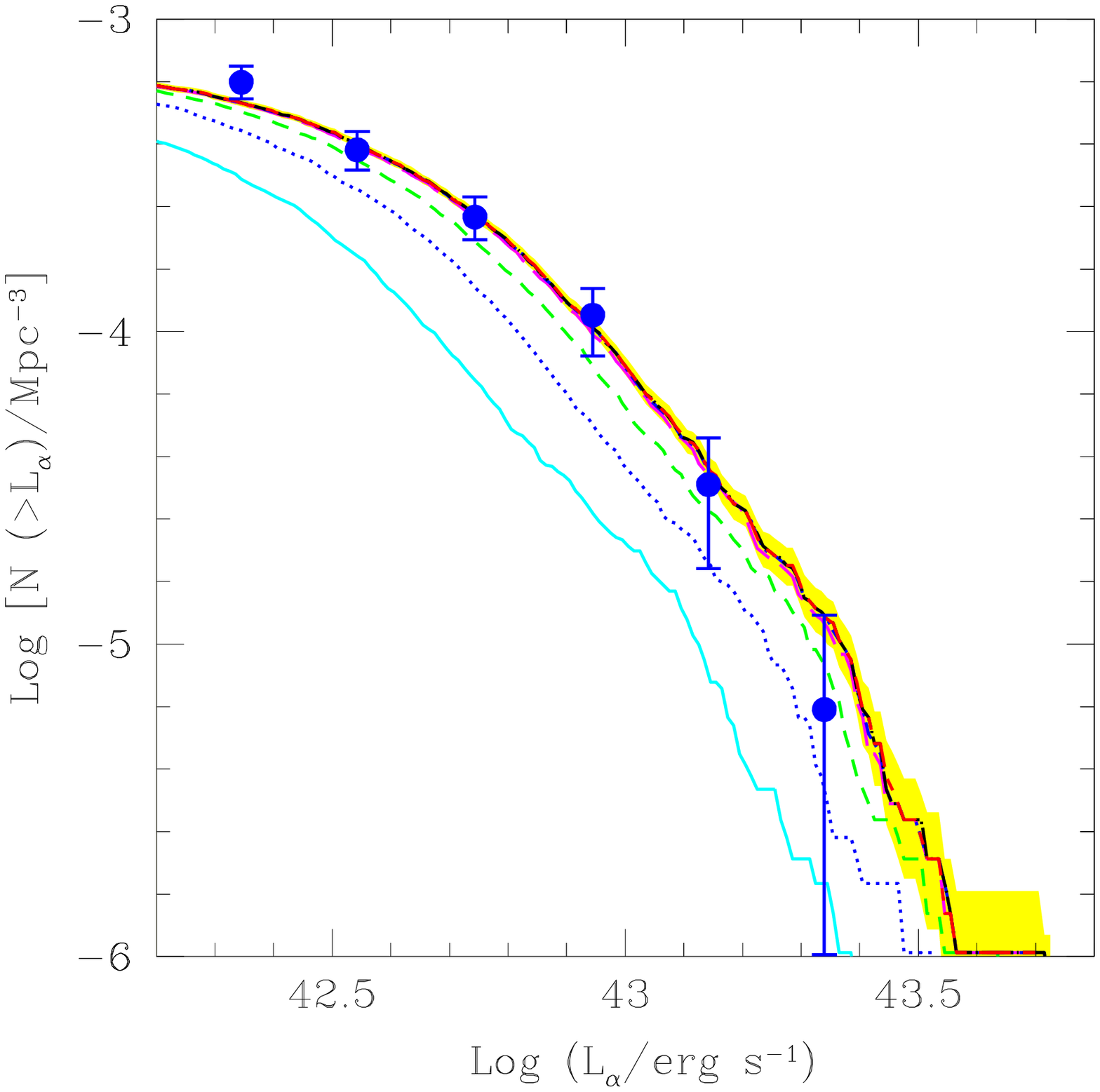}} 
  \caption{Cumulative Ly$\alpha$ LF for galaxies identified as LAEs from the simulation snapshot including the full RT calculation, density fields, velocity fields and dust, corresponding to models ${\rm M1}$ to ${\rm M7}$ in Tab. \ref{table1_ch5}. The lines from bottom to top correspond to decreasing values of $\langle \chi_{HI} \rangle$ such that: $\langle \chi_{HI} \rangle \sim$ 0.29 (solid), 0.24 (dotted), 0.16 (short dashed), $4.5 \times 10^{-2}$ (long dashed), $1.1 \times 10^{-2}$ (dot-short dashed), $4.3 \times 10^{-3}$ (dot-long dashed) and $3.4 \times 10^{-3}$ (short-long dashed). The shaded region shows the poissonian error corresponding to $\langle \chi_{HI} \rangle \sim 3.4 \times 10^{-3}$, i.e. model ${\rm M7}$ (Tab. \ref{table1_ch5}) and points show the observed LF at $z\sim 5.7$ (Shimasaku et al. 2006).}
\label{zch5_lya_incvel_incdust} 
\end{figure}

Once the above framework is in place, we include and study the effect of the final component that can affect the observed luminosity, the presence of peculiar velocities (models $M1-M7$, Tab. \ref{table1_ch5}). Once this is done by consistently deriving the peculiar velocity field from the simulations, we again reproduce both the magnitude and slope of the observed UV LF as shown in Fig.~\ref{zch5_uv_incvel_incdust}. Of course this does not come as a surprise since velocity fields do not affect UV photons. However, as explained in Sec. \ref{observed_lum}, galactic scale outflows (inflows) from a galaxy, redshift (blueshift) the Ly$\alpha$ photons, thereby leading to a higher (lower) $T_\alpha$ value. It is quite interesting to see that we can again match the theoretical Ly$\alpha$ LFs to the observed ones by a simple scaling between $f_\alpha$ and $f_c$ for $\langle \chi_{HI} \rangle \leq 0.04$ (models $M4-M7$, Tab. \ref{table1_ch5}), as seen from Fig.~\ref{zch5_lya_incvel_incdust}. However, we require a much higher value of $f_\alpha/f_c \sim 3.7$, compared to the value of $1.3$ excluding velocity fields (models $S4-S7$, Tab. \ref{table1_ch5}). Again, $T_\alpha$, and hence the Ly$\alpha$ LF, get progressively more damped with an increase in $\langle \chi_{HI} \rangle$, as seen from Fig.~\ref{zch5_lya_incvel_incdust}, (models $M3-M1$, Tab. \ref{table1_ch5}).

We now discuss the reason for the higher $f_\alpha/f_c \sim 3.7$ value required to reproduce the Ly$\alpha$ LF when velocity fields are considered, as compared to the ratio of $1.3$, when they are not, for $\langle \chi_{HI} \rangle \leq 0.04$.
The galaxies we identify as LAEs have halo masses $M_h \sim
10^{10.4-12} \, M_\odot$, which correspond to $\geq 2\sigma$
fluctuations at $z \sim 6.1$. These LAEs are therefore subject to
strong inflows since they lie in dense regions. As these inflows
blue-shift the Ly$\alpha$ photons, even photons in the red part of the
line are attenuated, thereby reducing $T_\alpha$. This can be seen
clearly from Fig.~\ref{zch5_fractx_ms}, where the solid lines, which
represent the outcomes from the models including velocities, are
always significantly below the dotted ones, which represent the model
with no velocity field included. When velocity fields are included we find
$T_\alpha \sim 0.08-0.12$ ($0.16-0.18$) for $\langle \chi_{HI} \rangle \sim 0.29$ ($3.4\times
10^{-3}$) for $\dot M_* \sim 8-200 \, {\rm M_\odot \, yr^{-1}}$. Averaging over all the LAEs for $\langle \chi_{HI}\rangle \sim 3.4\times 10^{-3}$, we find the value of $T_\alpha$ $\sim 0.17$ ($0.44$) including (excluding) the effects of peculiar velocity fields. This requires that when peculiar velocities are included, a correspondingly larger fraction ($0.44/0.17 = 2.6$) of Ly$\alpha$ photons must escape the galaxy, undamped by dust to bring the observed luminosity up to the levels it would reach in the absence of these inflows. Although many galaxies do show outflows, powered by supernova explosions, these dominate at very small scales ($\leq 170$ physical Kpc). However, the small redshift boost imparted by these is negligible compared to the blue-shifting of the Ly$\alpha$ line because of large scale inflows; eventually, it is these dominant inflows that determine the value of $T_\alpha$.

Further, including velocity fields changes the slope of the LF; when
velocity fields are not included, $T_\alpha$ basically scales with $\dot M_*$; when these are included, $T_\alpha$ is the most damped for the
largest masses, since these see the strongest inflow velocities by
virtue of their largest potential wells. 
This can be seen clearly from Fig.~\ref{zch5_fractx_ms} where the slope of
$T_\alpha$ is visibly shallower for the solid curves (including
velocities) than for the dotted ones (neglecting velocities). This has
the effect of {\it flattening} the slope of the Ly$\alpha$ LF as
can be seen in Fig.~\ref{zch5_lya_incvel_incdust}. 

\subsection {The $f_\alpha-T_\alpha$ degeneracy and the dusty nature of LAEs}
\label{degen_ch5}

\begin{figure}[htb]
  \center{\includegraphics[scale=0.6]{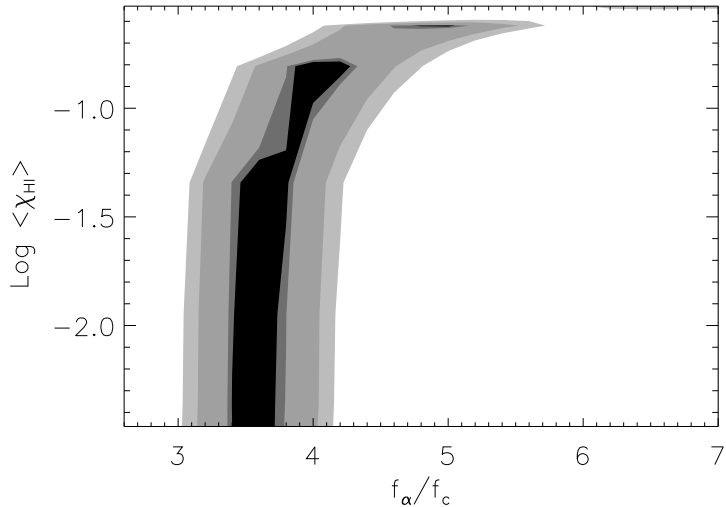}} 
 \caption{The $1-5\sigma$ probability contours (black to light gray respectively) for combinations of $\langle \chi_{HI} \rangle$ and $f_\alpha/f_c$ that fit the observed Ly$\alpha$ LF.  Within a $1\sigma$ ($5\sigma$) error, we can not distinguish an IGM with $\langle \chi_{HI} \rangle \sim 3.4 \times 10^{-3}$ from one where $\langle \chi_{HI} \rangle \sim 0.16$ (0.24) for $f_\alpha/f_c \sim 3.4-4.1$ ($3.0-5.7$). This implies a {\it degeneracy} between the ionization state of the IGM and dust clumping (or grain properties) inside high-redshift galaxies; the ionization state of the IGM cannot be tightly constrained unless the relative escape fraction of Ly$\alpha$ compared to the continuum photons is reasonably well understood. Refer Sec.~\ref{degen_ch5} for details. }
\label{zch5_contour_levels2}
\end{figure}

As an important result, the above analysis shows that there exists a degeneracy between the ionization state of the IGM and dust clumping (or grain properties) inside high-redshift galaxies, i.e., a high (low) $T_\alpha$ can be compensated by a low (high) $f_\alpha$. This is shown in Fig.~\ref{zch5_contour_levels2}, where we find that within a $1\sigma$ error, for $f_\alpha/f_c \sim 3.4-4.1$, we can not distinguish an IGM with $\langle \chi_{HI} \rangle \sim 3.4 \times 10^{-3}$ from one where $\langle \chi_{HI} \rangle \sim 0.16$. Within the area under the $5\sigma$ error, $f_\alpha/f_c \sim 4.1- 5.7$ can also fit the Ly$\alpha$ LF for $\langle \chi_{HI} \rangle \sim 0.24$. This leads to the very interesting conclusion that the ionization state of the IGM cannot be tightly constrained unless the relative escape fraction of Ly$\alpha$ compared to the continuum photons is reasonably well understood.

We now discuss the dusty nature of the LAEs we identify from the SPH simulation. We require that for the complete  LAE model ($M1-M7$, Tab. \ref{table1_ch5}), the value of $f_\alpha/f_c$ must range between $3.4-4.1$ ($3-5.7$) for an average neutral hydrogen fraction of $\langle \chi_{HI} \rangle \leq 0.16$ ($\leq 0.24$). However, no single extinction curve gives a value of $f_\alpha/f_c >1$. One of the simplest ways of explaining this large relative escape fraction is to invoke the multiphase ISM model as proposed by Neufeld (1991), wherein the ISM is multiphase and consists of a warm gas with cold dust clumps embedded in it. This inhomogeneity of the dust distribution can then lead to a larger attenuation of the continuum photons relative to the Ly$\alpha$.

We also translate $f_c$ into the color excess, $E(B-V)$, for models $M1-M7$ (Tab. \ref{table1_ch5}) and compare this value to other high redshift LAE observations. At high redshifts, the observed properties of the most distant quasars (Maiolino et al. 2006) and gamma-ray bursts (Stratta et al. 2007) can be successfully interpreted using a SN extinction curve (Todini \& Ferrara 2001; Bianchi \& Schneider 2007). Using the same curve, we find the average value of the color excess, $E(B-V)= 0.2$, corresponding to an average $f_c=0.12$. When expressed in terms of the widely used Calzetti extinction law (Calzetti et al. 2000), the average value of $f_c=0.12$ corresponds to a color excess $E(B-V)=0.21$. This average inferred color excess value is in very good agreement with recent experimental determinations: by fitting the SEDs of 3 LAEs at $z = 5.7$, Lai et al. (2007) have inferred $E(B-V) < 0.225-0.425$; in a sample of 12 LAEs at $z = 4.5$, Finkelstein et al. (2009a) have found $E(B-V) = 0.04-0.37$. Further, using SPH simulations similar to those used here, Nagamine et al. (2008) have inferred a value of $E(B-V)\sim 0.15$; this difference can be attributed to the fact that these authors use a single value of the IGM transmission across all galaxies. Since the Supernova and Calzetti extinction curves yield similar values of E(B-V), refer Fig. \ref{zch4_fesc_clrexs} for a flavor of $f_\alpha$ $=(1.5,0.6) f_c$, shown with open circles and triangles respectively) as a function of E(B-V) (obtained using the supernova extinction curve), compared to the GALEX LAE sample (Deharveng et al. 2008), IUE local starburst sample (McQuade et al. 1995; Storchi-Bergmann et al. 1995) and the Atek et al. (2008) sample.

We now summarize our two main results: (a) we find that the Ly$\alpha$ LF can be well reproduced (to within a $5\sigma$ error) by an average neutral hydrogen fraction as high as 0.24 (an almost neutral IGM), to a value as low as $3.4 \times 10^{-3}$, corresponding to an ionized IGM, provided that the increase in the transmission is compensated by a decrease in the Ly$\alpha$ escape fraction from the galaxy, (b) we find that to reproduce the Ly$\alpha$ LF, for any ionization state of the IGM, we require $f_\alpha/f_c>1$, a value that cannot be obtained using any existing extinction curve; this raises the need to invoke a multiphase ISM model, in which dust clumps are embedded in a highly ionized ISM, to facilitate the Ly$\alpha$ photon escape relative to that of continuum photons.

\section{Physical properties of LAEs}
\label{phy_prop_ch5}

We are now in a position to discuss the physical properties of the galaxies we identify as LAEs, including the stellar and gas mass, metallicity, stellar ages, dust mass and escape fraction of Ly$\alpha$ photons from the ISM. As mentioned in Sec.~\ref{post_proc_ch5}, while the UV LF is independent of the ionization state of the IGM, the Ly$\alpha$ LF can be well reproduced if a low (high) Ly$\alpha$ escape fraction from the galaxy is compensated by a high (low) transmission through the IGM. Hence, by scaling up $f_\alpha/f_c$ for increasing values of $\langle \chi_{HI} \rangle > 0.04$ (models M1-M3, Tab. \ref{table1_ch5}), we would broadly always identify the same galaxy population as LAEs. We now show the physical properties for the LAEs identified using model M7 (Tab. \ref{table1_ch5}), which includes the full RT calculation, density/velocity fields and dust, with $f_\alpha/f_c = 3.7$. 

\begin{figure*}[htb] 
  \center{\includegraphics[scale=0.9]{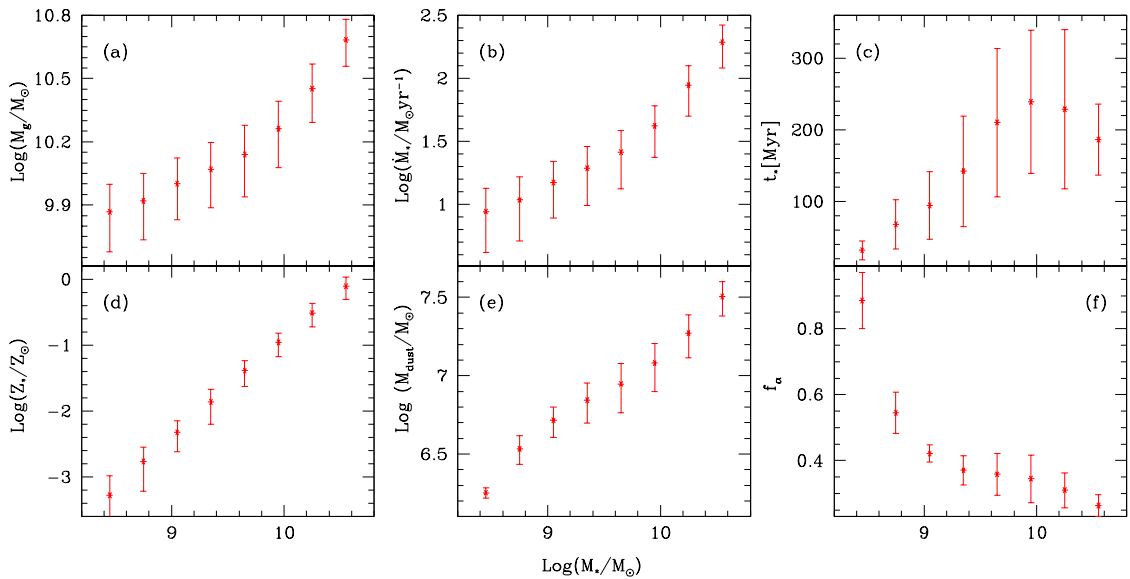}} 
  \caption{Physical properties of the galaxies identified as LAEs using model ${\rm M7}$ in Tab. \ref{table1_ch5}. This model includes the full RT calculation, density, velocity fields, dust and has $\langle \chi_{HI} \rangle \sim 3.4 \times 10^{-3}$. The panels show (a) the gas mass, $M_g$ (b) SFR, $\dot M_*$, (c) mass weighted stellar age, $t_*$ and (d) the mass weighted stellar metallicity, $Z_*$, (e) the total dust mass, $M_{dust}$ and, (f) the escape fraction of Ly$\alpha$ photons, $f_\alpha$,  plotted as a function of the stellar mass, $M_*$. The values of each quantity are shown averaged over $M_*$ bins spanning 0.2 dex, with the error bars representing the $1\sigma$ error in each bin. }
\label{zch5_phy_prop6} 
\end{figure*}      

We find that there is the direct correlation between $M_g$ and $M_*$;
LAEs with a larger $M_*$ are also more gas rich. However, the ratio
$M_g/M_*$ is the largest (smallest) for LAEs with the smallest
(largest) $M_*$, as one can deduce from panel (a) of
Fig.~\ref{zch5_phy_prop6}. If we assume the halo mass to scale with the
total baryonic mass ($M_g+M_*$), according to the cosmological ratio,
this implies that more massive galaxies are more efficient in turning
their gas into stars. This trend is as expected, since even a small amount of star formation activity in low mass galaxies can lead to large outflows of gas, thereby suppressing further star formation. This situation however, does not occur in galaxies with larger masses, which do not witness large, galactic scale outflows, by virtue of their much larger potential wells (Mac Low \& Ferrara 1999). 

LAEs with $M_* \leq 10^{9.7} M_\odot$ have $\dot M_* \sim 8-25 \, {\rm
  M_\odot \, yr^{-1}}$, i.e. most of the LAEs have a sustained but not
exceptionally large star formation activity. However, galaxies with
larger stellar masses are much more efficient in converting their gas
into stars, leading to SFR as large as $200 \, {\rm M_\odot\,
  yr^{-1}}$, as seen from panel (b), Fig.~\ref{zch5_phy_prop6}, which is
just a positive feedback of the $M_g-M_*$ relation mentioned above.

Although the average age of the galaxies are calculated as $t_* = M_* / \dot M_*$, it is interesting to see that all the galaxies we identify as LAEs have $t_*\geq 10$ Myr, as shown in panel (c) of Fig.~\ref{zch5_phy_prop6}; further the standard deviation on the smallest ages are the smallest. Since we calculate the ages assuming $\dot M_*$ to have been constant over the entire star formation history, it is entirely possible that many of the LAEs are actually older (younger) if $\dot M_*$ in the past was smaller (larger) than the final value at $z \sim 6.1$. It is important to see that the age distribution is in good agreement with that shown in Panel a1, Fig. \ref{zch3_phy1}, based on an accurate modelling of the star formation history. However, we are unable to comment on this further in absence of the complete star formation history for each galaxy. Using these average ages, we find that LAEs are intermediate age galaxies, instead of being extremely young ($<$ 10 Myr) or extremely old ($\sim 1$ Gyr) objects. 

The mass weighted stellar metallicity of LAEs scales with $M_*$ as shown in panel (d) of Fig.~\ref{zch5_phy_prop6}; LAEs with a higher $\dot M_*$ are more dust enriched, which is only to be expected since the metals have a stellar origin. The metallicity values for the smallest halos show the largest dispersion, possibly arising due to the differing values of feedback in these low mass halos. Compared to the analogous mass-metallicity relation observed at lower redshifts (Tremonti et al. 2004; Panter et al. 2008; Maiolino et al. 2008), we do not see the sign of a flattening of metallicity towards larger masses which might imply that at high redshift galaxies are not massive enough to retain their metals and behave as close-boxes.

In our model, we have assumed SNII to be the primary dust factories and the dust amount is regulated solely by stellar processes: dust is produced by SNII, destroyed in the ISM shocked by SNII and astrated into stars, as mentioned before in Sec.~\ref{post_proc_ch5}. The dust amount, therefore, scales with $\dot M_*$, with the  most star forming galaxies being
most dust enriched, as shown in panel (e) of Fig.~\ref{zch5_phy_prop6}. A caveat that must be mentioned here is that as outflows tend to occur on smaller scales with respect to inflows, they are only marginally resolved by our RT simulations. Lacking this information, we have not included their destructive impact into the computation of the dust mass,
which could therefore be somewhat overestimated. 

As expected, we find that $f_\alpha$ decreases with increasing dust enrichment of the galaxy; galaxies with larger $M_*$, and hence, $\dot M_*$, have a smaller Ly$\alpha$ escape fraction. The value of $f_\alpha$ decreases by a factor of about 3, going from 0.9 to 0.3 as $M_*$ runs from $10^{8.5-10.7} \, {\rm M_\odot}$, as shown in panel (f), Fig.~\ref{zch5_phy_prop6}.

\section{The LAE EWs}
\label{ew_ch5}                                                                                                    
 
\begin{figure}[htb]
  \center{\includegraphics[scale=0.5]{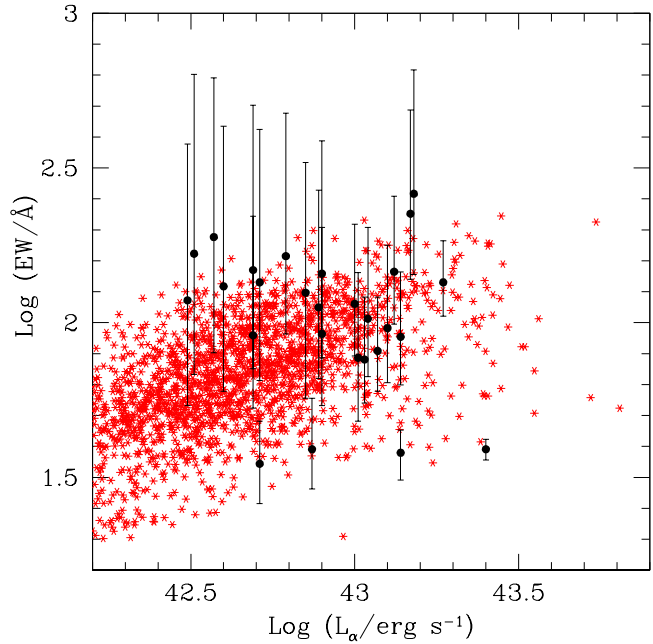}} 
\caption{Observed EWs from Shimasaku et al. (2006) (circles) and model values of the observed EWs (astrexes) as a function of the observed Ly$\alpha$ luminosity. The theoretical model includes the full RT calculation, density, velocity fields, dust and has $\langle \chi_{HI} \rangle \sim 3.4 \times 10^{-3}$, corresponding to model ${\rm M7}$, Tab. \ref{table1_ch5}.}
\label{zch5_ew} 
\end{figure}

We finally compare the EW distribution calculated from our model to that obtained from observations, as shown in Fig.~\ref{zch5_ew}. 

There is a trend of increasing EW in more luminous objects which cannot be yet solidly identified in the available (and uncertain) observational data. In addition, the observed EWs show a large dispersion, varying by as much as a factor of 3 for the same $L_\alpha$ value; this suggests that the observed EW depends on several structural parameters of LAEs, namely $\dot M_*$, peculiar velocities and dust clumping, to name a few. By virtue of using a LAE model which depends on the intrinsic galaxy properties, includes a dust calculation and takes peculiar velocities/ density inhomogeneities into account, we also obtain a large spread (20-222 \AA) in the observed EWs. This is a clear improvement on our own previous work (see Sec. \ref{match_obs_ch4}) in which some of these ingredients were not yet accounted for.  

Therefore, contrary to the results of Malhotra \& Rhoads (2002), we find that we do not need to invoke extremely young LAEs to be responsible for the large equivalent widths ($EW \geq 200$ \AA) observed for a number of LAEs at $z \sim 4.5, 5.7$ by Dawson et al. (2007) and Shimasaku et al. (2006) respectively.

\section{Conclusions }
\label{conc_ch5}

In this chapter, we have built a coherent model for LAEs, which includes: (a) cosmological SPH simulations run using {\tt GADGET-2}, to obtain the galaxy properties ($\dot M_*$, $Z_*$, $M_*$), (b) the Ly$\alpha$ and continuum luminosities produced, calculated using the intrinsic properties of each galaxy, accounting for both the contribution from stellar sources and cooling of collisionally excited \HI in their ISM, (c) a model for the dust enrichment, influencing the escape fraction of Ly$\alpha$ and continuum photons, and (d) a complete RT calculation, carried out using {\tt CRASH}, including the effects of density and velocity fields, used to calculate the transmission of Ly$\alpha$ photons through the IGM. The main results obtained using this model are now summarized:

\begin{itemize}

\item We find that assuming all galaxies to be dust-free and ignoring gas peculiar velocities, none of the ionization states of the IGM with $\langle \chi_{HI} \rangle \sim 0.29 $ to $3.4\times 10^{-3}$, can reproduce either the slope or the galaxy number density of the observed UV or Ly$\alpha$ LF. 

\item We find that we can reproduce the Ly$\alpha$ LF using $f_\alpha/f_c \sim 1.3$ (3.7) for all ionization states such that $\langle \chi_{HI} \rangle \leq 0.04$ excluding (including) peculiar velocity fields, since the Ly$\alpha$ LF settles to a constant value at this point due to a saturation in $T_\alpha$, as explained in Sec.~\ref{visibility_ch5}. The higher $f_\alpha/f_c$ value required to reproduce the observations when velocity fields are included, arise because LAEs reside in $\geq 2 \sigma$ fluctuations; LAEs are therefore subject to strong inflows which blue-shift the Ly$\alpha$ photons, thereby reducing $T_\alpha$. This decrease in $T_\alpha$ must therefore be compensated by a larger escape fraction from the galaxy to fit the observations. 

\item The above {\it degeneracy} between the ionization state of the IGM and
the dust distribution/clumping inside high-redshift galaxies has been
quantified (see Fig.~\ref{zch5_contour_levels2}); the Ly$\alpha$ LF can be
well reproduced (to within a $5\sigma$ error) by $\langle \chi_{HI}
\rangle \sim0.24$, corresponding to a highly neutral IGM, to a value
as low as $3.4 \times 10^{-3}$, corresponding to an ionized IGM,
provided that the increase in $T_\alpha$ is compensated by a decrease
in the Ly$\alpha$ escape fraction from the galaxy. This leads to the
very interesting conclusion that the ionization state of the IGM can
not be constrained unless the escape fraction of Ly$\alpha$ versus
continuum photons is reasonably well understood. 

\item As for the dusty nature of LAEs, we again find that $\langle f_c \rangle \sim 0.12$, averaged over all the LAEs in any snapshot, which corresponds to a color excess, $E(B-V) \sim 0.21 (0.2)$, using a Calzetti (SN) extinction curve. Secondly, since no single extinction curve (Milky Way, Small Magellanic Cloud, supernova) gives a value of $f_\alpha/f_c >1$; this larger escape fraction of Ly$\alpha$ photons relative to the continuum can be understood as an indication of a multi-phase ISM model (Neufeld 1991) where dust clumps are embedded in a more diffuse ionized ISM component.

\end{itemize}

At the very last, we discuss a few caveats in our model. 

\begin{itemize}

\item First, we have used a constant value of the escape fraction of \HI ionizing photons, $f_{esc}=0.02$ in all our calculations. However, the value of $f_{esc}$ remains poorly constrained both observationally and theoretically, as described in Sec. \ref{cosmic_reio_ch1}. A larger (smaller) value of $f_{esc}$ would lead to larger (smaller) \HII regions, thereby affecting the progress of the reionization process and hence $T_\alpha$; a change in $f_{esc}$ would also affect $L_\alpha^{int}$. However, this problem is alleviated to a large extent by our result that shows a degeneracy between $f_\alpha$ and $T_\alpha$; a constant higher (lower) value of $f_{esc}$ would lead to (a) a decrease (increase) in the intrinsic Ly$\alpha$ luminosity, $L_\alpha^{int}$, and (b) an increase (decrease) in $T_\alpha$. However, by scaling $f_\alpha/f_c$ appropriately, we would always be able to reproduce the Ly$\alpha$ LF.

\item Secondly, the average age of the galaxies are calculated as $t_* = M_* / \dot M_*$. Since we calculate the ages assuming a constant value of $\dot M_*$, it is entirely possible that some of the LAEs could be slightly older (younger) if $\dot M_*$ in the past was smaller (larger) than the final value at $z \sim 6.1$. Galaxies younger than $10$ Myr would have higher EWs as compared to the values shown here; however, this could be possible only for the the smallest galaxies, which could assemble due to mergers and undergo star formation in a time as short as 10 Myr.

\item Third, in the dust model explained in Sec.~\ref{post_proc_ch5}, we have considered dust destruction by forward sweeping SNII shocks. However, Bianchi \& Schneider (2007) have shown that reverse shocks from the ISM can also lead to dust destruction, with only about 7\% of the dust mass surviving the reverse shock, for an ISM density of $10^{-25} {\rm gm \, cm^{-3}}$.  Since we neglect this effect, we might be over-predicting the dust enrichment in LAEs. This, however, does not affect our results because the dust optical depth depends on the surface density of the dust distribution as shown in Eq. \ref{tauc_ch4}; a large (small) dust mass can be distributed in a large (small) volume to obtain identical values of the optical depth.

\end{itemize}

\chapter{LAEs: clues on the MW infancy}\label{ch6_mw_lae}
LAEs have by now been used extensively as probes of both the ionization state 
of the IGM and probes of high redshift galaxy evolution, as mentioned in Chapter \ref{ch1_intro} and shown in Chapters \ref{ch2_lya_sam}-\ref{ch5_lya_rt}. However, in spite of the growing data sets, there has been no effort to establish a link between the properties of these early galaxies to observations of the local Universe, {\it in primis} the Milky Way (MW). Our aim in this chapter is to investigate the possible connection between the Galactic building blocks and LAEs at a time when the Universe was $\approx 1$ Gyr old. This will allow us to answer to questions such as: are the progenitors of MW-like galaxies visible as LAEs at high redshifts? How can we discriminate amongst LAEs which are 
possible MW progenitors and those that are not? What are the physical 
properties of these Galactic building blocks?

To this end, we adopt a novel approach (Sec. \ref{model_ch6}) of coupling the semi-analytical code {\tt GAMETE}, which traces the hierarchical build-up of the Galaxy, successfully reproducing most of the observed MW and dwarf satellite properties at $z=0$, to the LAE model previously developed in Chapters. \ref{ch2_lya_sam}-\ref{ch4_lya_cool}, that reproduces a number of important observational data sets accumulated for high-z LAEs. The results are shown in Sec. \ref{results_ch6} and we end with a discussion that includes a mention of the caveats in our model, in Sec. \ref{conc_ch6}.

\section{The theroretical model}
\label{model_ch6}
We start by describing the principal features of {\tt GAMETE} in Sec. \ref{gamete_ch6} and how this is used to derive the properties of the MW progenitors, for different merger histories. We then use the model described in 
Sec. \ref{identifying_lae_ch6} to compute the intrinsic Ly$\alpha$/continuum luminosity for each progenitor, its dust content and the IGM Ly$\alpha$ transmission, which are used to obtain the observed luminosities, and identify the progenitors in each merger history that would be visibile as LAEs.

\subsection{Obtaining the MW progenitors}
\label{gamete_ch6}
We start by summarizing the main features of the code {\tt GAMETE} (Salvadori, Schneider \& Ferrara 2007, hereafter SS07; Salvadori, Ferrara \& Schneider 2008, hereafter SFS08, Salvadori \& Ferrara 2009), which stands for GAlaxy MErger Tree \& Evolution. This code is used to build-up 80 possible hierarchical merger histories and to derive the properties of the MW progenitors at $z=5.7$. We specifically choose 
$z=5.7$ for all our calculations as it represents the highest redshift for 
which a statistically significant sample of confirmed LAEs is available. 
 
First, the possible hierarchical merger histories of a MW-size DM halo are reconstructed up to $z=20$ via a Monte Carlo algorithm based on the extended Press-Schechter theory (see Salvadori, Schneider \& Ferrara 2007 for more details). The evolution of gas and stars is then followed along each hierarchical tree by assuming that: 
(a) the initial gas content of DM haloes is equal to the universal cosmological 
value $(\Omega_b/\Omega_m)M_h$, (b) at any redshift, there exists a minimum halo mass to form stars, $M_{sf}(z)$, whose evolution accounts for the suppression of SF in progressively more massive objects due to radiative feedback effects 
(see Fig.~1 of Salvadori \& Ferrara 2009), (c) the gradual accretion of cold 
gas, $M_c$, into newly virializing haloes is regulated by a numerically 
calibrated infall rate (Kere\v{s} et~al. 2005), (d) the SFR, $\dot M_*$, 
is proportional to the mass of cold gas inside each galaxy, 
$\dot M_*=\epsilon_* M_c/t_{ff}$, where $\epsilon_*$ is the SF efficiency 
and $t_{ff}$ the halo free fall time, (e) in halos with a virial temperature, 
$T_{vir}<10^4$~K (minihalos), the SF efficiency is reduced as 
$\epsilon = \epsilon_*[1+(T_{vir}/2\times 10^4{\rm K})^{-3}]^{-1}$ due to 
ineffective cooling by H$_2$ molecules. The chemical enrichment of gas, both in 
the proto-Galactic halos and in the MW environment is followed simultaneously 
by taking into account the mass-dependent stellar evolutionary timescales and 
the effects of mechanical feedback due to SN energy deposition 
(see Salvadori, Ferrara \& Schneider 2008 for more details).

The two free parameters of the model (star formation and wind efficiencies) 
are calibrated by reproducing the global properties of the MW (stellar/gas 
mass and metallicity) and the Metallicity Distribution Function (MDF) of 
Galactic halo stars (Salvadori, Schneider \& Ferrara 2007, Salvadori, 
Ferrara \& Schneider 2008); $M_{sf}(z)$ is fixed by matching 
the observed iron-luminosity relation for dwarf spheroidal galaxies 
(Salvadori \& Ferrara 2009). They are assumed to be the same for all 
progenitors in the hierarchical tree. 
\subsection{Identifying LAEs}
\label{identifying_lae_ch6}

By using {\tt GAMETE} we obtain the total halo/stellar/gas masses ($M_h$, $M_*$, $M_g$), the instantaneous SFR ($\dot M_*$), the mass weighted stellar metallicity ($Z_*$) and the mass-weighted stellar age ($t_*$) of each MW progenitor, in each of the 80 realizations considered. These outputs are used to calculate the total values of $L_\alpha^{int}$ and $L_c^{int}$, which include both the contribution from stellar sources and from the cooling of collisionally excited \HI in the ISM (see Sec. \ref{lum_coolh1_ch4} for details).

As in Sec. \ref{intrinsic_lum_ch3}, the intrinsic Ly$\alpha$ luminosity is again translated into the observed luminosity considering both the attenutation due to the dust inside the ISM and the transmission through the IGM, while the observed continuum luminosity only depends on the ISM dust attenuation. We now briefly summarize these calculations: the Ly$\alpha$ IGM transmission is calculated as explained in Sec. \ref{intrinsic_lum_ch3}, taking into account the effect of clustered sources explained in Sec. \ref{clus_ch3}. The value of $f_c$ is calculated using the same parameters as explained in Sec. \ref{dust_model_ch4}, we use $f_\alpha=1.5 f_c$ needed for reproducing the observations at $z \sim 5.7$, as shown in the same section.

Progenitors are then identified as LAEs based on the currently used 
observational criterion: $L_\alpha \geq 10^{42} \, {\rm erg \, s^{-1}}$ 
and the observed equivalent width, $L_\alpha/L_c \geq 20$~\AA. We use a 
comoving volume of the Milky Way environment ($30 {\rm Mpc}^3$) to 
calculate the number density of the progenitors visible as LAEs and the 
observed Ly$\alpha$ LF, shown in Fig. \ref{zch6_LF}.

\section{Results}
\label{results_ch6}
\begin{figure}
\begin{center}
\psfig{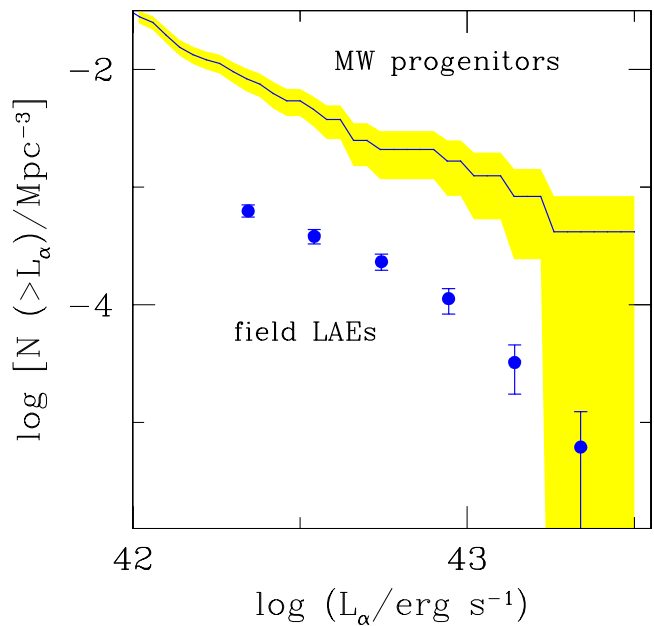}
              \caption{Cumulative Ly$\alpha$ luminosity function at 
       redshift $z \approx 5.7$. The solid line shows the mean Ly$\alpha$ 
       LF of the MW progenitors averaged over 80 realizations of the merger 
       tree; the dashed area represents the $\pm 1 \sigma$ spread among 
       different realizations. For comparison only, the points show the 
       field LAE LF (Shimasaku et~al. 2006).}
  \label{zch6_LF}
\end{center}
\end{figure}

We start by comparing the number density of field LAEs observed at 
$z\approx 5.7$ with the average LF of the Milky Way progenitors at the same 
epoch (Fig.~\ref{zch6_LF}). 

The number density of MW progenitors decreases with 
increasing Ly$\alpha$ luminosity, reflecting the higher abundance of the least 
massive/luminous objects in $\Lambda$CDM models. The MW progenitors cover the 
entire range of observed Ly$\alpha$ luminosities, $L_{\alpha}=10^{42-43.25} 
{\rm erg~s^{-1}}$. We then conclude that {\it among the LAEs observed at 
$z\approx 5.7$ there are progenitors of MW-like galaxies}. For any given 
$L_\alpha$ the number density of MW progenitors is higher than the observed 
value because the MW environment is a high-density, biased region. As discussed 
in Sec. \ref{conc_ch6}, uncertainties on the treatment of dust may also play a role.

\begin{figure}
  \centerline{\psfig{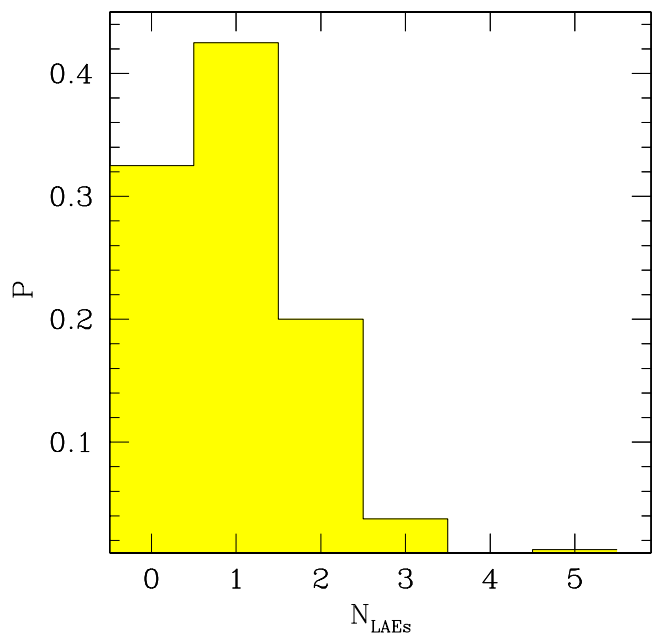}}
  \caption{Probability of finding a number, $N_{LAEs}$, of LAEs in a single
        MW realization, averaged over 80 hierarchical merger histories at
        $z \approx 5.7$.}
  \label{zch6_Prob}
\end{figure}

In Fig.~\ref{zch6_Prob} we show the probability distribution function (PDF), $P$, of 
finding a number $N_{LAEs}$ of LAEs in any given MW realization. The PDF has a 
maximum ($P=0.42$) at $N_{LAEs}=1$; while in any single MW realization, there 
are $\approx 50$ star-forming progenitors with stellar masses $M_* \gsim 10^7\Msun$, 
on average only one of them would be visible as a LAE. We note that $P=0.32$ for 
$N_{LAEs}=0$, while it rapidly declines ($P<0.2$) for $N_{LAEs}>1$. We conclude 
that the MW progenitors that would be observable as LAEs at $z\approx 5.7$ are rare 
($\approx 1/50$), but the probability to have {\it at least} one LAE in any 
MW hierarchical merger history is very high, $P=68\%$. 

Let us now consider the physical properties of the building blocks of the MW. 
$\dot M_*$ represents the dominant physical factor to determine whether a 
progenitor would be visible as a LAE, since it governs  the intrinsic 
Ly$\alpha$/continuum luminosity,  the dust enrichment (and hence absorption)
and $T_\alpha$, as both the size of the ionized region around each source 
and the \HI ionization fraction inside it scale with $\dot M_* $ (Sec. \ref{sfr}). 

\begin{figure*}
\centerline{\psfig{figure=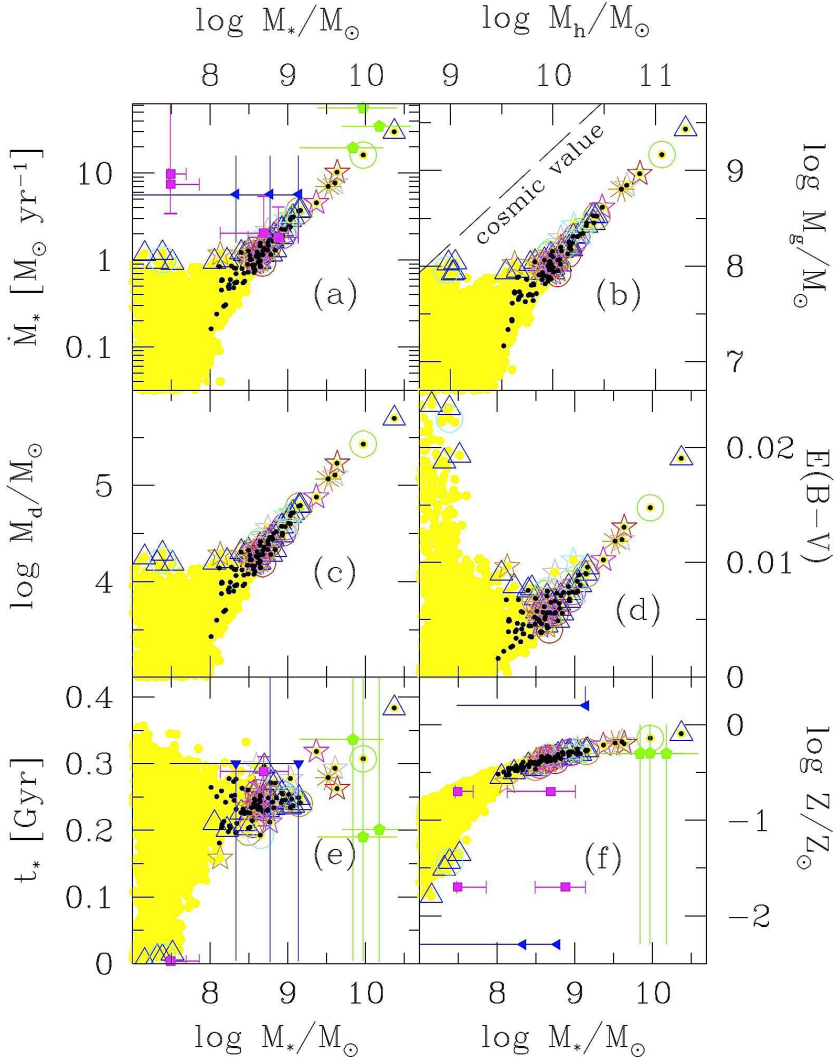,width=11.0cm,angle=0}}
\caption{Physical properties of $z \approx 5.7$ MW progenitors in 80 
different hierarchical merger histories. We show: (i) all the progenitors 
(yellow circles), (ii) the major branches of each hierarchical tree (black 
filled points) and (iii) the progenitors identified as LAEs (colored open 
symbols). LAEs pertaining to the same (different) realizations are shown 
with the identical (different) colored symbols (see the text and Fig. 
\ref{zch6_Prob}). Triangles are used for those realizations in which there is 
only one LAE. As a function of the total stellar mass $M_*$ the various 
panels show: (a) the instantaneous star formation rate, $\dot M_*$; (c) 
the dust mass, $M_{dust}$; (d) the color excess $E(B-V)$; (e) the average 
stellar age, $t_*$; (f) the average stellar metallicity $Z_*$. Panel (b) 
shows the relation between the halo and gas mass, with the cosmic value 
$(\Omega_b/\Omega_m) M_h$ pointed out by the dashed line. Points with 
errorbars are the observational LAE data collected by Ono et~al. 2010 
(magenta squares, 1 LAE, four different models), Pirzkal et~al. 2007 
(blue triangles, 3 LAEs) and Lai et~al. 2007 (green circles, 3 LAEs).}
\label{zch6_Fig3}
\end{figure*}

This implies the existence of a SF rate threshold 
for MW progenitors to be visible as LAEs which is 
$\dot M^{min}_* \approx 0.9\Msun$yr$^{-1}$ (panel (a) of Fig.~\ref{zch6_Fig3}). 
Since $\dot M_* \propto M_g$ (see Sec.~\ref{gamete_ch6}), such a lower limit can be translated into a threshold gas mass: $M^{min}_g\approx 8\times 10^7\Msun$ (panel (b)). 
In turn, the gas content of a (proto-) galaxy is predominantly determined by 
the assembling history of its halo and the effects of  SN (mechanical) 
feedback. While the most massive MW progenitors ($M_h > 10^{10}\Msun$) display 
a tight $M_g-M_h$ correlation, the least massive ones are highly scattered, 
reflecting the strong dispersion in the formation epoch/history of recently 
assembled halos (panel (b)). As a consequence, {\it LAEs typically correspond 
to the most massive progenitors of the hierarchical tree}, i.e. the major 
branches (black points in the panels). In particular,  we find that {\it all} 
haloes with $M_h \geq 10^{10}\Msun$ (40 haloes) are LAEs. At decreasing $M_h$, 
instead, the progenitors can be visible as LAEs only by virtue of a high gas 
mass content or extremely young ages; there are 5 such objects, with 
$M_h \approx 10^9 \Msun$, as seen from panel (b).

By comparing the panels (b) and (c) of Fig.~\ref{zch6_Fig3}, we can see that the dust mass, $M_{dust}$, of the Galactic building blocks closely tracks $M_g$. Since the gas mass 
content of the $z\approx 5.7$ MW progenitors is reduced by $\approx 1$ order 
of magnitude with respect to the initial cosmic value, due to gas (and dust) 
loss in galactic winds, their resulting dust mass is relatively small: 
$M_{dust}\approx 10^{4-5.7}\Msun$. As a consequence all the progenitor
galaxies have a color excess E(B-V)$<0.025$. In panel (d) we see that as expected,  
E(B-V) increases with $M_*$, i.e. more massive galaxies are redder. However 
the trend is inverted for $M_*< 10^8 \Msun$; even though the dust masses in 
these low mass objects ($M < 10^9 \Msun$) is $M_{dust} \lsim 10^{4.2}\Msun$, due 
to their small virial radius, both the gas distribution scale, and hence the 
dust distribution scale are very small (see Sec. \ref{dust_model_ch4}). The concentration 
of the dust in a small area leads to a large dust attenuation and hence, a 
large value of the color excess.

MW progenitors visible as LAEs are generally intermediate age objects, 
$t_* \approx 150 -400$ Myr, as seen from panel (e); the largest progenitors 
tend to be the oldest ones, an expected feature of standard hierarchical 
structure formation scenarios. However, the ages show a large scatter, 
especially at decreasing $M_*$, reflecting the great variety of assembling 
(and SF) histories of recently formed halos. Interestingly the five 
$M_h\approx 10^9\Msun$ newly formed ($z < 6$) progenitors visible as LAEs 
have a very young stellar population, $t_*\leq 5$~Myr. The high $\dot M_*$ 
induced by the large mass reservoir and the copious Ly$\alpha$ production 
from these young stars makes them detectable.

In panel (f) we can see that these newly virializing galaxies are
metal-poor objects, with an average stellar metallicity  $Z_*\approx 0.016-0.044 
\Zsun$. As these galaxies host a single and extremely young stellar population, 
such low $Z_*$ values reflect the metallicity of the MW environment at their 
formation epoch, $z\approx 5.7$ (see the middle panel in Fig.~1 of Salvadori, 
Ferrara \& Schneider 2008). 
The more massive MW progenitors visible as LAEs, instead, are more metal rich, 
$Z_*\approx 0.3-1 \Zsun$; their intermediate stellar populations form during a 
long period (see Fig.~4) and from a gas that is progressively enriched by 
different stellar generations.

The fact that the physical properties of the LAEs progenitors obtained with our
model are consistent with those inferred from the observations of field LAEs 
(Fig.~\ref{zch6_Fig3}) is a notable success of our model.

\begin{figure}[htb]
  \centerline{\psfig{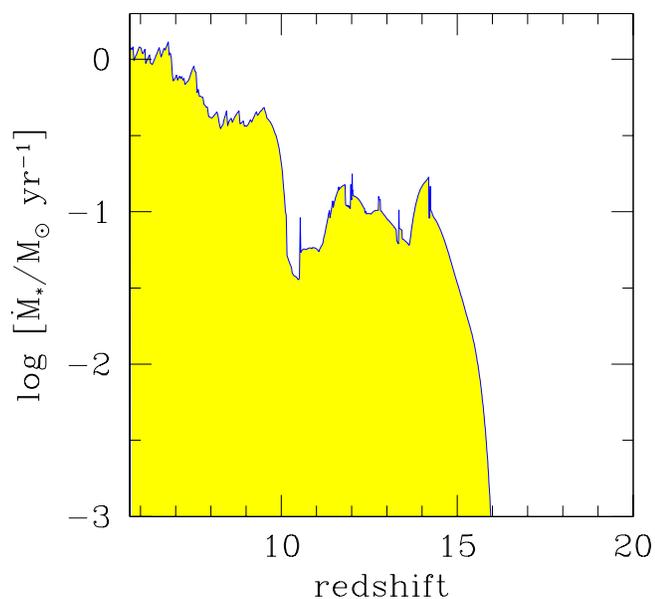}}
  \caption{Star formation history of a typical MW progenitor 
        identified as a LAE at $z\approx 5.7$. }
  \label{zch6_sfh}
\end{figure}

As mentioned above, the scatter in the LAE ages originates from different 
assembling and SF histories (SFH) of the MW progenitors. This can be better 
understood by considering the SFH of a typical MW progenitor 
identified as a LAE, as shown in Fig.~\ref{zch6_sfh}; we define as 
``typical'' a LAE whose properties match the average values: 
$M_h\approx 10^{10}\Msun$, $M_g \approx 10^8 \Msun$, 
$ \dot M_* \approx 2.3$~$\Msun $${\rm yr^{-1}}$, $t_*\approx 230$~Myr.

We find that since most 
LAEs typically correspond to the major branch, their progenitor seeds are 
associated with high-$\sigma$ peaks of the density field virializing and 
starting to form stars at high redshifts ($z\approx 16$). The SF rapidly 
changes in time, exhibiting several bursts of different intensities and 
durations, which follow merging events refueling gas for SF. During the 
``quiet'' phase of accretion, instead, SN feedback regulates SF into a more 
gentle regime. The duration and intensity of the peaks depends on the 
effectiveness of these two competitive physical processes. At high redshifts 
($z>10$) the peaks are more pronounced because (i) the frequency of major 
merging is higher and (ii) mechanical feedback is stronger given the shallower 
potential well of the hosting halos ($M\approx 10^{8.5}\Msun$, Salvadori,
Ferrara \& Schneider 2008). 

\begin{figure}[htb]
 \centerline{\psfig{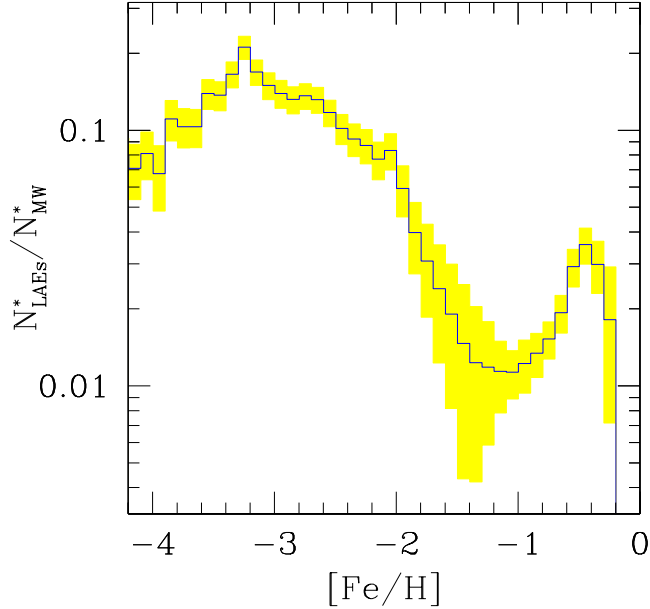}}
 \caption{Fraction of LAE MW halo relic stars as a function of their iron 
abundance. The histogram shows the average fraction among 80 realizations 
of the hierarchical tree; the shaded area shows the $\pm 1\sigma$ dispersion 
among different realizations.}
 \label{zch6_mdf}
\end{figure}

We finally turn to the last question concerning the contribution of old and 
massive progenitors seen as LAEs, to the very metal-poor stars ([Fe/H]$<-2$) 
observed in the MW halo. In Fig.~\ref{zch6_mdf} we show the fractional contribution 
of long-living stars from LAEs to the MDF at $z=0$. 

LAEs provide $> 10\%$ of 
the very metal-poor stars; the more massive the LAE, the higher is the number 
of [Fe/H]$<-2$ stars they contribute. This is because such metal-poor stellar 
fossils form at $z>6$ in newly virializing halos accreting pre-enriched gas 
out of the MW environment (see Salvadori, Schneider \& Ferrara 2007 and 
Salvadori et~al. 2010a). By $z \approx 5.7$, many of these premature building 
blocks have merged into the major branch, i.e. the LAE. Because of the gradual 
enrichment of the MW environment, which reaches [Fe/H]$\approx -2$ at $z\approx 5.7$ 
(see Fig.~1 of Salvadori, Ferrara \& Schneider 2008), most of [Fe/H]$>-2$ stars 
form at lower redshifts, $z < 5.7$, thus producing the drop at [Fe/H]$>-2$. 
Note also the rapid grow of $N^*_{LAEs}/N^*_{MW}$ at [Fe/H]$>-1$, which is 
a consequence of the self-enrichment of building blocks resulting from 
{\it internal} SN explosions. 

\section{Conclusions}
\label{conc_ch6}
We now summarize the main results obtained in this chapter.

\begin{itemize}
\item According to our results, the progenitors of MW-like galaxies cover a wide 
range of observed Ly$\alpha$ luminosity, $L_{\alpha} =10^{39-43.25}$ erg~s$^{-1}$, 
with $L_{\alpha}$ increasing with $M_*$ (or, equivalently, $M_h$); {\it hence some
of them can be observed as LAEs.}

\item In each hierarchical merger history we find 
that, on average, {\it only one} star-forming progenitor (among $\approx 50$) 
is a LAE, usually corresponding to the major branch of the tree ($M_h\approx 
10^{10}\Msun$). Nevertheless, the probability to have {\it at least}
one visible progenitor in any merger history is very high ($P=68\%$).  

\item Although rare, some LAEs (5 out of 80) are hosted 
by small DM halos, $M_h\approx 10^9\Msun$, which are yet visible as (faint) 
LAEs ($L_{\alpha}\approx 10^{42.05}$~erg~s$^{-1}$) by virtue of their high 
star formation rate and extremely young stellar population, $t_* < 5$~Myr. 

\end{itemize}

The main uncertainity in this work concerns the treatment of dust in calculating the LF and observed properties of the MW progenitor LAEs identified here, especially 
at the low luminosity end of the LF. Several aspects require additional study. 
As gas, metal and dust are preferentially lost from low mass halos, pushing 
the mass resolution of simulations to even lower masses would be important. 
Also, the amount of dust lost in SN-driven winds remains very poorly understood, 
an uncertainty that propagates in the evaluation of the continuum and Ly$\alpha$ 
radiation escaping from the galaxy. Finally, Ly$\alpha$ photons could be affected 
considerably by the level of dust clumping. It is unclear to what extent these 
effects may impact the visibility of LAE. Progress on these issues is expected when high-resolution FIR/sub-mm observations of LAEs with ALMA will become available in the near 
future, as shown by Dayal, Hirashita \& Ferrara (2010), Finkelstein et al. (2009b) (see also Sec. \ref{pred_submm_ch4}).


\chapter{Simulating high-redshift galaxies}\label{ch7_reio_sources}
\label{introduction_ch7}
As mentioned in Chapter \ref{ch1_intro}, the search for the most distant galaxies, located at the beginning of the cosmic dawn, is now entering its maturity. An enormous amount of galaxy data has been collected upto $z \sim 10$, which has been made possible by a combination of new instruments and selection techniques. Although the most robust information that is obtained concerns the UV LF, its interpretation has been complicated by our lack of understanding of galaxy evolution. In this chapter, we present results using a cosmological SPH simulation that can even resolve dwarf galaxies at high-$z$, in addition to having an excellent treatment of metals; the latter is important to study the PopIII to PopII transition. We start by describing some of the data sets and the problems in interpreting them, in Sec. \ref{quest_highz_ch7}, this is followed by a description of the simulations used, in Sec. \ref{simulations_ch7}. We then build the UV LFs, to compare to observations in Sec. \ref{lf_ch7}. Once this is done, we discuss the properties of these high-$z$ galaxies, including their stellar ages, stellar metallicities, and the stellar mass-SFR relation, in Sec. \ref{prop_1gal_ch7}. We study the contribution from these early galaxies to reionization in Sec. \ref{Rei_sources_ch7} and follow this with a study on the importance of PopIII stars, compared to PoPII, in Sec. \ref{pop3_ch7}. We then discuss the role of early GRBs as probes of both the stellar metallicity and PopIII stars, in Sec. \ref{early_grb_ch7} and end by summarizing the main results, in Sec. \ref{summary_ch7}.

Throughout the paper, we adopt a $\Lambda$CDM cosmological model with parameters $\Omega_M = 0.26$, $\Omega_{\Lambda} = 0.74$, ${\rm h}=0.73$, $\Omega_b=0.041$, $n=1$ and $\sigma_8=0.8$, in agreement with the 3-yr WMAP results (Spergel et al. 2007).

\section{The quest for high redshift galaxies}
\label{quest_highz_ch7}

 The last few years have witnessed a tremendous increase in the data available, and the number of candidates at redshifts as high as $z=10$, corresponding to only half a billion years after the Big Bang. This has been made possible by instruments like the Hubble Space Telescope (HST) and surveys like the Hubble Deep Field, and Hubble Ultra Deep Field. Follow-up experiments performed with the newly installed Wide Field Camera (WFC3), yielding sky images in the F105W (Y-band, $1.05 \mu$m), F125W (J-band, $1.25 \mu$m) and F160W bands (H-band, $1.60 \mu$m), have allowed to push the exploration to very faint (e.g. AB mag = 28.8 in the above bands) galaxies as remote as $z=10$. In addition, the WFC3 crafted filters have considerably alleviated the contamination problem due to interlopers and provided more precise photometric redshift estimates. The standard selection method applied to these survey data sets is based on the dropout technique introduced by Steidel et al. (1996) and later constantly refined and improved by several authors (e.g. Giavalisco et al. 2004, Bouwens et al. 2007). Though this method has proved to be very solid in identifying high-redshift sources, it has the drawback that the exact source redshift cannot be determined with complete confidence.

What have we learned from this wealth of experimental results ? The most solid piece of information that can be determined from the data appears to be the UV LF and, less robustly, its evolution. It is useful to briefly recap the present observational situation marching towards increasing redshift. Bouwens et al. (2007) presented a comprehensive view of galaxy candidates from the UDF/ACS/GOODS fields using NICMOS in the redshift range $z=4-6$. They identified 1416 (627) V-dropouts ($i$-dropouts) corresponding to $z\approx 5$ ($z\approx 6$) down to an absolute UV magnitude of $M_{UV}\approx -17$ with a LF described by a Schechter function with a characteristic luminosity and faint-end slope given by $M_{UV}^* = -20.64 \pm 0.13$ and $\alpha=-1.66\pm 0.09$ ($M_{UV}^* = -20.24 \pm 0.19, \alpha=-1.74\pm 0.16$) respectively. The same group (Bouwens et al. 2008) has extended the data analysis to include $z\approx 7$ $z$-dropouts (8 candidates at $z=7.3$) and J-dropouts (no candidates at $z \approx 9$). More recently, the installation of WFC3 on board the HST has triggered a new series of searches. Oesch et al. (2010) used data collected during the first-epoch WFC3/IR program  (60 orbits) in the Y, J, H bands reaching a magnitude limit of AB$\approx 29 (5\sigma)$. They identified 16 $z$-dropouts in the redshift range $z=6.5-7.5$ from which they obtained a LF with ($M_{UV}^* = -19.91 \pm 0.09, \alpha=-1.77\pm 0.20$), essentially confirming the previous findings while extending it to fainter luminosities ($M_{UV} \approx -18$). Bouwens et al. (2010) pushed the investigation to $z=8.0-8.5$ by using 5 Y-dropouts; finally, Bouwens et al. (2009) were able to identify three J-dropouts. If confirmed, these sources would be the most distant objects detected so far. Similar studies using the same data has been performed by Bunker et al. (2009), who find a comparable number of $z$ and $Y-$dropouts. McLure et al. (2009) did not apply specific color cuts as in the previous works, thus finding a larger number of candidates; however, they pointed out that about 75\% of the candidates at $z>6.3$ (100\% at $z>7.5$) allow a $z<2$ interloper solution. 

Besides the LF, tentative information on the physical properties of these sources can be extracted from their SED, exploiting available {\it Spitzer} data (Eyles et al. 2005; Yan et al. 2006; Stark et al. 2009). In a recent study Labb\'e et al (2009), based on follow-up {\it Spitzer/IRAC} observations, analyzed the SED of 12 $z$-dropout and 4 Y-dropout candidates. None of them is detected in the {\it Spitzer/IRAC} 3.6 $\mu$m band to a magnitude limit of AB=26.9 $(2\sigma)$, but a stacking analysis reveals a robust detection for the $z$-dropout sample and a strong upper limit for the Y-dropout one. The stacked SEDs are consistent with a stellar mass of about $10^9 M_\odot$, no dust reddening, sub-solar metallicity, and best-fit ages of about 300 Myr, implying a formation epoch $z\approx 10$. These results for the stacked sample should be compared with those obtained by Finkelstein et al. (2009a) who performed an object-by-object analysis and found similar ages but with a considerable spread, allowing ages as low as a few Myr.       

One of the major triggers to look for very high-$z$ galaxies is the quest for the reionization sources. The ionizing photon budget provided by the candidate high-$z$ galaxies is often estimated by extrapolating their LF to lower luminosities, a step that introduces a considerable uncertainty in the final determination. Having this in mind, it is still interesting to note that most studies tend to agree on the fact that the integrated UV specific luminosity for the detected galaxies at $z=7-8$ falls short of accounting for the ionizing power required to reionize the intergalactic medium. Of course, this conclusion is subject to at least two major unknown factors, these being the gas clumping factor (affecting its ability to recombine), and the escape fraction of ionizing photons (affected by dust and neutral hydrogen absorption within galaxies). Additionally, the effect of poorly constrained ages and metallicities (including the presence of metal-free, massive Pop III stars), further complicate the calculation.

In spite of the large experimental effort, surprisingly little attention has been devoted by modelers to the very high redshift universe. Most of the work has so far concentrated on a 
semi-analytical approach (Stiavelli, Fall \& Panagia 2004; Schneider et al. 2006; Bolton \& Haehnelt 2007a; Mao et al. 2007; Samui, Subramanian, Srianand 2009) to compute the luminosity function, number counts and emissivity evolution of high-$z$ galaxies. Albeit quite fast and versatile, these methods cannot provide detailed information on the properties of the galaxies, often being based on simplified assumptions. Numerical dedicated simulations attempting to model galaxy populations beyond $z=5-6$ are also very scarce, with the partial exceptions constituted by the works by Nagamine et al. (2006) and Finlator, Dav\'e \& Oppenheimer (2007). 

Our approach is novel and different in spirit from all the previous theoretical ones. As our main aim is to model very high redshift reionization sources, we can afford smaller simulation boxes, thereby reaching the high resolutions required to resolve the dominant reionization sources - dwarf galaxies. Most importantly, though, we have implemented a careful treatment of metal enrichment and of the transition from Pop III to Pop II stars, along with a careful modelling of supernova feedback. Here we are interested in deriving the LF plus other observables from the simulations and to cast them in a form that can be compared directly with the available data or used to make new predictions for the James Webb Space Telescope (JWST).

\section{Numerical simulations}
\label{simulations_ch7}

For the present study we have performed a set of cosmological simulations using the publicly available code GADGET\footnote{www.mpa-garching.mpg.de/galform/gadget/} (Springel 2005) with an improved treatment of chemical enrichment as detailed in Tornatore, Ferrara \& Schneider (2007, TFS07). A unique feature of the computation concerns the IMF of stars, which is taken to
be different for Pop III and Pop II stars and depends on gas metallicity, $Z_g$.  In brief, if $Z_g < Z_{\rm cr}$, the adopted IMF is a Salpeter 
law with lower (upper) limit of $100M_\odot$ ($500 M_\odot$) and only pair-instability supernovae  ($140 M_\odot < M < 260 M_\odot$) contribute to metal enrichment
(Heger \& Woosley 2002). If $Z_g \ge Z_{\rm cr}$, the
above limits are shifted to $0.1 M_\odot$ ($100 M_\odot$),
respectively; stars above $40 M_\odot$ end their lives as black holes
swallowing their metals. According to the canonical choice, we fix $Z_{\rm
  cr} = 10^{-4}Z_\odot$. These two populations, to which we will refer
to as Pop III and Pop II stars respectively, differ also for their
metal yield and explosion energy. Complete details of the simulation can be found in TFS07. The simulation follows the production and transport of six different metal species, namely:
C, O, Mg, Si, S, Fe, based on which the locally appropriate IMF is selected.  
The simulated volume has a linear (comoving) size $L=10 {\rm h^{-1}}$~Mpc with 
$N_p=2\times 256^3$ (dark+baryonic) particles, corresponding to a DM 
(baryonic) particle mass of $M_p=3.62 \times 10^6 {\rm h^{-1}} M_\odot$ ($6.83 
\times 10^5 {\rm h^{-1}} M_\odot$); the corresponding force resolution is $2~$kpc. Our
resolution does not allow us to track the formation of mini-halos, whose stellar 
contribution remains very uncertain due to radiative feedback effects (Haiman 
\& Bryan 2006; Susa \& Umemura 2006; Ahn \& Shapiro 2007, Salvadori \& Ferrara 2009).
The computation is initialized at $z=99$ and carried on until $z=2.5$. The gas photo-ionization and heating rates are calculated at equilibrium with a background ionizing radiation due to the combined contribution of galaxies and quasars, taken from Haardt 
\& Madau (1996), shifted so that the intensity at 1 Ryd is $J_\nu= 0.3\times
10^{-21}$~erg ~s$^{-1}$~Hz$^{-1}$, in agreement with Bolton et al.
(2005). Gas cooling and wind treatment details are the same as in TFS07.

\section{The LFs}
\label{lf_ch7}

As a first check of the simulation results, we compute the evolution of the 
LF of the simulated galaxies and compare it with available data. To this end we make a number of physical assumptions that are discussed in the following. 

For each galaxy at a redshift $z$, the rest-frame luminosity at a wavelength, $\lambda$, is the sum of the contribution of its Pop II and Pop III stars. The SED of Pop II stars is computed by running the {\tt STARBURST99} code (Leitherer et al. 1999) using 
the metallicities and stellar ages appropriate for the galaxy under consideration as obtained from the cosmological simulation. The SEDs of massive (metal-free) Pop III stars have been computed by Schaerer (2002) including the effect of nebular emission. For both populations the IMF is taken according to the same prescription used in the simulation and described in Sec. \ref{simulations_ch7}. The total luminosity of a galaxy is then given by

\begin{equation}
L_\lambda=L_{\lambda}^{\rm II}+L_{\lambda}^{\rm III}=
l_{\lambda}^{\rm II}(t_*^{\rm II},Z_*)\dot{M_*}^{\rm II}+l_{\lambda}^{\rm III}\dot{M_*}^{\rm III}t_*^{\rm III}
\end{equation}

\noindent
where $l_{\lambda}^{\rm II}(t_*^{\rm II},Z_*)$ is the SED template for Pop II stars with mean age $t_*^{\rm II}$ and stellar metallicity $Z_*$ corresponding to a continuous star formation rate of 1 $\Msun$ yr$^{-1}$; $\dot{M_*}^{\rm II}$ is the Pop II star formation rate. The SED template of Pop III stars, whose star formation rate is $\dot{M_*}^{\rm III}$, is $l_{\lambda}^{\rm III}$; 
$t_*^{\rm III} = 2.5\times 10^6$ yr is the mean lifetime of massive Pop III stars (Schaerer 2002). We have implicitly assumed that the emission properties of Pop III stars are constant during the short lifetime of these massive metal-free stars. Next, the luminosity is converted into absolute AB magnitudes, $M_{UV}$, where the suffix UV refers to the wavelength $\lambda= (1600, 1350, 1500, 1700, 1500, 1600)$ \AA~for $z=(5, 6, 7, 8, 9, 10)$, respectively. Though this wavelength is chosen to be consistent with the observational data we compare with, a different value would not affect the results in any sensible way, given the flatness of the spectrum in this short wavelength range. 

\begin{figure*}[htb]
\center{\includegraphics[scale=0.75]{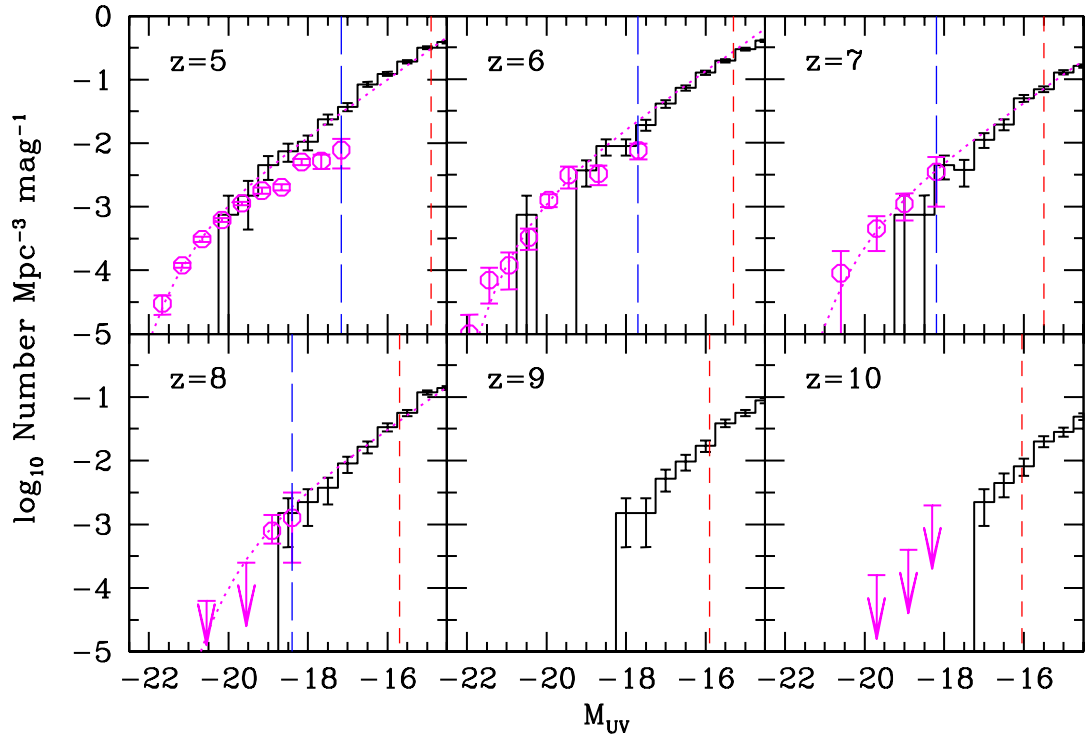}}
\caption{The UV LF of galaxies at the different redshifts shown in 
each panel. Observational data (and upper limits) from HUDF are taken from Bouwens et al. (2007) for $z=5,6$, Oesch et al. (2010) for $z=7$, Bouwens et al. (2010) for $z=8$ and Bouwens et al. (2009) for $z=10$; they are shown as circles (arrows). The histograms show the 
simulated LF with error bars representing Poisson errors. Dotted lines are the
Schechter function fits to the LF;  the vertical short(long)-dashed lines mark the sensitivity limit of JWST (HST/WFC3).}
\label{zch7_highlf_fig1}
\end{figure*}

The LF of galaxies at any redshift $z$ is obtained by counting the galaxies with a given absolute magnitude in each magnitude bin and dividing the final result by the total volume of the
simulation and bin size (0.5 mag). We perform this procedure for the following six redshifts $z=(5, 6, 7, 8, 9, 10)$.  The results are shown in Fig.~\ref{zch7_highlf_fig1} as solid histograms, where the error bars represent the Poisson error on the number of galaxies in each magnitude bin. These theoretical LFs are then compared to the experimental ones collected from the various analyses of the HUDF. For $z=10$ we show the upper limits on the LF obtained from the three available candidates identified by Bouwens et al. (2009). 

Let us now analyze the results shown in Fig. \ref{zch7_highlf_fig1} in more detail. It is clear that the luminosity range sampled by the observations and our predictions is only partially overlapping. This is because on one hand, even the exquisite sensitivity of WFC3 is not sufficient to properly sample the faint-end of the LF ($M_{UV}\simgt -18$); on the other hand, our simulations, which are specifically designed to properly resolve the very first galactic units in a relatively small volume, lack the most massive, rare objects which comprise the bright end of the LF. In spite of these shortcomings, we consider it a rewarding success that the amplitudes of the theoretical and experimental LFs match almost perfectly, and at the same time, have quite similar slopes at all redshifts for which data are available. This is even more striking as no attempts have been made to fit or adjust the theoretical curves to the observed LF, i.e. they have been computed directly from the simulation output with no free extra parameters.

\begin{figure*}[htb]
\begin{tabular}{lr}
\includegraphics[scale=0.3]{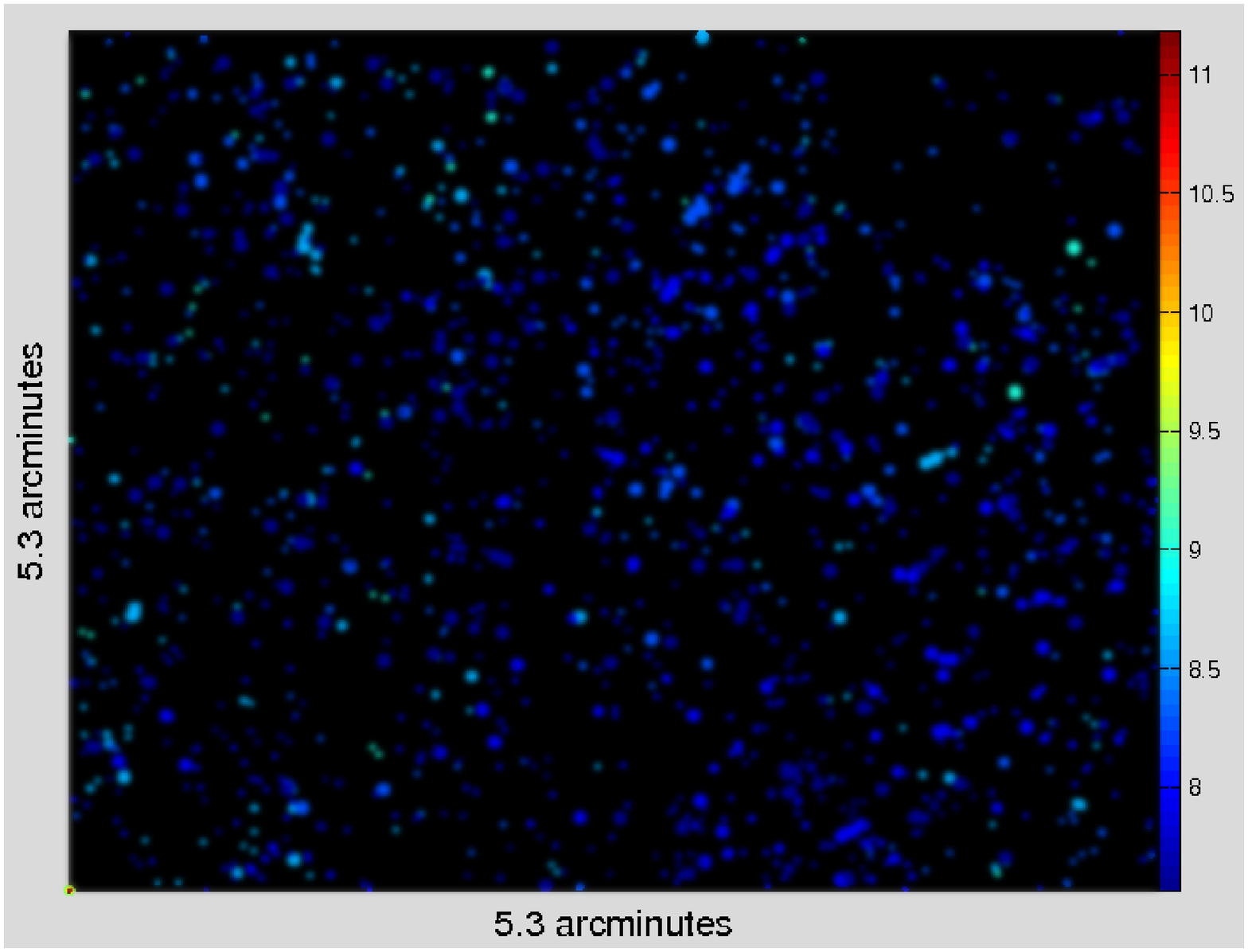}& \includegraphics[scale=0.3]{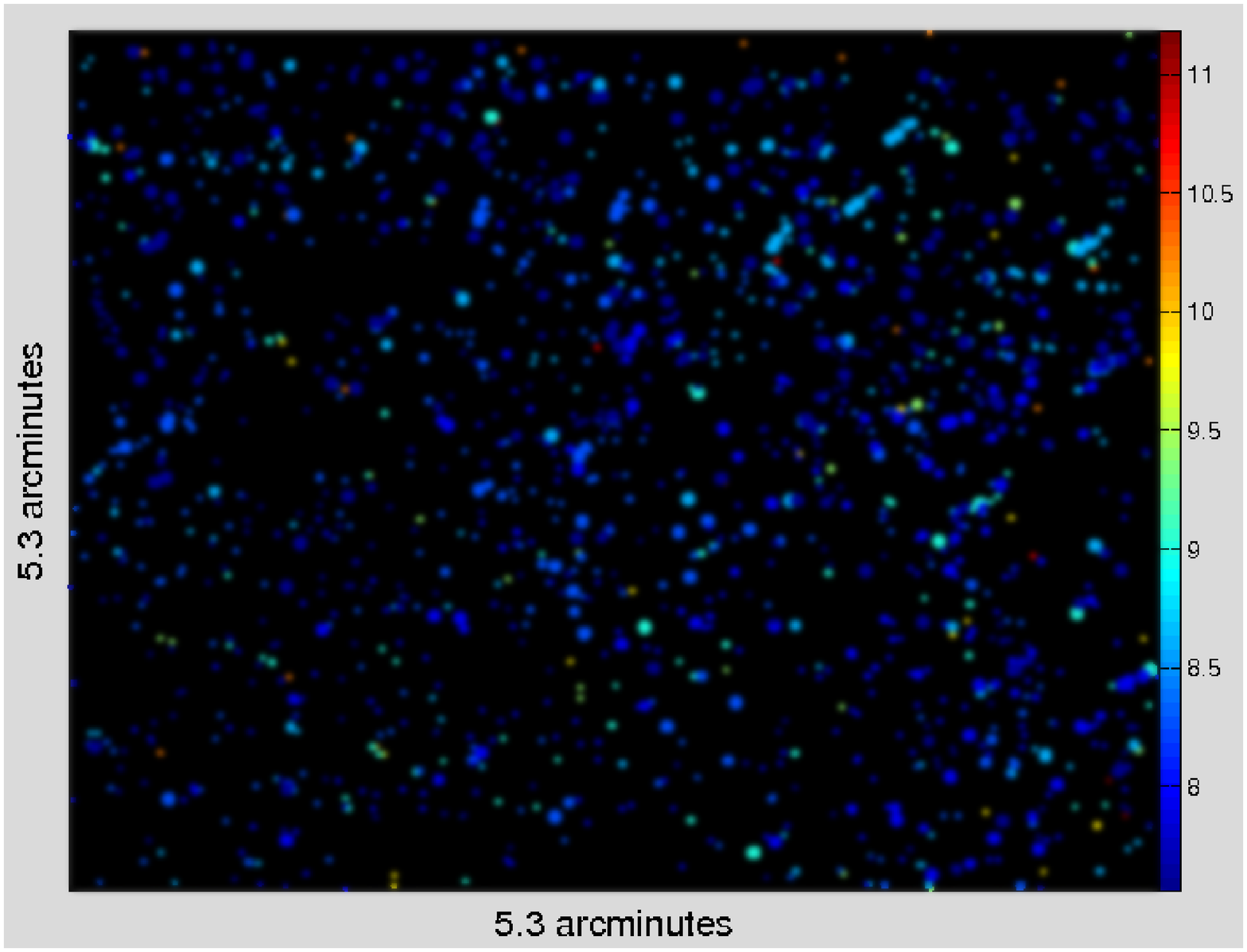}
\end{tabular}
\caption{Maps showing the distribution of galaxies with an observed flux larger than the JWST sensitivity limit of 1 nJy in the J-band (left panel) and H-band (right). The maps are a 2D cut of a 3D image produced by stacking simulation snapshots between $z \sim 7.6-11.6$. The vertical color bar gives the redshift of the object, the size of the galaxies scales with their flux in the range 1 nJy $< {\cal F} <$ 24.4 nJy.}
\label{zch7_highlf_fig2}
\end{figure*}

Our results suggest two clear trends. First, the LFs shift towards fainter luminosities with increasing redshift, mimicking a pure luminosity or density evolution. This is quite consistent with the trend of an increasing $M^*_{UV}$ with redshift, found in the data by several groups (see the extended discussion in Ouchi et al. 2009), preferring a pure luminosity evolution. Second, the faint-end slope of the simulated LF does not vary (within errors) from $z=5$ to $z=10$, maintaining an almost constant value of $\alpha=-2.0$. This value is slightly larger than the one derived from the data (Bouwens et al. 2007; Oesch et al. 2010) which suggest $\alpha$ in the range $[-1.57,-1.97]$. This behavior is most likely produced by a combination of the halo mass function evolution, feedback effects and evolving stellar populations.   

Clearly the faint-end slope will be better constrained by forthcoming facilities such as the JWST; the above successful test of our model allows us to make reliable predictions for surveys that will be performed with such instruments. For a deep exposure of $10^6$ s, JWST is expected to reach a photometric sensitivity of about $1$~nJy ($10\sigma$) allowing an investigation of the faint-end of the LF predicted by our simulations at least up to $z=9$ (see Fig. \ref{zch7_highlf_fig1}). This will be particularly exciting because it will very likely unveil the physical properties of these objects which are now thought to be the main reionization sources, as we will discuss in detail in Sec. \ref{Rei_sources_ch7}. To give a visual idea of the high-$z$ ($z > 7.5$) galaxy population in a typical JWST field (of size 5.3 arcmin) we have produced the maps shown in Fig. \ref{zch7_highlf_fig2} which demonstrate the wealth of objects that JWST will be able to detect up to $z \simgt 11$. The maps are a 2D cut of a 3D image produced by stacking all the available simulation snapshots between $z \sim 7.6-11.6$, and filling in the gaps between snapshots by randomly rotating the orientation axes in all the three-dimensions simultaneously.

\begin{figure}[htb]
\center{\includegraphics[scale=0.5]{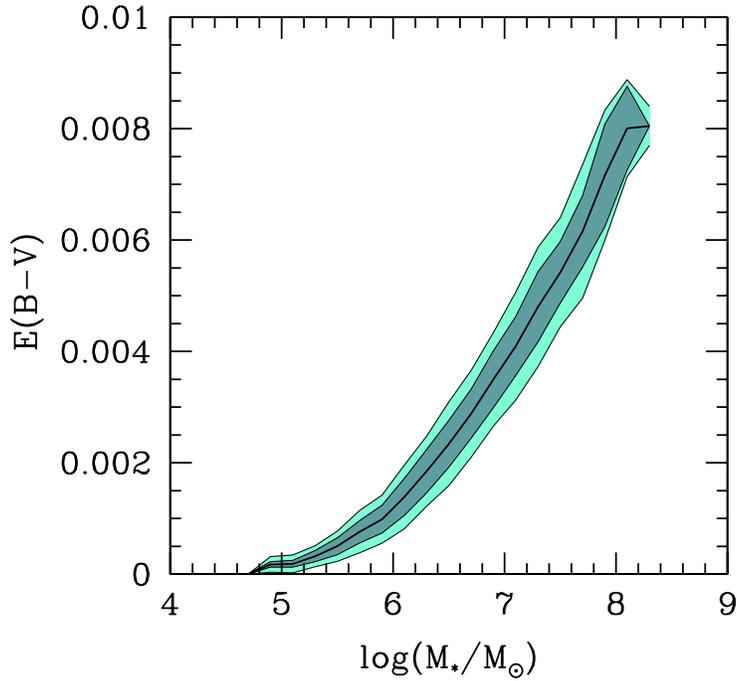}}
\caption{the color excess, $E(B-V)$, as a function of the stellar mass for $z\approx 7$ galaxies obtained from the simulation snapshot. The dark (light) region shows the color excess for 65\% (95\%) of the galaxies. }
\label{zch7_highlf_fig3}
\end{figure} 

As it has emerged from the observational results (see also Labb\'e et al. 2009), dust seems to have only a minor effect on shaping the SED of faint high-redshift galaxies. This is easily understood, as we will see later on, as a result of the relatively low stellar masses and metallicities characterizing these objects. Dust can however become more important in luminous LBGs and LAEs as many studies have shown (Lai et al. 2007; Atek et al. 2008; Nagamine, Zhang \& Hernquist 2008; Dayal et al. 2008, 2009, 2010a, 2010b; Finkelstein et al. 2009b). 

The dust enrichment of galaxies can be followed self-consistently by post-processing our simulation through the model introduced in Sec. \ref{dust_model_ch4}, using which, we can quantitatively check the previous statement. In brief, the model assumes that dust is produced by supernovae  and taking into account three processes: (a) dust forms in the expanding ejecta with a yield per SNII of  $0.54 M_\odot$ (Todini \& Ferrara 2001; Nozawa et al. 2003; Bianchi \& Schneider 2007), (b) SNII destroy dust in the ISM they shock to velocities $> 100 $ km s$^{-1}$, with an efficiency of 0.12 (McKee 1989), and (c) a homogeneous mixture of gas and dust is assimilated into star formation (astration). Once the dust mass is calculated for each galaxy in the simulation, we translate this into a value of $E(B-V)$ using the appropriate SN dust extinction curve given by Bianchi \& Schneider (2007) as explained in the same section. The resulting values of $E(B-V)$ for galaxies at $z\approx 7$ are shown in Fig. \ref{zch7_highlf_fig3} as a function of the stellar mass. Many galaxies, especially the smallest ones, are almost dust-free, and none of them shows a dust reddening value larger than  $E(B-V)=0.009$. This evidence allows us to safely neglect the effects of dust on the UV LF. 


\section{Properties of the first galaxies}
\label{prop_1gal_ch7}

\begin{figure*}[htb]
\center{\includegraphics[scale=0.75]{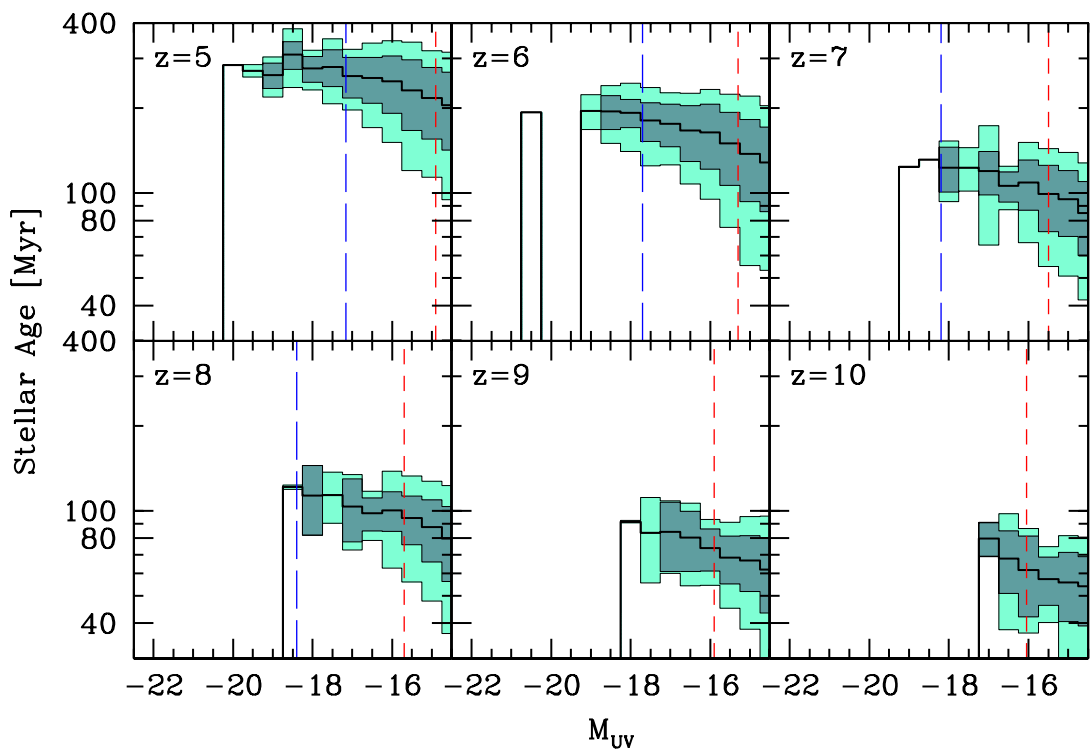}}
\caption{Mean stellar age of galaxies located at different redshifts 
(see labels) as a function of their absolute UV magnitude. The dark (light) shaded 
area show 68\% (95\%) of the galaxies in the magnitude bin.  The vertical short (long)-dashed lines mark the sensitivity limit of JWST (HST/WFC3).}
\label{zch7_highlf_fig4}
\end{figure*}

In addition to predicting the global evolution of the LF, a major strength of our study is that it makes it possible to extract the physical properties of the high-$z$ galaxies which are a part of the  faint-end of the LF. We start by concentrating on the stellar ages, reported in Fig. \ref{zch7_highlf_fig4} as a function of the UV magnitude, from which the following main conclusions can be drawn. On average, fainter (and less massive) objects tend to be younger at all redshifts, with typical ages for observable objects in the range 200-300 Myr at $z=5$ and 80-130 Myr at $z=7-8$, with the caveat that a considerable age spread exists at all luminosities. These ages imply that these galaxies started to form stars as early as $z=9.4$, clearly suggesting that their UV light might have influenced the cosmic reionization history. From a different perspective, we note that stellar ages of $\sim 100-150$ Myr are expected for galaxies at the sensitivity limit of WFC3 for $z=7-8$, in agreement with the observational estimates obtained by Labb\'e et al. (2010) and Finkelstein et al. (2009a).

\begin{figure*}[htb]
\center{\includegraphics[scale=0.75]{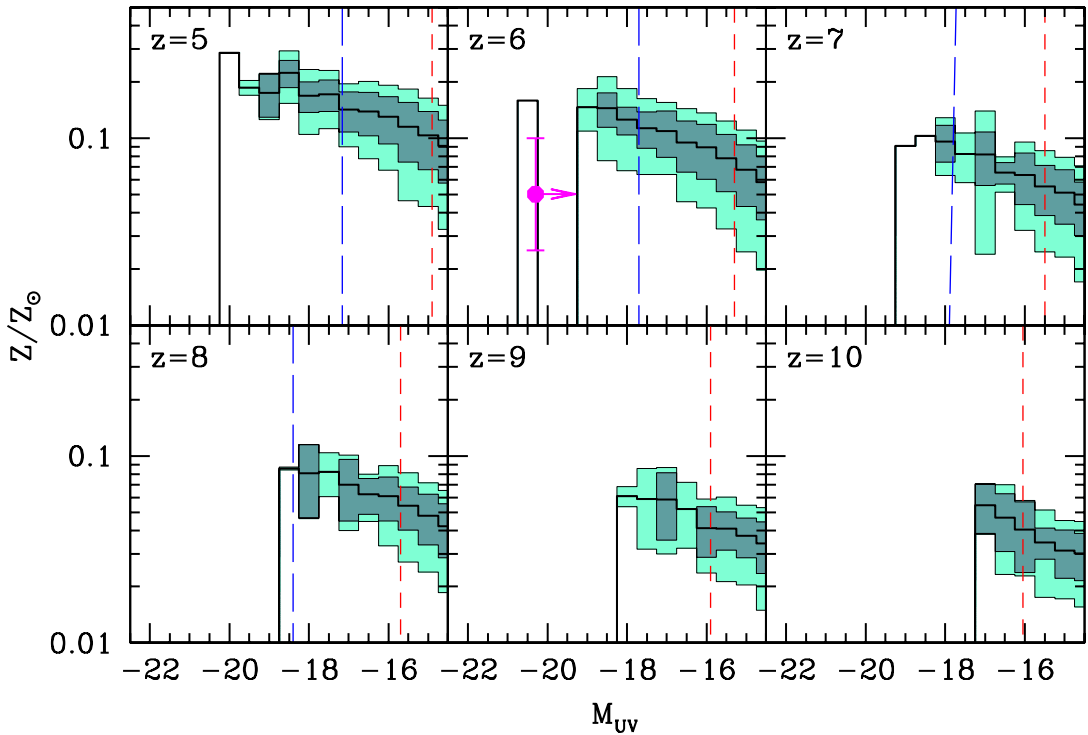}}
\caption{Mean metallicity of galaxies located at different redshifts 
(see labels) as a function of their absolute UV magnitude. The dark (light) shaded 
area show 68\% (95\%) of the galaxies in the magnitude bin. The vertical short (long)-dashed lines mark the sensitivity limit of JWST (HST/WFC3).
 The filled circle refer to the GRB~050904 host galaxy at $z=6.3$.}
\label{zch7_highlf_fig5}
\end{figure*}

A second important key physical parameter of pristine galaxies is their stellar and gas metallicity. In the following we will only discuss the stellar metallicity, $Z_*$, keeping in 
mind that the two closely match each other. Somewhat surprisingly (but not unexpectedly) 
even the faintest galaxies appear to be already enriched to remarkable levels: at the JWST sensitivity threshold, we find $Z_* > 0.03 Z_\odot$ at all redshifts; galaxies identified 
by HST ($M_{UV} < -18$) systematically show metallicities in excess of $Z_*=0.1 Z_\odot$ even at $z=7-8$. Thus we come to the interesting conclusion that even at these early epochs, the self-enrichment, due to the metals produced, following the first star formation episodes is able to increase the metal abundances of such small objects to levels comparable to their present-day counterparts (e.g. the Magellanic Clouds). In addition, such high mean metallicites could in principle preclude the formation of Pop III stars according to the critical metallicity scenario (Schneider et al. 2002, 2003; Schneider \& Omukai 2010) which predicts $Z_{cr}=10^{-5\pm 1} Z_\odot$ as the upper limit 
for the formation of Pop III stars. However, as we will discuss in more detail later, inside these early structures, small pocket of (quasi) pristine gas may survive in which a relatively tiny amount of Pop III stars continue to form as pointed out by TFS07 and Jimenez \& Haiman (2006).

Taken together, Fig.~\ref{zch7_highlf_fig4} and \ref{zch7_highlf_fig5} allow a disentanglement of the known metallicity--age degeneracy which arises when fitting the SEDs of high-$z$ galaxies. In fact,
once the absolute UV magnitude of a galaxy is known, our plots directly provide  
a good guess for its stellar age and metallicity. 

\begin{figure*}[htb]
\center{\includegraphics[scale=0.45]{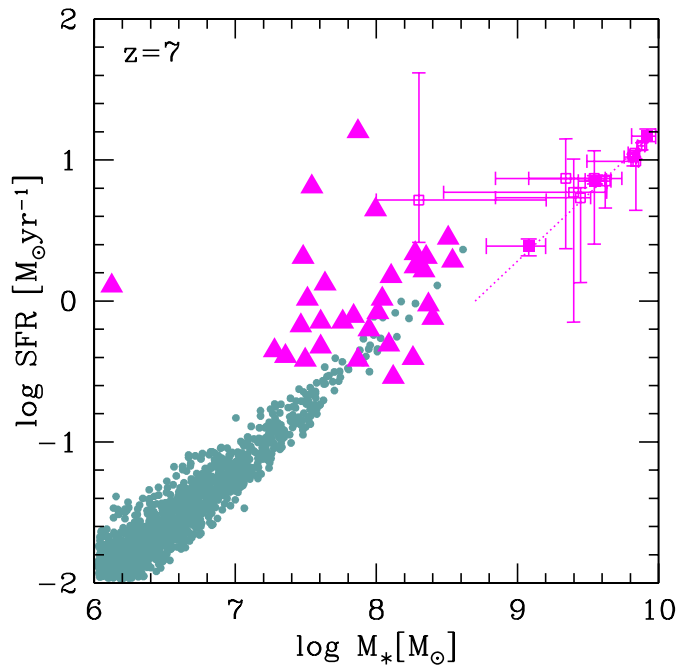}}
\caption{Stellar mass--SFR relation for simulated 
galaxies at $z=7$ (black points) compared with recent estimates for observed
$z$-dropout as derived 
by different authors: triangles are from Schaerer \& de Barros (2010), 
open squares from Gonzales et al. (2009), and filled squares for the mean
values derived by Labb{\'e} et al. (2009). The dotted line shows the empirical
relation between the two quantities obtained by Labb{\'e} et al. (2010).}
\label{fig:schaerer}
\end{figure*}

Fig.~\ref{fig:schaerer} shows the relation between the SFR and stellar mass of
galaxies at $z=7$. Simulated objects follow an almost linear relation with 
significant spread, i.e. an almost constant specific star formation rate. 
This trend closely matches the one found in the
analysis of stacked SEDs of WFC3 $z$-dropout by Labb{\'e} et al. (2009), 
although a single object analysis reveals large errors in the determination
of both the stellar mass and the SFR for these objects (Gonzalez et al. 2009). 
We note that simulated galaxies tend to have a slightly higher SFR (for a fixed 
stellar mass) than expected by the extrapolation of the relation to
smaller objects. However, in a re-analysis of the 
observed $z=7$ sample including the effect of nebular continuum and line 
emission along with that of dust absorption, Schaerer \& de Barros (2010) find smaller stellar masses
and larger SFR with respect to previous works. Our simulated galaxies lie 
somewhat in between these two observational estimates.

\begin{figure*}[htb]
\center{\includegraphics[scale=0.75]{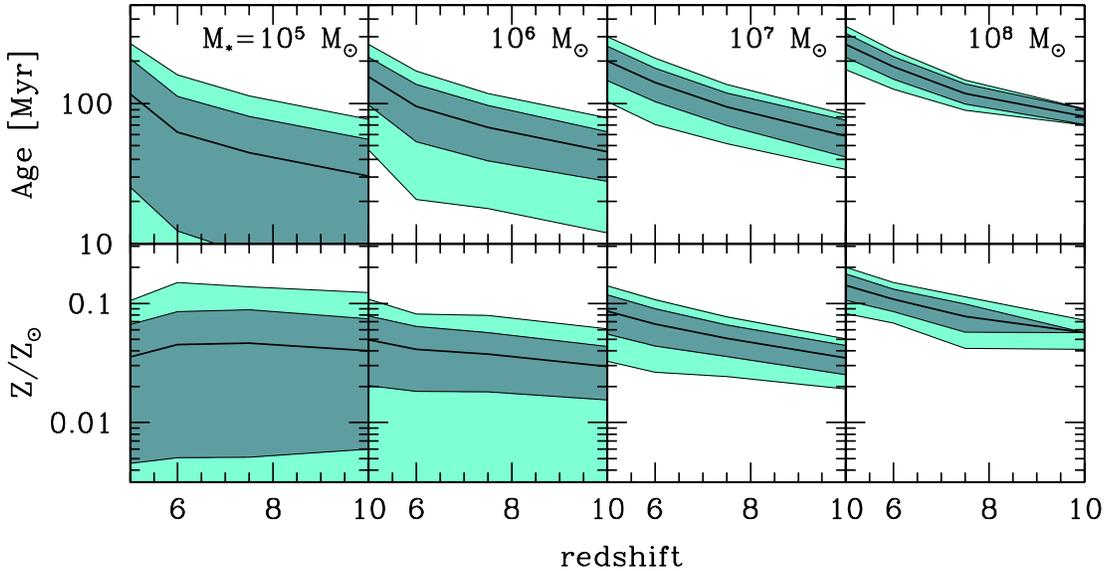}}
\caption{Redshift evolution of the mean stellar age and 
metallicity for galaxies in given stellar mass ranges (see labels). 
The dark (light) shaded area show 68\% (95\%) of the galaxies in the redshift bin. 
}
\label{zch7_highlf_fig7} 
\end{figure*}

Fig.~\ref{zch7_highlf_fig7} shows the evolution of the galaxy properties as a
function of redshift for galaxies with different stellar masses. 
The mean age of stellar population for all galaxies in the simulation is found
to decrease with redshift $\propto (1+z)^{-2}$; however, the age spread increases 
for less massive galaxies at any redshift. The average stellar metallicity is also 
growing with time for $M_*>10^7 M_\odot$, but show a much flatter (almost constant) 
evolution in the smallest objects, leveling at about 1/25 $Z_\odot$ for the tiniest star-forming galaxies. However, even for dwarf galaxies with $M_* \approx 10^5 M_\odot$ one 
can find individual objects enriched up to $1/10$ of solar metallicity already at $z=10$.    
This mass-dependent metallicity evolution is probably caused by the different ability  
to retain metals deposited by supernova explosions of the massive and dwarf populations. 
As the potential wells of the latter one are shallower, metals escape easily into the intergalactic medium thus setting an upper limit to the amount of metals than can be 
kept in their main body (Mac Low \& Ferrara 1999). An obvious implication is that IGM 
metals preferentially come from small and common objects, thus resulting in a more 
homogeneous enrichment (Ferrara, Pettini \& Shchekinov 2000; Madau, Ferrara \& Rees 2001).

\section{Reionization sources}
\label{Rei_sources_ch7}
Having obtained the luminosity function and characterized some of the physical properties of the current high-$z$ candidate galaxies, we now turn to an analysis of their role in the reionization process. 

The rate of ionizing photons from the $j$-th galaxy is given by

\begin{equation}
\dot{n}_{ion,j}=f_{esc}\left[Q^{\rm II}\dot{M}^{\rm II}_{j}+Q^{\rm III}\dot{M}^{\rm III}_{j}t_*^{\rm III}\right]
\end{equation}

\noindent
where $Q^{\rm II}$ ($Q^{\rm III}$) is the ionizing photon flux for Pop II (Pop III) stars, which of course is a function of the IMF, metallicity and stellar age of any given galaxy. Finally $f_{esc}$ is the escape fraction of ionizing photons. For simplicity, we assume a redshift-independent value of $f_{esc}$ for the two stellar populations. The ionization rate provided by galaxies in the simulation box per unit comoving volume at redshift $z$, $\dot{N}_{ion}(z)$, is then given by the sum over all galaxies at that redshift divided by the volume of the simulation. 

\begin{figure}[htb]
\center{\includegraphics[scale=0.42]{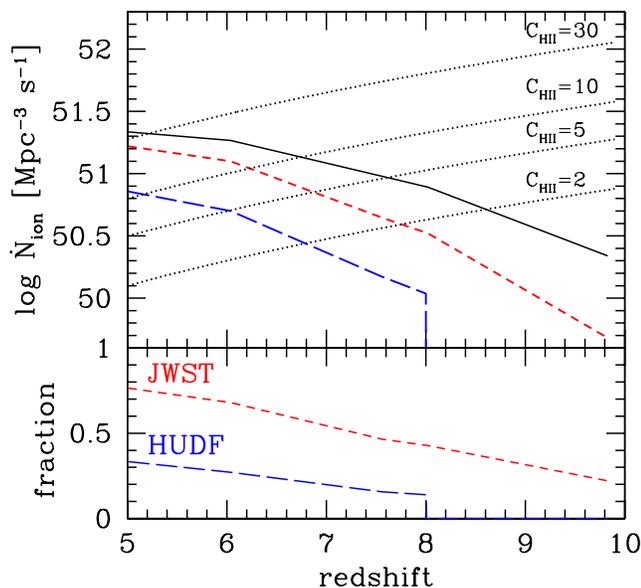}}
\caption{{\it Upper panel:} Redshift evolution of the total specific ionization rate (solid line). The short (long)-dashed line corresponds to galaxies detectable by JWST(HST/WFC3); dotted lines show the specific IGM recombination rate IGM for different values of the clumping factor $C_{\rm HII}$. {\it Bottom panel:} Fraction of ionizing
photons coming from galaxies identified by JWST and in the HUDF. An escape fraction $f_{esc}=0.2$ has been assumed.}
\label{zch7_highlf_fig8}
\end{figure}

The actual ionization rate must also include galaxies that are too rare to be caught in our relatively small simulation volume. To account for this correction, we add the ionization rate due to bright galaxies by integrating the observed LF (or the upper limits) from the luminosity of the brightest galaxy in the output down to very low magnitudes ($M_{UV}=-25$).   
The ionizing photon flux is obtained using a SED derived assuming that rare galaxies have the same stellar age and metallicity as the brightest simulated one. The extra contribution of the unaccounted bright-end of the LF is found to be at most $10$\% of the total ionizing photon emission. 

The evolution of the total specific ionization rate, $\dot{N}_{ion}$, is plotted in the top panel of Fig.~\ref{zch7_highlf_fig8} ($f_{esc}=0.2$) along with the same rate due to galaxies detected in the HUDF, $\dot{N}_{ion}^H$, or detectable by JWST, $\dot{N}_{ion}^J$. The ratios
between ($\dot{N}_{ion}^H$, $\dot{N}_{ion}^J$) and $\dot{N}_{ion}$ are also shown for clarity in the bottom panel of the same Figure. HST is now resolving the sources that provide $\approx 1/3$ of the ionizing photon budget at $z=5$ and $\sim 20$\% at $z=7-7.5$.  
At the sensitivity limit of JWST, it will be possible to detect the bulk of ionizing sources up to $z\sim 7.3$, but at higher redshifts most of the ionizing photons will still be produced by sources that are too faint to be detected even by JWST. 

The total ionization rate density $\dot{N}_{ion}(z)$ should then be compared 
with the recombination rate density of the IGM, $\dot{N}_{rec}(z)$, given by (e.g. Madau, Haardt \& Rees 1999)

\begin{equation}
\dot{N}_{rec}=\frac{\langle n_H \rangle}{\langle t_{rec} \rangle}=10^{50.0}C_{HII}\left(\frac{1+z}{7}\right)^3 {\rm s}^{-1} {\rm Mpc}^{-3},
\end{equation}

\noindent
where $\langle n_H \rangle$ is the mean comoving hydrogen density in the Universe and 
$\langle t_{rec} \rangle$ is the volume-averaged recombination time for ionized hydrogen
with an effective H$_{\rm II}$ clumping factor $C_{HII}=\langle n^2_{HII}\rangle/\langle n_{HII}\rangle^2$. The recombination rate density is shown in the top panel of Fig.~\ref{zch7_highlf_fig8} with dotted lines for different value of the clumping factor $C_{HII}$. For $f_{esc}=0.2$ and $C_{HII}=10$, the balance between ionization and recombination is obtained 
at $z\sim 6.8$. For $f_{esc}=0.1$ $\dot{N}_{ion}=\dot{N}_{rec}$ at $z=6$
assuming $C_{HII}=10$.

\section{Light from Pop III stars}
\label{pop3_ch7}

Another piece of useful information than can be extracted from the simulation outputs is the relative fraction of normal (Pop II) and massive, metal-free (Pop III) stars.  There are several questions to which we can provide quantitative answer: (i) do some of the current
candidates contain Pop III stars ? (ii) in that case, what fraction of their UV luminosity is powered by them ? (iii) how is this fraction dependent on their $M_{UV}$ luminosity ? (iv) is there a clear observational signature imprinted by Pop III stars ? 

The answer to the first question is straightforward: having analyzed the stellar populations  of the simulated galaxies present at four observationally relevant redshifts, $z=(5, 6, 8, 10)$, we find that a fraction 0.07-0.19 (depending on $z$) of the galaxies contain at least some Pop III stars. We should not emphasize too much on the exact values of this Pop III/Pop II galaxy ratio as fluctuations in galaxy mass and star formation rate might introduce a very large dispersion. The most robust physical quantity to understand the relative importance of the two populations is the ratio of Pop III-to-Pop II star formation rates which is a decreasing function of time (see Fig. 1 of TFS07) never exceeding $10^{-3}$ below $z=10$.

\begin{figure}[htb]
\center{\includegraphics[scale=0.5]{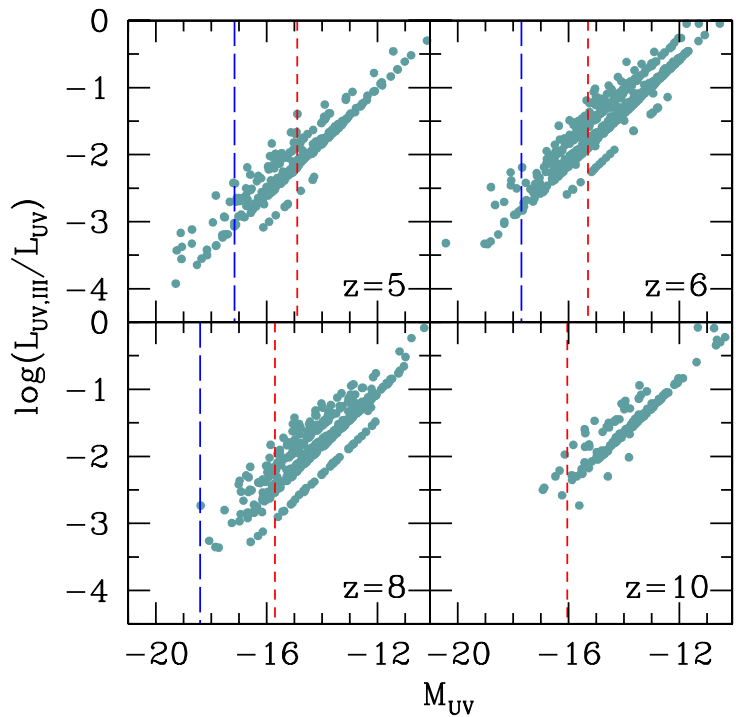}}
\caption{Fraction of the total 
luminosity $L_{\rm UV}$ due to Pop III stars $L_{UV,III}$ as a function of the
absolute UV magnitude $M_{UV}$. The vertical short(long)-dashed lines mark 
the sensitivity limit of JWST (HST/WFC3). Absolute UV magnitudes are 
computed as in Fig.~\ref{zch7_highlf_fig1}.} 
\label{zch7_highlf_fig9}
\end{figure}

More relevant is question (ii) above, whose answer can be obtained by inspecting Fig. \ref{zch7_highlf_fig9},  showing the ratio, $L_{UV,III}/L_{UV}$ of UV luminosities contributed by Pop III and Pop II stars as a function of $M_{UV}$. Light from Pop III stars becomes progressively more important towards fainter objects. This confirms previous findings (e.g. Schneider et al. 2006) that Pop III stars preferentially form in low-$\sigma$ peaks rather than in larger galaxies whose gas has already been enriched by several stellar generations. At the same time, we do not find galaxies containing only Pop III stars: this occurrence is made unlikely by the short lifetimes of such very massive stars. At $z=6$ ($z=10$) the contribution of Pop III stars to the total luminosity is always less than 5\% for $M_{UV}<-17$ ($M_{UV}<-16$). 

Among the candidate galaxies detected so far, we find that Pop III stars contribute less than a few percent to the total galaxy luminosity. Even at the detection limit of JWST, no Pop III-dominated galaxies ($\simgt 50$\% of total luminosity) will be found, due to their extreme faintness ($M_{UV}\simgt -14.$ at $z=6$). However, some objects having $\approx 10$\% of their luminosity powered by Pop IIIs are present at $z=6$ at the 1 nJy sensitivity level reachable by the JWST. 

Recently, many authors (Shapley et al. 2003, Nagao et al. 2008, di Serego Alighieri et al. 2008) have searched for the HeII 1640\AA\, emission line in the spectra of the so-called {\it dual emitters}, i.e. high-$z$ galaxies showing strong emission in both Ly$\alpha$ and HeII lines. The HeII line is usually taken as a well-defined signature of  Pop III stars. Up to now, these searches have given negative results. Motivated by these attempts, we have computed the rest-frame equivalent width of the HeII line for the objects in our simulation box. In particular, we focus here on $z=6$ which might be more easily accessible to present or future observations. Regretfully, the perspectives of direct detection of PopIII stars through this technique do not appear as very promising. For objects detectable in the JWST (HUDF) deep field survey, the expected HeII rest-frame equivalent width is $<0.5$ \AA~($<0.1$ \AA). Such small EWs will be very difficult to detect. The EW increases to more accessible values of about $10$ \AA~only if much fainter objects ($M_{UV} = -13$) could be observed. We have to underline that the above discussion implies that we cannot set limits on the total cosmic SFR in Pop III stars using the results of dual emitters searches given that the bulk of Pop III stars may be 'hidden' in galaxies much fainter than any present (and probably future) survey can detect (Scannapieco, Schneider \& Ferrara 2003). 

In conclusion, Pop III stars are found to form essentially at any redshift and
in $ < 20$\% of the galaxies. However, their contribution to the total galaxy 
luminosity is very low, apart from the very faint objects and their detection 
could be extremely difficult even with the next generation of space telescopes.


\section{Early Gamma-Ray Bursts}
\label{early_grb_ch7}

Long GRBs are powerful flashes of $\gamma$-rays that are observed with a frequency of about one per day over the whole sky. The $\gamma-$ray emission is accompanied by a long-lasting tail, called afterglow, usually detected over the whole electromagnetic spectrum. Their extreme brightness easily over-shines the luminosity of their host galaxy and
makes them detectable up to extreme high redshifts, as shown by the discovery of GRB~090423 at $z=8.2$ (Salvaterra et al. 2009b, Tanvir et al. 2009). Metal absorption lines can often be identified in their afterglow spectra, allowing a study of the metal (and dust) content of the environment in which they blow. Finally, once the afterglow has faded, follow-up searches of the GRB host galaxy become possible. At low redshifts,
GRBs are typically found in blue, low-metallicity dwarf galaxies with stellar masses $M_*\sim 10^{8-9}\;\Msun$ and high specific star formation rates (Savaglio et al. 2009). These objects closely resemble the properties of high-$z$ galaxies identified in our simulations, whose 
mean specific star formation rates (SSFR) are $\approx 8-10$ Gyr$^{-1}$, albeit associated with a large spread (for $M_*\ge 10^8$, SSFR=1.5-7 Gyr$^{-1}$ at $z=6-8$). This suggests that 
high-$z$ GRBs can be used as signposts of the same faint galaxies that provide the bulk of the ionization photons in the early Universe. Moreover, the study of their afterglows can provide new hints about the metal (and dust) content of the parent galaxy. Finally, it has been proposed that even Pop III stars may eventually blow as powerful GRBs (Fryer, Woosley \& Heger 2001; Yoon, Langer \& Norman 2006; Hirschi 2007; Komissarov \& Barkov 2010). The observation of these Pop III-GRBs may provide a valuable way to detected the elusive, short living first stars. We will discuss these points in the following.

\subsection{GRBs as metallicity probes}
Fig.~\ref{zch7_highlf_fig5} shows the mean stellar metallicity of the simulated sample of 
high-$z$ galaxies as a function of their absolute UV magnitude at different redshifts $z\ge 5$, i.e. the luminosity--metallicity relation for such objects. Although the experimental determination of such a relation would be of the utmost importance in testing theoretical predictions, in practice such effort is hampered by the extreme difficulty in inferring $Z_*$ from lines in the galaxy spectrum, with available facilities. 

This problem can be considerably alleviated if a GRB could be found inside of one of these
remote galaxies. In this case absorption features produced by heavy elements dispersed in the ISM surrounding the GRB would leave a characteristic and recognizable imprint on top of the GRB afterglow spectrum. Moreover, there is now a good agreement on that GRBs are usually hosted in relatively small, star-forming galaxies (Savaglio et al. 2009): these requirements would make the dwarf high-$z$ galaxies which, according to our findings, are the dominant population during the first cosmic billion year, optimal candidate hosts. 

The only high-quality spectra burst at $z>6$ available so far, GBR~050904 at $z=6.3$ (Kawai et al. 2006), shows the expected metal absorption features, witnessing the presence of metals at the same redshift of the GRB (Kawai et al. 2006). From this data, and with the further assumption that the measured sulfur ([S/H]=$-1.3\pm 0.3$) is a good proxy for metallicity, one can determine the metal content of the host galaxy. Berger et al. (2007) attempted a detection of the GRB host galaxy with HST and {\it Spitzer}, and were able to set an upper limit to the host luminosity of 
$M_{UV}\simgt -20.3$, i.e. $M_{UV}\simgt M_{UV}^*$ at $z=6$. By combining the HST and {\it Spitzer} upper limits, they set an upper limit to the stellar mass of the galaxy, $M_*\simlt {\rm few}\times 10^9\;\Msun$.  This constraint is consistent with the predictions presented in Fig. \ref{zch7_highlf_fig5} (notice the data point corresponding to the Berger et al. 2007 determination), which might indicate that at $z\approx 6$, a galaxy with 
$Z_*\approx 0.05 Z_\odot$ is on average 2-3 magnitudes fainter than the detection limit of the experiment; in addition, such metal abundance level is quite typical for galaxies located at that epoch. 

GRB~090423 (Salvaterra et al. 2009b, Tanvir et al. 2009) has been recently detected at $z=8.2$. The intrinsic properties of the burst (both of the prompt and afterglow phases) are similar to those observed at low/intermediate redshifts, suggesting that the progenitor and the medium in which the burst occurred are not markedly different from those of low-$z$ GRBs 
(Salvaterra et al. 2009a, Chandra et al. 2009). Chary et al. (2009) observed the field of GRB~090423 for 72 hours using {\it Spitzer}/IRAC at 3.6 $\mu$m, looking for its host galaxy. 
The observation was performed $\sim 46$ days after the GRB trigger, (corresponding to about 5 days in the burst rest-frame). A weak source was detected at the location of the GRB
afterglow with  $L_{AB}=27.2\pm 0.3$. The observed flux is consistent with the power-law decay of the GRB afterglow, suggesting that the source is still contaminated by the GRB emission and implying a limit on the absolute magnitude of the host of $M_{3900{\rm \AA}}>-19.96$. 
From Fig.~3, we expect that the host galaxy of GRB~090423 should be enriched at a level of a few percent solar. This relatively high metallicity may explain the high equivalent neutral hydrogen column density measured in the X-ray afterglow, although a wide range of systematic effects or the contamination by low-$z$ intervening absorption systems may be a more valuable, 
alternative explanation for the observed absorption (Chandra et al. 2009).

These examples nicely show how GRBs represent an unique tool to study the high-$z$ Universe. In particular, the observation of their optical-NIR afterglow may allow to eventually study the evolution of the mass-metallicity relation up to very high-$z$. To achieve this goal, 
high resolution and good signal-to-noise afterglow spectra are required as soon as possible 
after the GRB detection in $\gamma$-rays. As nicely demonstrated by GRB~090423, GRBs are 
easily detectable well beyond any other astrophysical object. Indeed, up to $\sim 5$\% of 
all GRBs detected by the {\it Swift} satellite are expected to be at $z>6$ (Salvaterra \& Chincarini 2007, Salvaterra et al. 2009). Future missions (e.g. EXIST, XENIA, SVOM) will rapidly increase the high-$z$ GRB sample (Salvaterra et al. 2008) allowing a statistical study of GRB hosts in the high-$z$ Universe and a direct check of the galaxy metal enrichment history at those early epochs. In particular, EXIST, thanks to its 1.1m optical-NIR 
telescope, will be able to take the GRB afterglow spectrum only 300 s after the trigger, allowing an on-board direct measure of the redshift and the identification of metal  absorption lines when the afterglow is still sufficiently bright even for high-$z$ bursts (Grindlay et al. 2009).

\subsection{GRBs from Pop III stars}

It has been widely discussed (see Bromm \& Loeb 2007 for a review) whether GRBs can arise from the collapse of  massive, metal-free stars. While the large envelopes of these objects may suppress the emergence of relativistic jets out of their surface, the weak winds expected for 
low-metallicity stars can prevent angular momentum loss during their evolution, producing the rapidly rotating central configurations needed to produce a GRB. In the following we will assume that massive Pop III stars produce GRBs.

Excluding Pop III progenitors in the mass range [140,260] $\Msun$, that will explode as Pair Instability Supernovae (PISN, Heger \& Woosley 2002) leaving no remnant, we have two possible channels through which GRBs can occur: (i) 100-140 $\Msun$ (Yoon et al. 2006, Hirschi 2007) and (ii) 260-500 $\Msun$ (Fryer et al. 2001, Komissarov \& Barkov 2010). In the latter case, since both the luminosity and the duration are thought to be proportional to the black hole mass, an extremely bright and long GRB is expected (Fryer et al. 2001), while more typical luminosities and durations may be expected in the case of smaller progenitors (Hirschi 2007).

We can estimate a strong upper limit for the rate of Pop III-GRB detections as follows. Suppose a fraction $f_{GRB}$ of all Pop III stars with masses in the ranges discussed above produce GRBs with a typical beaming angle $\theta$, and that all Pop III-GRBs are detectable by present-day satellites given the their extreme brightness. Then the observed rate of Pop III-GRB, $R_{GRB}(>z)$, is given by
\begin{equation}
R_{GRB}(>z)=5.5\times 10^{-3} \left(\frac{\theta}{6^\circ}\right)^2 \int_z dz\, f_{GRB}\, \eta\,
\frac{\dot{\rho}^{III}(z^\prime)}{1+z^\prime} \frac{dV}{dz}
\end{equation}
where $\dot{\rho}^{III}(z)$ is the total Pop III star formation rate at redshift $z$. Further, 
\begin{equation}
\eta=\int_{M_{min}}^{M_{max}} \phi(m_*) dm_*/\int_{100}^{500} m_* \phi(m_*) dm_*,
\end{equation}
 is the number of Pop III stars with masses in the range $[M_{min},M_{max}]$ given the assumed IMF $\phi$, and $V$ is the cosmological volume per unit solid angle. The factor $(1+z)^{-1}$ accounts for time dilation effects due to redshift. The computed rate of Pop III-GRB at $z>6$ is then 
$R_{GRB}\sim 7.5 (2.5) f_{GRB}  (\theta/6^\circ)^2$ yr$^{-1}$ sr$^{-1}$ for GRB progenitors in the mass range 100-140 $\Msun$ (260-500 $\Msun$). This rate is of the same order of that expected for normal (i.e. Pop II/PopI progenitors) GRBs (Salvaterra \& Chincarini 2007, Salvaterra et al. 2009). However, we stress here that very likely $f_{GRB}\ll 1$, resulting in lower detection rates. The detection of one of such Pop III-GRBs might represent the most 
promising way to directly detect the very first stars to have formed.

\section{Conclusions}
\label{summary_ch7}

By using high resolution simulations specifically crafted to include the relevant physics of galaxy formation, along with a novel treatment of the metal dispersion that allows us to follow the PopIII-PopII transition as dictated by the critical metallicity scenario, we have been able to reproduce the observed UV LFs over a wide redshift range, $5 < z < 10$. It is then useful to schematically summarize the main findings of the present work:

\begin{itemize}

\item{The simulated high-$z$ galaxy UV LFs match remarkably well with the amplitude and
slope of the observed LFs in the redshift range $5 < z < 10$.}
\item{The LF shifts towards fainter luminosities with increasing
redshift, mimicking a pure luminosity (or density) evolution. The faint-end
slope of the LF does not vary from $z=5$ to $z=10$, keeping an almost constant
slope value of $\alpha=-2$.}
\item{Many galaxies at $z\approx 7$, especially the smallest ones, are virtually dust-free, and none of them shows dust extinctions larger than  $E(B-V)=0.009$. This evidence allows us 
to safely neglect the effects of dust on the UV LF.}
\item{The stellar population of high-$z$ galaxies shows typical ages in the range 
100-300 Myr at $z=5$ and 40-130 Myr at $z=7-8$, implying that they started to form 
stars as early as $z=9.4$. These objects are enriched rapidly with metals and galaxies identified by HST/WFC3 show metallicities $\approx 1/10 \Zsun$ even at $z=7-8$. 
The trend of decreasing metallicity (and increasing spread) towards low mass halos indicates that small galaxies are more affected by supernova feedback and loose a larger fraction of the heavy elements they produce.}
\item{The relation between the star formation rate and stellar mass of simulated follows an almost linear relation with significant spread towards the lowest masses, implying an almost constant specific star formation rate.}
\item{The bulk of the ionizing photons is produced by objects populating the 
faint-end of the LF. These galaxies are beyond the capabilities of current
survey, but JWST will be able to resolve them up to $z=7.3$.}
\item{Massive PopIII stars continue to form essentially at all redshifts and
in $<20$\% of the galaxies. However, their contribution to the total galaxy 
luminosity is negligible ($<$ 5\%) for all objects with the marginal exception of the 
extremely faint ones; their detection will be tremendously difficult even for
the next generation of space telescopes.}
\item{The typical high-$z$ galaxies closely resemble the GRB host population observed 
at lower redshifts. This fact suggests that GRBs can be used to detect and study these objects, providing unique information about the first stages of structure formation. In particular, they can be used to extend the study of the mass-metallicity and its evolution 
to very high redshifts. Moreover, if PopIII stars end their lives in a GRB explosion, the
detection of one of these bursts might represent the most promising way to directly detect 
the very first stars.}

\end{itemize}


\chapter{Conclusions}\label{conc_ch8}
In the CDM model of galaxy formation, the epoch of reionization is the last major event left to be uncovered in the history of the Universe. Reionization begins when the first sources of \HI ionizing photons form within dark matter halos and begin to carve out a region of ionization around themselves, the so-called Str\"omgren Sphere. Hence, reionization marks the second major change in the ionization state of the Universe; from neutral to ionized. However, the astrophysical sources of \HI ionizing photons and their relative contributions to cosmic reionization are still controversial. This is because the reionization process depends on a number of unknown parameters including the IMF of these first sources, their SFR, the escape fractions of \HI ionizing photons from the galaxies and the clumping of the IGM, to name a few.  This is one of the primary reasons for the debate regarding the reionization history and the redshift at which it ends.

To this end, a new class of high-redshift galaxies, called Lyman Alpha Emitters (LAEs) have rapidly been gaining popularity as probes of both reionization and early galaxy evolution for three primary reasons. Firstly, specific signatures like the strength and width of the Ly$\alpha$ line (1216 \AA) make the detection of LAEs unambiguous. Secondly, since Ly$\alpha$ photons are highly sensitive to the presence of neutral hydrogen, their attenuation can be used to put constraints on the ionization state of the IGM. Thirdly, surveys like the Hubble Ultra Deep Field (HUDF), Subaru Deep Field (SDF), Large Area Lyman Alpha (LALA) and Multiwavelength Survey by Yale-Chile (MUSYC) have enabled the confirmation of hundreds of LAEs between $z\sim2. 25 - 6.6$, with tentative detections at $z \sim 7.7$. Since these are also amongst the earliest galaxies to have formed, LAEs represent superb probes of the properties and evolution of early galaxy populations. This is important since mechanical, chemical and radiative feedback from these galaxies determine the properties of the subsequent generations of structures.

As presented in this thesis, our aim has been to build a physically motivated, self-consistent model to probe reionization and galaxy evolution using LAEs. We started with a semi-analytic model to calculate the Ly$\alpha$/continuum production/transmission in Chapter \ref{ch2_lya_sam}. This was was then coupled to cosmological SPH simulations ({\tt GADGET-2}) and additional mechanisms of luminosity production, as shown in Chapters \ref{ch3_lya_sim} and \ref{ch4_lya_cool}. These were used to obtain a number of results concerning the ionization state of the IGM, the intrinsic properties of LAEs, and their dust enrichment and detectability. We coupled the LAE model, SPH simulations and an RT code ({\tt CRASH}) to build the first complete LAE model, as shown in Chapter \ref{ch5_lya_rt}. Using the LAE model in conjunction with a code that builds the MW merger tree ({\tt GAMETE}), we were able to link the properties of high-$z$ LAEs to the local Universe, as shown in Chapter \ref{ch6_mw_lae}. Finally, we used cosmological SPH simulations ({\tt GADGET}) to study the nature of the earliest galaxies that have been detected as of yet, in Chapter \ref{ch7_reio_sources}. The main results obtained from all these works, which have formed my doctorate thesis, are now summarized. \\

\begin{center}
{\underline {\bf {\Large LAEs and reionization}}}
\end{center}
We first summarize the constraints we have obtained on reionization from the available LAE data, and then discuss the contribution of LAEs to reionization.\\

\begin{center}
{\bf {\large (a) LAEs as probes of reionization}}
\end{center}

Using our semi-analytic model presented in Chapter \ref{ch2_lya_sam}, we showed that a model wherein reionization ends at $z \sim 7$ (Early Reionization Model, ERM) explains the observed Ly$\alpha$ LF data at $z \sim 5.7, 6.6$ more consistently as compared to a model where reionization ends at $z \sim 6$ (Late Reionization Model, LRM). Hence, the IGM must already have been highly ionized at the redshifts considered and the evolution of the Ly$\alpha$ LF can be attributed solely to an evolution in the underlying mass function. However, the situation changes remarkably when a full RT calculation is included. As shown in Chapter \ref{ch5_lya_rt}, there exists a {\it degeneracy} between the ionization state of the IGM and dust clumping (or grain properties) inside high-redshift galaxies; the Ly$\alpha$ LF can be well reproduced (to within a $5\sigma$ error) by $\langle \chi_{HI}
\rangle \sim0.24$, corresponding to a highly neutral IGM, to a value
as low as $3.4 \times 10^{-3}$, corresponding to an ionized IGM,
provided that the increase in $T_\alpha$ is compensated by a decrease
in the Ly$\alpha$ escape fraction from the galaxy. 

We also quantified the effects of clustering on the visibility of LAEs in Chapter \ref{ch3_lya_sim}: for an almost neutral IGM ($z \sim 7.6, \chi_{HI}=0.16$), if clustering effects are ignored, only those few galaxies that are able to carve out a large enough \HII region are visible. However, about 4 times more galaxies become visible when clustering effects are included. In a highly ionized IGM ($z\sim 6.6, 5.7$, $\chi_{HI} \sim 10^{-4}, 10^{-5}$), however, the effect of clustering is indiscernible on the Ly$\alpha$ LF.\\

\newpage
\begin{center}
{\bf {\large (b) LAE contribution to reionization}}
\end{center}

As shown in Chapter \ref{ch3_lya_sim}, the SFR density provided by LAEs is $\dot\rho_* = 1.1 \times 10^{-2}\, {\rm M_\odot\, yr^{-1}\, Mpc^{-3}}$ at $z \sim 5.7$, and this decreases to $\dot\rho_* = 4.9 \times 10^{-3}\, {\rm M_\odot\, yr^{-1}\, Mpc^{-3}}$ at $z \sim 6.6$. Corresponding to this SFR density and $f_{esc}=0.02$, it is easy to derive that the \HI ionizing photon rate density contributed by LAEs is $3.1\times 10^{49} \, {\rm s^{-1}\, Mpc^{-3}}$ at $z=6.6$. When compared to the  photon rate density of $1.24 \times 10^{50} \, {\rm s^{-1}\, Mpc^{-3}}$, necessary to balance recombinations for a homogeneous IGM (clumping factor of 1) given by Madau, Haardt and Rees (1999), we find that LAEs on the Ly$\alpha$ LF at $z\sim 6.6$ can contribute at most 25\% 
of the \HI ionizing photons needed to balance recombinations at this redshift. This value decreases to 0.8\% 
as the clumping factor increases to 30. Since the IGM is expected to be clumped, our results seem to indicate that LAEs are passive tracers of reionization. \\
\begin{center}
{\underline {\bf {\Large The physics of LAEs}}}
\end{center}

We now discuss our main results concerning the sources of luminosity in high-$z$ LAEs, and their physical nature. \\
\begin{center}
{\bf {\large (a) Sources of the Ly$\alpha$ luminosity}}
\end{center}

Traditionally, stellar sources have been assumed to be the main sources of both Ly$\alpha$ and continuum luminosity in LAEs. However, following the calculations shown in Dijkstra (2009), we included a new source of luminosity, that from cooling of collisionally excited \HI in the galaxy. We presented two important results regarding this in Chapter \ref{ch4_lya_cool}: while cooling of collisionally excited \HI in the ISM on average contributes about 20\% of the Ly$\alpha$ radiation produced by stars, the continuum luminosity is almost completely dominated by stellar sources. \\

\begin{center}
{\bf {\large (b) The physical nature of LAEs}}
\end{center}

Using our LAEs model coupled to SPH simulations, we have been able to pin down the elusive properties of LAEs including the ages, SFR and metallicity for the first time. We found that at $z \sim 5.7$, the dark matter halo mass of LAEs ranges between $10^{9}-10^{12} \rm{M_\odot}$, with the halo mass range narrowing towards higher redshifts. At the same $z$, the SFR lie in the range $1-120\, {\rm M_\odot \, yr^{-1}}$; LAEs residing in low mass dark matter halos are inefficient in forming stars as compared to those in larger halo masses. The mass weighted ages for LAEs are $t_* > 20$ Myr at all redshifts, while the mean stellar metallicity increases from $Z_*=0.12\,{\rm Z_\odot}$  at $z \sim 7.6$ to $Z_*=0.22\,{\rm Z_\odot}$ at $z \sim 5.7$. The brightest LAEs are all characterised by large SFR and intermediate ages ($\approx 200$ Myr), while objects in the faint end of the Ly$\alpha$ LF show large age and star formation rate spreads.\\


\begin{center}
{\underline {\bf {\Large The dusty nature of LAEs}}}
\end{center}

Dust grains absorb both Ly$\alpha$ and continuum photons and hence, play a very important role in determining the Ly$\alpha$ and continuum luminosity that is observed. Further, the IGM transmission-dust attenuation degeneracy found in Chapter \ref{ch5_lya_rt} makes it imperative to understand the dust enrichment of LAEs.\\


\begin{center}
{\bf {\large (a) The dust content and its detectability}}
\end{center}

Using a semi-analytic model that calculates the dust enrichment of LAEs based on the intrinsic galaxy properties (including the SFR, age and gas mass) as shown in Chapter \ref{ch4_lya_cool}, we have found that, on average, only (33\%, 23\%) of the Ly$\alpha$ and (23\%, 38\%) of the continuum photons escape the galaxy, undamped by dust. We have also presented a clean way of looking for these dusty galaxies (see Chapter \ref{ch4_lya_cool}): we calculated the sub-millimeter flux of LAEs to make predictions for ALMA, where we showed that the brightest LAEs would be detectable with ALMA for a 5$\sigma$ integration time as short as 1 hour. \\

\begin{center}
{\bf {\large (b) The dust distribution}}
\end{center}

As shown in Chapters \ref{ch2_lya_sam}, \ref{ch3_lya_sim}, \ref{ch4_lya_cool}, while the relative escape fractions of Ly$\alpha$ and continuum photons, $f_\alpha/f_c <1 $ for $z\sim 6.6$, the situation reverses itself for $z \leq 5.56$, i.e. $f_\alpha/f_c >1 $. However, no single extinction curve gives a value of $f_\alpha/f_c >1$. One of the simplest ways of explaining this large relative escape fraction is to invoke the multiphase ISM model as proposed by Neufeld (1991), wherein the ISM is multiphase and consists of a warm gas with cold dust clumps embedded in it. This inhomogeneity of the dust distribution can then lead to a larger attenuation of the continuum photons relative to the Ly$\alpha$. Thus, the evolution of the relative escape fractions of Ly$\alpha$ and continuum photons hints at the dust content of the ISM becoming progressively inhomogeneous/clumped with decreasing redshift. The result that dust is clumped at $z \sim 5.7$ has been reinforced using a model that includes a full RT calculation, shown in Chapter \ref{ch5_lya_rt}; we find that the value of $f_\alpha/f_c$ is always greater than 1 and must range between $3.4-4.1$ ($3-5.7$) for an average neutral hydrogen fraction of $\langle \chi_{HI} \rangle \leq 0.16$ ($\leq 0.24$). \\

\newpage
\begin{center}
{\underline {\bf {\Large Clues of the infant MW from high-z LAEs}}}
\end{center}

We have linked the properties of high-$z$ LAEs to the Local Universe by coupling the semi-analytical code {\tt GAMETE} to our LAE model. We find that the MW progenitors cover the entire range of observed Ly$\alpha$ luminosities, $L_{\alpha}=10^{42-43.25} {\rm erg~s^{-1}}$; we thus conclude that among the LAEs observed at 
$z\approx 5.7$ there are progenitors of MW-like galaxies. We find that that the MW progenitors that would be observable as LAEs at $z\approx 5.7$ are rare ($\approx 1/50$), but the probability to have at least one LAE in any MW hierarchical merger history is very high, $P=68\%$. The LAEs generally correspond to the major branch of the merger tree and their physical properties can be summarized as follows: $\dot M_* \geq 0.9 \, M_\odot \,{\rm yr^{-1}}$, $M_h \geq 10^{10} \,M_\odot$, $E(B-V) < 0.025$, $t_*=150-400$ Myr and $Z_*=0.016-0.044\, Z_\odot$ at $z \sim 5.7$. At decreasing $M_h$, instead, the progenitors can be visible as LAEs only by virtue of a high gas mass content or extremely young ages; there are 5 such objects, with $M_h \approx 10^9 \Msun$.

\begin{center}
{\underline {\bf {\Large Simulating high-z galaxies}}}
\end{center}

We use cosmological SPH simulations ({\tt GADGET}) to study the properties of high-$z$ galaxies. The simulation boxes are optimized for our study in the following ways: they reach the high resolutions required to resolve the dominant reionization sources - dwarf galaxies. Most importantly, they are implemented with a state-of-the-art treatment of metal enrichment and of the transition from Pop III to Pop II stars, along with a careful modelling of supernova feedback.

\begin{center}
{\bf {\large (a) The physical properties of the first galaxies}}
\end{center}

We find that on average, fainter (and less massive) objects tend to be younger at all redshifts, with typical ages for observable objects in the range 200-300 Myr at $z=5$ and 80-130 Myr at $z=7-8$, with the caveat that a considerable age spread exists at all luminosities. These ages imply that these galaxies started to form stars as early as $z=9.4$, clearly suggesting that their UV light might have influenced the cosmic reionization history. Even the faintest galaxies appear to be already enriched to remarkable levels: at the JWST sensitivity threshold, we find $Z_* > 0.03 Z_\odot$ at all redshifts; galaxies identified 
by HST ($M_{UV} < -18$) systematically show metallicities in excess of $Z_*=0.1 Z_\odot$ even at $z=7-8$.

\begin{center}
{\bf {\large (b) The light from PopIII stars}}
\end{center}

Having analyzed the stellar populations of the simulated galaxies present at four observationally relevant redshifts, $z=(5, 6, 8, 10)$, we find that a fraction 0.07-0.19 (depending on $z$) of the galaxies contain at least some Pop III stars. We should not emphasize too much on the exact values of this Pop III/Pop II galaxy ratio as fluctuations in galaxy mass and star formation rate might introduce a very large dispersion. The most robust physical quantity to understand the relative importance of the two populations is the ratio of Pop III-to-Pop II star formation rates which is a decreasing function of time (see Fig. 1 of TFS07) never exceeding $10^{-3}$ below $z=10$. We also find that although the light from Pop III stars becomes progressively more important towards fainter objects (although it is never greater than a few percent), we do not find galaxies containing only Pop III stars: this occurrence is made unlikely by the short lifetimes of such very massive stars. 

\begin{center}
{\underline {\bf {\Large Open questions}}}
\end{center}

In spite of all the physical ingredients implemented into the model, and its successes, there remain a number of open issues. The first of these concerns the value of the escape fraction of \HI ionizing photons, $f_{esc}$. As discussed in Sec. \ref{cosmic_reio_ch1}, the value of $f_{esc}$ is largely unconstrained and ranges between $f_{esc}=0.01-0.73$ ($0.01-0.8$) observationally (theoretically). A larger (smaller) value of $f_{esc}$ would lead to larger (smaller) \HII regions, thereby affecting the progress of the reionization process and hence $T_\alpha$; a change in $f_{esc}$ would also affect $L_\alpha^{int}$, Sec.~\ref{intrinsic_lum}. 

The second problem concerns the gas topology and distribution within the galaxy. For our works, we have used the results of Ferrara, Pettini \& Shchekinov (2000) to calculate the gas distribution radius, and assumed the gas to be homogeneously distributed in this radius. However, in reality, the ISM consists of dense, star forming, molecular clouds embedded in the ISM. Considering the gas to be clumped in these clouds could boost up the luminosity from the cooling of collisionally excited \HI by many factors. Due to the large simulation volumes required for our work ($\sim 100$ Mpc), such galactic scales of a few Kpc can not be resolved with our simulations. To solve this problem, simulations of individual galaxies must be used to link the gas clumping and distribution scales to the physical properties of the galaxy.

Thirdly, in the dust model explained in Sec.~\ref{dust_model_ch4}, we have considered dust destruction by forward sweeping SNII shocks. However, Bianchi \& Schneider (2007) have shown that reverse shocks from the ISM can also lead to dust destruction, with only about 7\% of the dust mass surviving the reverse shock, for an ISM density of $10^{-25} {\rm gm \, cm^{-3}}$. Since we neglect this effect, we might be over-predicting the dust enrichment in LAEs. However, the problem arises that the amount of dust destroyed by the reverse shocks depends sensitively both on the type of the dust grains considered, and the density of the ISM; both of these are only poorly known for the high-$z$ galaxies we are interested in.

The last issue concerns the dust topology and distribution in the ISM of galaxies. There are a number of questions regarding this which remain unanswered as of now, such as: is the dust clumped inside these galaxies, or is there an alternative explanation for the relative fractions of Ly$\alpha$ and continuum photons? Where are the sites in the ISM where the dust is most clumped/concentrated? How does this clumping depend on the intrinsic properties of the galaxy?

\clearemptydoublepage
\newpage
\begin{center}
{\LARGE
\bf{Citations to previously published studies}}
\end{center}
Part of the content of this Thesis has already appeared in the following publications:\\

\begin{center}
{\raggedleft \textbf{\large { \underline {Refereed papers}}}}
\end{center}
{\bf Pratika Dayal}, Andrea Ferrara \& Simona Gallerani, 2008, {\it Signatures of reionization on Lyman Alpha Emitters}, MNRAS, 389, 1683 \\
\\
{\bf Pratika Dayal}, Andrea Ferrara, Alexandro Saro, Ruben Salvaterra, Stefano Borgani \& Luca Tornatore, 2009, {\it Lyman alpha emitter evolution in the reionization epoch}, MNRAS, 400, 2000 \\
\\
{\bf Pratika Dayal}, Andrea Ferrara \& Alexandro Saro, 2010, {\it The cool side of Lyman alpha emitters}, MNRAS, 402, 1449 \\
\\
{\bf Pratika Dayal}, Hiroyuki Hirashita \& Andrea Ferrara, 2010, {\it Detecting Lyman alpha emitters in the submillimeter}, MNRAS, 403, 620 \\
\\
{\bf Pratika Dayal}, Antonella Maselli \& Andrea Ferrara, 2010, {\it The visibility of Lyman alpha emitters during reionization}, arXiv:1002.0839, accepted for publication in MNRAS \\ 
\\
Stefania Salvadori, {\bf Pratika Dayal} \& Andrea Ferrara, 2010, {\it High redshift Lyman alpha emitters: clues on the Milky Way infancy}, arXiv:1005.4422, accepted for publication in MNRAS 
\\

\begin{center}
{\raggedleft \textbf{\large {\underline {Submitted papers}}}}
\end{center}
Ruben Salvaterra, Andrea Ferrara \& {\bf Pratika Dayal}, 2010, {\it Simulating the sources of cosmic reionization}, arXiv:1003.3873


\chapter*{}
\begin{center}
{\large
{\bf I acknowledge that in the construction of this sphinx, by my side were:}}
\end{center}
Professor Andrea Ferrara, the chief architect, who with his undying enthusiasm, faith and a vision, made my tentative drawings a reality. SISSA which gave me the tools, without which the construction could never have begun. Profs. Annalisa Celotti, Gigi Danese, John Miller, Carlo Baccigaluppi, Stefano Liberati, Paolo Salucci, who gave me so much precious mortar of courage and endurance. Bhanu sir, Prof. Stefano Borgani, Hiroyuki Hirashita, Antonella Maselli, Stefania Salvadori, Ruben Salvaterra, Alex Saro, for the artillery to clear the land and build the walls with me. Prof. Sangeeta Malhotra, for the provisions of resilience when mine ran dry.  To my referees, Profs. Daniel Schaerer and Adriano Fontana for approvinf the groundwork and their positivity. The DAVID group, where I found so many unexpected friends, so much laughter, so many beautiful gems of ideas.

Giulia, Melo, Barbara, my family in Trieste; for being constantly by my side, for facing each landslide, flood and earthquake with me. Goff and raffo, for all the laughter and psychedellic mime that brightened our dark days. Ernazar for being my solid rock, unperturbed even when walls collapsed around us. Fabio and Jasmina, for the rivers of laughter and the kindness that feed this land. Munia and Moira who showed me worlds different to what I have ever known. Robert and Nakul, who came from distant lands and made me appreciate what I was building. Ghiido, luca lepori and 105 for keeping our army of workers on the move. Emiliano, a rediscovered friend who brought with him beautiful memeories of home. Lucia and Eleonora for the abundant gifts of food and drink, that kept us alive in the desert. Angelo, my informer, without whom the real world would have been lost to me. Lydia, who reminded me to look from afar at what I was building. All those I met on my journeys to other lands; China, Japan, Paris, Jaime, Mark, Sherry, Peter, Luca, Cecelia, Sebastian, who gave me precious jewels, from the stars, without which this Sphinx would not shine. Appu, Mansi, Josh, my eternal friends, who always welcomed me with love and open arms.

To my family, nanu, mom, dad, anna for all the love, strength and support; even thought they were far, they infused my work with soul and felt joy in every brick I put in. And Anu, the constant of my life, who so painstakingly painted every stroke of the heiroglyphs with me. Without whom, none of this would have had the soul, and golden brightness it does now.


\end{document}